\documentclass[12pt,preprint]{aastex}

\slugcomment{to appear in ApJ, January 20, 2007, v655}

\shorttitle{Ages and Metallicities of Globular Clusters}
\shortauthors{Wolf et al.}

\begin{document}

\title{Ages and Metallicities of Extragalactic Globular Clusters from Spectral and Photometric Fits of Stellar Population Synthesis Models}

\author{Marsha J. Wolf\altaffilmark{1}, Niv Drory, Karl Gebhardt, and Gary J. Hill}
\affil{University of Texas at Austin, Department of Astronomy, 1
University Station, \\ Austin, TX 78712}

\altaffiltext{1}{Current address:  University of Wisconsin - Madison, Department of Astronomy, 475 N Charter St, Madison, WI 53706, mwolf@astro.wisc.edu}

\begin{abstract}

Spectra of galaxies contain an enormous amount of information about
the relative mixture of ages and metallicities of constituent
stars. We present a comprehensive study designed to extract the
maximum information from spectra of data quality typical in large
galaxy surveys. These techniques are not intended for detailed stellar
population studies that use high quality spectra. We test techniques
on a sample of globular clusters, which should consist of single
stellar populations and provide good test cases, using the
Bruzual-Charlot 2003 high resolution simple stellar population
synthesis models to simultaneously estimate the ages and metallicities
of 101 globular clusters in M31 and the Magellanic Clouds by fitting
their integrated spectra and photometry. The clusters cover a wide
range of ages and metallicities, 4~Myr~$<$~t$_{age}$~$<$~20~Gyr and
$-$1.6~$<$~[Fe/H]~$<$~+0.3, estimated by other methods in the
literature. We compare results from model fits to both the spectra and
photometry and find that fits to continuum-normalized spectra over the
entire range available, typically 3500~\AA~to 1~$\mu$m for this
sample, provides the best results. For clusters older than 1~Gyr we
agree with literature ages to 0.16~dex (35\%) and [Fe/H] to 0.12
dex. For younger clusters we agree with literature ages to 0.3~dex
(63\%), but cannot constrain the metallicity. It is particularly
important to use the entire continuum-normalized spectrum to avoid
problems with model continua for young objects and to break
age-metallicity degeneracies of broadband photometry. Our required S/N
is 15-30~\AA$^{-1}$ for 20\% age uncertainties and 30-55~\AA$^{-1}$
for 10\% uncertainties over spectral resolutions of
$\Delta\lambda$~=~5-25~\AA. This technique should work well for the
age-metallicity parameter space expected for early-type galaxies at
z$\sim$1, although individual galaxy spectral S/N may require the
coaddition of a few like objects. Lack of accurate flux calibration in
large surveys is not an issue for the continuum-normalized spectra.

\end{abstract}


\section{Introduction}

Understanding when and how different types of galaxies formed in the
universe is an age old problem. In recent years the hierarchical
merging scenario \citep{white78,peebles80,white91}, where large
galaxies are built up over time through mergers of smaller objects,
has been the leading theory for the mechanism by which structure
formed in the universe. Large galaxy surveys have been undertaken to
provide observational evidence for how galaxies formed. From their
observations, we are compiling a snapshot of galaxies througout time.
At z$\sim$3 we find both massive starbursting galaxies in highly
clustered regions \citep{neri03,blain04,chapman04} and actively star
forming galaxies elsewhere \citep{papovich01}. At z$\sim$2 we see
galaxies that are still forming stars, but have redder colors and
larger masses from continued star formation since z$\sim$3
\citep{steidel04}. At z$\sim$1-2 a population of extremely red objects
(EROs) are seen by near infrared surveys such as FIRES
\citep{labbe03}, K20 \citep{cimatti02}, and MUNICS \citep{drory01},
many of which appear to contain older stellar populations that have
experienced passive evolution \citep{forster04,daddi02,saracco03}. At
z$\sim$1 we see a decrease in global star formation rates
\citep{lilly96,madau96,madau98,steidel99,barger00,thompson01,cwolf05},
an increase of stellar mass densities in early-types
\citep{hogg02,bell03,dickinson03,kauffmann03b,rudnick03,cimatti04,cwolf05},
and a change in galaxy morphologies to more ordered systems
\citep{strateva01,dickinson03}. At z$\lesssim$1 we see early-type
galaxies that appear to have formed at z$\sim$1.5-2 in the field
\citep{im02,gebhardt03} and at z$\sim$2-3 in clusters
\citep{vandokkum98,thomas05}, possibly from those that were seen as
star forming galaxies at z$\sim$2 and starbursts at z$\sim$2-3.

This observed galaxy demography, however, is not matched by the
semi-analytic hierarchical galaxy formation models. Although the
models do well on predicting the global star formation rate and
stellar mass densities, they do not match observations on the correct
proportions of galaxy types as a function of redshift
\citep{somerville04}. In particular, not enough EROs or submm galaxies
are formed at z$\sim$1-3 in the models. We do not yet understand the
details of how galaxies assembled their mass, although the epoch of
z$\sim$1-2 continues to stand out as an important transitional
period. Detailed studies of individual galaxies around this epoch will
uncover essential clues about how they assembled into the distribution
of galaxies that we see today.

One such detail for study is determining ages or formation redshifts
of galaxies in this epoch. Various techniques have been used in the
past to determine an overall age from integrated galaxy light,
including using colors to estimate age from the amounts of old (red)
and young (blue) stars, using the equivalent widths (EWs) of specific
spectral lines or breaks that are known to be sensitive to stellar age
\citep{trager00a,trager00b,kauffmann03a,kauffmann03b}, matching galaxy
spectra to stellar spectra \citep{stockton95,dunlop96,spinrad97},
using fits of stellar population synthesis models to spectral energy
distributions (SEDs) obtained from broadband or narrowband photometry
\citep{papovich01,cimatti04,drory04,mccarthy04}, and fitting galaxy spectra
with model spectra
\citep{yi00,nolan03,cimatti04,ferreras04,mccarthy04}.  We will refer
to photometric SEDs throughout the paper as ``SEDs,'' not to be
confused with spectra.

Photometry has been widely available and is a good way to cover a
larger spectral range than typically available with spectroscopy to
get information about different stellar populations, but it lacks
spectral details and suffers from more degeneracy between age,
metallicity, and dust content. Recent surveys with spectra of hundreds
or thousands of galaxies, in addition to the photometry, provide more
information for detailed analyses. Furthermore, the release of
population synthesis models at high spectral resolution \citep{bc03}
allows one to take advantage of the detailed information contained in
individual lines over the entire observed spectral range, rather than
relying on only a few lines or on more degenerate broadband colors.

In this work, we evaluate age and metallicity estimation techniques on
globular clusters by comparing results from \citet{bc03} simple
stellar population (SSP) model fits to the spectra, broadband
photometry, and line indices of the clusters. Globular clusters
provide a simpler test case than galaxies since their stars likely
provide a coeval and nearly homogeneous metallicity population that
has no internal dust extinction. We test the utility of using these
models to estimate ages and metallicities of high redshift galaxies by
first applying them to a sample of extragalactic globular clusters
from the \citet{santos02} public database that have integrated
spectra, photometry, and age or metallicity estimates obtained by
other methods. Galaxies will be discused in a future paper.

This paper is arranged as follows. Globular cluster spectra and
photometry are described in \S 2, comparison ages and metallicities
from the literature are described in \S 3, the stellar population
synthesis models are described in \S 4, our model fitting procedures
are explained in \S 5, age and metallicity results are presented in \S
6, a discussion of issues is given in \S 7, and conclusions
are summarized in \S 8.

\section{The Cluster Data}

The \citet{santos02} online spectral database provides our sample of
extragalactic globular clusters that cover a significant range of age
and metallicity, making their results applicable to evolving
galaxies. We chose those clusters that also have photometry available
in the literature, which includes 79 clusters in the Large Magellanic
Cloud (LMC), 4 clusters in the Small Magellanic Cloud (SMC), and 18
clusters in M31. The spectra were obtained with the 1.52 and 2.2-meter
telescopes at ESO in La Silla
\citep{bica86a,bica87a,bica87b,bica90,bica94}, the 2.15-m CASLEO
telescope in Argentina \citep{santos95}, the 3.6-m CFHT in Hawaii
\citep{jablonka92}, and the 4.2-m William Herschel Telescope in La
Palma \citep{jablonka98}.

The data include near ultraviolet, optical, and near infrared spectra,
covering a total wavelength range of 3200-10000 \AA~for some clusters,
and varying in spectral resolution from 6 to 23~\AA.  We should note
that for all clusters except a few in M31, this resolution element is
larger than the $\Delta\lambda$~=~10~\AA~required for Lick/IDS line
indices. All available spectral segments for each object are used to
utilize maximum wavelength coverage, but since they originate from a
number of different sources, flux calibration between regions may not
be consistent. Table \ref{res_tbl} lists each object with its specific
wavelength coverage, spectral resolution, and source references.

We compile photometry of the clusters from various sources in the
literature. \citet{bica96} provide UBV photometry for the LMC clusters
with $\sim$20\% of the objects taken from \citet{vdb81} and the rest
re-observed by the authors at CTIO or CASLEO. \citet{persson83}
provide VJHK photometry for LMC clusters from observations on 1-2.5
meter telescopes at Las Campanas and Cerro Tololo observatories. For
the SMC clusters, \citet{persson83} provide VJHK and UBV are from
\citet{santos95} who use photometry from \citet{vdb81} and a few other
sources. For the M31 clusters, \citet{battistini93} supply BVRI from
observations on the 152-cm telescope in Loiano, Bologna and on the 4-m
KPNO telescope. \citet{barmby00} provide UBJHK from observations on
the 1.2-m telescope at the Fred L. Whipple Observatory.

\section{Comparison Ages and Metallicities}

Even for a given globular cluster, a wide range of age and metallicity
estimates typically exists in the literature. This is not completely
surprising given the different methods employed to derive the
parameters, some based on broadband colors and others on very
narrowband spectral line strengths, but the dispersion in estimates is
quite large for many clusters. We will compare our results to age and
metallicity values from a number of sources and different techniques
to show where our values lie in the total spread. Literature ages and
metallicities for specific clusters can be found in Table
\ref{age_tbl}, along with our derived values.

One of the oldest techniques for measuring the age of a star cluster
is using its integrated UBV colors. When placed on color-color plots,
globular clusters follow a sequence that can be related to age. We use
MC cluster U-B color ages that were determined by \citet{bica90} and
\citet{santos95}. When photometry is available for individual stars in
a cluster, the location of the main sequence turnoff (MSTO) on a
color-magnitude diagram (CMD) can be used to determine the age of the
cluster. We use a large compilation of MSTO ages for MC clusters from
\citet{hodge83}.  Refinement of the MSTO technique is achieved by
fitting theoretical isochrones to the CMDs. We take CMD fit ages from
\citet{elson88} and \citet{girardi95,girardi98}. We also add
comparison ages determined from photometric fits to GISSEL96 model
SEDs \citep{bc93} in the BATC filter system, using 15 intermediate
bandwidth passbands, by \citet{jiang03}, and from the line index
ratios H$\delta$~/~Fe~I $\lambda$4045 and Ca~II~H/K by
\citet{leonardi03}, which were developed to work better on younger
objects than the typical Lick line indices for use on older
objects. We consider the Leonardi \& Rose line index ratio ages to be
some of the most secure, given that they are determined from high
quality spectra and are tailored for the younger ages of many clusters
in this sample. \citet{santos04} utilize diagnostic diagrams
constructed from EW sums of metal (Ca~II~K, G band, Mg~I) and Balmer
(H$\delta$, H$\gamma$, H$\beta$) lines, which are calibrated against
literature values and placed on homogeneous age and metallicity
scales, to determine ages and metallicites of MC clusters. We also
include these in our comparison values of age and metallicity.

Fitting theoretical ishchrones to cluster CMDs is generally accepted
as the most secure age determination possible when using photometry,
however the results do vary between sets of ishochrones. The absolute
derived age depends on model zeropoints and uncertainties.
\citet{schiavon02} find that Padova isochrones with AGB stars included
underpredict the luminosity function of giants by 0.2-0.3 dex when
compared to observations of 47~Tuc, resulting in an overprediction in
the spectroscopic age of 2-3 Gyr for old stellar populations. Young
stellar populations are much less affected.  Additionally, the Padova
isochrone uncertainty in T$_{eff}$ of 75~K leads to a 1~Gyr age
uncertainty or 0.1 dex uncertainty in the [Fe/H]
scale. \citet{vazdekis01} also find discrepancies between CMD and
spectroscopic ages of 47~Tuc due to the exclusion of
$\alpha$-enhancement and atomic diffusion in evolutionary models,
causing a zeropoint offset. If 47~Tuc is representative of other old
stellar populations, for clusters in the 10-14~Gyr age range these
model isochrone issues could lead to spectroscopic based age
discrepancies of 20-40~\% when compared to CMD based values.

Large homogeneous samples of globular cluster metallicities are harder
to find, since metallicity cannot be as accurately determined from
photometry and typical methods require high quality spectra. However,
some photometric techniques have been calibrated to spectra. Of these
we use [Fe/H] values from VJK colors by \citet{cohen94}, fits to CMDs
\citep{seggewiss89,sagar89,suntzeff92,girardi95,dirsch00,johnson01,piatti02},
and fits to GISSEL96 model SEDs by \citet{jiang03}. Our more secure
comparison metallicites come from spectra-based techniques. For these
we use [Fe/H] from the \citet{leonardi03} line index ratios and their
comparison literature values that were derived from EWs of Ca, Fe, Mg,
and Na lines \citep{cohen82}, and from EWs of the Ca II triplet at
$\lambda$8500 \citep{olszewski91,dacosta98}. We incorporate the
comparison literature values used by \citet{santos04}, which include
metal line abundances \citep{jasniewicz94,hill00}, line indices
\citep{beasley02}, and an infrared index at 1.6 $\mu$m
\citep{oliva98}. And finally we use a combination of line indices and
spectral breaks from \citet{huchra91}: D(4000), CNB $\lambda$3883 \AA,
G-band $\lambda$4300 \AA, MgH, Mg \textit{b}, and Fe $\lambda$5270
\AA.

\section{Stellar Population Synthesis Models}

We use the high spectral resolution (R$\sim$2000) SSP ``standard
model'' from Bruzual \& Charlot (2003; hereafter BC03) that utilizes
the STELIB/BaSeL 3.1 spectral library, the Padova 1994 evolutionary
tracks, and the \citet{chab} initial mass function (IMF) with stellar
mass limits of 0.1 and 100 M$_{\sun}$. They provide model spectra at
[Fe/H]~= $-$2.25, $-$1.65, $-$0.64, $-$0.33, +0.093, and +0.56, which
we linearly interpolate at each wavelength point to a grid of spectra
that spans [Fe/H] of $-$2.2 to +0.5 in increments of 0.1 dex, and ages
of 1 Myr to 20 Gyr in increments of 2 Myr for 1$<t_{age}<$10 Myr, 20
Myr for 10$<t_{age}<$100 Myr, 200 Myr for 100$<t_{age}<$1000 Myr, and
1 Gyr for 1$<t_{age}<$20 Gyr.

The BC03 models include thermally-pulsing asymptotic branch stars
(TP-AGB), making use of the multi-metallicity models of
\citet{vassiliadis93} that have been calibrated on stars in the
Galaxy, LMC, and SMC. The importance of these stars, which have a
strong influence on the integrated near infrared light from star
clusters of certain ages, will become apparent later in the
paper. This phase can dredge up carbon in the stellar atmospheres,
leading to carbon-rich stars, which are also very red and can dominate
the near infrared light from some clusters. Although no simple
prescription can be expected to match all clusters, these stars must
somehow be included. \citet{bc03} achieve this by defining the
transition to carbon stars and the duration of this phase with the
models of \citet{groenewegen93} and \citet{groenewegen95}. This
semi-empirical prescription for TP-AGB and carbon stars has been
tested on and provides good agreement with observed colors of
Magellanic Cloud clusters and with optical and near infrared surface
brightness fluctuations of metal-poor Galactic globular clusters and
more metal-rich nearby elliptical galaxies \citep{liu00}.

\section{Data Preparation and Model-Fitting Procedures}

Our approach is to use the maximum amount of information possible for
each object. Broadband photometry often covers a much wider spectral
range than spectroscopy, providing better constraints on the possible
mixes of stellar types, but suffers from age-metallicity
degeneracy. However, by combining broadband information with detailed
spectral analysis, we hope to better constrain model fits and thus
derived ages and metallicities. We investigate the ability of both
broadband and detailed spectral features of the models to estimate age
and metallicity of the globular clusters by fitting models to the full
spectrum, to the continuum-normalized spectrum, to the continuum
shape, to photometry, and to spectral line indices of the
clusters. Each of these fitting procedures is described in detail in
the following sections.

\begin{itemize}

\item \textit{Full spectrum fits} utilize information from both the line
strengths and the continuum shape.

\item \textit{Continuum-normalized (CN) spectrum fits} use only the
information contained in the lines, losing important continuum
information, but also removing any adverse effects of inaccurate flux
calibration in the data or continuum shape errors in the models. Model
continua problems are particularly possible in the near infrared due
to the inability to properly account for the number of TP-AGB or
carbon stars that can contribute much of the light at these
wavelengths.

\item \textit{Continuum fits} use only the continuum shape that was
removed from the spectrum.

\item \textit{Photometry fits} cover a much broader wavelength range, but lose
detailed spectral information and may also be affected by continuum
shape problems in the data or models.

\item \textit{Line index fits} move from broad to narrow and focus only on
specific spectral lines that are known to be sensitive to metallicity
or age. We evaluate these specific lines in the models by calculating
the indices directly from model spectra and comparing those to line
indices calculated from the cluster spectra in the same way. 

\end{itemize}

\subsection{Spectra}

The spectra for each object in the \citet{santos02} database consist
of near ultraviolet, optical, and near infrared segments from multiple
sources at spectral resolutions of 6-23~\AA. We splice all spectra
together to cover the widest possible range for each cluster. No noise
or sky spectra are included in the database, so we must estimate the
noise from the spectra. To do this we calculate the standard deviation
of the spectrum in 100~\AA~bins and scale the resultant
signal-to-noise ratio (S/N) spectrum to match the average S/N value
quoted in the source publications for some objects. For those objects
that do not have quoted S/N values, the standard deviation spectrum
provides the noise. This likely overestimates the noise since
absorption lines will increase the standard deviation of a bin, but
the required scaling is typically less than 10\%.

We prepare the data for fitting by marking regions that contain
emission lines, noise spikes, and sky background residuals, as well as
two regions for which the STELIB stellar library used in the models
has problems with telluric features (6850-6950~\AA~and
7550-7725~\AA). These regions are masked and deweighted for each
cluster by significantly lowering its S/N spectrum over the affected
wavelengths. We smooth the grid of model spectra, which have an
intrinsic resolution of 3 \AA, to the wavelength dependent resolution
of each cluster and fit its spectrum using two free parameters, age
and metallicity, and a S/N-weighted normalization constant,
$\alpha_{n}$, which is uniquely determined for each model. We select
the best fitting model by calculating the summed $\chi^2$ for each
model spectrum of a given age and metallicity, relative to the cluster
spectrum, and selecting the model in the grid with the minimum
value. In the $\chi^2$ calculation, we use the previously determined
cluster noise spectrum as the $\sigma_{i}$ for the flux,
\textit{f}$_{i}$, at each wavelength bin, \textit{i}, for each model,
\textit{n}, as given in Equations 1 and 2. Once the best fitting model
is determined for a cluster, we inflate its noise until the
$\chi^2_{reduced}$=1 to include any additional noise sources and to
better estimate confidence contours and error bars. If, after this
process, the estimated errors are smaller than the distance to the next
model grid point, an error bar of half this distance is assigned to
the cluster for both age and metallicity.

\begin{equation}
\alpha_{n} = \frac{\sum_{i} f_{data_{i}} \left (
\frac{S}{N} \right )_{i} }{\sum_{i} f_{model_{i}} \left (
\frac{S}{N} \right )_{i} } \label{alpha}
\end{equation}
\begin{equation}
\chi ^{2}_{n} = \sum_{i} \left ( \frac{f_{data_{i}} - \alpha_{n}
f_{model_{i}}}{\sigma_{i}} \right ) ^{2} \label{chi2}
\end{equation}

\begin{figure}
    \includegraphics[scale=.40]{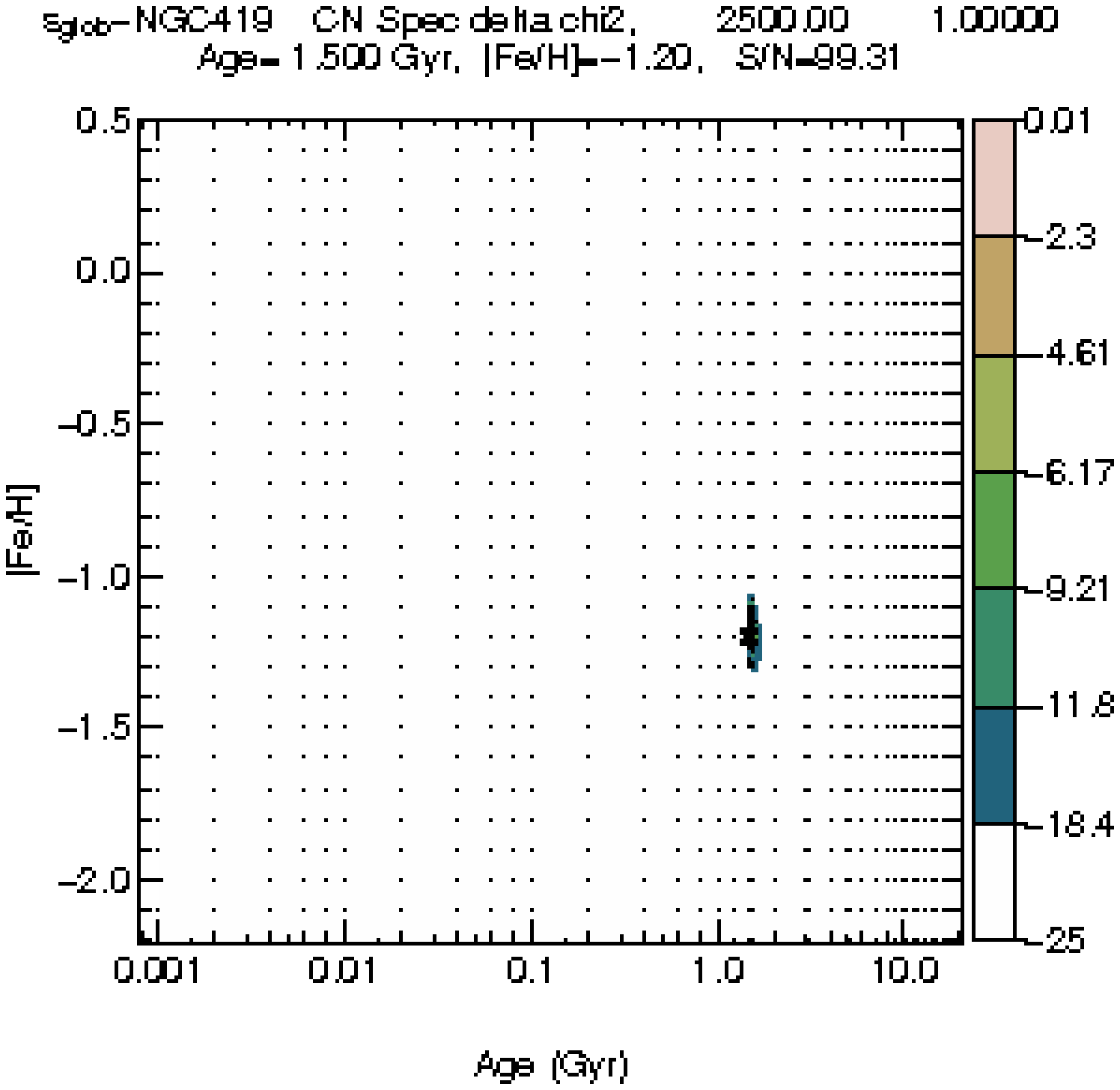}
    \includegraphics[scale=.40]{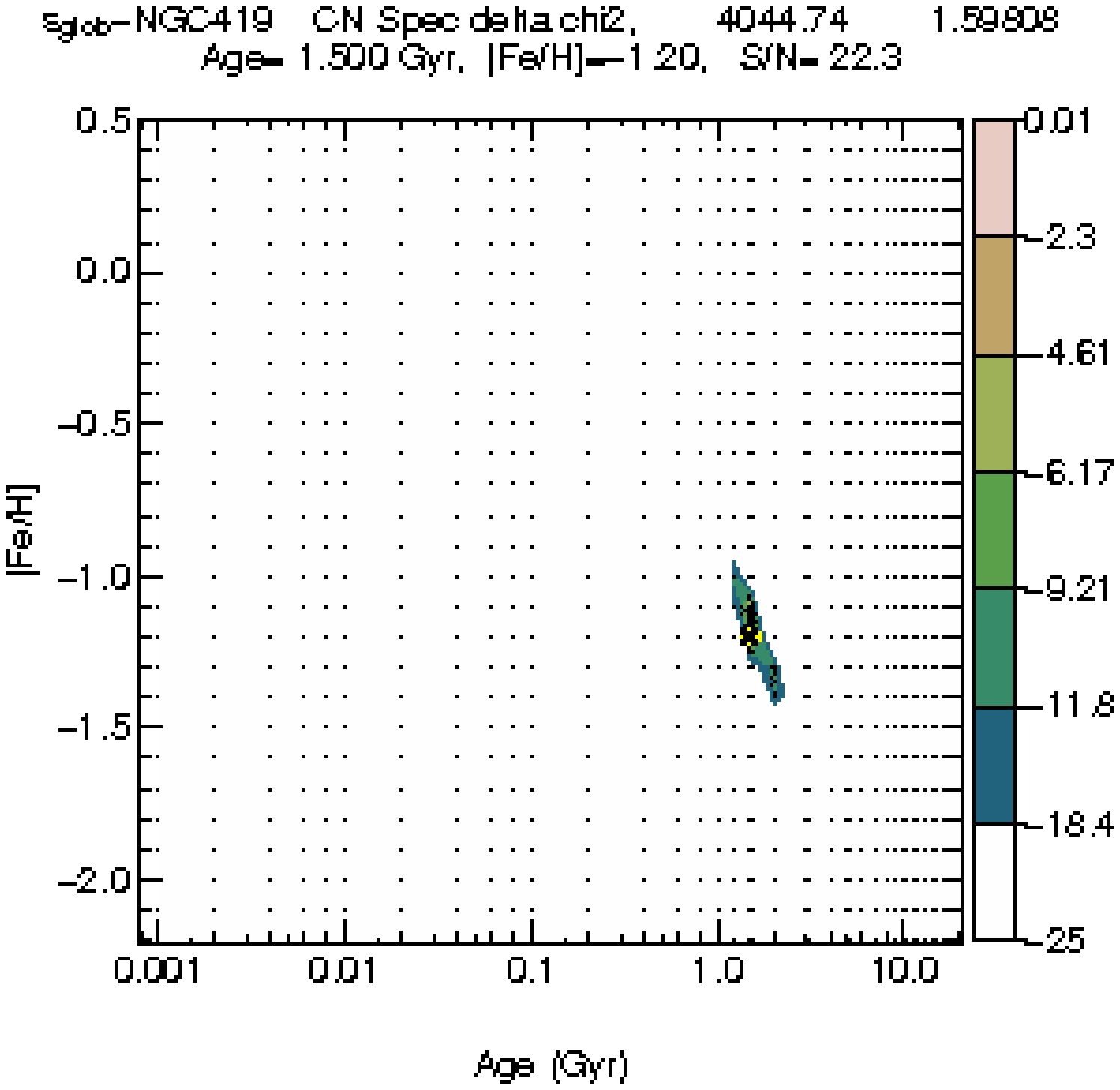}
    \includegraphics[scale=.40]{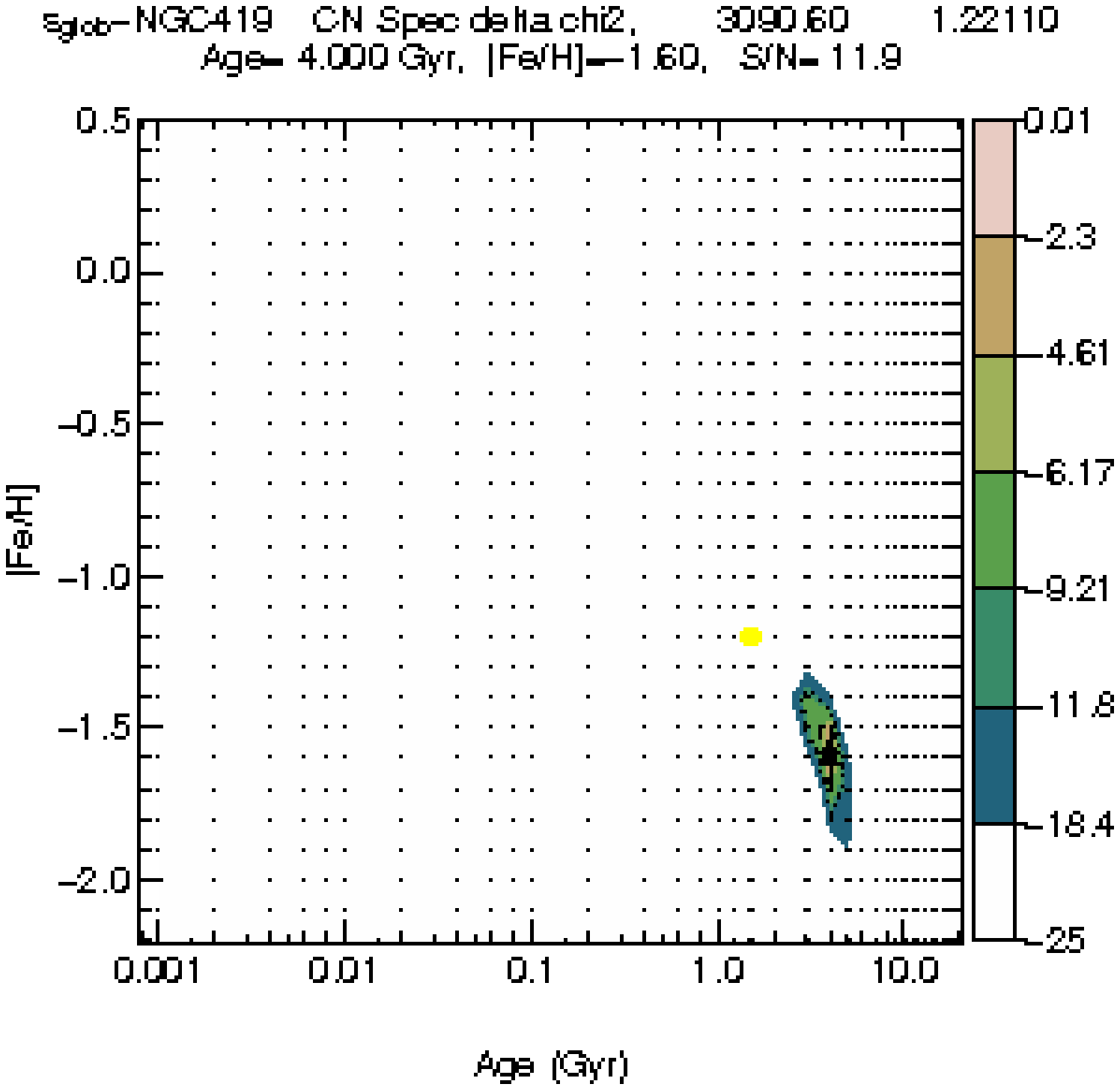}
    \includegraphics[scale=.40]{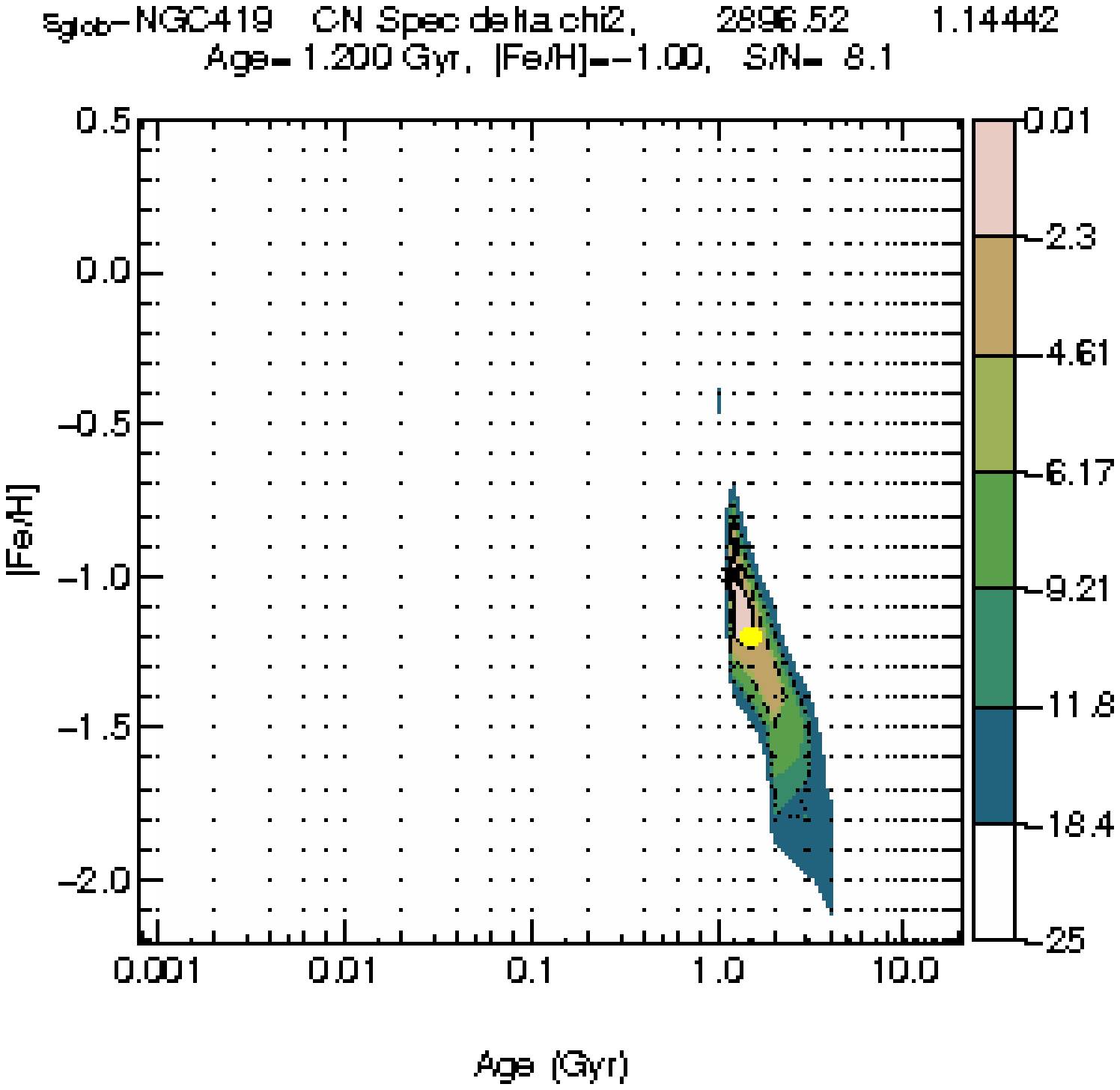}
    \includegraphics[scale=.40]{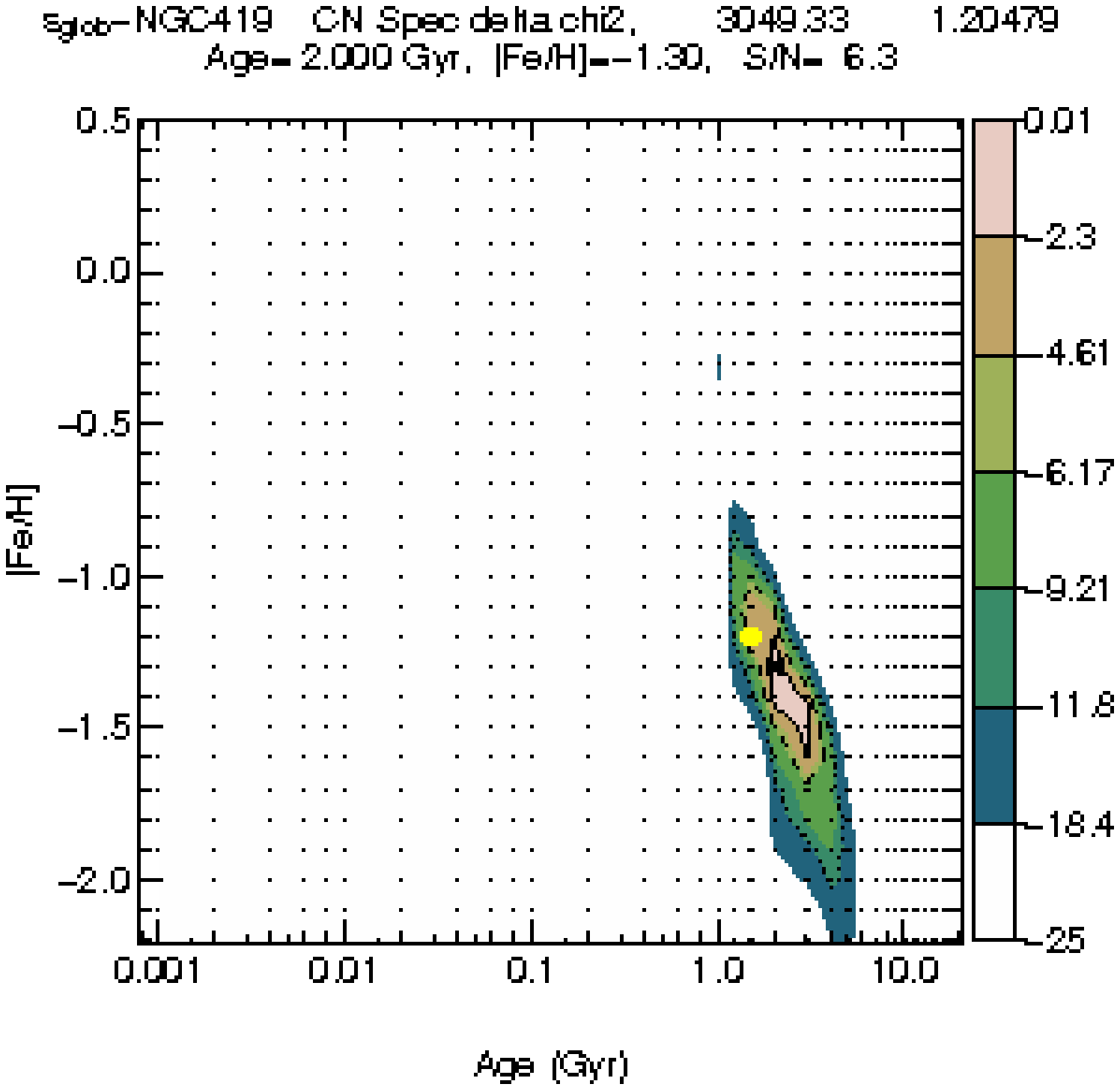}
    \includegraphics[scale=.40]{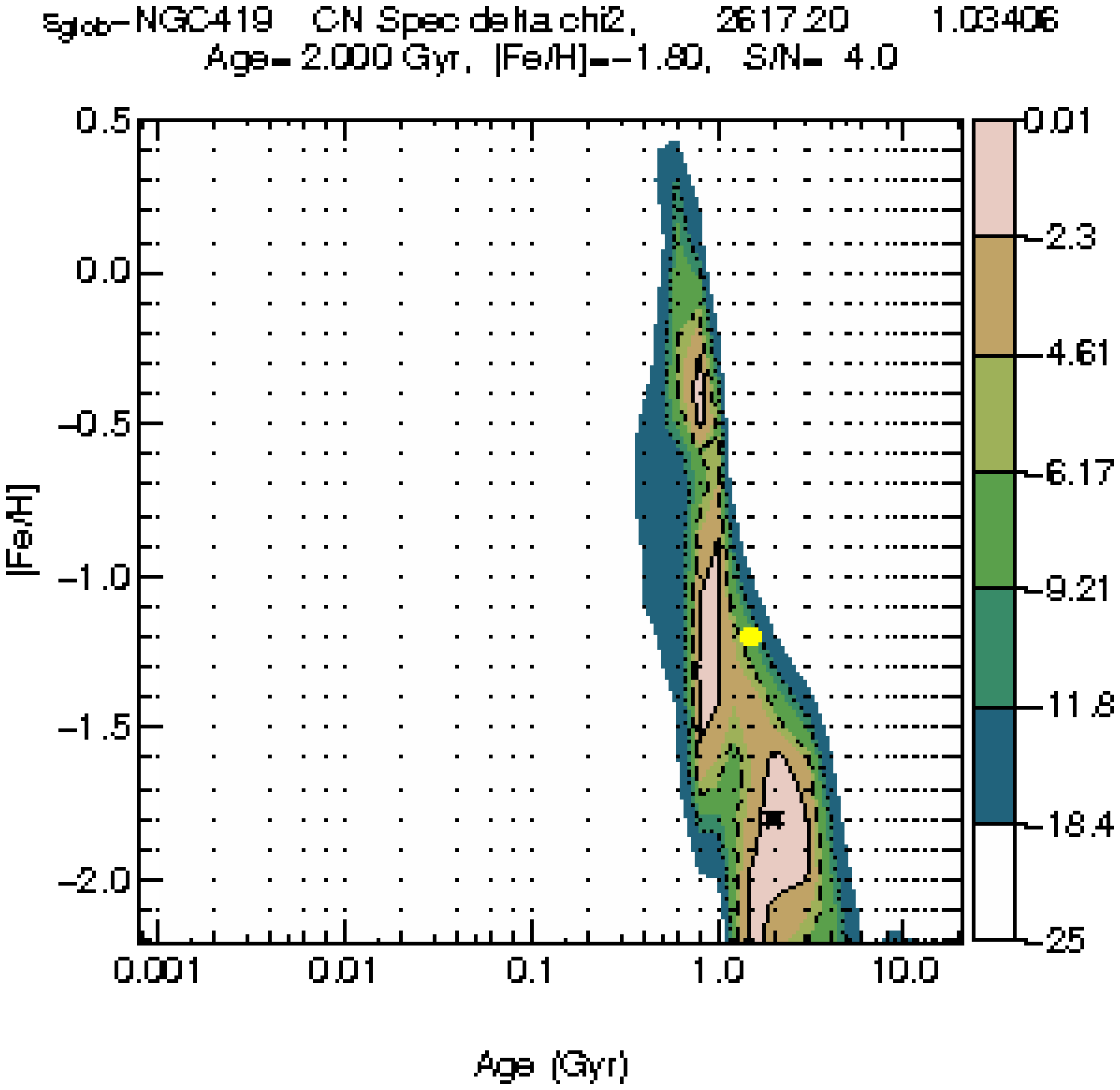}
\caption{The effect of S/N on $\Delta\chi^2$ contours of age and
metallicity for CN spectrum fits to the SMC globular cluster NGC
419. The S/N per resolution element from top left to lower right is
99, 22, 12, 8, 6, and 4. The best-fitting age and [Fe/H] is given in
the plot titles and marked by the black crosses. The yellow circles mark
the age and [Fe/H] from the highest S/N fit. The solid, dashed, and
dotted lines are 1, 2, and 3$\sigma$ error contours,
respectively. Color bars give the $\Delta\chi^2$ values for 68.3\%
($\Delta\chi^2$=2.3), 90\%, 95.4\%, 99\%, 99.73\%, and 99.99\%
($\Delta\chi^2$=18.4) confidence levels for two degrees of
freedom. (The highest level is converted to white for plot clarity.)
Black dots mark the locations of models in the age-metallicity
grid. \label{fig1}}
\end{figure}

Errors on age and metallicity estimates are determined from the
$\Delta\chi^2$ contours. The S/N of the data controls the size of
these contours, and thus the confidence of the derived
parameters. Typical levels for the globular clusters are S/N$\sim$100
per resolution element. Figure~\ref{fig1} shows a simulation of the
effect of S/N on the confidence of the derived age and metallicity for
NGC 419 in the SMC, achieved by artificially adding different levels
of gaussian noise to its spectrum and fitting models to each of the
noisy continuum-normalized spectra. The simulated S/N per resolution
element from top left to lower right in Figure~\ref{fig1} is 99, 22,
12, 8, 6, and 4. Solid, dashed, and dotted lines represent 1, 2, and
3$\sigma$ contours. The color bars give the $\Delta\chi^2$ values for
confidence levels of 68.3\% ($\Delta\chi^2$=2.3), 90\%, 95.4\%, 99\%,
99.73\%, and 99.99\% ($\Delta\chi^2$=18.4) for two degrees of
freedom. Black dots mark the locations of models in the
age-metallicity grid. The best fitting model is marked by a black cross
in each panel. The age and metallicity from the best fitting model in
the highest S/N case is marked by a yellow circle in each subsequent
panel for comparison.

Two things become apparent from these plots. First, as the confidence
contours grow with decreasing S/N, metallicity becomes harder to
constrain than age. In some low S/N conditions (e.g.~the last panel)
multiple islands of age-metallicity combinations arise as equally
likely answers. The best fitting model in each panel does not always
overlap the 1$\sigma$ contours of all other panels, however, this
mostly happens in the [Fe/H] dimension and those values are less
secure than the ages. This is our first indication that metallicity is
poorly constrained, compared to ages. The second point is that we find
a limiting S/N below which the estimated age and metallicity become
unreliable. This happens somewhere between S/N = 22 and 12 in these
plots, at which point the contours in the low S/N cases do not overlap
the estimates from the highest S/N case. We find that the S/N must be
$\gtrsim$15 per resolution element (S/N$\sim$5~\AA$^{-1}$) for
$\Delta\lambda$=13-23~\AA~to adequately constrain the age. Metallicity
is harder to constrain under all conditions. A more detailed analysis
of limiting S/N and spectral resolution can be found in \S
\ref{snr_sec}.

\subsection{Continuum-Normalization}

Errors in flux calibration of the spectra can induce errors in derived
ages and metallicities. Additionally, model continua can be off in the
near infrared due to stochastic contributions from TP-AGB stars in
$\sim$ 0.1 to 1 Gyr old populations and from carbon stars in
populations of age 0.3 to 2.5 Gyr \citep{frogel90,marigo96,girardi98}.
These variations can result in differences of nearly 2 magnitudes in
V-K (Bruzual \& Charlot 2003, Figure~8). AGB stars in the thermally pulsing phase
can contribute over 80\% of the K-band light \citep{maraston05}, the
exact amount of which is hard to predict since this phase involves
ejection of the outer stellar envelopes and geometry-dependent
obscuration of the central stars. The BC03 models include TP-AGB and
carbon stars semi-empirically, but in any given cluster the K-band
light can vary depending on the numbers of stars it actually has in
these phases.

To avoid these issues, we normalize out the continua with a
median-binning routine that uses bin sizes tailored to work for
different classes of objects. The young clusters of the Magellanic
Clouds have strong Balmer discontinuities and need to be sampled by
small bins to correctly trace this sharp continuum break, while older
objects in M31 require larger bins to smooth over absorption troughs
in the red part of the spectrum. The bins that we use are 100 \AA~wide
blueward of 4000~\AA~and 300-700~\AA~wide redward of 4000~\AA. The
points in each bin are sorted by flux values, the lower 1/2 of the
points are ignored to reduce the weight of absorption lines, the
highest few points are rejected to ignore noise spikes, and the median
of the remaining flux values is assigned to that bin. The medianed bin
fluxes are linearly connected and the object spectrum is divided by
the result, providing a flat continuum. The same binning parameters
used for an object are also used on each model spectrum in the
grid. 

Removal of the continuum shape should not affect the values of line
indices measured on the spectrum, providing a good check of our
procedure. We applied this test, as suggested by the referee, by
choosing a random sample of 25 globular clusters spanning the entire
age range and measured line indices before and after continuum
normalization. The indices used were Mg\textit{b}, Fe5270, Fe5335,
H$\beta$, H$\delta_A$, H$\delta_F$, H$\gamma_A$, H$\gamma_F$, G4300,
$<$Fe$>$, and [MgFe]$^\prime$. Only the positive index values that
could be reliably measured were used, as in \S \ref{line_ind}. On
average the values change by less than 1\%, and all line index median
differences for the 25 clusters are below 3\%~with no correlation to
the derived cluster age.

The continuum shape that we divide out is also fit by model continua
and referred to throughout the paper as ``continuum fits.''  Different
types of spectral fits are illustrated in Figure~\ref{fig2} for an old
cluster, G177 in M31, and a young cluster, NGC 1711 in the LMC.

\subsection{Photometry}

To test ages and metallicities derived from broadband features over a
wide baseline, we fit models to the UBVJHK photometry from the
literature, using all of the bands that are available for each
cluster. Our model fitting procedure requires scalable fluxes, so we
convert cluster magnitudes to fluxes using the UBV zeropoints from
\citet{bessell79} and the JHK zeropoints from \citet{wamsteker81}. We
obtain model fluxes for each band by convolving filter transmission
curves with the model spectra. These fluxes are scaled to those of the
object with an overall S/N-weighted normalization parameter during the
fit. The best fitting model is selected by $\chi^2$ minimization,
using Equations \ref{alpha} and \ref{chi2} with \textit{i} now
representing each photometric band. Errors are not given for all
literature photometry, so we initially assume a 5\% photometric flux
error in all bands for each cluster, which is then adjusted to make
$\chi^2_{reduced}$=1 for the best-fitting model. Fitting parameters
are age and metallicity.  Photometric fits are also illustrated in
Figure~\ref{fig2}.

\subsection{Line Indices}

Lick line indices have been used for some time
\citep{faber85,burstein86,gorgas93,worthey92,worthey94a,worthey94b,worthey97,trager98}. They
are based on a very specific set of data and their application to new
objects is somewhat complicated.  To properly use Lick/IDS indices,
some of the same stars must be observed with the instrumental setup of
the new objects and the spectra of the new objects must be convolved
to exactly the same wavelength dependent resolution as the original
Lick data, which varied from run to run during the development of the
indices. Very small differences in the wavelengths can cause large
errors in the indices. To investigate the robustness of these specific
lines in the model spectra, we calculate line indices directly from
the models, similar to the calibrations done in \citet{bc03}. Direct
application of the models for this purpose would avoid the step of
convolving the object spectra to the estimated Lick resolutions. If
the models are smoothed to the same spectral resolution as the cluster
data, indices can be calculated directly from both spectra in the same
manner. However, the bandpasses of the Lick indices were originally
chosen for data with $\Delta\lambda$~=~8-12~\AA~\citep{worthey97}.
Specific lines in lower resolution spectra may begin to lose age and
metallicity information. Nevertheless, given the long history of the
use of line indices, we wanted to attempt a variation of this
technique on data with resolution and S/N representative of that from
typical galaxy surveys.

First, we broaden the model spectra to match the resolution of each
object, which may vary for wavelength regions obtained on different
instruments. Then, we calculate indices for the lines in the Lick
system using the passbands defined in \citet{trager98} and
\citet{worthey97}. To these we add $<$Fe$>$, an average of Fe
$\lambda$5270 and Fe $\lambda$5335; D$_n$(4000), the 4000 \AA~break
strength using the narrow passbands defined in \citet{balogh99}; CNB
$\lambda$3883 \AA, MgH, and Ca II H+K, as defined by \citet{brodie90};
and [MgFe]$^\prime$ = $\sqrt{{\rm Mg} b (0.72 \times {\rm Fe}5270 +
0.28 \times {\rm Fe}5335)}$, as defined by \citet{thomas03}. The
[MgFe]$^\prime$ index has been found to be a good tracer of total
metallicity and to be insensitive to $\alpha$/Fe enhancement (for
spectral resolution near 10~\AA), which may be important for some of
the globular clusters in this sample.

We calculate the indices as described in \citet{trager98}, by
connecting the average flux values of the pseudocontinua sidebands
with a straight line and integrating the flux under that line over the
index passband. Indices (or break strengths) are calculated for CNB,
H$\delta_{A}$, H$\delta_{F}$, CN$_{1}$, CN$_{2}$, Ca4227, G4300,
H$\gamma_{A}$, H$\gamma_{F}$, Fe4383, Ca4455, Fe4531, C$_{2}$4668,
H$\beta$, MgH, Fe5015, Mg$_{1}$, Mg$_{2}$, Mg\textit{b}, Fe5270,
Fe5335, Fe5406, Fe5709, Fe5782, Na~D, TiO$_{1}$, TiO$_{2}$,
[MgFe]$^\prime$, $<$Fe$>$, and D$_n$(4000). Indices for model spectra
are calculated in the same manner. We select the model with the
smallest sum of absolute value residuals between its index values and
the corresponding indices derived from a cluster's data. The age and
metallicity of this closest matching model are adopted as the
estimates for the cluster; no interpolation is performed between model
grid points. To facilitate comparisons with the literature, we
repeated the model selection using different subsets of the indices
listed above: all indices simultaneously; Huchra et al.'s (1991) group
of 6, CNB, D$_n$(4000), G4300, MgH, Mg\textit{b}, and Fe5270; and
H$\beta$-[MgFe]$^\prime$ as in \citet{thomas03}.

\section{Results \label{results}}

\begin{figure}
\figurenum{2}
    \includegraphics[scale=.35,angle=-90]{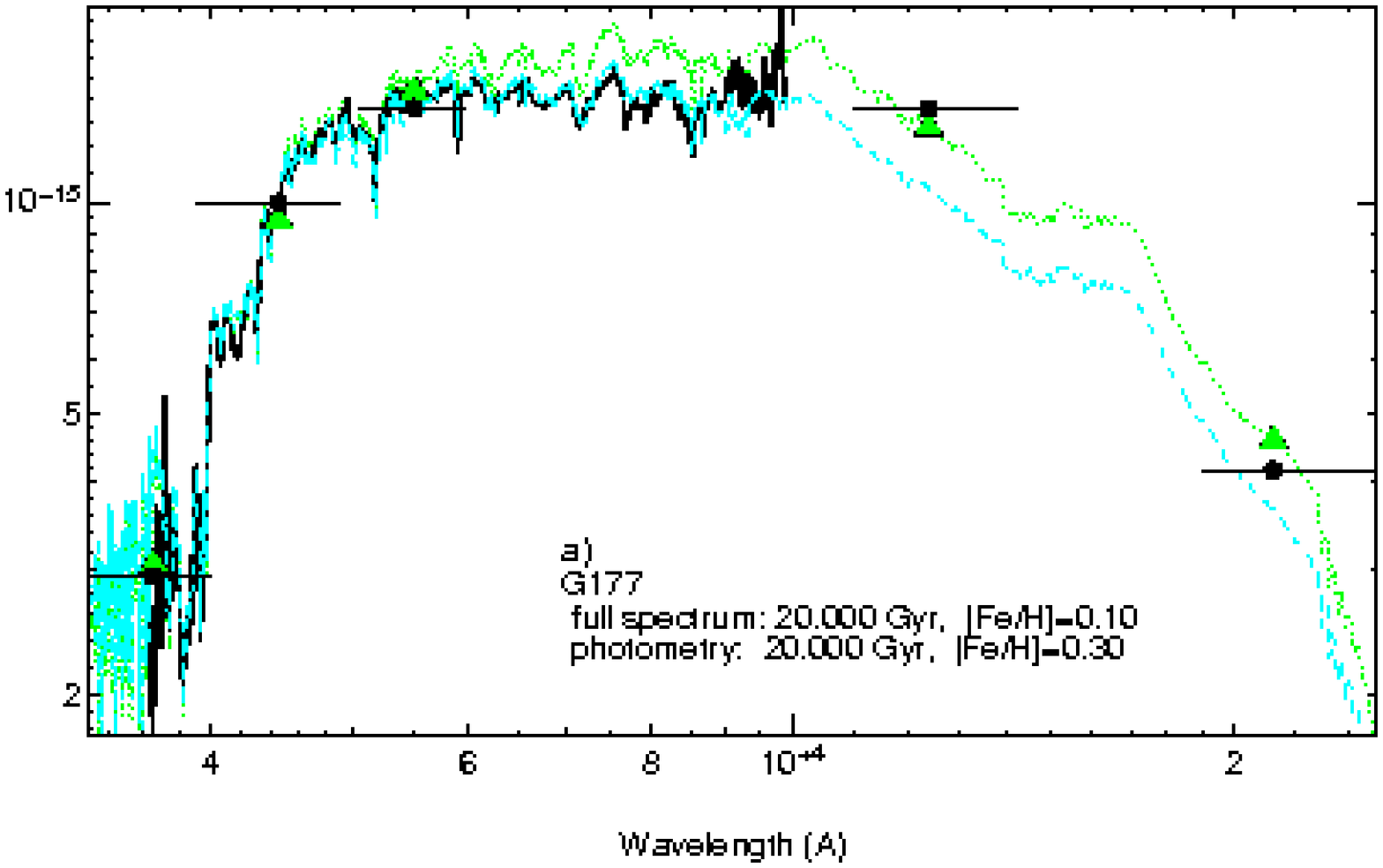}
    \includegraphics[scale=.35,angle=-90]{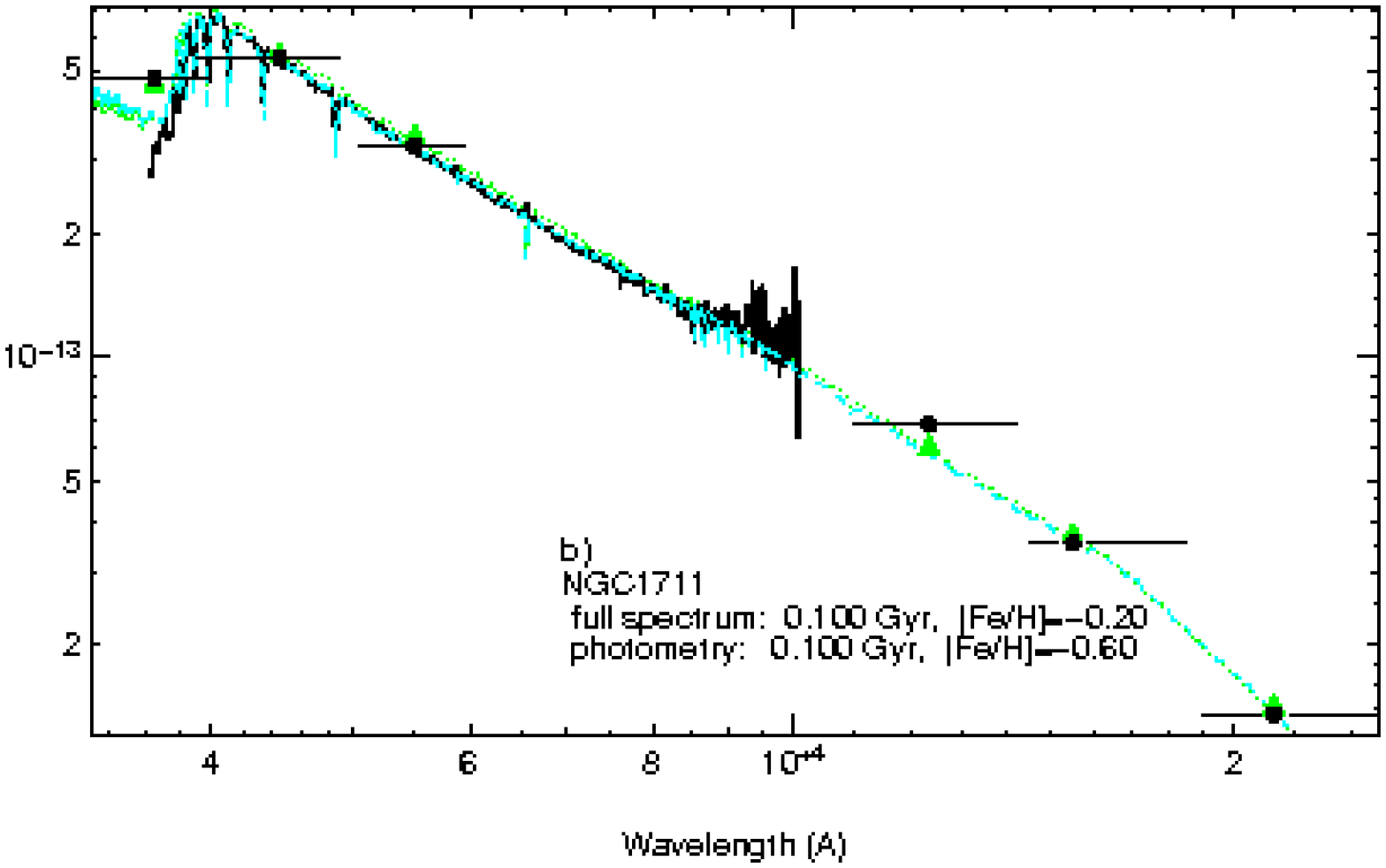}
    \includegraphics[scale=.35,angle=-90]{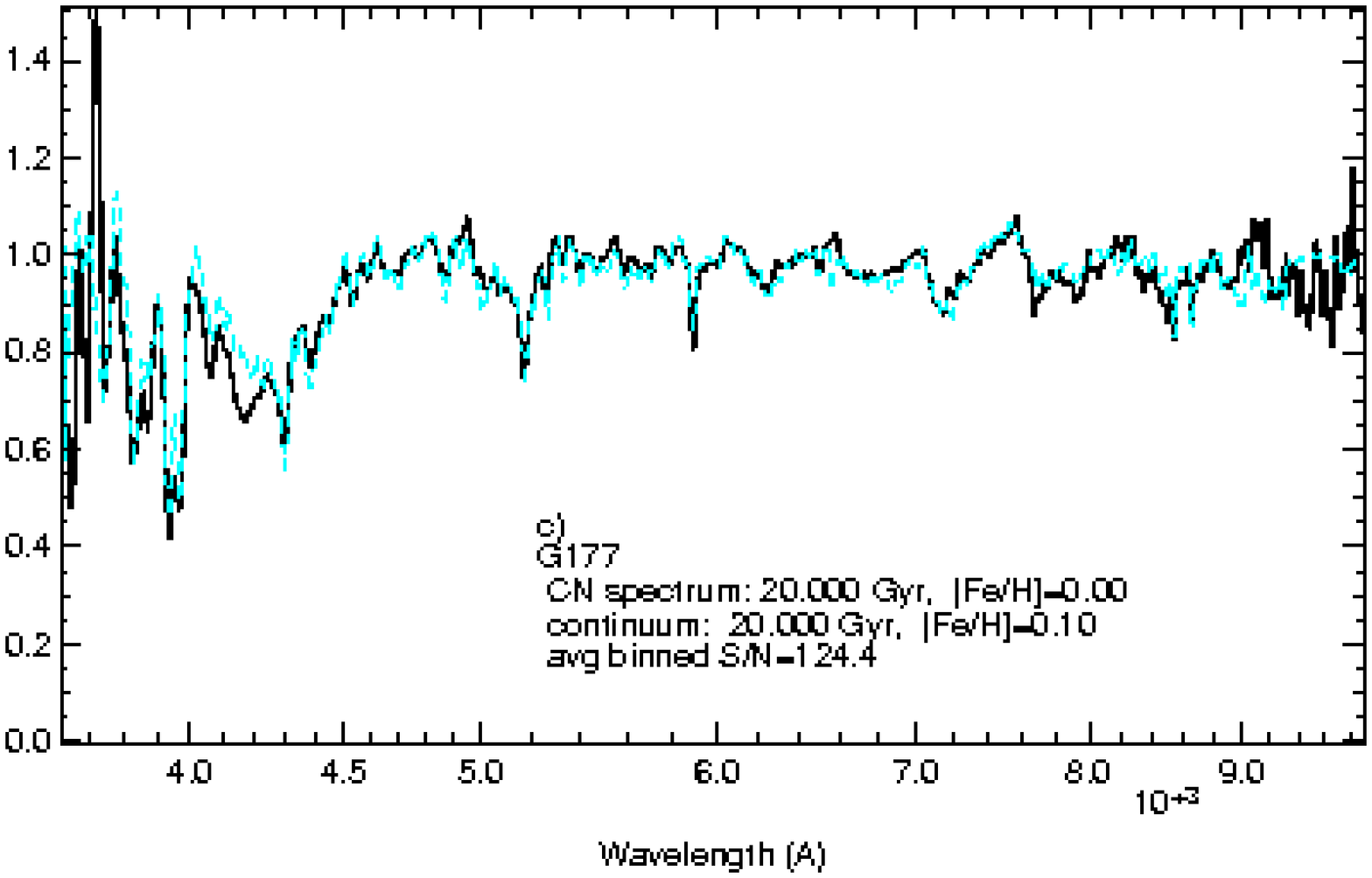}
    \includegraphics[scale=.35,angle=-90]{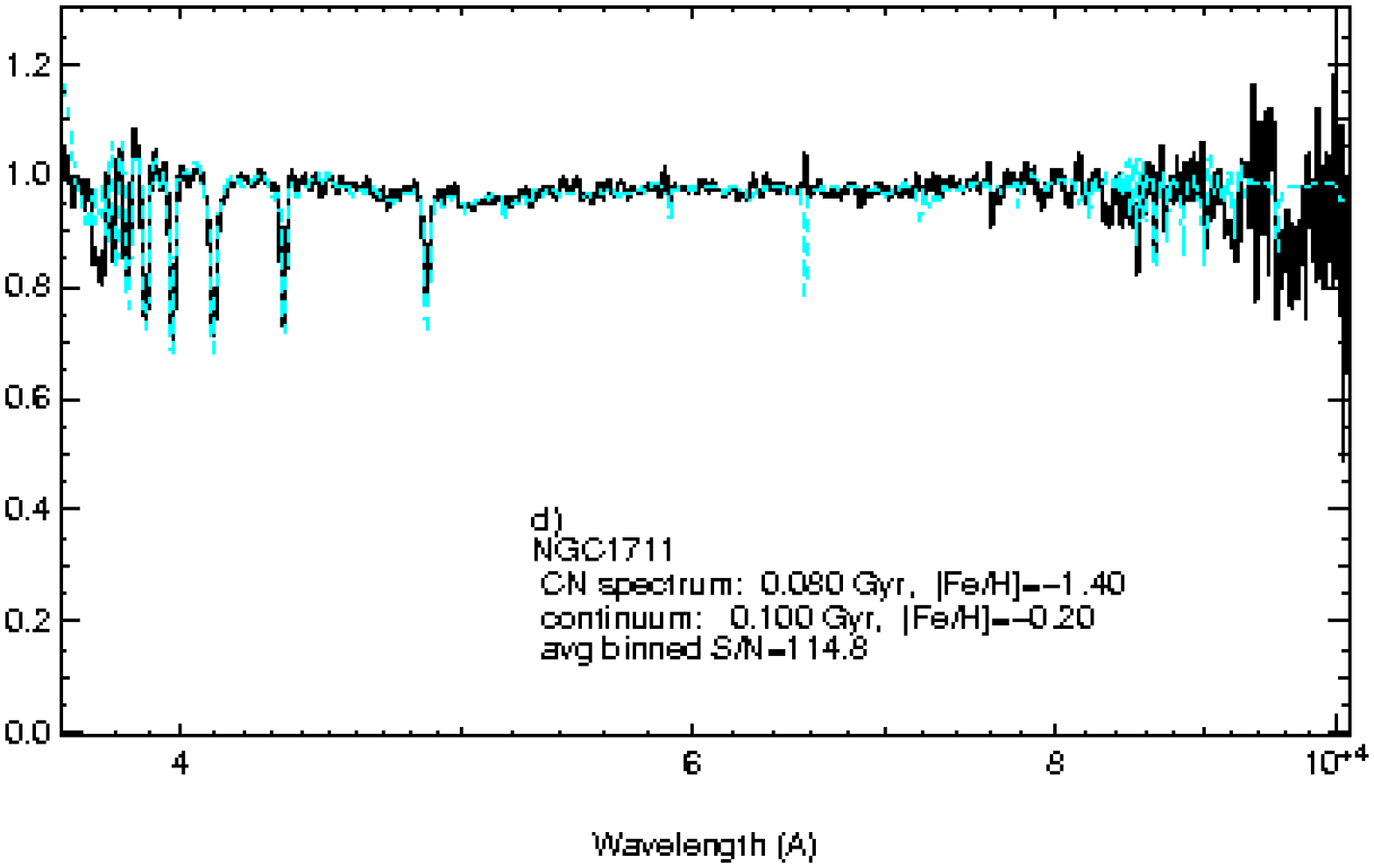}
    \includegraphics[scale=.35,angle=-90]{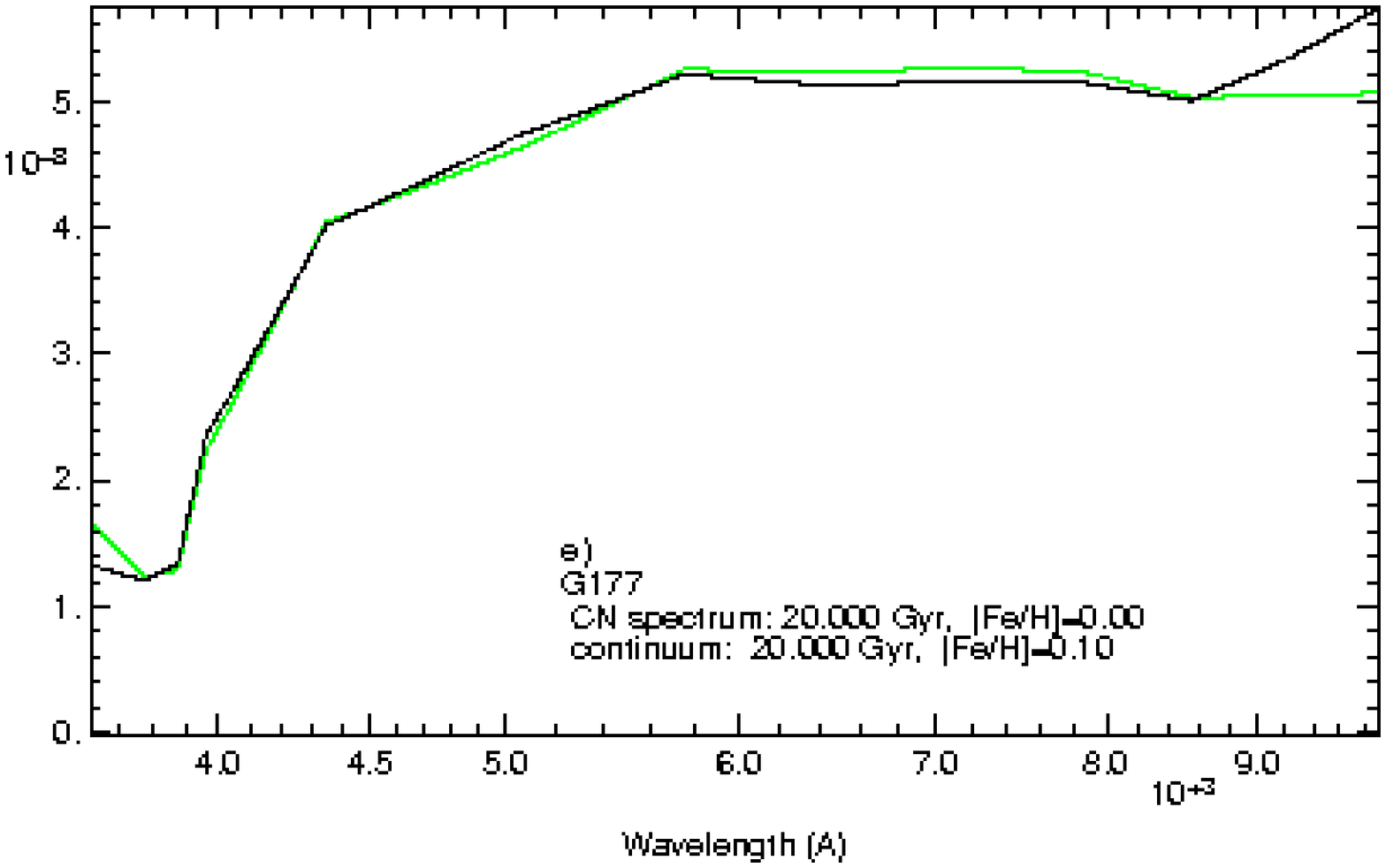}
    \includegraphics[scale=.35,angle=-90]{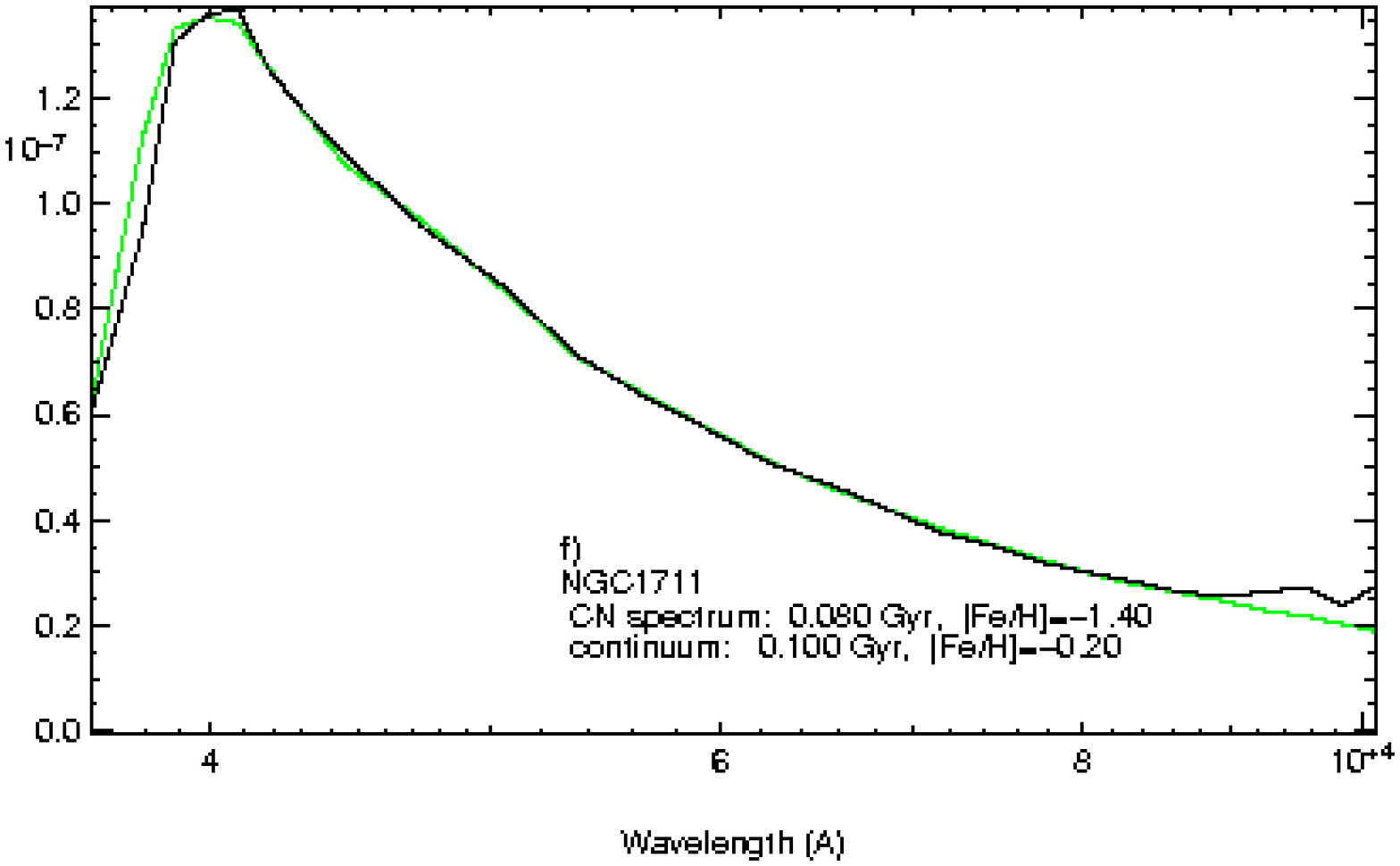}
\caption{Example model fits to two clusters of very different ages,
G177 in M31 and NGC 1711 in the LMC. a \& b are the full spectrum
fits, c \& d are the continuum-normalized spectrum fits, and e \& f
are the continuum fits. Solid black lines are the cluster spectra,
dashed cyan lines are model spectra corresponding to the best
spectroscopic fit, dotted green lines are the model spectra
corresponding to the best photometric fit, filled circles are the
cluster photometry, triangles are photometry from the best photometric
fit models, and horizontal bars mark the filter bandwidths. Photometry
for G177 includes U,B,V,J,K bands and NGC 1711 includes U,B,V,J,H,K
bands. Best-fitting ages and metallicities for each method are given
in the plot labels. In the continuum fits, black lines are the cluster
continua and green lines are the best-fitting model
continua. \label{fig2}}
\end{figure}

Examples of model fits are shown in Figure~\ref{fig2} for old and young
clusters. Panels a,c,\&e are the cluster G177 in M31 and panels
b,d,\&f are NGC 1711 in the LMC. The full spectrum fits are shown in
a\&b, the continuum-normalized spectrum fits are shown in c\&d, and
the continuum fits are shown in e\&f. Photometry fits are also shown
in a\&b with U, B, V, J, and K bands for G177 and U, B, V, J, H, and K
bands for NGC 1711. The circles are the object photometry and
triangles are model fluxes from the best photometric fit. The solid
lined black spectra are the clusters, the dashed cyan lines are best
fitting models from the spectral fits, and the dotted green lines are
spectra corresponding to the best fitting models from the photometry
fits. The model spectra corresponding to the best fitting photometry
are normalized to the cluster spectra at V band in these plots. In the
continuum fits, black lines are the data and green lines are the
models. The resulting ages and metallicities from the different types
of fits are given in the plot labels.

\begin{figure}
\figurenum{3}
    \includegraphics[scale=.35,angle=-90]{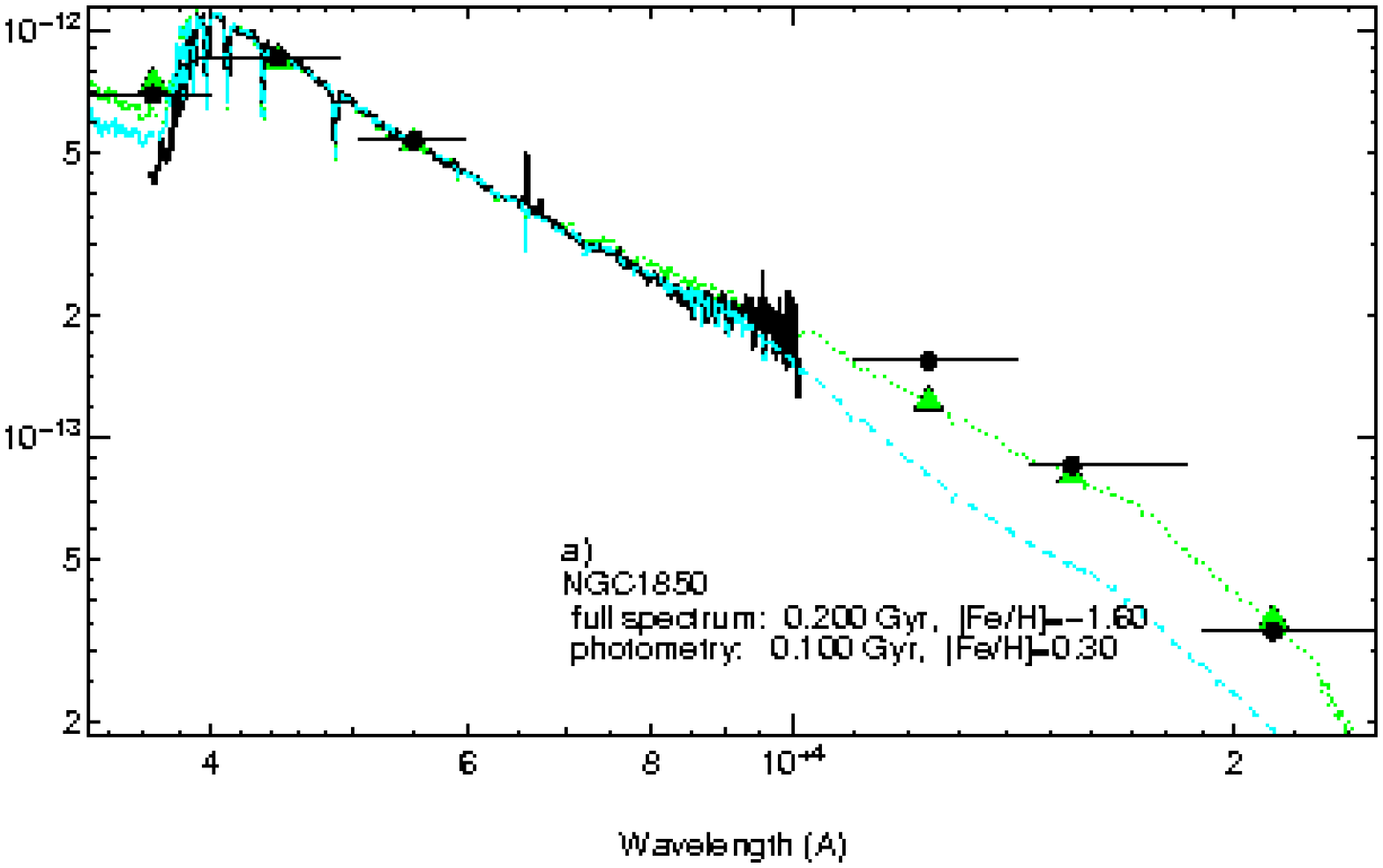}
    \includegraphics[scale=.35,angle=-90]{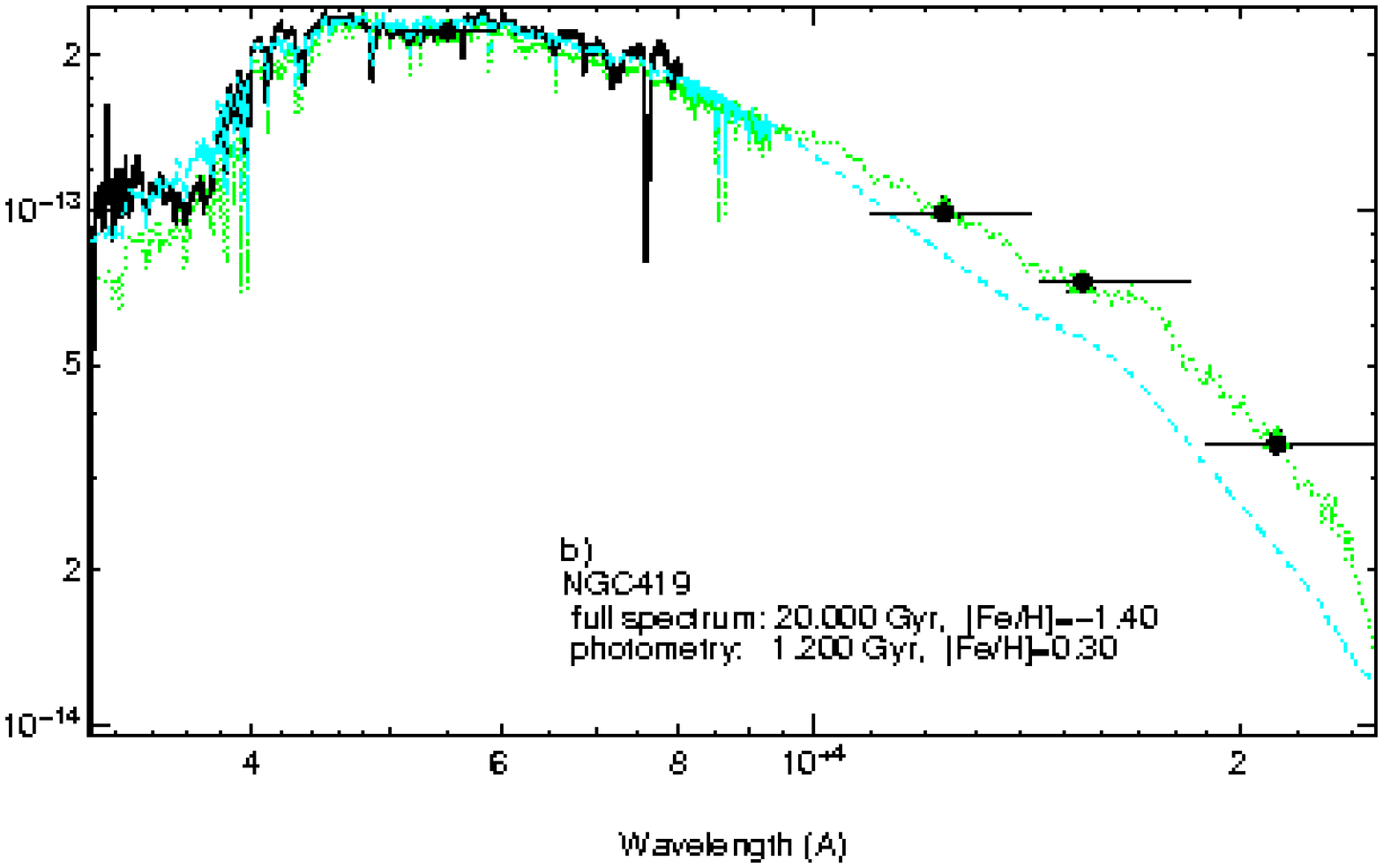}
    \includegraphics[scale=.35,angle=-90]{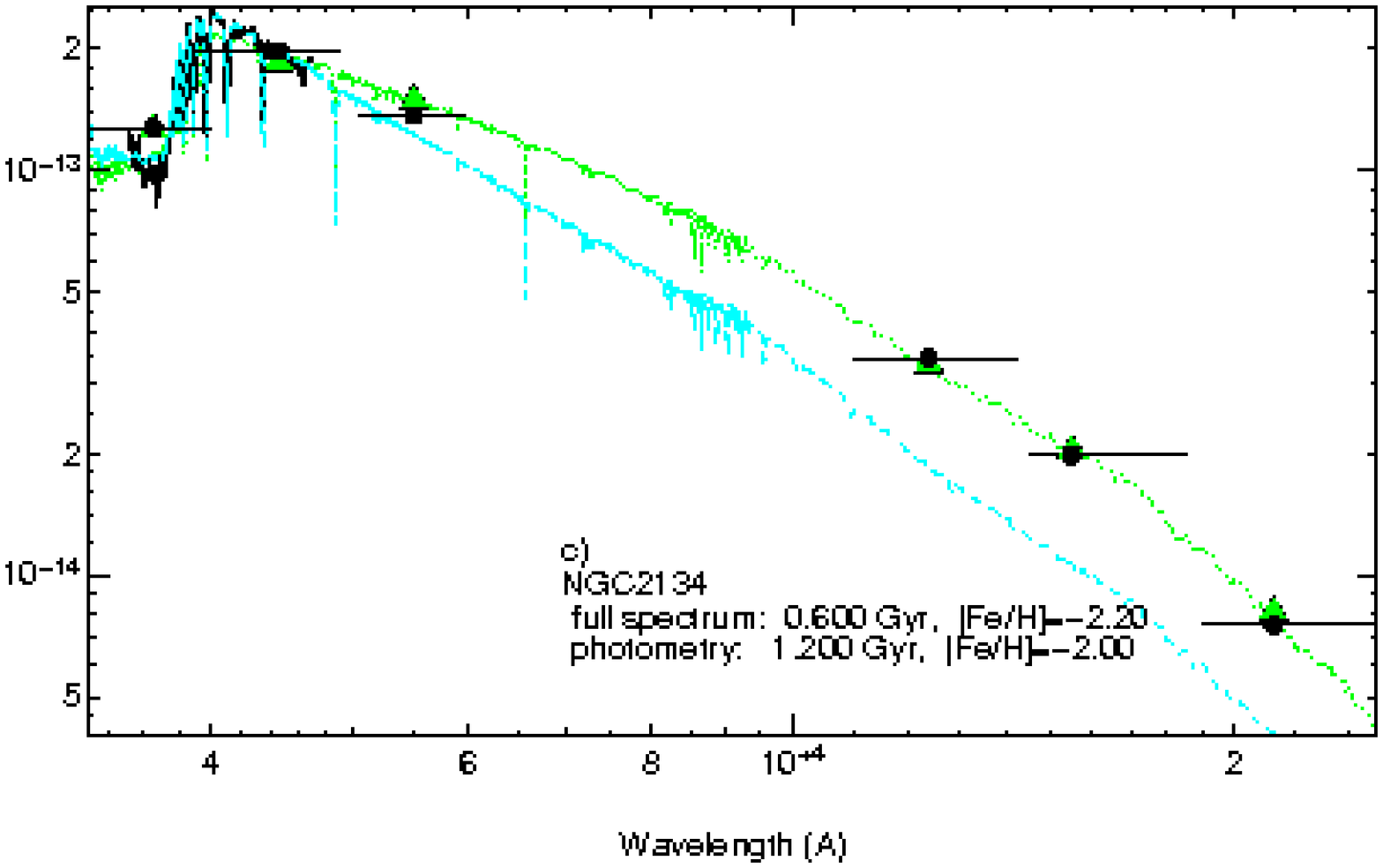}
    \includegraphics[scale=.35,angle=-90]{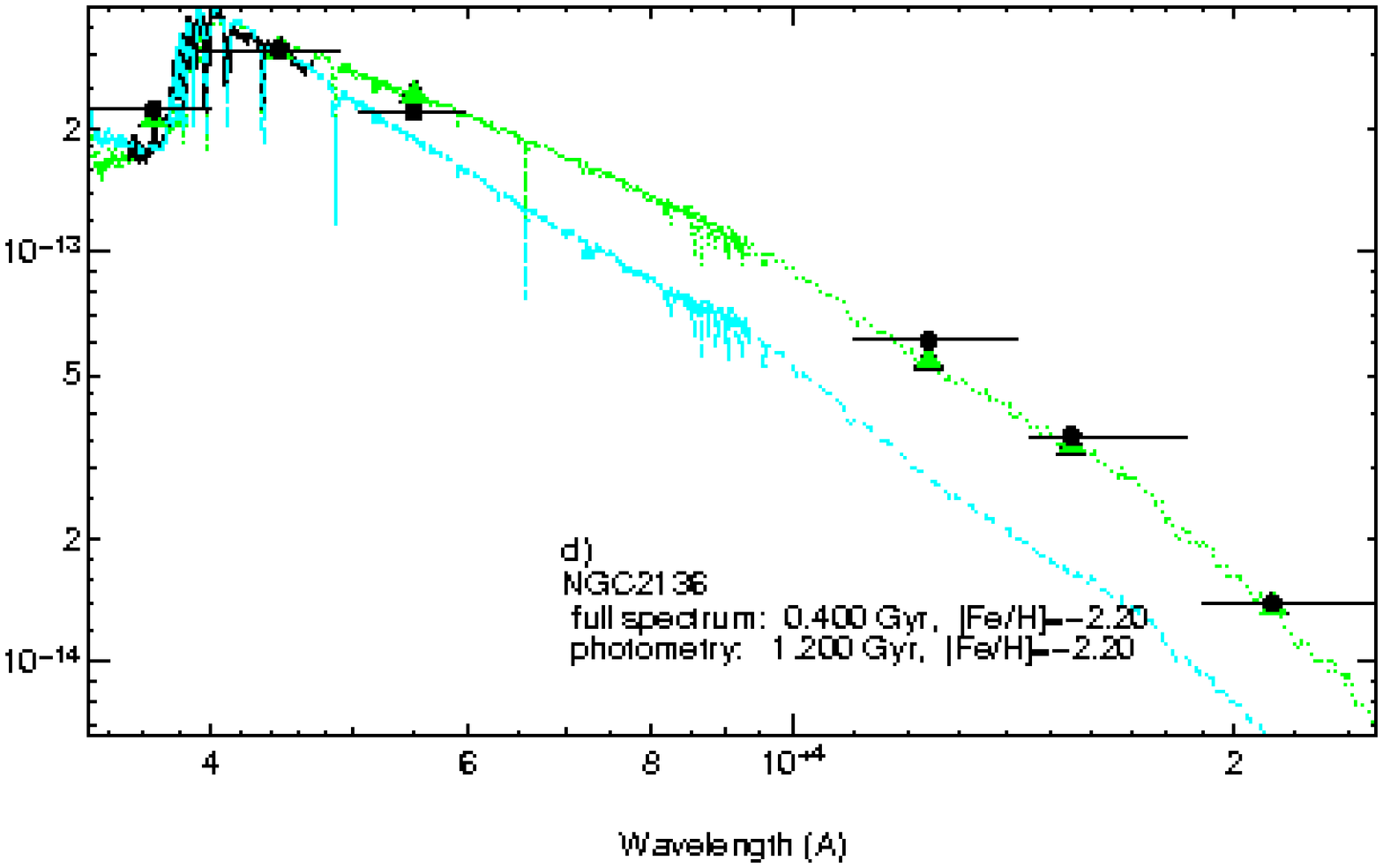}
\caption{Spectra with possible signs of TP-AGB or carbon stars because
  their near infrared photometry points show an excess over the best
  fitting model to the spectrum. The legend is the same as in
  Figure \ref{fig2}. The ages of these clusters all fall within the
  range where TP-AGB and carbon stars are important, 0.1-2 Gyr. Our CN
  spectrum fits give ages of 0.1 Gyr for NGC 1850, 1.5 Gyr for NGC
  419, 0.2 Gyr for NGC 2134, and 0.1 Gyr for NGC 2136. 
  \label{fig3}}
\end{figure}

For these two clusters the ages derived from the different types of
model fits nearly agree at 20 Gyr for G177 and about 0.1 Gyr for NGC
1711. The agreement between photometric and spectroscopic fits seen
here is not always the case. G177 in Figure~\ref{fig2}a illustrates a
case where the photometry through K band reveals a redder SED than the
optical spectrum. In this case, the best fitting photometry model has
a higher metallicity than the model that best fits the spectrum to
make the object redder, an effect of age-metallicity degeneracy. A
similar effect is seen in some of the young clusters, presumably due
to a near infrared excess from TP-AGB or carbon stars in that
case. Some examples of such clusters within the affected age range of
0.1-2 Gyr are shown in Figure~\ref{fig3}.

\subsection{Different Model Fitting Techniques \label{compare}}

To evaluate the different model fitting techniques, we first compare
their results to each other. In the following sections we compare our
results to the literature. Figure~\ref{fig4} compares the ages derived
from CN spectrum, full spectrum, photometry, and continuum
fits. Clusters with average literature ages $\geq$~1~Gyr are maked
with yellow circles and $<$~1~Gyr with cyan diamonds. Although the
general trends of the ages agree, there are differences in the
results. When compared to CN spectrum fits, the methods that contain
the continuum overestimate the ages of many clusters younger than 1
Gyr. We believe that this is due to red supergiant stars for clusters
in the age range of a few to tens of Myrs and TP-AGB stars for
clusters in the age range of 0.1 to 1~Gyr. Both of these types of
stars can have stocastic effects in star clusters because of their
extremely high luminosities. If more light from them appears in the
integrated spectra of the cluster than is included in the models, the
cluster SED will appear redder than the model, forcing an older redder
model SED as the best fit. All fitting methods that contain model
continua will have this same bias. This effect is discussed further in
comparison with literature ages in \S \ref{age_results}.

For clusters with average literature ages $<$~1~Gyr our tightest
estimated age correlation is seen between the full spectrum and
continuum fits (Figure 4c), both of which contain the spectral
shape. The median fractional age offset, relative to the average
literature age, between these fits is 0.04 for 63 clusters. The fact
that the continuum-full spectrum correlation is tighter than the
CN-full spectra correlation (fractional offset of 2.40) suggests that
the continuum shape has a larger effect than the spectral lines in
constraining the ages of these young globular clusters for our
employed fitting methods. Therefore, any problems in the continuum
shape of the data or models will have grave effects on the derived
ages. For the clusters older than 1~Gyr our tightest estimated age
correlation is between the CN spectrum and photometry fits (Figure 4b),
with a median fractional offset of 0.18 for 12 clusters. 

For young clusters the metallicities obtained from our different
techniques show virtually no correlation, with a tendency for the
continuum to give higher metallicities than the full spectrum. For
clusters older than 1~Gyr there is a very weak correlation between
metallicities obtained with continuum and full spectrum fits, possibly
indicating that the derived metallicity has a stronger dependence on
the continuum shape than the spectral lines for our fitting method.

\subsection{Ages \label{age_results}}

In Figure~\ref{fig5} we compare our globular cluster ages to
those obtained by other methods in the literature. The thick cyan
horizontal bars connect points that represent the same cluster, but
that have ages from multiple sources in the literature.  These
illustrate the large spread in previously derived ages. The thin
horizontal bars are the quoted errors on the literature ages and the
thin vertical bars are our 1$\sigma$ errors on ages derived from model
fits. Note that the errors are much larger for the photometry fits,
reflecting more degeneracy when using broadband colors than when using
spectra or the continua shapes.

Over the entire age range spanned by the clusters and considering all
literature age estimation techniques, our CN spectrum fit ages best
match those from the literature. The average offsets and rms scatter
our ages from groups of literature values based on different
techniques are given in Table \ref{rms_age_tbl}, where column 1 gives
values for all clusters in the sample that have literature ages and
column 2 excludes two outliers (for reasons explained in \S
\ref{weird_ones}) and the clusters that only have literature values
from photometric fits to BC96 models (because these used only a few
fixed metallicity values in the model grid, which could skew the
derived ages, and they are based on old models). Excluding these
points, our CN spectrum fits compared to the averages of the
literature values for each object have an average offset of 0.29~dex
(0.19~Gyr) with a dispersion of 0.36~dex (1.00~Gyr).

Our photometric ages have an average offset of 0.59~dex (0.48~Gyr) and
dispersion of 0.76~dex (1.19~Gyr). There are some systematic
differences between the two sets of results. The scatter of the CN
spectrum fit points is more uniformly distributed around the
literature values, unlike the photometric points that tend to
overestimate the age, particularly for the younger clusters. The full
spectrum fits, which contain the continuum, also overestimate the ages
below 1 Gyr, with an average offset of 0.69~dex (0.54~Gyr) and a
dispersion of 0.9~dex (3.2 Gyr). This is the same effect that was seen
when comparing our fits with and without the continua included.

We now return to stars that can strongly affect the continua
shapes. Detailed studies of such stochastic effects in star clusters
can be found in \citet{bc03},
\citet{cervino00,cervino01,cervino02,cervino06}, and
\citet{maraston05}. Our photometric fits, in particular, show two
different age ranges where our best-fit ages depart from literature
ages on the high side (Figure~\ref{fig5}c): 4 to 100~Myr and 0.1 to
1~Gyr.  The age range of 0.1 to 1~Gyr would be affected by TP-AGB
stars. If the model prescription for these stars does not exactly
match those in the clusters, then the older ages from fitting methods
that include the continuum could be explained by older, redder spectra
giving the lowest $\chi^2$ for clusters that have this near infrared
excess. Our derived ages also scatter to older values for clusters
younger than 0.1 Gyr when using models that include the continuum
shape. This effect is likely due to stocastic red supergiants, which
have been observed with estimated masses up to 120 M$_{\sun}$ in the
LMC \citep{massey05}. These stars will affect clusters younger than a
few tens of Myrs. Either the number of red supergiants is
underestimated in the BC03 models, or the upper mass cutoff of 100
M$_{\sun}$ for the Chabrier IMF is not quite high enough for the
LMC. See \S \ref{weird_ones} for a discussion of two LMC clusters that
are dominated by supergiants, which are also our extreme outliers in
the CN spectrum fits of Figure~\ref{fig5}a.

Because of the large age range covered by the clusters, it is perhaps
useful to look at the behavior of the fractional age errors relative
to the literature values, which are plotted in
Figure~\ref{fig6}. Overall, the CN spectrum fits provide ages with
fractional offsets from the literature of 61\%, while the photometry
does much worse at 1048\%. If we consider younger and older clusters
separately, the CN spectrum fits do better than photometry on the
average age offsets for both young and old clusters, but the
photometry does better on fractional errors for the older objects. For
clusters with ages $<$~1~Gyr, the CN spectrum fits have an average
offset of 63\% (0.30~dex or 0.09~Gyr) and dispersion of 0.37~dex
(0.35~Gyr), while the older clusters have an offset of 35\% (0.16~dex
or 1.63~Gyr) and dispersion of 0.17~dex (3.78~Gyr). The dispersion is
higher below ages of 0.1 Gyr even though the CN spectra are not
affected by the continua effects mentioned earlier. This increased
dispersion is likely due to the degeneracy of many spectral lines in
this age range (see Figure~\ref{fig10} and \S\ref{line_ind}).  The
photometry fits do much worse fractionally on young clusters with an
average offset of 1115\% (0.62~dex or 0.38~Gyr) and dispersion of
0.78~dex (1.02~Gyr), and better on the fractional errors of older
clusters with an offset of 23\% (0.10~dex or 2.06~Gyr) and dispersion
of 0.11~dex (2.67~Gyr).

The lowest fractional errors for the CN spectrum fits occur for the
\citet{leonardi03} line index ratios and for the clusters older than
1~Gyr, both at 35\%. Photometry fits do slightly better on the older
objects with errors of 23\%, but these errors rise substantially for
the young objects to 1115\%. The largest fractional age offsets occur
in the clusters less than 1 Gyr old for the full spectrum (3450\%),
continuum (4432\%), and photometry fits. This is likely due to AGB
stars. Furthermore, the full spectrum and continuum fits have larger
age offsets in the older clusters (425\% and 480\%) than do the CN
spectrum or photometry fits, which may be due to flux calibration
issues, since both of these methods include the observed spectral
continuum shape. To take care of both data and model continua issues,
it appears that continuum-normalized spectral fits are the most robust
in producing accurate age estimates for the clusters of all ages.

The age errors do not correlate with metallicity, so they seem to be
dominated by the age regime of the cluster. Although photometry does a
decent job on older objects, it should be noted that after excluding
the BC96 photometry literature points, we only have 4 clusters in the
$>$ 1 Gyr bin at ages of 1.2, 2.2, 12.4, and 13.1 Gyrs. It could just
be that the two younger ones have no problems with boosted near
infrared emission. In general, since there is not a good way to
determine whether the 1-2~Gyr old objects are affected by the AGB star
problem, using the CN spectrum fits would avoid this issue. Some
combination of CN spectrum and photometry fits will likely provide the
most robust answer. This will be explored further in future work.

Other studies have compared results from their cluster age estimation
techniques to the literature. \citet{rafelski04} use colors formed
from UBVI photometry along with Starburst99 \citep{leitherer99} and
GALEV \citep{anders03} models to derive ages and compare to those in
the literature from integrated colors \citep{vdb81,hunter03} and
isochrone fitting \citep{pietrzynski99,deoliveira00,mighell98,rich00}
in Figure~8 of their paper. They find an age correlation with dispersion
of 0.76 Gyr, which drops to 0.49 Gyr when considering only the more
secure literature ages derived from CMDs. Over a similar age range our
CN spectrum fits provide ages with an overall dispersion of 1.0 Gyr
about literature comparison values, with 0.12 Gyr for UBV colors and
1.06 Gyr for CMD fits. Our photometry fits have dispersions of 1.04
Gyr for UBV colors and 0.85 Gyr for CMD fits. Although
\citet{rafelski04} use clusters in the SMC, only two overlap with our
sample. Our literature ages come from different sources as well, so we
cannot make a direct comparison to this work.

To summarize our results for globular cluster age estimation, the CN
spectrum model fits best match the entire range of age estimates from
the literature (Figure~\ref{fig5} and Table~\ref{rms_age_tbl}),
with average errors of 0.16 dex (35\%) for older clusters and 0.3 dex
(63\%) for younger clusters. The full spectrum, continuum, and
photometry fits overestimate the ages of many clusters below 1~Gyr,
apparently suffering from an excess of AGB stars making the cluster
spectra redder than the models and forcing older aged models to
provide the best fits. Although photmetric fits seem to do better for
clusters older than 1 Gyr (0.1 dex or 23\% error), we will show in \S
\ref{mets} that CN spectrum fits are superior for simultaneously
providing both age and metallicity for these older
objects. Furthermore, the uncertainty of whether the clusters will
have boosted near infrared emission due to AGB stars makes using the
photometry alone a less reliable technique than using
spectra. Comparing the fitting methods (Figure~\ref{fig4}), we find
a tighter correlation between ages derived from the full spectrum and
continuum fits than from the CN and full spectrum fits, suggesting
that the ages are more strongly driven by the continuum shape than by
the spectral lines. Because the derived ages are so strongly
influenced by the continua and we have seen signs of problems in the
models matching the cluster continua, we conclude that the best method
for deriving accurate ages for globular clusters of all ages,
especially when simultaneously determining metallicity, is fitting
models to their continuum-normalized spectra.

\subsubsection{The Outliers \label{weird_ones}}

There are two extreme outliers in the CN spectrum age plot of
Figure~\ref{fig5}a. The spectra of these outlying objects are shown in
Figure~\ref{fig7}. The upper leftmost outlier in the CN spectrum age
plot is NGC~2092 in the LMC (Figure~\ref{fig7}a,b), with a literature
age of 4-12 Myr (from MSTOs and UBV colors) and for which we derive an
age of 10 Gyr from both the CN and full spectrum fits. Its spectral
coverage is only over the optical range of 3500-5870~\AA. The shape of
its spectrum is fairly flat with emission lines and few absorption
features. The emission lines include [OII]$\lambda$3727,
[OIII]$\lambda\lambda$4959,5007, and [NeIII]$\lambda$3869 nebular
emission. The emission lines are masked out in our fits, which leaves
very few distinguishing features. The cluster's UBV photometry is also
essentially flat. Our best fit metallicity of this object is bottomed
out at the minimum in the model grid, [Fe/H]=$-$2.2, for both the CN
and full spectrum fits, likely because nothing would fit well and the
lowest metallicity model that has very few metal lines in the red
produced the lowest, although high, $\chi^2$. \citet{santos95} list
this object as a cluster embedded in a star forming complex with a
flat continuum probably caused by red supergiants, which are stocastic
in nature even for large star clusters. This appears to be a case that
cannot be fit well with either spectral or photometric techniques
using the BC03 SSP models.

The other outlier in the upper lefthand corner of the CN spectrum age
plot in Figure~\ref{fig5}a is NGC~2096 in the LMC. This object has a
literature age of 49~Myr from its U-B color. Its spectrum
(Figure~\ref{fig7}c,d) contains a moderate sized 4000\AA~break of
D$_{n}$(4000)=1.28, which is consistent with an age of at least 1~Gyr
from index plots in \citet{bc03}, and shows no evidence of the
characteristic spectral shape of a young object with a strong Balmer
discontinuity that would be expected for an age of only 49~Myr. For
this object, our age is 1.2~Gyr (full spectrum and photometry fits) to
6~Gyr (CN spectrum fit). However, \citet{santos95} note this object as
one where a few luminous intermediate temperature supergiants dominate
the cluster's integrated spectrum and claim that its age is even
younger at 6-12~Myr. This may be a stocastic case where the supergiant
dominated spectrum is not properly matched by the models.

Two out of 101 globular clusters in this sample could not be properly
fit by the BC03 models. Both are very young clusters that show
evidence of supergiant dominance in their integrated spectra. The
stocastic nature of these stars make it difficult for us to derive
accurate ages for clusters younger than a few tens of Myrs.

\subsection{Metallicites \label{mets}}

We compare our derived cluster metallicites to the literature values
in Figure~\ref{fig8}. There are fewer points on these plots than
for the ages in Figure~\ref{fig5} because fewer of the clusters
had metallicity estimates in the literature. The magnitudes of the
metallicity errors relative to literature values are a stronger
function of cluster age than [Fe/H]. This is illustrated in
Figure~\ref{fig9} where the [Fe/H] errors are plotted against average
literature ages. Table \ref{rms_met_tbl} gives the average [Fe/H]
offsets and dispersions for different groups of clusters.

Metallicity is harder to constrain than age. This was first seen in
Figure~\ref{fig1} when the $\Delta\chi^2$ contours grew primarily in
the metallicity dimension as noise was added to the cluster
spectrum. Nevertheless, our CN spectrum fits do well on metallicity
estimates for the older clusters. In Figure~\ref{fig8} we see a
tight correlation of metallicity from these fits with the literature
for clusters older than 1 Gyr. The large symbols mark literature
estimates that are based on spectra. There is an overall [Fe/H] scale
offset between M31 (\textit{blue pluses}) and the MC (\textit{red
circles, black crosses, black inverted triangles}). If this offset of
0.45 dex is removed from the MC cluster literature metallicities, our
metallicites agree to 0.12~dex with those from the \citet{huchra91}
line indices (\textit{blue pluses}), literature values on a
homogeneous scale from \citet{santos04} (\textit{black inverted
triangles}), and \citet{leonardi03} line index ratios (\textit{red
circles}) and their comparison literature values (\textit{black
crosses}). 

The small symbols have literature metallicity estimates that are based
on colors. There is more scatter in these values. The weaker
correlation with VJK colors (\textit{green triangles}) can be seen, as
well as its break down at low metallicities of
[Fe/H]~$\lesssim-$0.7. The \citet{jiang03} metallicities
(\textit{magenta squares}) are for older clusters in M31 and were
derived from photometric fits to BC96 models using only three
metallicities, [Fe/H] = 0.0, $-$0.7, $-$1.7. Their discrete
metallicity steps and any differences between the 1996 and 2003 models
are the reasons for the large dispersion of these points.

For the clusters younger than 1 Gyr, our metallicites from CN spectrum
fits show no correlation with literature metallicity values. We
believe this is because continuum information is also necessary for
determining metallicities of young clusters, particularly since
younger clusters likely have higher metallicity with more line
blanketing. A hint that the continuum information might be helping to
determine the metallicity of some young clusters can be seen in the
plots of our full spectrum fits in Figs. \ref{fig8} and \ref{fig9}. In
Figure \ref{fig8}c, if we ignore the points below the 1:1 correlation
line for the moment, there does seem to be a metallicity trend with
the literature for many of the other points. All of the clusters that
we ignored below the line show possible signs of a near infrared
excess in their photometry over the model that best fits their optical
spectra. Some of these spectra are shown in Figure \ref{fig3}.  These
clusters are in the correct age range, 0.1-2 Gyr, to be affected by
TP-AGB stars. It is not clear why our derived metallicities of these
objects appear too low, since it seems that redder spectra would be
better fit by models with higher metallicities, but maybe it is more
an indication that these results cannot be trusted. Fits to the
continua shape alone (Figure~\ref{fig8}e) result in much less
correlation to the literature, however, 5 of the 6 outliers below the
1:1 correlation line in this case are common to those in the full
spectrum plot. Therefore, although it seems that correct continuum
information might allow metallicity estimates to be made for young
objects, the uncertainty in proper modeling of the continua shapes
makes the full spectrum fits unreliable.

We expect our metallicity estimates that are based on broadband
photometry to be less well constrained than those based on
spectra. Three of our comparison literature sources make cluster
metallicity estimates based on photometry (denoted in our plots by
smaller sized symbols): the \citet{cohen94} VJK colors, the
\citet{jiang03} BC96 colors, and two of the \citet{santos04}
homogenized literature values. Our photometrically derived
metallicities (Figure~\ref{fig8}d) do not correlate with these
literature values, except for possibly a very weak relation with the
VJK colors (\textit{green triangles}). As expected, our metallicities
derived from broadband photometry and the continuum shape show no
correlation to any values from the literature that are based on
spectral lines.

\citet{santos04} use empirical relationships between the sums of EWs
of Balmer and metal lines to the literature age and metallicity values
to estimate ages and metallicities of clusters in the Magellanic
Clouds and in the Galaxy. They find that
EW(H$\delta$+H$\gamma$+H$\beta$) and EW(CaK+Gband+Mg) are both
sensitive to age for clusters younger than 10 Gyr, while
EW(CaK+Gband+Mg) is sensitive to [Fe/H] only for clusters older than
10 Gyr. Neither of these EW sums correlate with the ages of old
clusters (t$_{age} >$ 10 Gyr) or with the metallicities of young
(t$_{age} <$ 10 Gyr) clusters. Our CN spectrum fits are similar to
this technique in that they use line strengths, but over the entire
spectral range of the data. Our fits produce metallicities that
agree with literature values only for the older clusters, supporting
the \citet{santos04} result, but extending the minimum age from 10 Gyr
down to $\sim$~1~Gyr for valid metallicity estimates. 

To summarize our results for metallicity estimates, we agree very well
with the literature metallicities that are based on spectra for
clusters older than 1 Gyr. Those based on colors have much more
scatter, which suggests that our CN spectrum fits are more robust in
estimating [Fe/H] than methods using colors. For clusters younger than
1 Gyr, metallicity is hard to constrain. Our full spectrum fits hint
that correct continua shapes would aid in metallicity estimates of
younger clusters, but given the uncertainties in modeling this shape,
we cannot accurately derive metallicity estimates for clusters younger
than $\sim$~1~Gyr.

\subsection{Using Model Line Indices \label{line_ind}}

We compare line indices measured on the BC03 model spectra to those
measured on the globular cluster spectra in Figure~\ref{fig10} for
H$\beta$-[MgFe]$^\prime$ and $<$Fe$>$-Mg\textit{b}. Panels
a-d show model grids calculated at the spectral resolutions
of the data, 6~\AA, 12~\AA, 16~\AA, and 23~\AA~ respectively. Symbols
mark the indices that could be measured for the clusters in each
resolution group (indices that came out negative due to noise or
emission lines were ignored). The 6 and 23 \AA~resolution groups
(\textit{squares and crosses}) are older clusters in M31, while the
rest are younger clusters in the Magellanic Clouds. The
H$\beta$-[MgFe]$^\prime$ index grids clearly show that the model
indices become degenerate at $t_{age} <$ 100 Myr for $-$1.0 $<$ [Fe/H]
$<$ +0.5, and at $t_{age} \lesssim$ 1 Gyr for [Fe/H] $<$ $-$2.0. Most
of the young MC clusters fall in the degenerate regions of the
grids. This degeneracy also occurs for the traditional Lick indices
(tabulated with the BC03 model package) in this region of parameter
space, as shown by the grids in Figure~\ref{fig10}e. This plot only
includes points from clusters that have spectra with resolutions near
that of Lick indices (11 and 12~\AA).

\citet{bc03} investigated how well line indices calculated directly
from a library of their models containing complex stellar populations
with a range of star formation histories could match galaxy spectra
from the Sloan Digital Sky Survey (SDSS). They concluded that those
models could simultaneously fit observed strengths of H$\beta$,
H$\gamma_{A}$, H$\delta_{A}$, [MgFe]$^\prime$, [Mg$_{1}$Fe],
[Mg$_{2}$Fe], and D$_{\textit{n}}$(4000) in high quality galaxy
spectra with S/N$_{med} >$ 30 per pixel. Their sample of moderately
low redshift SDSS galaxies probes a different region of
age-metallicity space than does our sample of globular clusters. Local
galaxies are generally old with considerably higher metallicity than
these clusters, a region of parameter space where the models do
well. In contrast, the M31 globular clusters are old and metal poor
and the MC globular clusters are young and metal poor. Bruzual \&
Charlot note that their SSP models using the STELIB/BaSeL 3.1 library
do poorer on line strengths for [Fe/H]~$<$~$-$0.7 and our index plots
add model degeneracy for $t_{age} <$ 100 Myr, both of which are parts
of parameter space occupied by globular clusters in this sample.

Figure \ref{fig10}f shows $<$Fe$>$ vs. Mg~\textit{b} with model grids
included for the minimum and maximum spectral resolutions. Some of the
clusters depart from the model grid such that [Mg/Fe] may be enhanced
\citep{maraston03}. Since we see signs of $\alpha$-enhancement and the
BC03 models use scaled solar abundance ratios, the preferred metal
index here is [MgFe]$^\prime$, which was shown by \citet{thomas03} to
be sensitive to metallicity while insensitive to $\alpha$-enhanced
abundance ratios at the Lick/IDS spectral resolution. However, when
[MgFe]$^\prime$ is plotted against Balmer indices, which are affected
by $\alpha$-enhancement because of metal lines in the index passbands,
many of the clusters fall outside of the model grids with lower Balmer
indices than the models (Figure \ref{fig10}a-d). One might suspect that
the lower H$\beta$ indices are caused by the lines being partially
filled in by emission, however, many of these clusters are old ones
from M31 (\textit{squares and crosses}) and unlikely to have
emission. The younger MC clusters, on the other hand, could be
affected by emission. Nevertheless, since our purpose in calculating
line indices is for comparison of using the model indices to using the
entire spectrum, and the affected MC clusters fall in the degenerate
region of the model grid with unusable indices, we do not correct for
emission.

Further possible causes of the H$\beta$ index discrepancy could be
$\alpha$-enhancement or the presence of blue horizontal branch (BHB)
stars in the clusters. The M31 clusters do show signs of
$\alpha$-enhancement in Figure~\ref{fig10}f. Line indices are changed by
$\alpha$-enhancement, but in the wrong sense to explain the high
H$\beta$ in the BC03 scaled solar abundance ratio model
grids. H$\beta$ increases with [$\alpha$/Fe] \citep{tantalo04a,
tantalo04b, thomas03}, due to extra absorption in the blue
pseudo-continuum and some in the central passband. Therefore, if the
clusters are $\alpha$-enhanced their H$\beta$ indices should be higher
than the model grid. The second possible culprit, BHB stars in old
clusters with low metallicities, also affect H$\beta$, but in the
wrong sense. H$\beta$ should increase if a BHB exists in the
cluster. We cannot explain the H$\beta$ behavior of the clusters
relative to the models by either of these causes.

This same situation is seen by \citet{leborgne04} in their Figs.~12
and 13 where they calculate Lick indices directly from the PEGASE-HR
model spectra, which use the same stellar library as the BC03 models,
and compare to globular clusters of intermediate to old ages in
Andromeda, M31, M33, M81, and M87. They claim that this behavior of
H$\beta$ is due to the clusters being extremely metal-poor, a regime
where the interpolated stellar libraries suffer from large
uncertainties due to lack of sufficient numbers of stars. This same
regime of [Fe/H]$<-$0.7 was noted as a less reliable region by Bruzual
\& Charlot. The metallicites of our M31 clusters have literature
values of $-$1.35$<$[Fe/H]$<$0.29, with an average of $-$0.73, so many do
fall in the unreliable metallicity regime of the models. That, coupled
with the index degeneracy below ages of 100 Myr, make indices
calculated directly from the models unusable for most of the globular
clusters in our sample. 

Because calculating line indices from the models is troublesome for
the age-metallicity parameter space occupied by many of the clusters
in this sample, we take a different approach to investigating the
importance of these spectral features. We fit model spectra to the
clusters only over regions that include and exclude some of the
typically used line indices and breaks. These features are
$D_n$(4000), H$\delta$, G4300, H$\gamma$, H$\beta$, Mg \textit{b},
Fe5270, and Fe5335. The index definitions for the extents of the
pseudocontinua passbands of lines, or for the flux ratio windows for
breaks, define our wavelength ranges over which to fit the models to
cluster spectra. Figure~\ref{fig11} shows the results of these fits,
where a\&b use only the regions including these features and c\&d
exclude these regions. It is clear that these features are very
important for deriving cluster parameters from the models, but also
that fitting the entire spectral range (refer to Figure~\ref{fig5} and
\ref{fig8}) does a better overall job in estimating age and
metallicity.

\subsection{S/N and Spectral Resolution \label{snr_sec}}

Spectra of z$\sim$1 galaxies from large surveys will most certainly
have lower S/N than the spectra of the globular clusters used in the
present study. We determine the lowest S/N value that can reliably be
used to estimate ages by adding gaussian noise of different levels to
cluster spectra via the \textit{noao.artdata.mknoise} task in IRAF and
fitting CN model spectra to them over the wavelength range of
3325-9000~\AA. We first use actual cluster spectra to make the test
realistic and avoid any biases that might be introduced by model
spectra. The clusters in this test were chosen for their high S/N
spectra and availability of both age and metallicity estimates in the
literature. Further tests including effects of spectral resolution
were then conducted on smoothed model spectra with added noise.

The resulting ages and metallicities as a function of S/N~\AA$^{-1}$
for each of four clusters, G158, NGC 419, NGC 2134, and NGC 1818, are
shown by different symbols in Figure~\ref{fig12}. The average spectral
resolutions of these data are 23\AA, 13\AA, 13\AA, and 13\AA,
respectively. The derived ages of the clusters hover around the same
values as S/N decreases until $\sim$5~\AA$^{-1}$, below which the
derived ages become unstable. The horizontal line associated with each
set of symbols is the average of the derived ages or metallicites for
that cluster from its S/N$\geq$5 spectra. For comparison, the spread
of literature ages and metallicities for each cluster are shown as
vertical lines of the same styles on the right-hand side of the
plots. Our derived parameters for these four clusters fall within the
literature age ranges. If the 0.45 dex [Fe/H] scale offset is
subtracted from the MC cluster literature values, then our average
[Fe/H] lines fall within the literature values for NGC 419 and G158,
which are the clusters older than 1~Gyr. We do not match the
literature [Fe/H] of the younger clusters. We find that the spectra
need to have S/N $\geq$~5~\AA$^{-1}$ to achieve stable age estimates
from CN spectral fits. Examples of fits to NGC 419 with added noise
and S/N~=~99, 22, and 12 RE$^{-1}$ (27, 6, 3~\AA$^{-1}$) are shown in
Figure~\ref{fig13}. Note that the last one is below the limiting S/N
value.

We have also begun a study of the effects of spectral resolution on
derived ages and metallicities. To accomplish this we smooth model
spectra to $\Delta\lambda$= 5, 10, 15, 20, 25, and 30~\AA, add noise
in 10 realizations at each S/N level up to 60~\AA$^{-1}$ for a total
of 300-500 simulations at each $\Delta\lambda$, and fit CN model
spectra to them. The lowest resolution of $\Delta\lambda$=30~\AA~was
chosen to match the lowest typically achieved in galaxy surveys. We
select two input models that are consistent with age and metallicity
combinations occupied by some of the globular clusters in this sample
and that may also be useful for high redshift galaxies. These
combinations are: 1~Gyr, [Fe/H]=0.0; and 10 Gyr, [Fe/H]=0.0 (similar
to G158 in Figure~\ref{fig12} at 18 Gyr, [Fe/H]$\sim-$0.1,
$\Delta\lambda$=23~\AA).

The results of these fits are presented in
Figures~\ref{fig14}-\ref{fig16}.  Plots in Figure~\ref{fig14} show the
best fit ages (top rows) and metallicities (bottom rows) as a function
of S/N~\AA$^{-1}$ for spectral resolutions of $\Delta\lambda$~=~5, 15,
25~\AA~(to conserve space $\Delta\lambda$~=~10, 20, and 30~\AA~are not
shown) for the 1and 10~Gyr, [Fe/H]~=~0 inputs. Input parameters are
marked by solid lines, dashed lines represent uncertainties of 10\% in
age and 0.1 dex in [Fe/H], and dotted lines mark age uncertainties of
20\%. Small filled circles show individual simulations, open circles
are the average derived parameters in S/N bins of width 2, and
vertical cyan bars mark the 1$\sigma$ variation within each bin.

Age-metallicity degeneracy in the derived parameters can clearly be
seen in Figure~\ref{fig15} for the two input models at the same
spectral resolutions presented in Figure~\ref{fig14}. The increased
difficulty of distinguishing older ages is also apparent from the much
larger spread in derived ages for this regime where age changes result
in small spectral differences. As spectral resolution degrades,
derived ages and metallicities for the lower S/N spectra tend to be
driven to the model grid limits more frequently.

We choose limiting S/N values for each spectral resolution
corresponding to the point at which the standard deviation in derived
ages for a S/N bin begins to increase beyond 10 or 20\% of the input
model age. Figure~\ref{fig16} summarizes the required S/N~\AA$^{-1}$ as
a function of $\Delta\lambda$ for the 1~Gyr case on the left and the
10~Gyr case on the right. Solid lines denote derived ages within 10\%
and dashed lines within 20\% of the input. Recall that the S/N is
calculated for these spectra by taking the standard deviation of the
flux in wavelength bins along the spectrum, which has the disadvantage
that spectral features increase the apparent ``noise'' in the
spectrum.  Because of this fact, there is a limit to the maximum S/N
calculated in this manner.  These artificial limits are plotted as
dotted lines in Figure~\ref{fig16}, and therefore the open upward
facing triangles are lower limits to the required S/N at those
resolutions. The 10~Gyr object requires higher S/N than the 1~Gyr
object at almost all resolutions, implying a strong dependence on the
object's age. However, because the age is unknown a priori, the 10~Gyr
model must be used to determine the minimum required S/N.

The expected trend is seen for ages within 10\% of the input: higher
S/N is required for lower spectral resolution. At the 20\% level,
however, somewhat puzzling behavior is seen for the 10~Gyr old
object. At $\Delta\lambda>$~15~\AA~the required S/N actually
decreases. This does not appear to be an aritfact of the simulations,
as we see no correlations between the number of points within S/N bins
and the standard deviation in derived ages. Interestingly, the
globular cluster G158 at $\Delta\lambda$~=~23~\AA~in
Figure~\ref{fig12} may demonstrate similar behavior. Although only one
noise realization was performed at each S/N level there, its correct
age is obtained at much lower S/N than the younger clusters in the
plot. Perhaps this trend is real. One could envision a scenario in
which with decreasing resolution model mismatches of narrow lines in
the spectrum become less important than broad features typical of old
spectra, such as the 4000~\AA~break and the Mg absorption trough. In
order to get the correct age, high S/N is required, but once the
allowed uncertainty is increased, the broader features can provide
enough information in lower S/N spectra. Clearly, this is only an
initial guide for required S/N and more investigation needs to be done
into the cause of this unexpected feature. At the high resolution end,
our results are consistent with those of \citet{mathis06} who also fit
continuum-removed spectra (via a data compression algorithm) and find
required S/N$\sim$11-17~\AA$^{-1}$ (20-30 per 3~\AA~pixel) at a
spectral resolution of 3~\AA. Extrapolating our 5~\AA~point down to
3~\AA~would give S/N$\sim$7-22~\AA$^{-1}$.

In general, at the spectral resolutions of the globular clusters in
Figure~\ref{fig12}, $\Delta\lambda \sim$~13-23~\AA, our simulations
indicate that a S/N~$\ga$~25~\AA$^{-1}$ is required for a 20\% age
uncertainty and $\ga$~45~\AA$^{-1}$ for a 10\% uncertainty. This is
higher than the S/N$\sim$5~\AA$^{-1}$ obtained from adding noise to
actual cluster spectra. This discrepancy is not currently understood
and will be investigated further in future work.

\section{Discussion}

\subsection{Different Spectral Regions}

In \S \ref{line_ind} we looked at the results of fits to specific
regions of the spectrum around typically used line indices. Here we
test the broader effects of the regions blueward and redward of the
4000~\AA~break on derived ages and metallicities. This will be
important for fitting galaxies at varying redshifts, since at z$\sim$1
optical spectra typically reach just beyond the 4000~\AA~break in the
galaxy's rest frame. For the clusters in our sample that include near
ultraviolet spectra, we analyze the range of 3200-5650~\AA, which is
the maximum extent common to all clusters in this group. Within this
range, we fit models to the blue ($\lambda<$~4000~\AA), the red
($\lambda>$~4000~\AA), and the region right around the break
(3750-4250~\AA, as defined in the empirical calibration of the break
by \citet{gorgas99}). The clusters in this study are listed in
Table~\ref{red_blue_id} with their ages derived from different parts
of the spectrum. Correlations of ages from the different spectral
regions are plotted in Figure~\ref{fig17}.

Fits to different regions give different results. For CN spectrum
fits, the red part of the spectrum drives the result. The red ages
correlate with the ages from the entire spectrum and are slightly
older than the values in the literature. The blue region gives younger
ages than both the red region and the literature. Ages from the break
region correlate to the red ages with about the same amount of scatter
as we saw in the red-entire spectrum correlation. Since the continuum
is normalized out for these fits, the break region is left with only
the signatures of strong features like the Ca H and K lines and
H$\delta$, similar to the D$_{n}$(4000)-H$\delta$ analysis done by
\citet{kauffmann03a} on SDSS galaxies. The rough agreement of ages
from this region with those from the entire spectrum emphasizes the
importance for the 4000~\AA~break to be in the spectrum used for
estimating the age of an object. The only correlation between [Fe/H]
is a weak one between the entire spectrum and the red spectrum (not
shown in Figure \ref{fig17}), not surprising since most metallicity
sensitive lines typically used are in the red part of the spectrum.

For full spectrum fits, the trends change and the blue region carries
more weight, demonstrating the importance of information in the
continuum shape of the blue part of the spectrum. This is especially
true for young objects with strong Balmer discontinuities. Ages from
the blue and red regions agree, but with some scatter. However, now
the entire spectrum agrees equally well with the blue and the red
regions. The 4000~\AA~break ages agree much better with the blue than
with the red ages. The red, blue, and break ages all correlate with
the literature ages, but the correlation is tightest for the break
region, even more so than for the full spectrum fits over the entire
range. There is no correlation between our [Fe/H] derived from the
different regions, but the red [Fe/H] has a weak correlation with the
literature.

Because of the importance of the continuum information blueward of the
4000~\AA~break, we also test hybrid fits that use the full spectrum
over $\lambda<$~4250~\AA~and the CN spectrum over
$\lambda>$~3750~\AA. Wavelengths below 4000~\AA~will not be affected
by the TP-AGB and carbon stars that dominate the near infrared, so as
long as there are no continuum problems with the data this type of fit
should be valid. We expected this approach to do better on very young
objects. However, the hybrid fits match the full spectrum fits that
used the entire wavelength range. The continuum shape dominates the
fits and spreads out the age correlation with the literature for
objects younger than $\sim$0.1 Gyr. This is the same problem seen
earlier with the continua of very young clusters being redder than the
models, causing older best-fitting ages.

In summary, the continuum shape contains a lot of age information. In
the blue part of the spectrum the continuum is more important than the
lines for deriving age, while in the red part of the spectrum the
lines alone are sufficient to derive the age. The most robust age
estimates are obtained by using the entire CN spectrum of an object
($\sim$ 3200-8200~\AA), but the 4000~\AA~break is an essential feature
to be contained within that spectrum. When using CN spectra to avoid
problems in model and data continua, ages from only the blue region
are younger than those obtained by the red region or the entire
range. Age biases introduced by using different regions of the
spectrum must be considered when combining objects that have different
spectral coverage or different redshifts that change the rest frame
wavelength coverage significantly. Additionally, for objects that
cannot be considered a single stellar population, the ages determined
from different spectral ranges may not be straighforward to
interpret. Nonetheless, near infrared spectra of high redshift
galaxies will be important to provide rest frame wavelengths around
and redward of the 4000~\AA~break, particularly when fitting their CN
spectra.

\subsection{TP-AGB and Carbon Stars}

The effects of TP-AGB stars on the near infrared luminosity of objects
in which they reside have long been known and discussed by groups
creating stellar population synthesis models
\citep{renzini92,bc93,girardi98,maraston98}. Recent models by
\citet{maraston05} incorporate TP-AGB stars through an empirical
spectral library of C- and O-type stars by \citet{lancon02}. The
effects of these stars have been reanalyzed and calibrated on MC
cluster colors in the new models. They show that over 80\% of the K
band light can be contributed by AGB stars between the ages of 0.1 and
2~Gyr for solar and lower metallicities. This effect is down to 20\%
in the V band.

Maraston fits photometric SEDs of these models to MC cluster SEDs. Two
objects in common with our sample, NGC~1783 and NGC~419, show
discrepancies between the Maraston models that best fit the
photometric SEDs and BC03 models that best fit the spectra. These
inconsistencies are the very point being address by the new models,
which show that younger ages may be obtained for objects that have a
near infrared excess due to TP-AGB stars if these stars are properly
included. However, analysis of the spectra may reveal a different
story. The CN spectra of the two clusters (shown in Figure~\ref{fig18}b)
are very similar to each other.  NGC~1783 has D$_{n}$(4000)=1.295 and
NGC~419 has D$_{n}$(4000)=1.246. We derive similar ages for the two
clusters by fitting their CN spectra: 1.5 Gyr and [Fe/H]=$-$0.8 for
NGC~1783, and 1.5 Gyr and [Fe/H]=$-$1.2 for NGC~419. The Maraston
results for NGC~1783 are 0.3 Gyr and [Fe/H]=$-$0.33, and for NGC~419
are 1 Gyr and [Fe/H]=$-$1.35. BC03 models are overplotted on the CN
spectra in Figure~\ref{fig18}c,d for all the best-fitting models. The
older age of 1.5 Gyr is a better fit to the CN spectra of both
clusters.

Our photometric fit age for NGC~1783 is 2.0 Gyr and [Fe/H]=$-$0.6, and
for NGC~419 is 20 Gyr and [Fe/H]=$-$1.4. This much older photometric
age for NGC~419 could be due to the presence of TP-AGB stars making
the continuum redder. The near infrared photometry of NGC~419 (see
Figure \ref{fig3}) does show an excess over best-fitting models to the
spectra. It is likely that the BC03 models do not include the enough
TP-AGB stars in this case. On the other hand, the presence of AGB
stars was detected in NGC~1783 by \citet{frogel90}, but apparently
their inclusion in the BC03 models is sufficient for this case. We
conclude that the detailed information contained in spectra is
necessary for breaking degeneracies that can occur when using only
broadband photometry through the near infrared. In a more recent paper
\citet{maraston06} apply the models with enhanced TP-AGB stars to high
redshift galaxies and find that some degeneracies, particularly
between TP-AGB stars and reddening due to dust, can be broken when
mid-infrared photometry is added from Spitzer. These new models look
very promising for young objects if infrared data are available,
however their spectral resolution is not high enough to fit to galaxy
spectra.

\subsection{Implications for Galaxies}

The ages expected for early-type galaxies at z$\sim$1 fall within the
range spanned by the globular clusters in this sample. The youngest
early-type galaxies at these redshifts will likely be field galaxies
that were the products of mergers. If the bulk of stars in these
galaxies were formed in mergers at redshifts of z$\sim$1.5-2, which
are the formation redshifts found for field ellipticals
\citep{im02,gebhardt03}, they would be at least 1.5~Gyr old by
z$\sim$1 (for $\Omega_{M}$~=~0.3, $\Omega_{\Lambda}$=0.7, $H_{0}$=70
km s$^{-1}$Mpc$^{-1}$). Our CN spectrum fits estimate ages to within
0.16 dex (35\%) of the literature values for globular clusters older
than 1 Gyr in our sample.

The metallicities of z$\sim$1 galaxies will likely be higher than many
of the clusters in this sample, however, a few clusters spanned most
of the range expected for galaxies. At redshifts of 0.3$<$z$<$1.0,
galaxies in GOODS-N have metallicities from 0.3 to 2.5 times solar, or
$-$0.5~$<$~[Fe/H]~$<$~+0.4 \citep{kobulnicky04}. We were able to
determine the [Fe/H] of globular clusters older than 1 Gyr to within
0.12 dex over the range of $-$1.6~$<$~[Fe/H]~$<$~+0.3. The only region
we have not tested is the highest metallicities of [Fe/H]~$>$~+0.3.

Spectra at a resolution of $\Delta\lambda$~=~15~\AA~(R~$\sim$~400 at
6000~\AA) must have S/N~$\sim$~30-45~\AA$^{-1}$ in order to derive
ages with 20 or 10\% uncertainties, respectively, for old galaxies and
S/N~$\sim$~10-20~\AA$^{-1}$ for young galaxies. This is higher than
typically achieved for many objects in modern galaxy surveys such as
MUNICS, DEEP, and TKRS. Our initial checks on TKRS spectra from DEIMOS
on Keck of GOODS-N galaxies at z$\sim$0.8-1.0 with magnitudes of
R=23-23.5 find S/N~$\sim$~5~\AA$^{-1}$ for individual
objects. Therefore, most galaxies will require coaddition of a few
similar objects to reach sufficient S/N.

For galaxies at z$\sim$1 typical optical spectra covering
$\sim$4000-9000~\AA~will provide rest frame wavelengths of
$\sim$2000-4500~\AA. Our tests have shown that when using CN spectra,
data to longer wavelengths will provide more robust age
constraints. This means that near infrared spectra of galaxies will be
important data to obtain for accurate age estimates.

Many early-type galaxies appear to have $\alpha$-element abundance
ratios enhanced above solar
\citep{trager00a,trager00b,thomas03,thomas05}. This enhancement is a
sign of short duration bursts of star formation that may well be typical
for early-type galaxies formed during mergers, as we expect to find at
z$\sim$1. We see indications of $\alpha$-enhancement in some of the
M31 globular clusters in this sample and we still derive ages within
35\%~of the literature values when fitting the entire CN spectra. This
suggests that results from the entire continuum-normalized spectrum,
effectively averaging over all the lines, may be less affected by
$\alpha$-enhancement than detailed line index methods that rely on
only a few affected lines in the spectrum. \citet{cassisi04} find that
broadband colors in the blue are significantly affected by
$\alpha$-enhancement, but removal of the continuum shape must mitigate
these effects to some extent.

Model fitting techniques will have to be modified for galaxies. Simple
stellar populations will no longer be sufficient descriptions of these
complex systems. Fits will require models with complex star formation
histories, multiple stellar populations, and dust. However, this work
has verified that the reliable regions of model parameter space cover
most of that expected to be occupied by galaxies. We have also proven
techniques by which model fits to galaxy spectra can be expected to
work. 

\section{Summary and Conclusions}

We have investigated the utility of using the \citet{bc03} high
resolution simple stellar population synthesis models for estimating
the ages and metallicities of globular clusters. We chose clusters
from M31 and the Magellanic Clouds that cover a wide range of age and
metallicity values, $\sim$ 0.004$<$t$_{age}$(Gyr)$<$20 and
$-$1.6$<$[Fe/H]$<$+0.3, to explore model performance in different
regions of parameter space. We tested various techniques of fitting
the models to data using spectra with and without their continua and
photometry. Results were compared to age and metallicity estimates in
the literature from other methods to evaluate our fitting techniques.

The models are reliable in regions of parameter space where
t$_{age}>$0.1 Gyr and [Fe/H]~$\gtrsim-$0.7. These regions are limited
by the incompleteness of the stellar libraries used in the models. For
objects between the ages of 0.1 and 2.5 Gyr, uncertainties in the
amount of near infrared light coming from TP-AGB and carbon stars make
it hard to match the continuum shape of individual globular
clusters. Red supergiant stars affect clusters younger than 0.1
Gyr. Continuum-normalized spectra can be used to get around the TP-AGB
star problem, as well as any flux calibration issues in the data, with
the requirement that high spectral resolution models are used. The red
supergiant star problem limits accurate age estimates to objects older
than $\sim$0.1~Gyr.

Ages from our CN spectrum model fits best match the entire range of
age estimates from the literature (Figure~\ref{fig5} and
Table~\ref{rms_age_tbl}), with average differences of 0.16 dex (35\%)
for clusters older than 1 Gyr and 0.3 dex (63\%) for younger
clusters. The methods that depend on the continuum shape -- the full
spectrum, continuum, and photometry fits -- overestimate the ages of
many clusters below 1~Gyr. We believe this to be due to a near
infrared excess from TP-AGB stars making the cluster spectra redder
than the models and forcing older aged models to provide the best fits
for cluster ages of 0.1-1~Gyr. Red supergiant stars similarly affect
clusters younger than 0.1 Gyr. Although photometric fits seem to do
well for clusters older than 1 Gyr (0.1 dex or 23\% age error relative
to the literature), the CN spectrum fits are better for simultaneously
providing both age and metallicity for these older objects. Fits to
photometry alone cannot reproduce the correct metallicities.

\begin{figure}
\figurenum{4}
    \plottwo{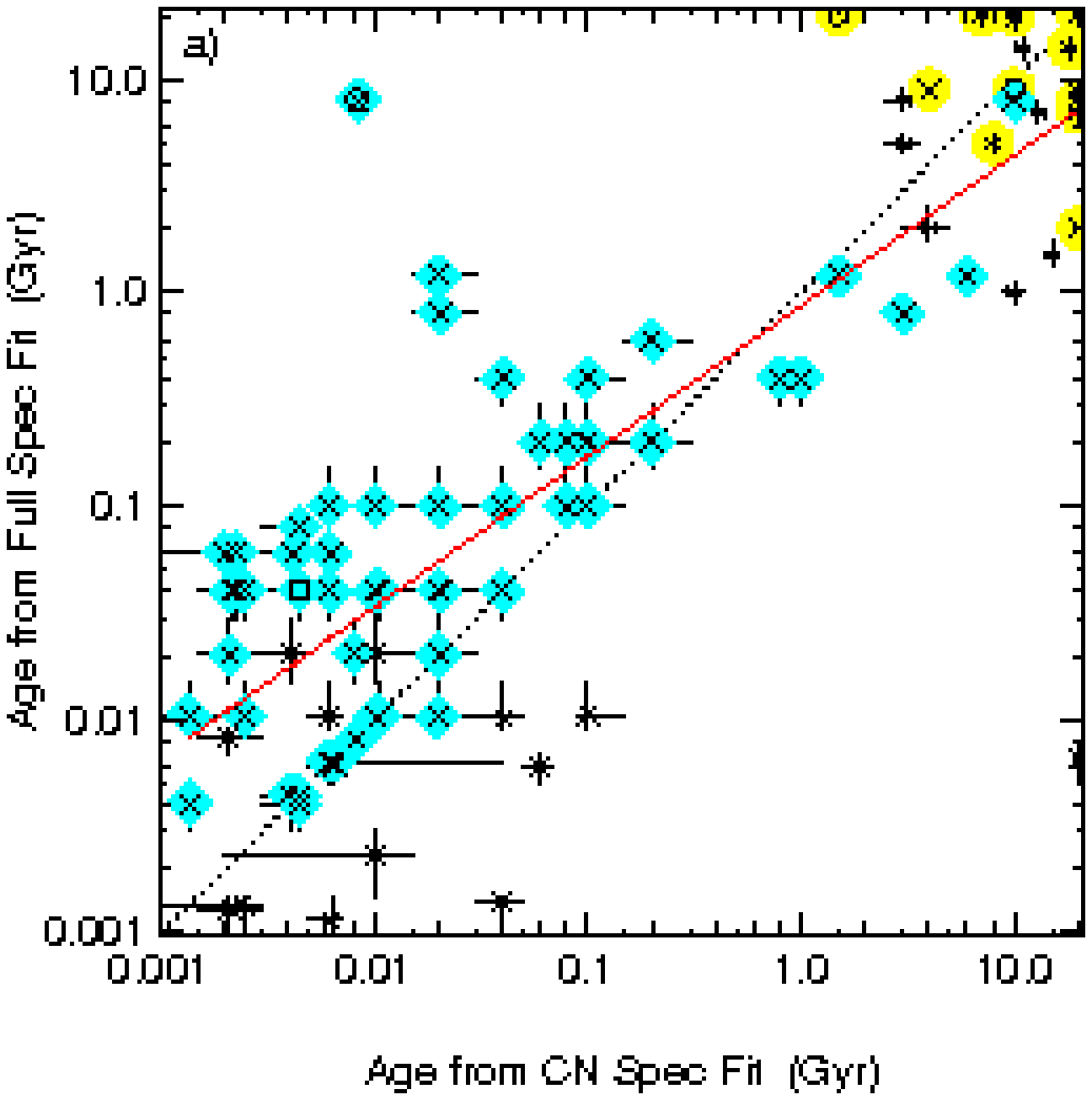}{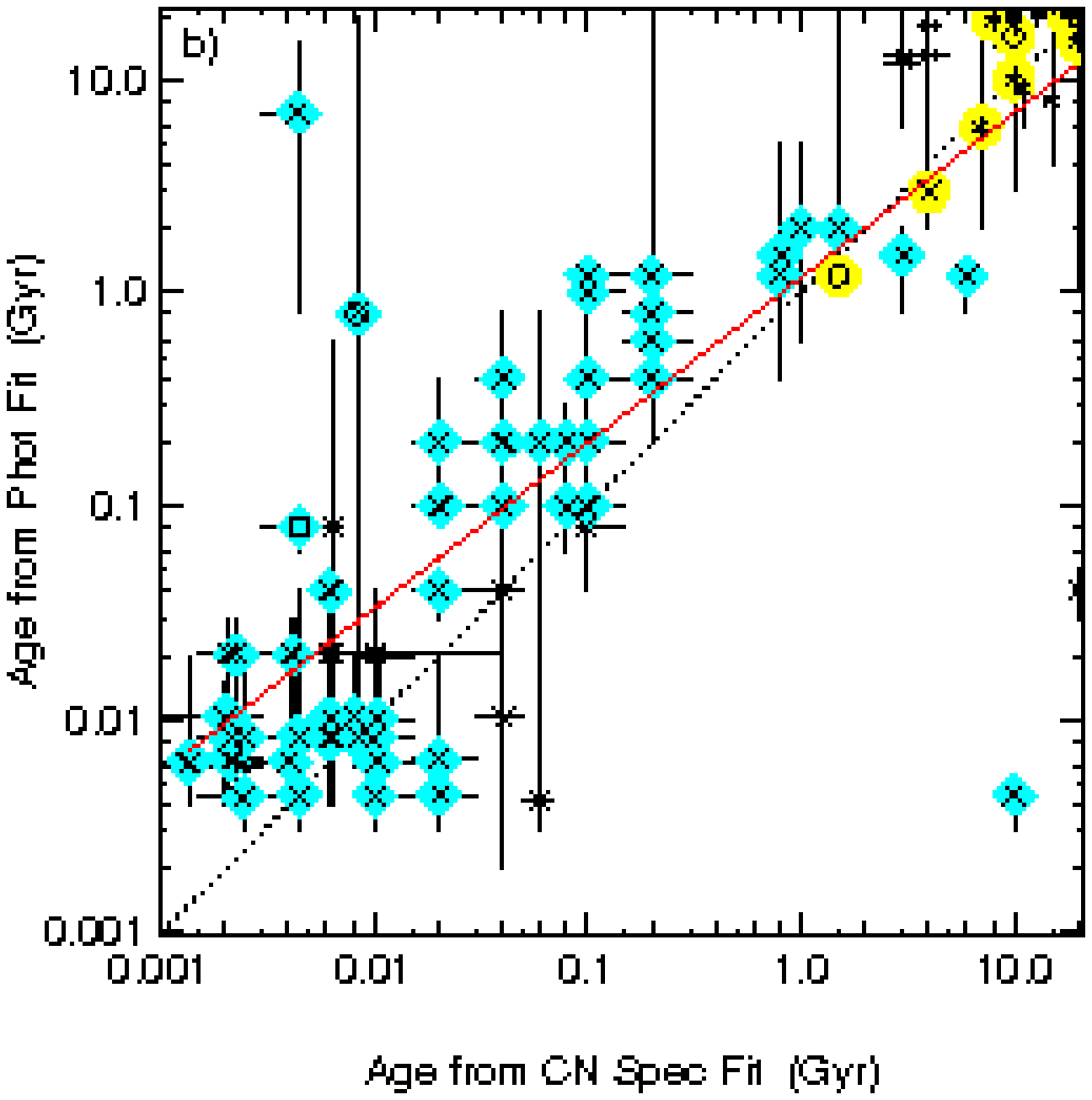}
    \plottwo{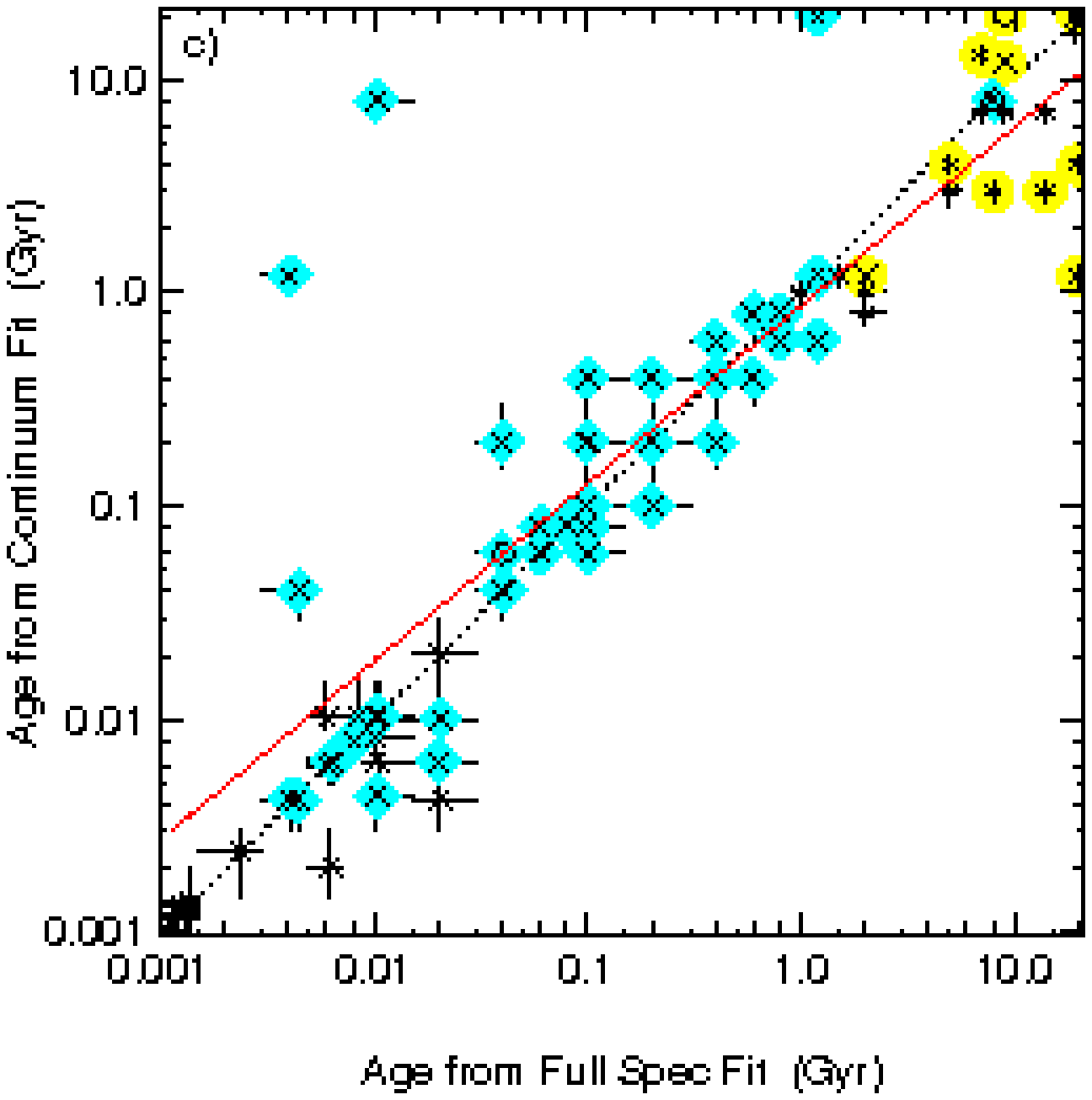}{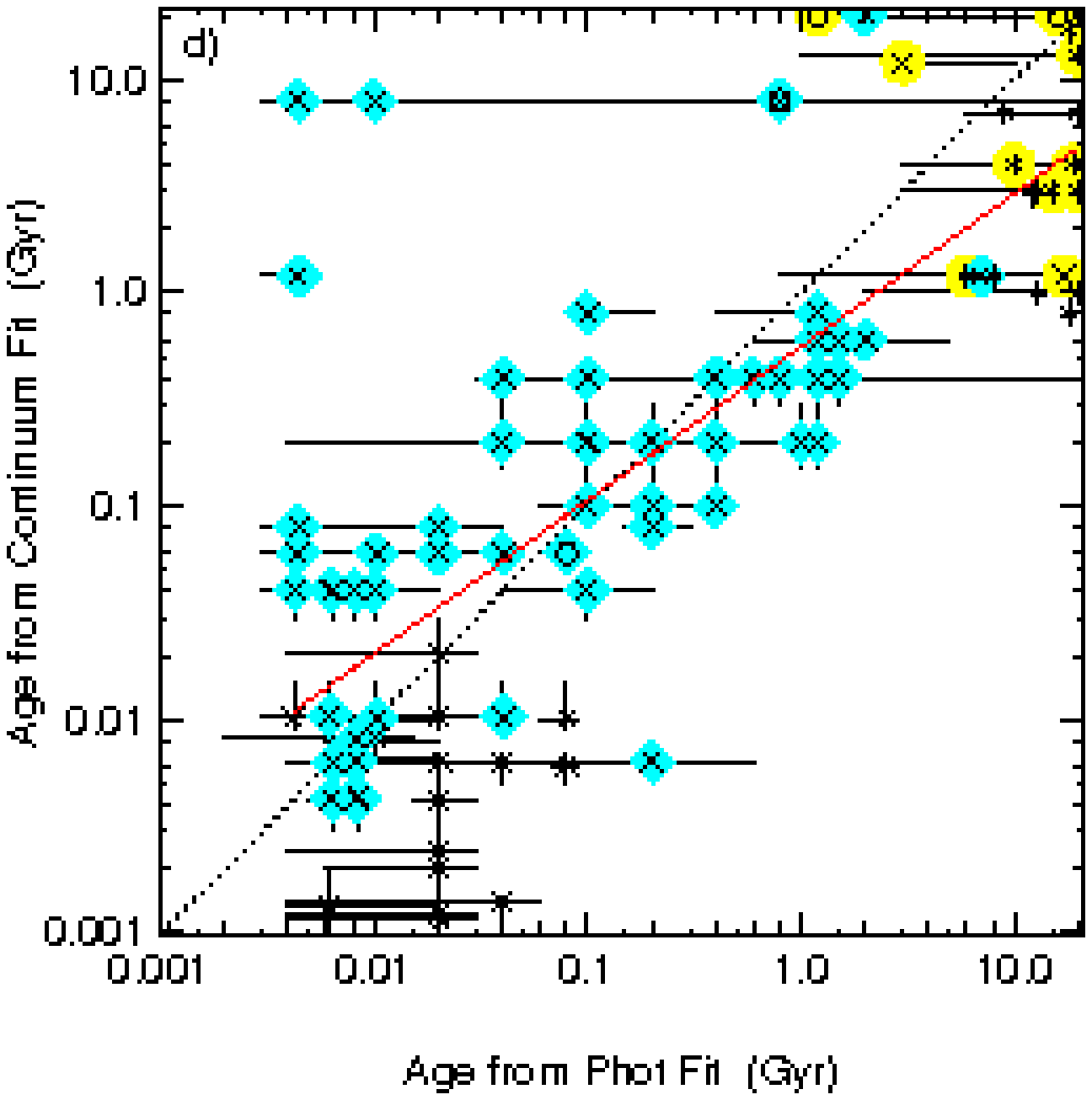}
\caption{Comparison of ages and metallicities derived by different
types of model fits for LMC clusters (\textit{crosses}), SMC clusters
(\textit{circles}), and M31 clusters (\textit{asterisks}) with linear
fits (\textit{solid red lines}) and lines of 1:1 correlation
(\textit{dotted lines}). Clusters with average literature ages
$\geq$~1~Gyr are marked by yellow circles and $<$~1~Gyr by cyan
diamonds. The error bars are 1$\sigma$. If no error bars are seen,
they are smaller than the symbols. Age comparisons are between a) the
full and CN spectrum fits, b) the photometry and CN spectrum fits, c)
the continuum and full spectrum fits, and d) the continuum and
photometry fits.
\label{fig4}}
\end{figure}

\begin{figure}
\figurenum{5}
    \plottwo{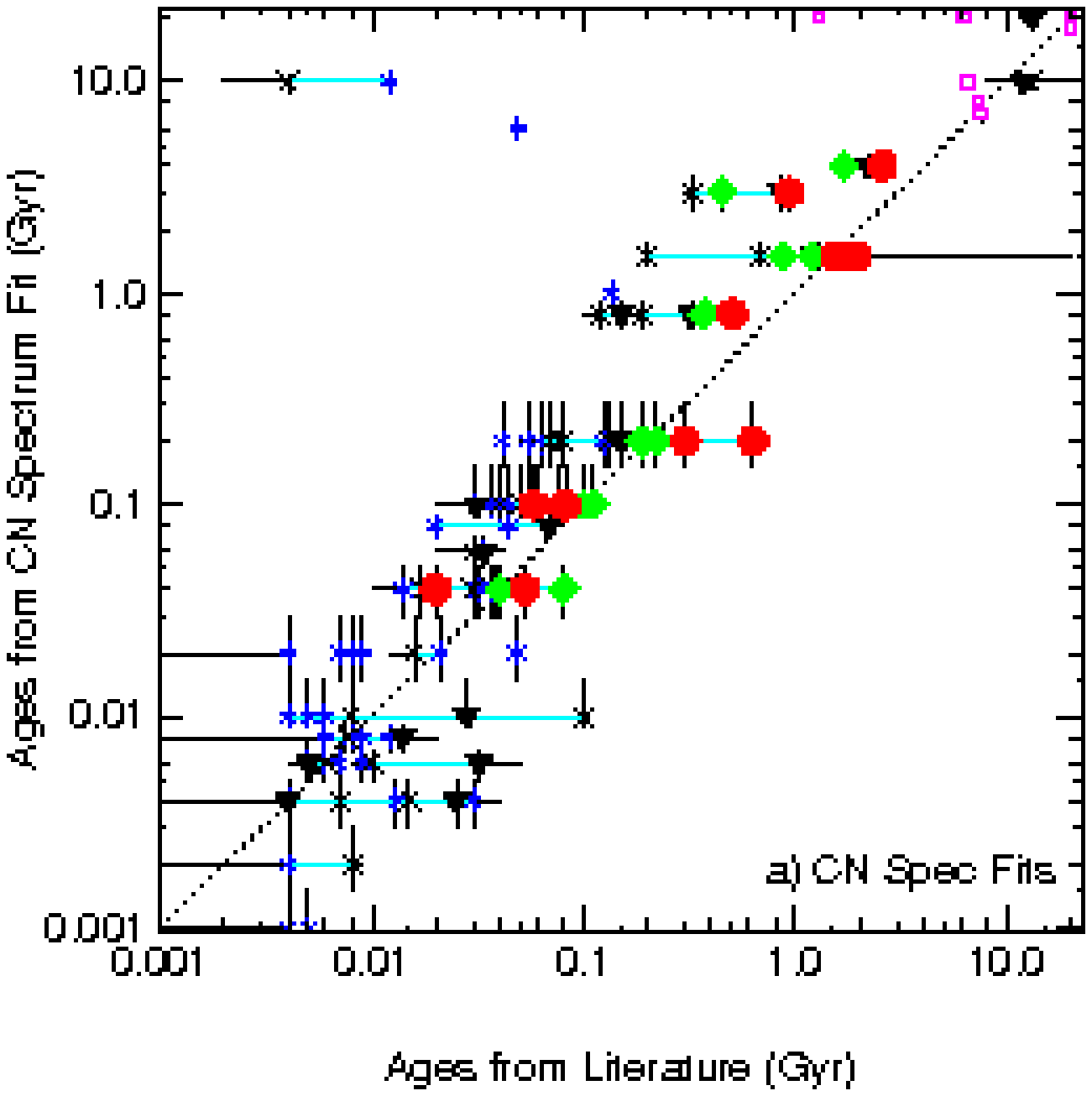}{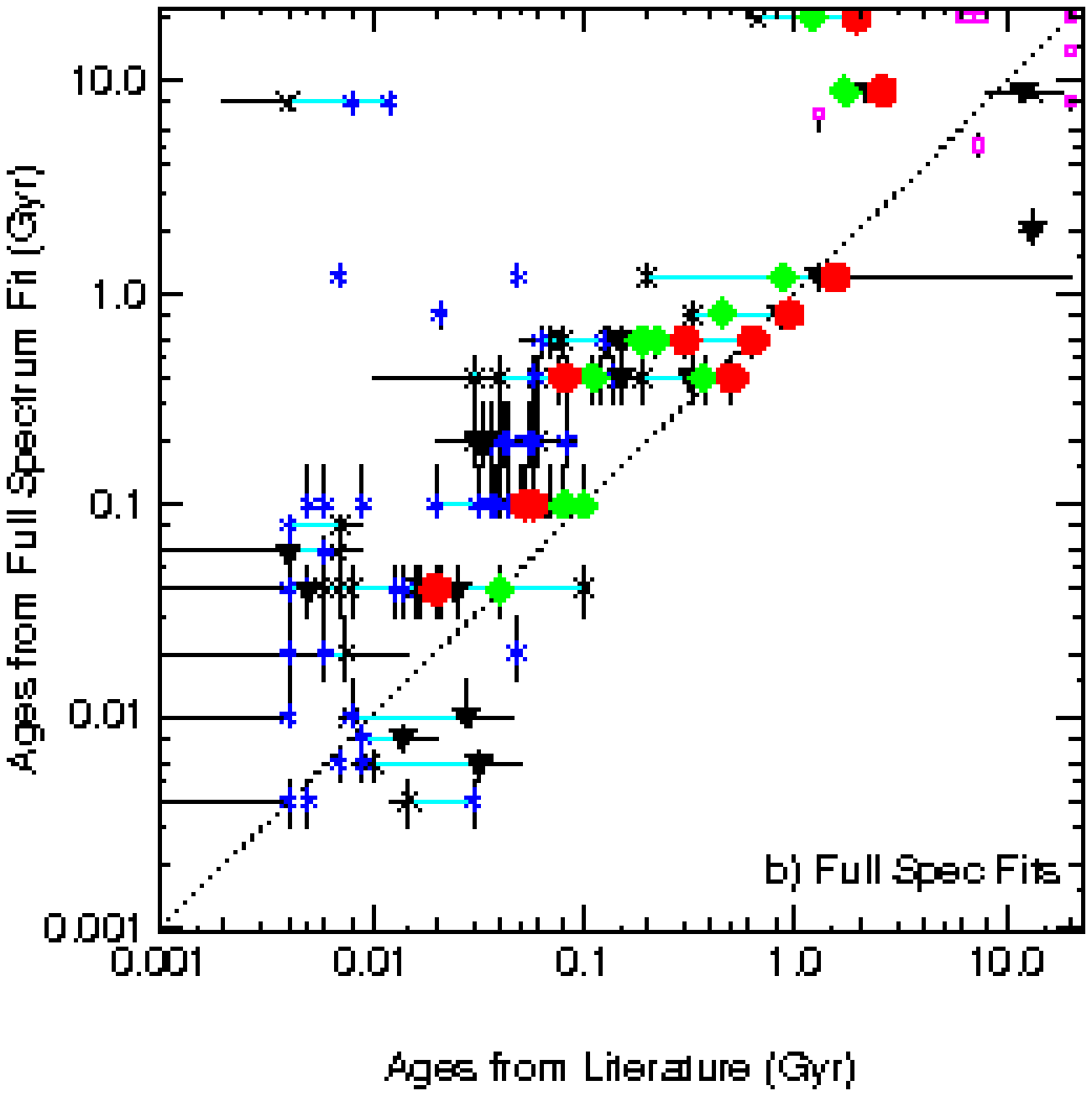}
    \plottwo{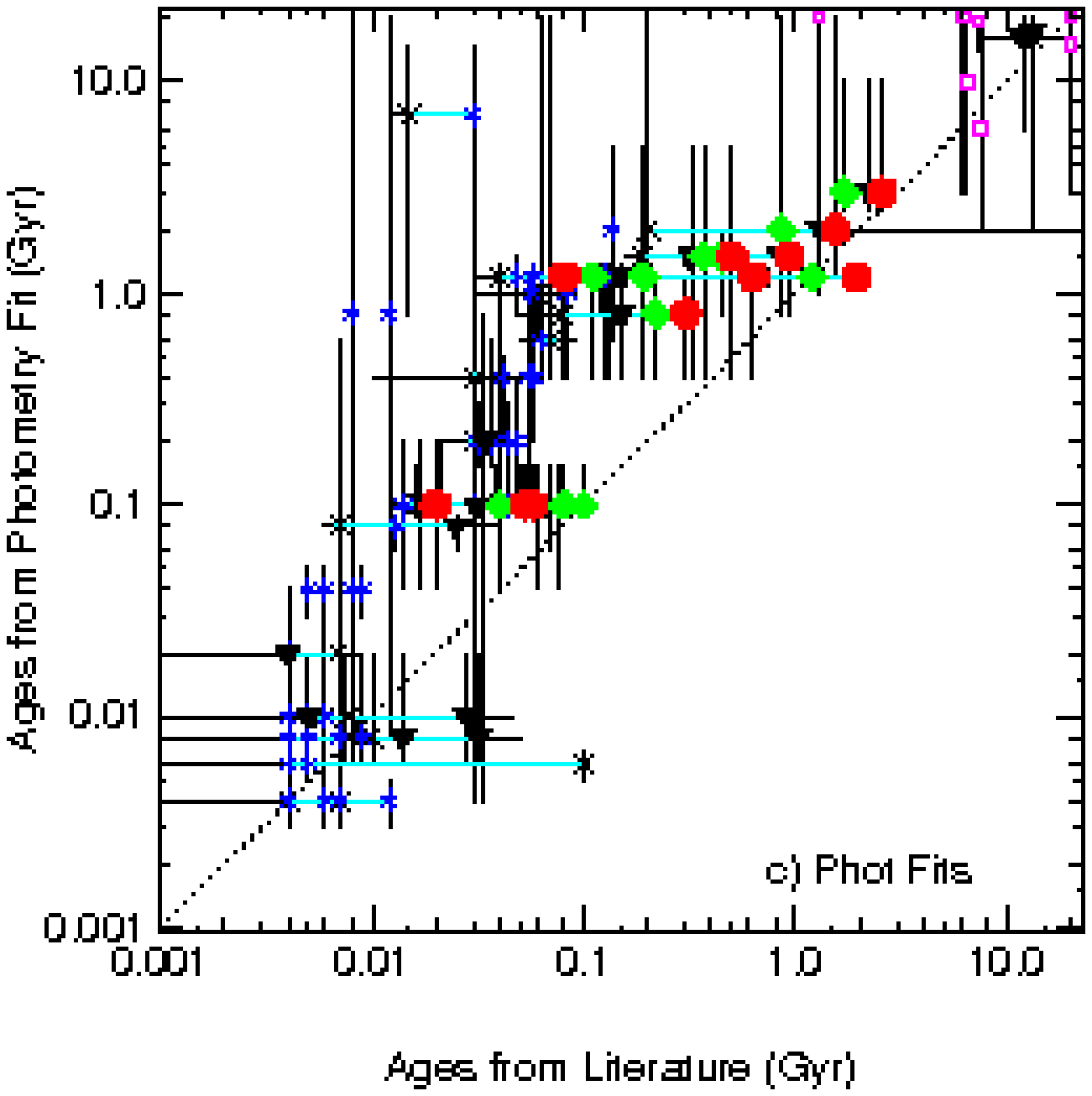}{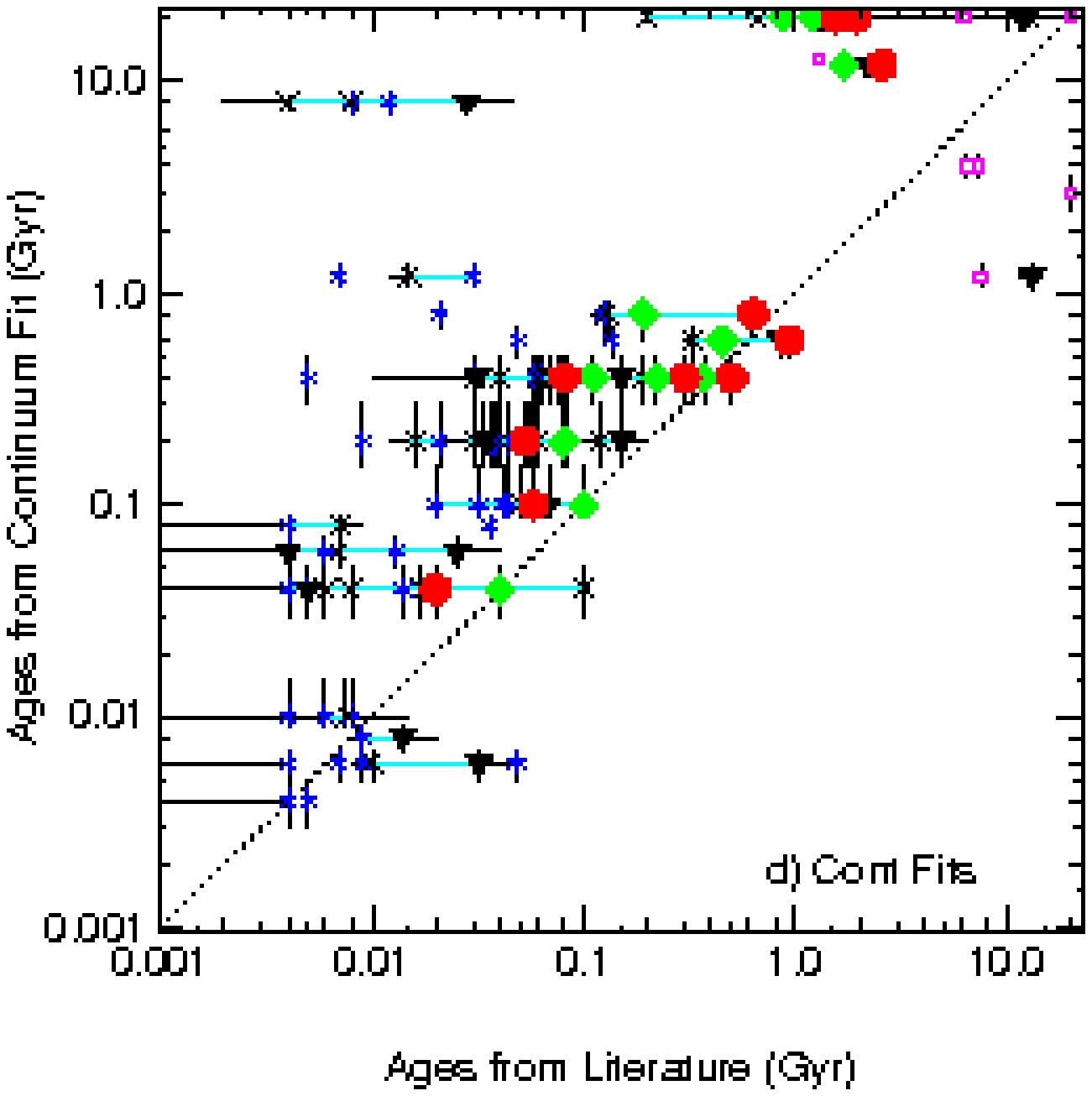}
  \caption{Cluster ages derived from model fits compared to ages from
  the literature. a) is the CN spectrum fits, b) is the full spectrum
  fits, c) is the photometry fits, and d) is the continuum
  fits. Literature ages are derived from main sequence turnoffs by
  \citet{hodge83} (\textit{black crosses}), UBV photometry by
  \citet{santos95} (\textit{blue asterisks}), CMDs by
  \citet{girardi95,girardi98} and \citet{elson88} (\textit{green
  diamonds}), photmetric fits to BC96 models by \citet{jiang03}
  (\textit{magenta squares}), line index ratios by \citet{leonardi03}
  (\textit{red circles}), and literature ages on a homogeneous scale
  by \citet{santos04} (\textit{black inverted triangles}). The thick
  horizontal cyan lines connect points that are the same cluster with
  multiple literature ages. Thin horizontal lines are quoted error
  bars on literature ages, thin vertical lines are 1$\sigma$ error
  bars on our derived ages. If no error bars are seen, they are
  smaller than the symbols. \label{fig5}}
\end{figure}

\begin{figure}
\figurenum{6}
    \plottwo{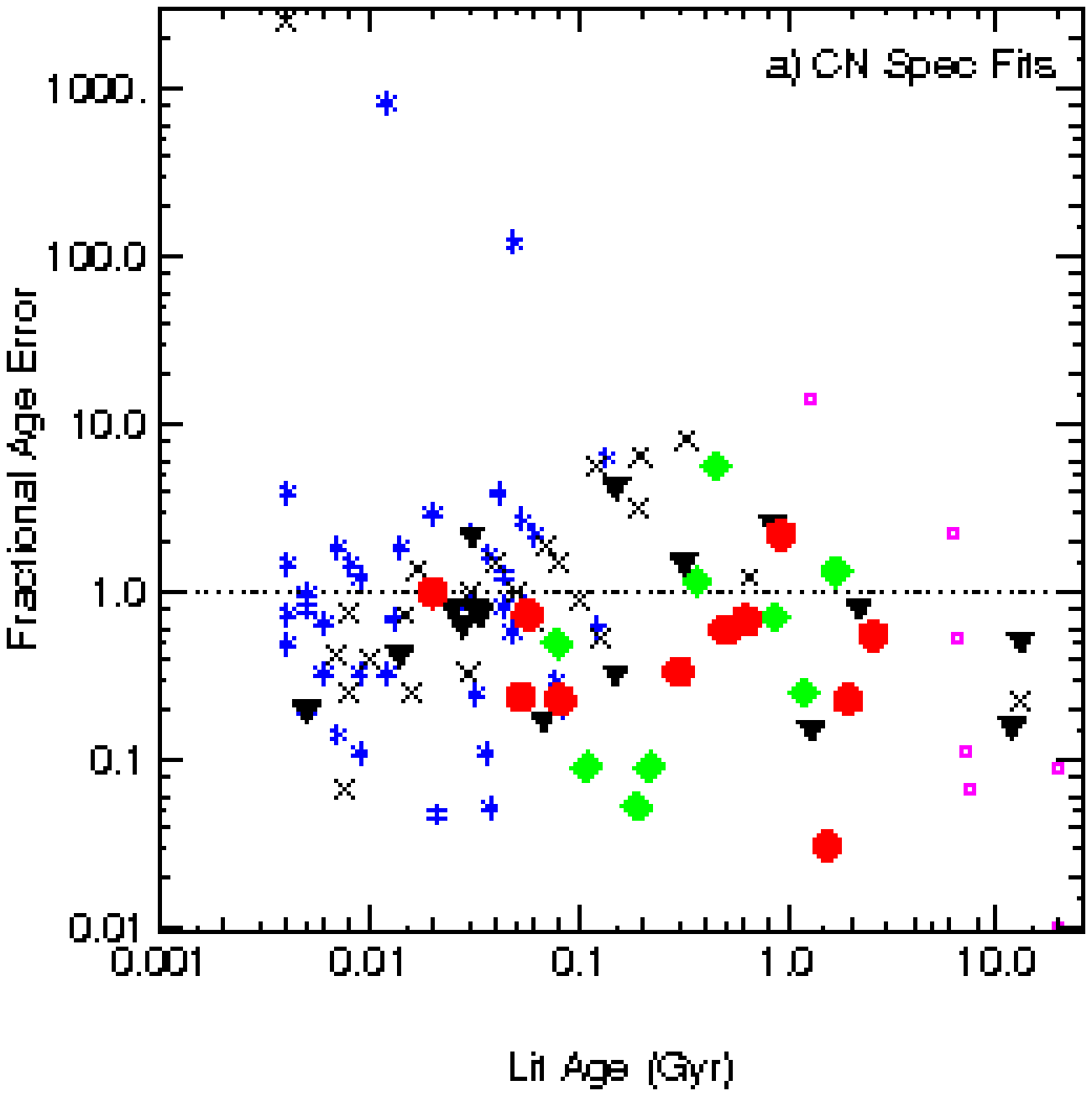}{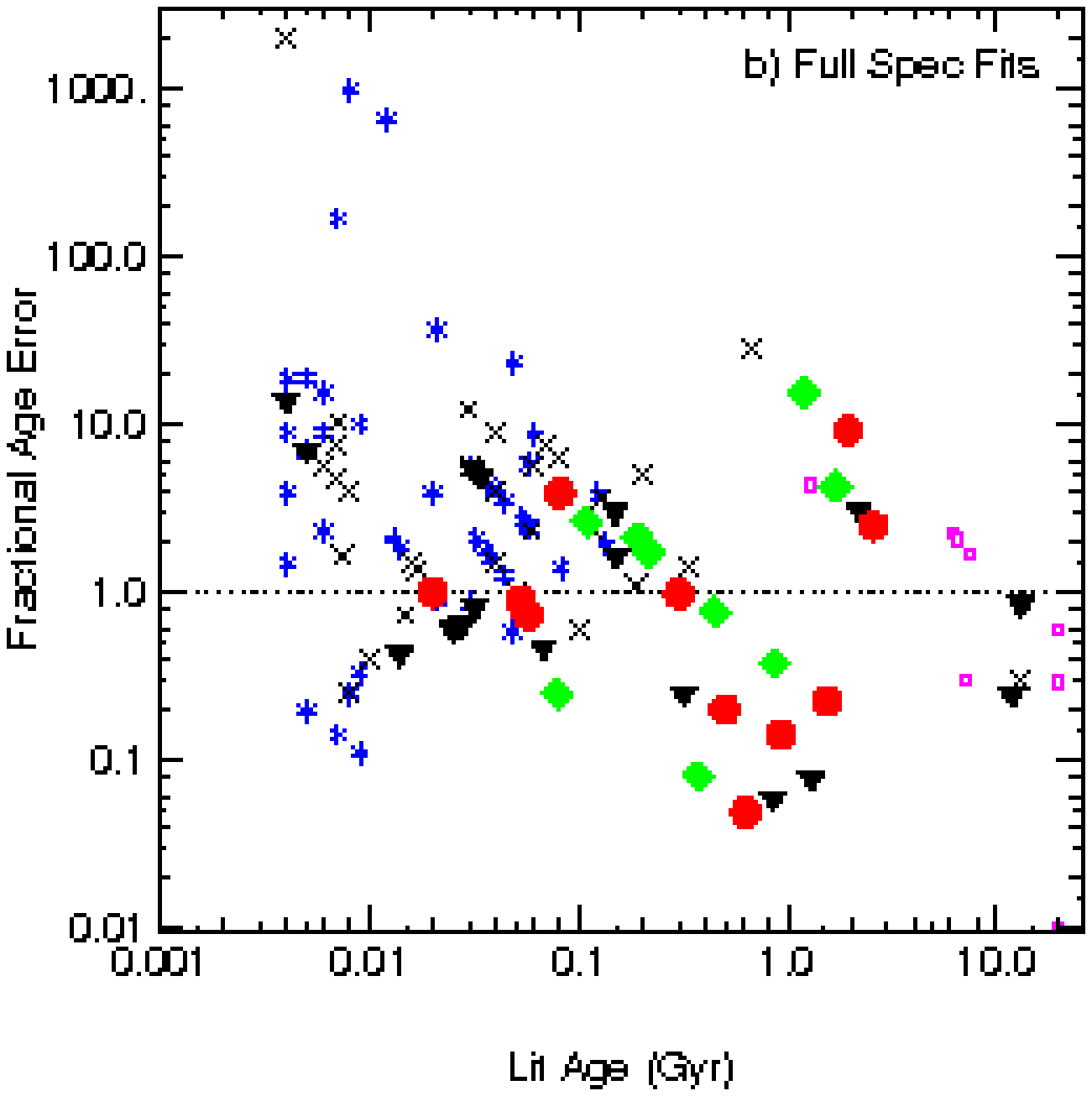}
    \plottwo{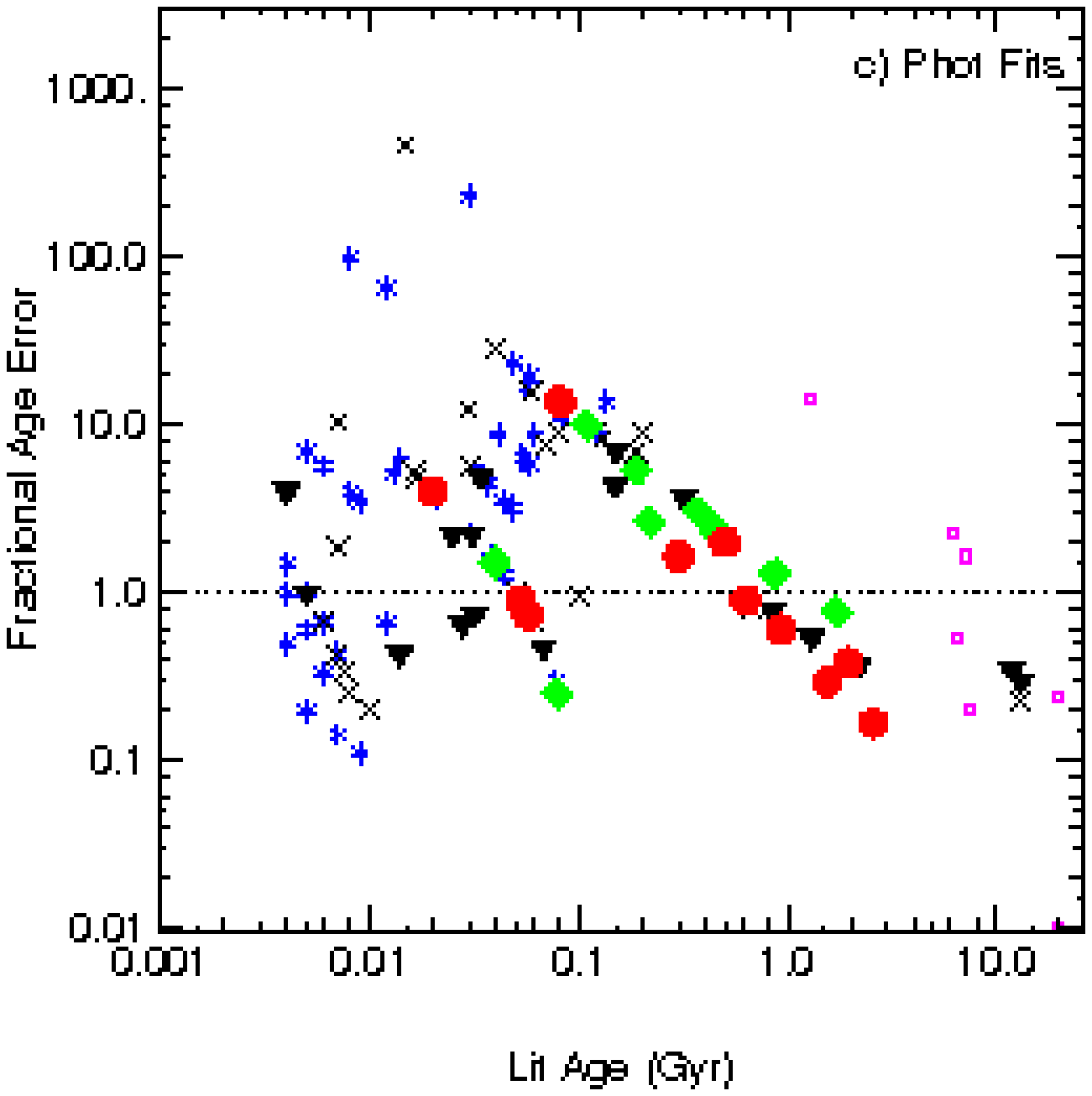}{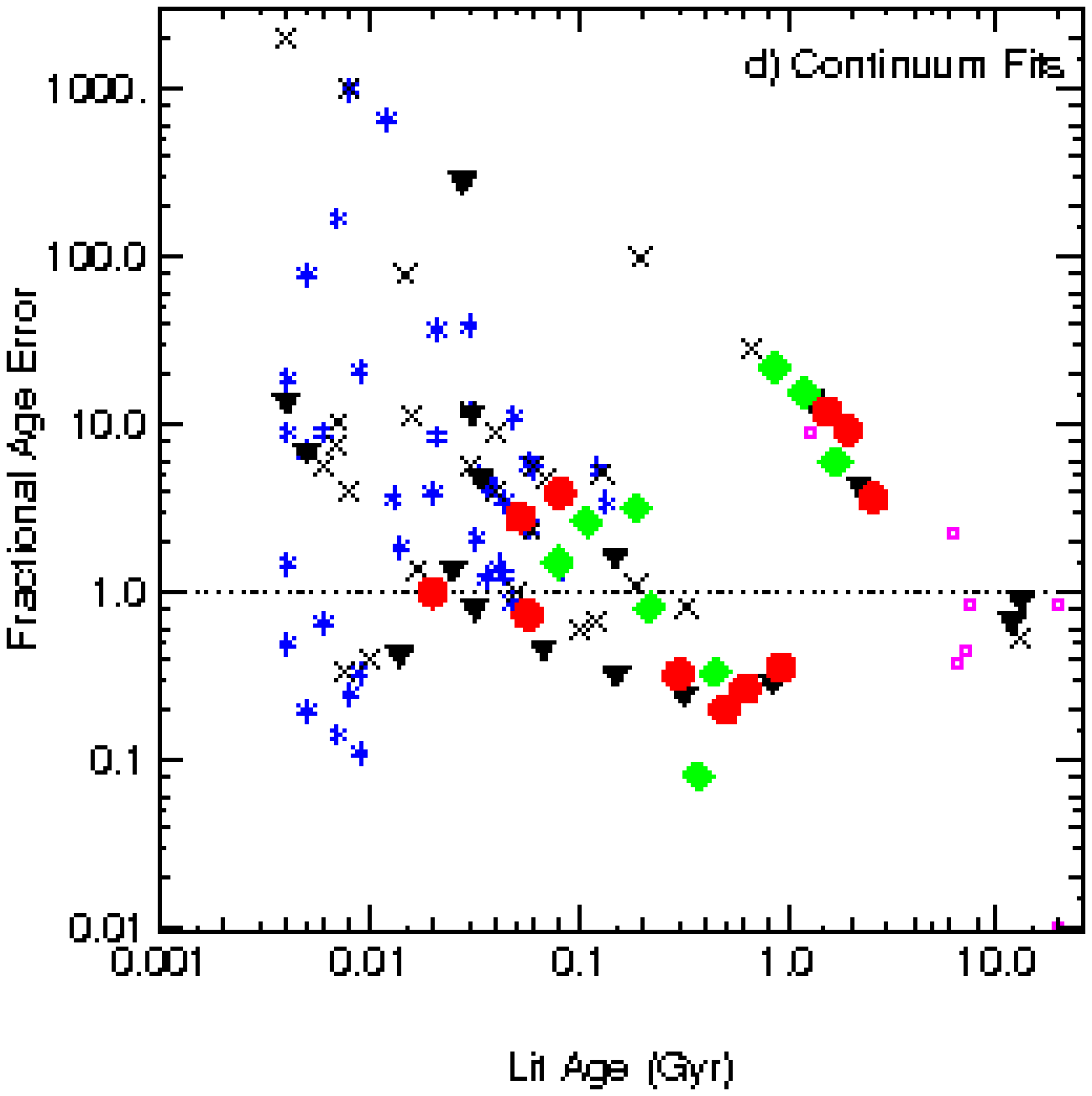}
  \caption{ Absolute values of fractional age errors relative to
  literature values for CN spectrum, full spectrum, photometry, and
  continuum fits. The dotted lines mark the 100\% error point. Symbols
  are the same as in Figure~\ref{fig5} with literature ages from
  main sequence turnoffs by \citet{hodge83} (\textit{black crosses}),
  UBV photometry by \citet{santos95} (\textit{blue asterisks}) and
  \citet{girardi95,girardi98} (\textit{green diamonds}), photmetric
  fits to BC96 models by \citet{jiang03} (\textit{magenta squares}),
  line index ratios by \citet{leonardi03} (\textit{red circles}), and
  literature ages on a homogeneous scale by \citet{santos04}
  (\textit{black inverted triangles}).
  \label{fig6}}
\end{figure}

\begin{figure}
\figurenum{7}
    \includegraphics[scale=.35,angle=-90]{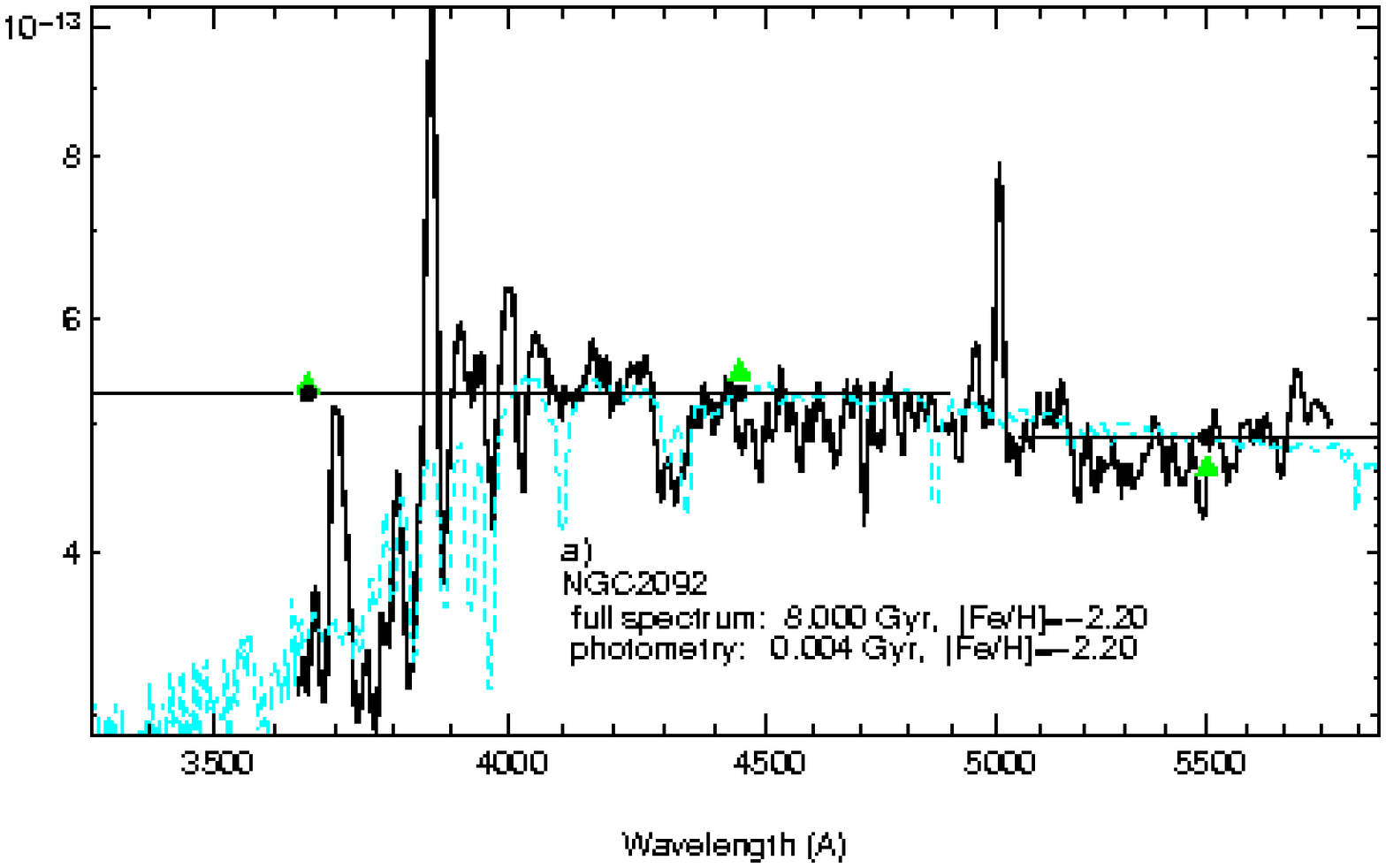}
    \includegraphics[scale=.35,angle=-90]{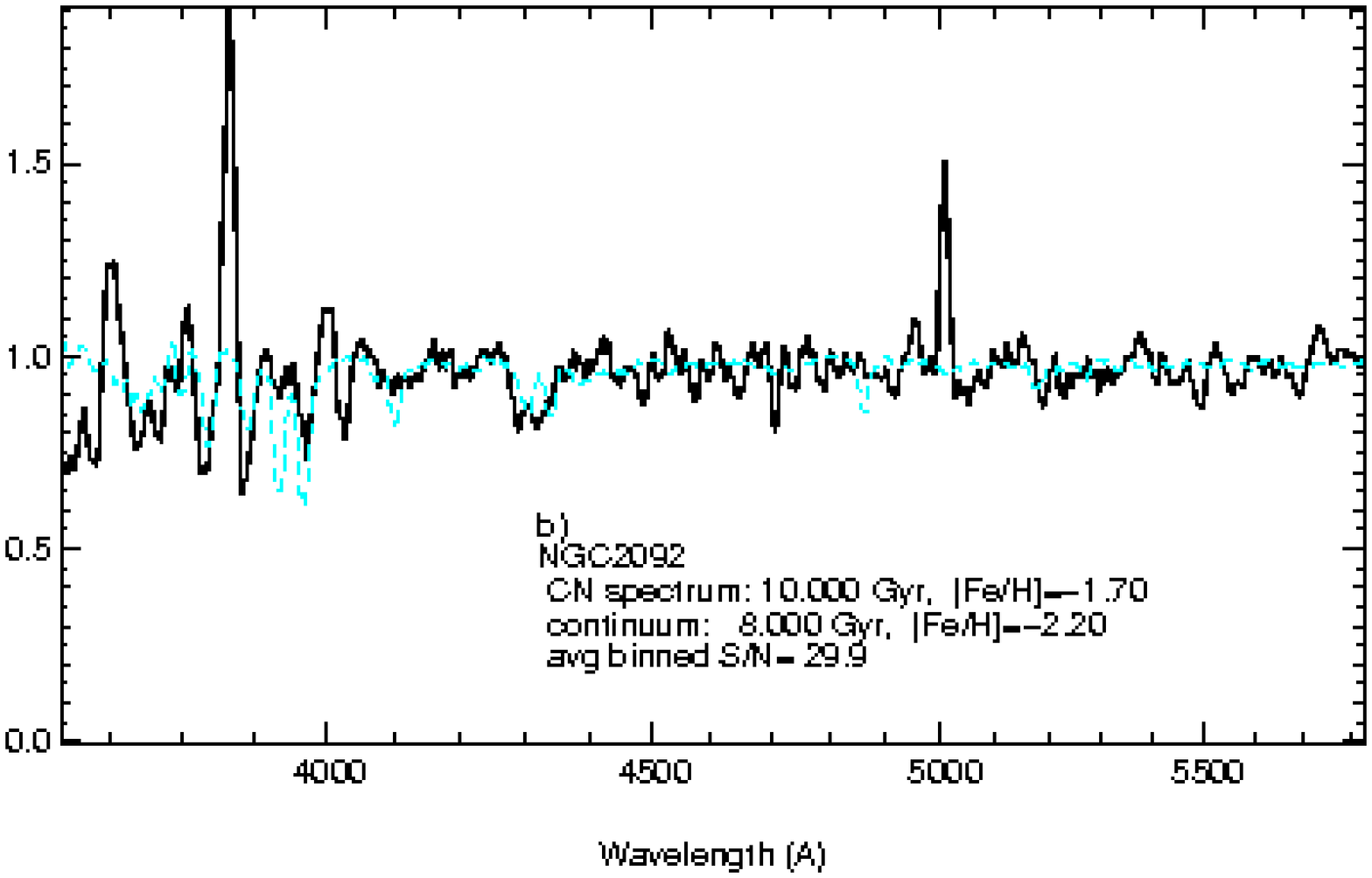}
    \includegraphics[scale=.35,angle=-90]{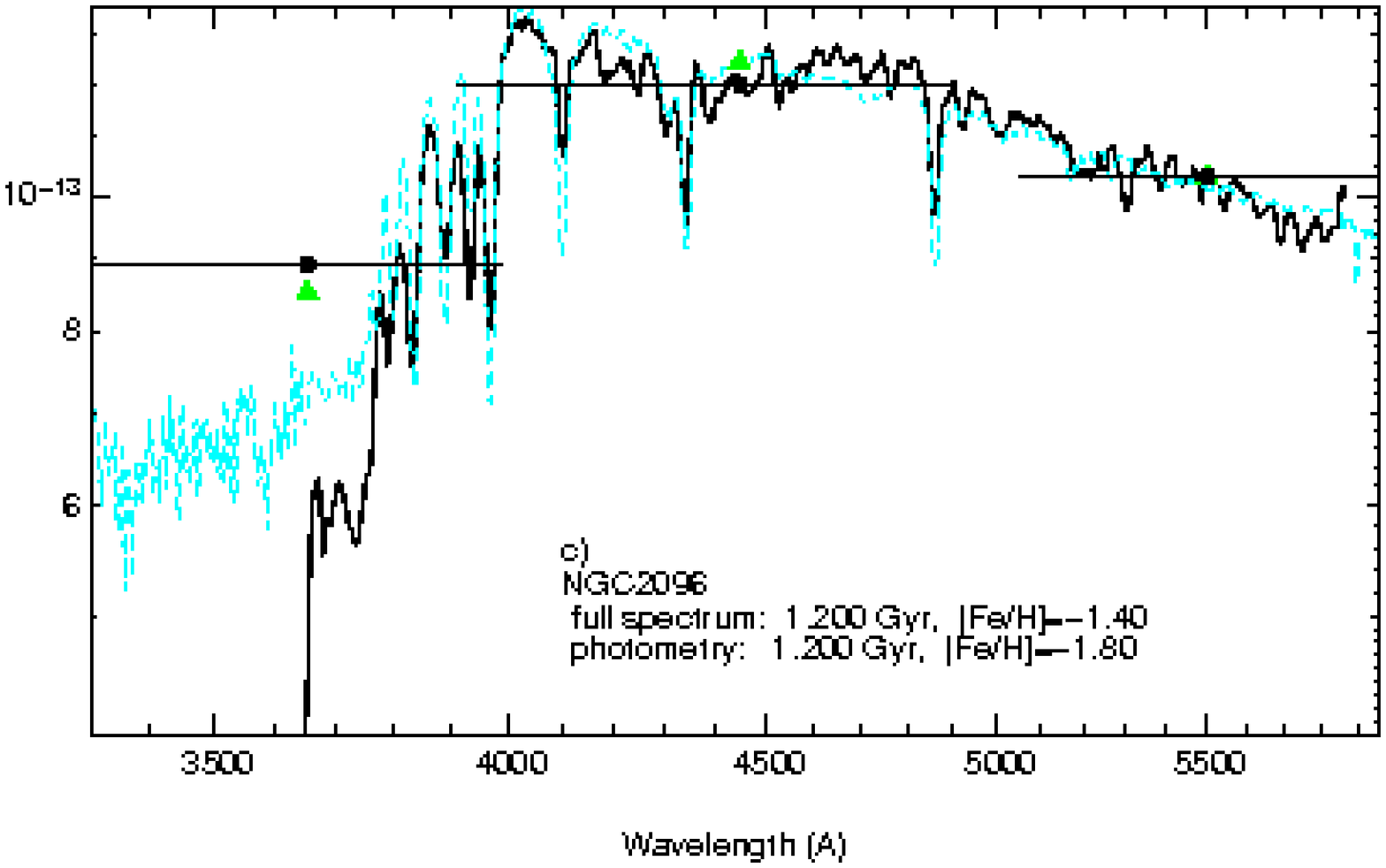}
    \includegraphics[scale=.35,angle=-90]{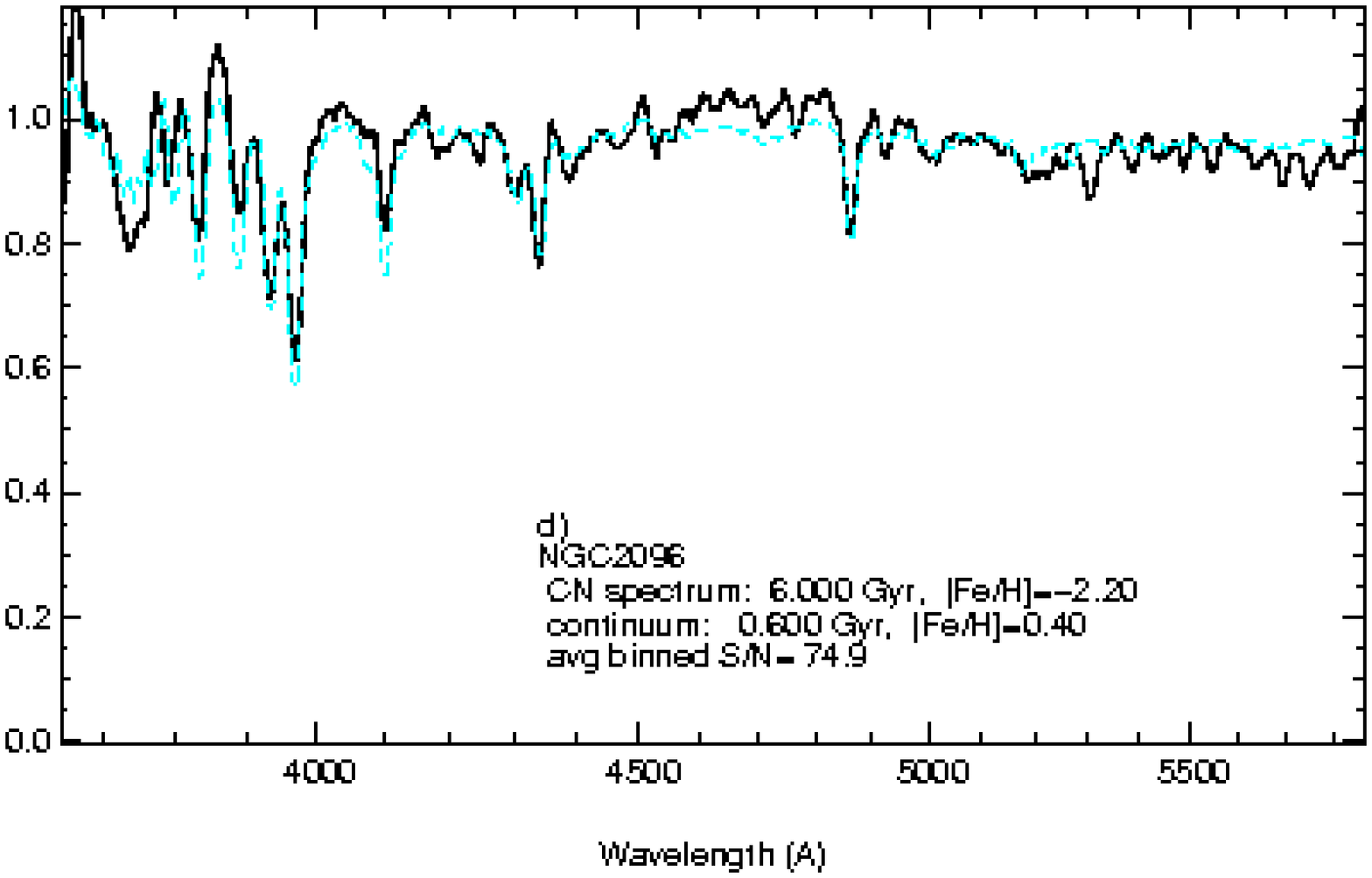}
\caption{Spectra of extreme outliers in the age correlation plot for
  CN spectrum fits in Figure~\ref{fig5}. Here, a\&b) are spectra of
  NGC~2092 and c\&d) are spectra of NGC~2096. a\&c) are the full
  spectrum fits to show the original spectral shapes, and b\&d) are
  the CN spectrum fits. After masking emission lines, NGC~2092 does
  not have enough distinguishing spectral features to obtain a good
  fit. NGC~2096 appears older than its literature age of 49 Myr from
  its 4000~\AA~break strength of D$_{n}$(4000)=1.28. These clusters
  are both known to contain red supergiants and NGC 2092 is embedded
  in a nebula. \label{fig7}}
\end{figure}

\begin{figure}
\figurenum{8}
    \includegraphics[scale=.38]{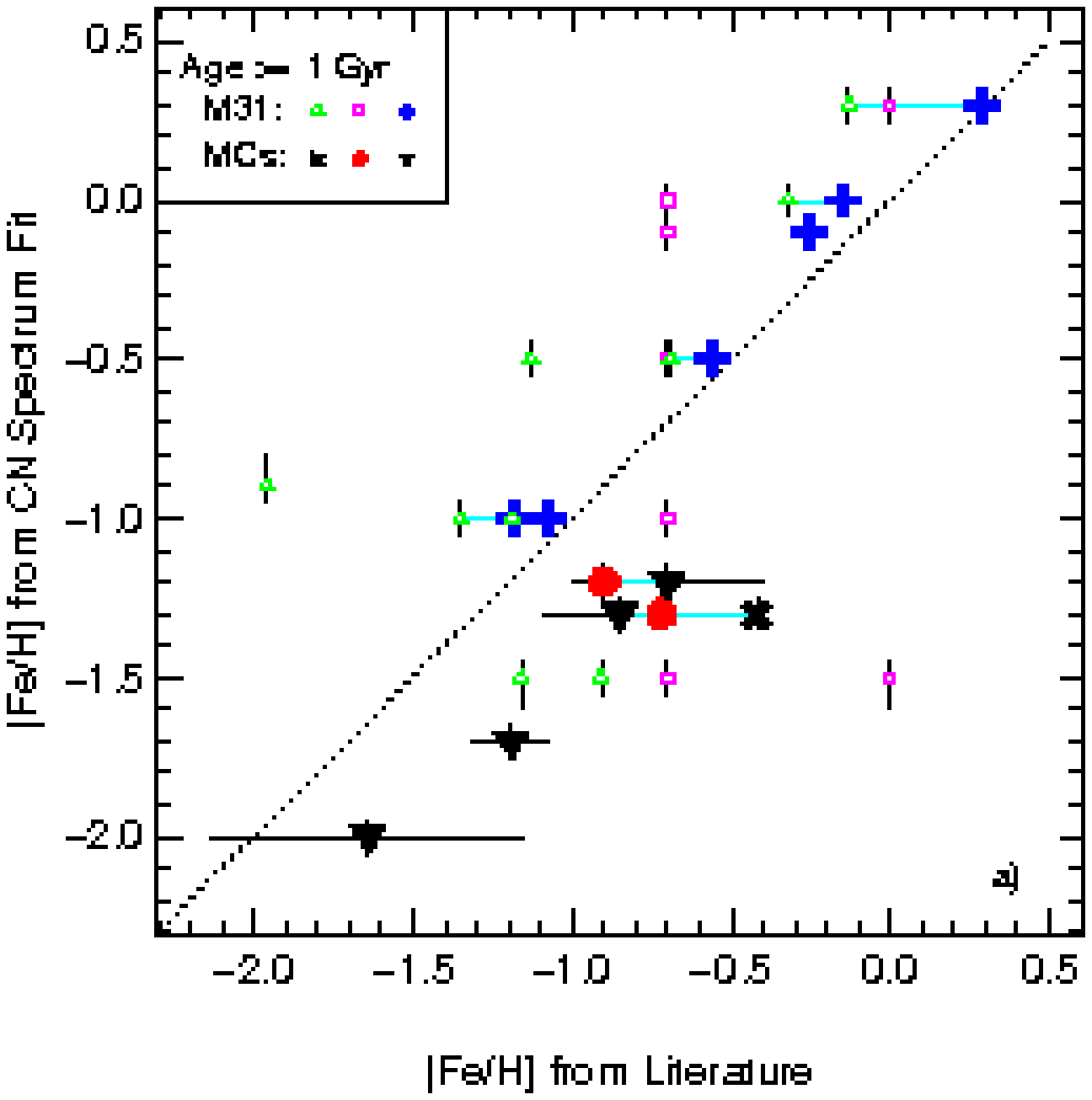}
    \includegraphics[scale=.38]{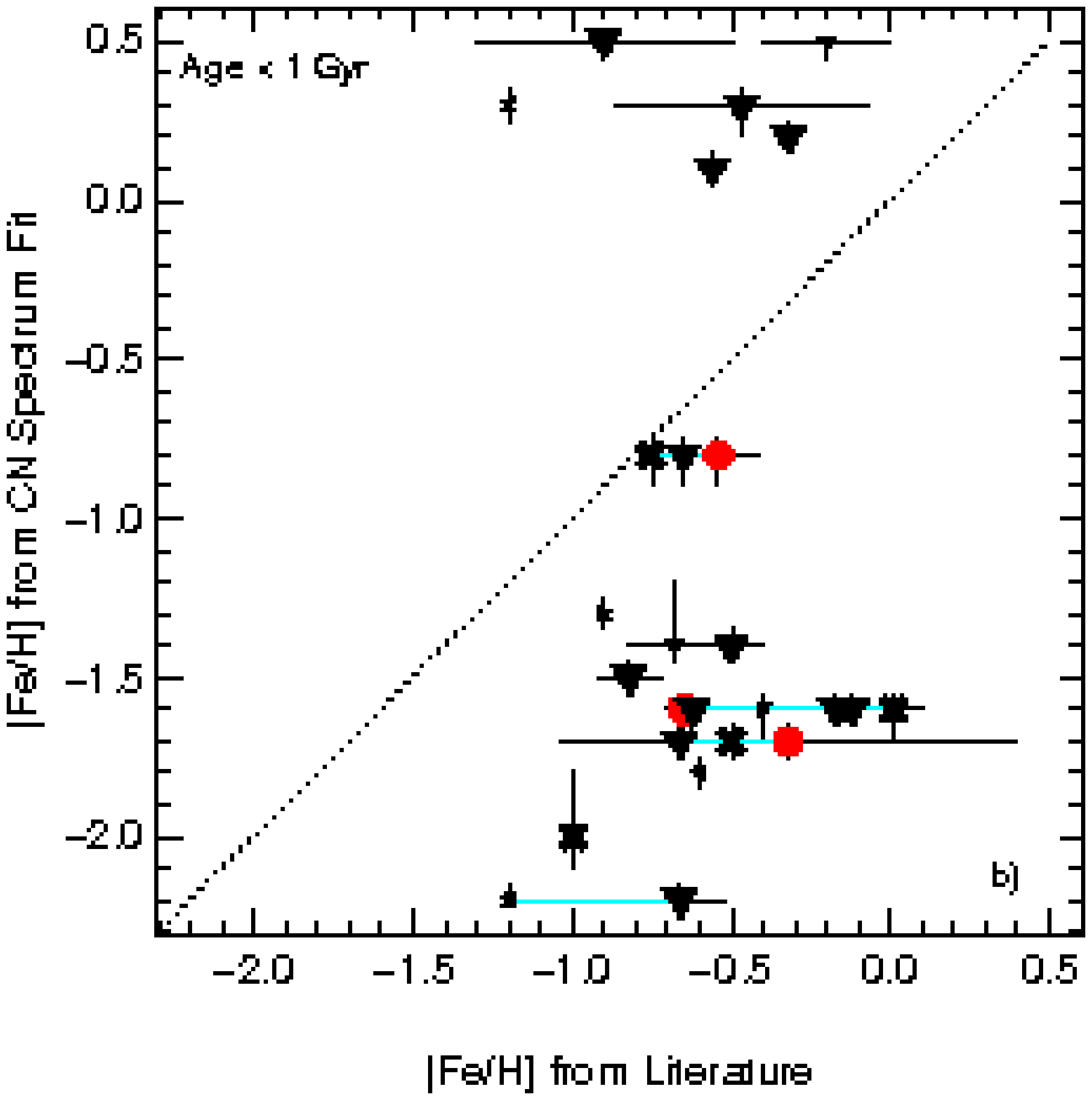}
    \includegraphics[scale=.38]{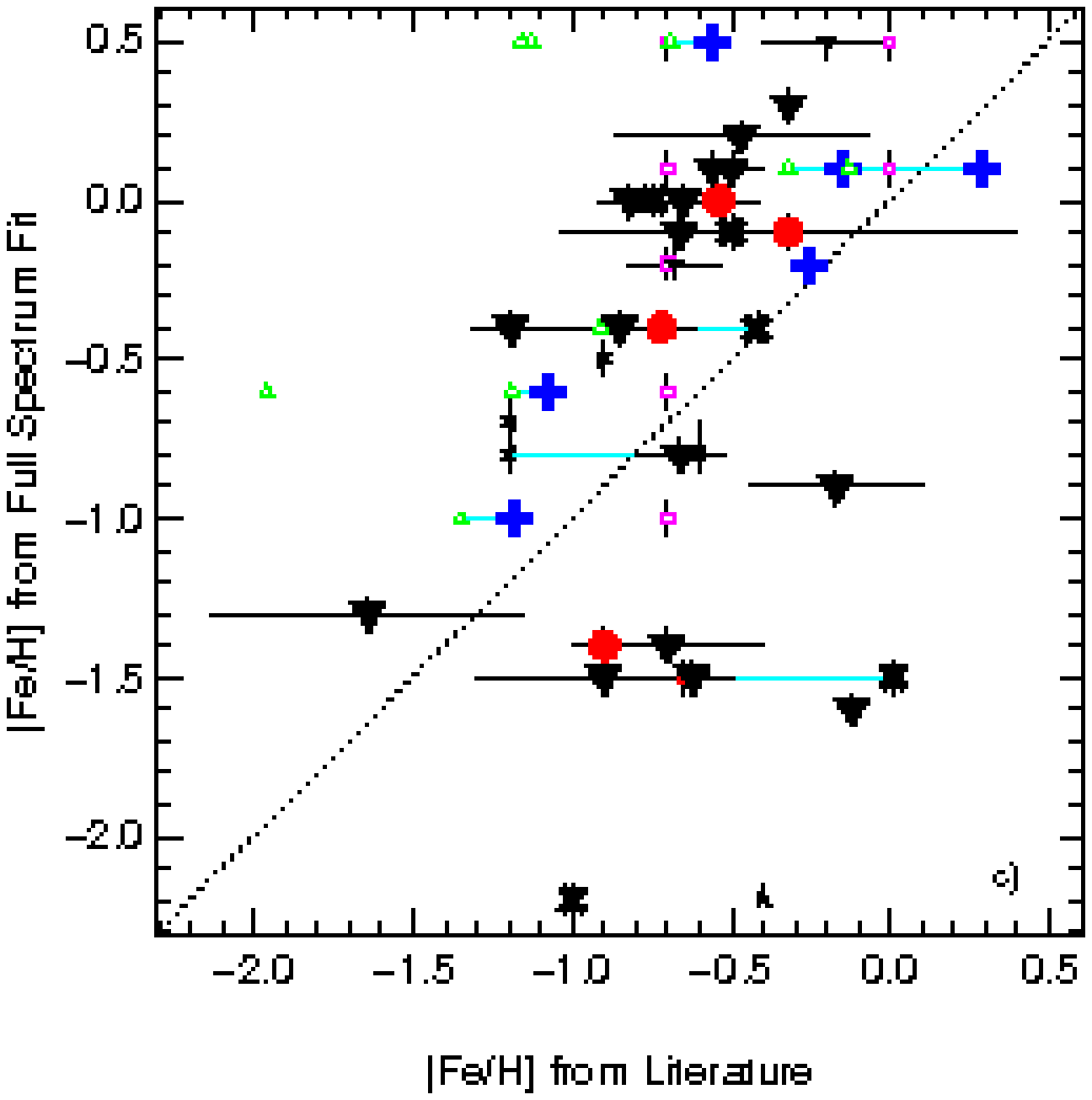}
    \includegraphics[scale=.38]{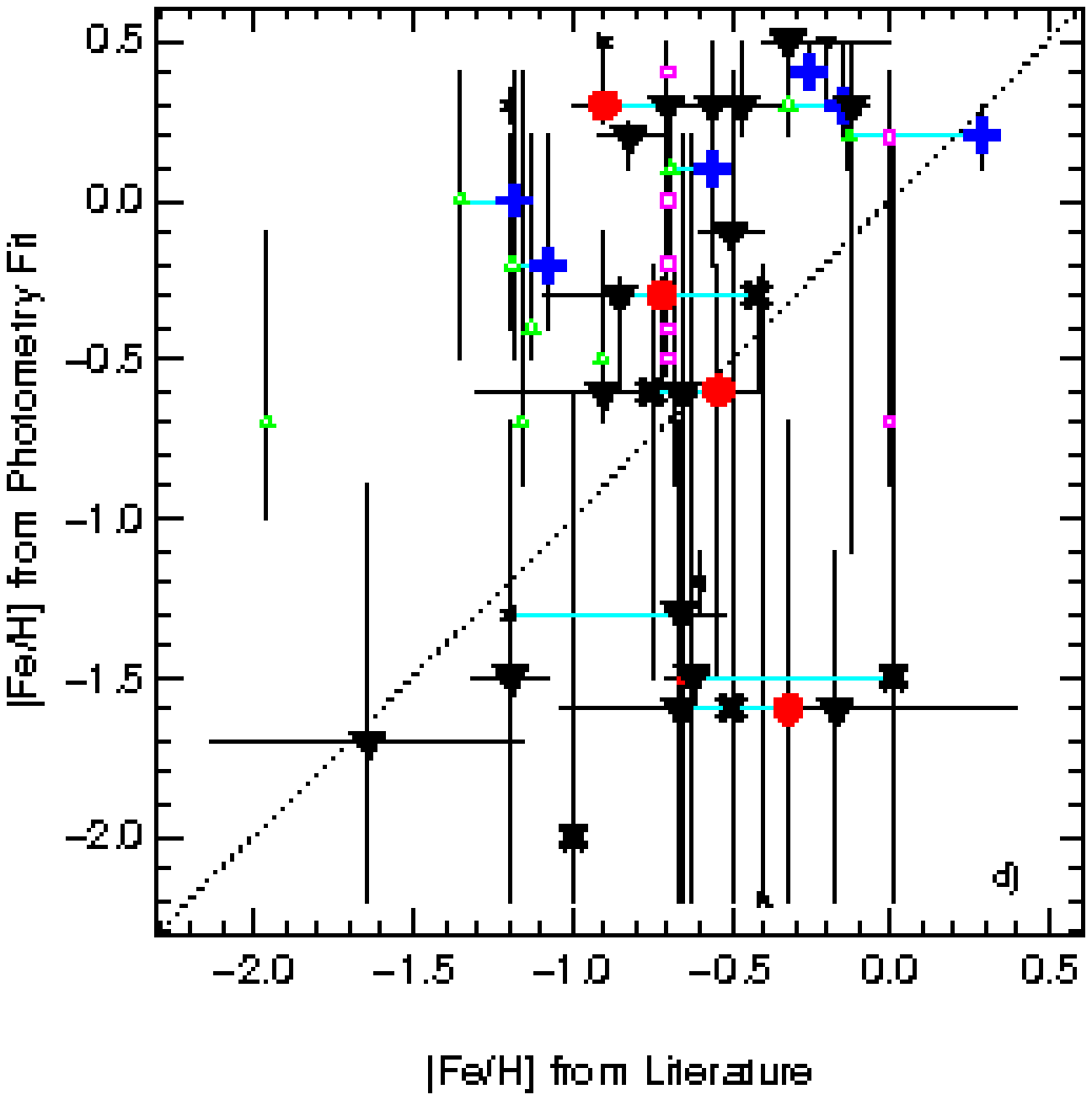}
    \includegraphics[scale=.38]{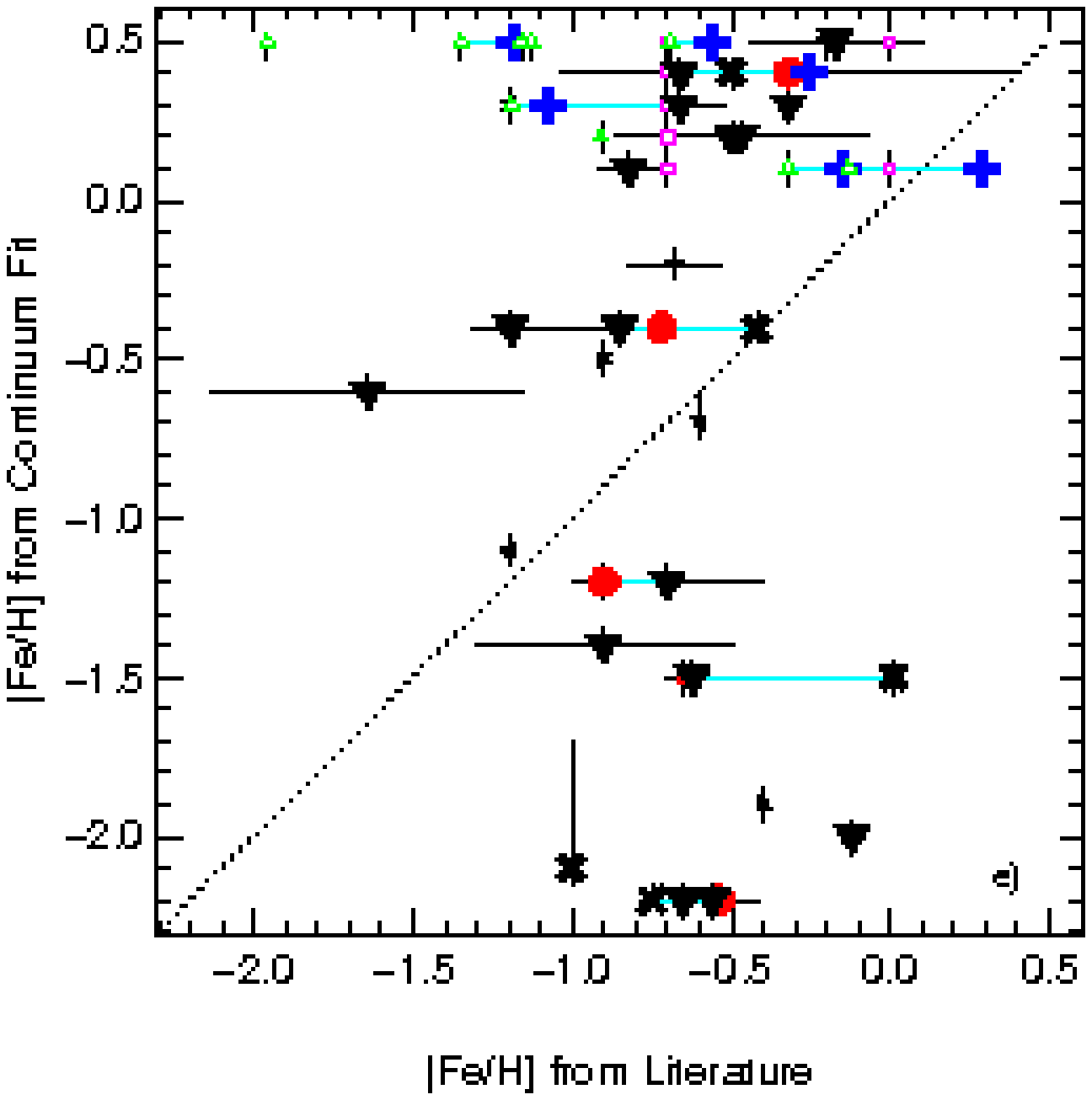}
  \caption{Cluster [Fe/H] derived from our different types of model
  fits compared to the literature. a\&b) are from the CN spectrum fits
  for old and young clusters separately, c) is from the full spectrum
  fits, d) is from the photometry fits, and e) is from the continuum
  fits. Literature values are from comparison values in
  \citet{leonardi03} (which come from \citet{olszewski91},
  \citet{cohen82}, \citet{seggewiss89}, \citet{sagar89}, and
  \citet{piatti02}) in the MC (\textit{black crosses}), line indices
  by \citet{huchra91} in M31 (\textit{blue pluses}), line index ratios
  by \citet{leonardi03} in the MC (\textit{red circles}), VJK colors
  by \citet{cohen94} in M31 (\textit{green triangles}), photmetric
  fits to BC96 models by \citet{jiang03} in M31 (\textit{magenta
  squares}), and literature [Fe/H] on a homogeneous scale by
  \citet{santos04} in the MC (\textit{black inverted
  triangles}). Large symbols denote literature values based on spectra
  and small symbols are values based on colors. The thick horizontal
  cyan lines connect points that are the same cluster with multiple
  literature metallicities. Thin horizontal lines are error bars on
  literature [Fe/H], thin vertical lines are 1$\sigma$ error bars on
  our derived values. The best agreement with literature metallicity
  occurs for spectra-based values of clusters older than 1 Gyr from
  our CN spectrum fits.
  \label{fig8}}
\end{figure}

\begin{figure}
\figurenum{9}
    \plottwo{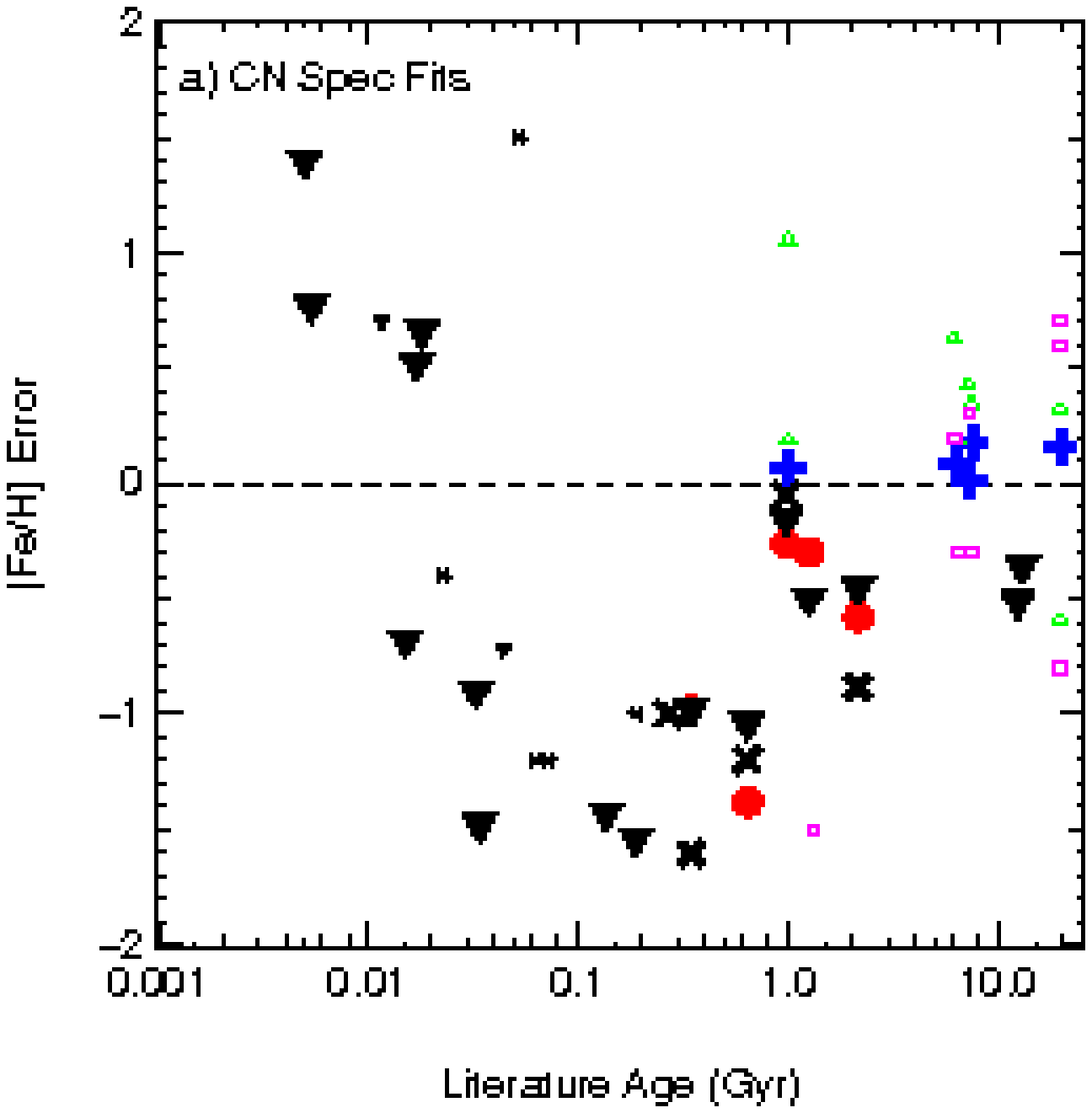}{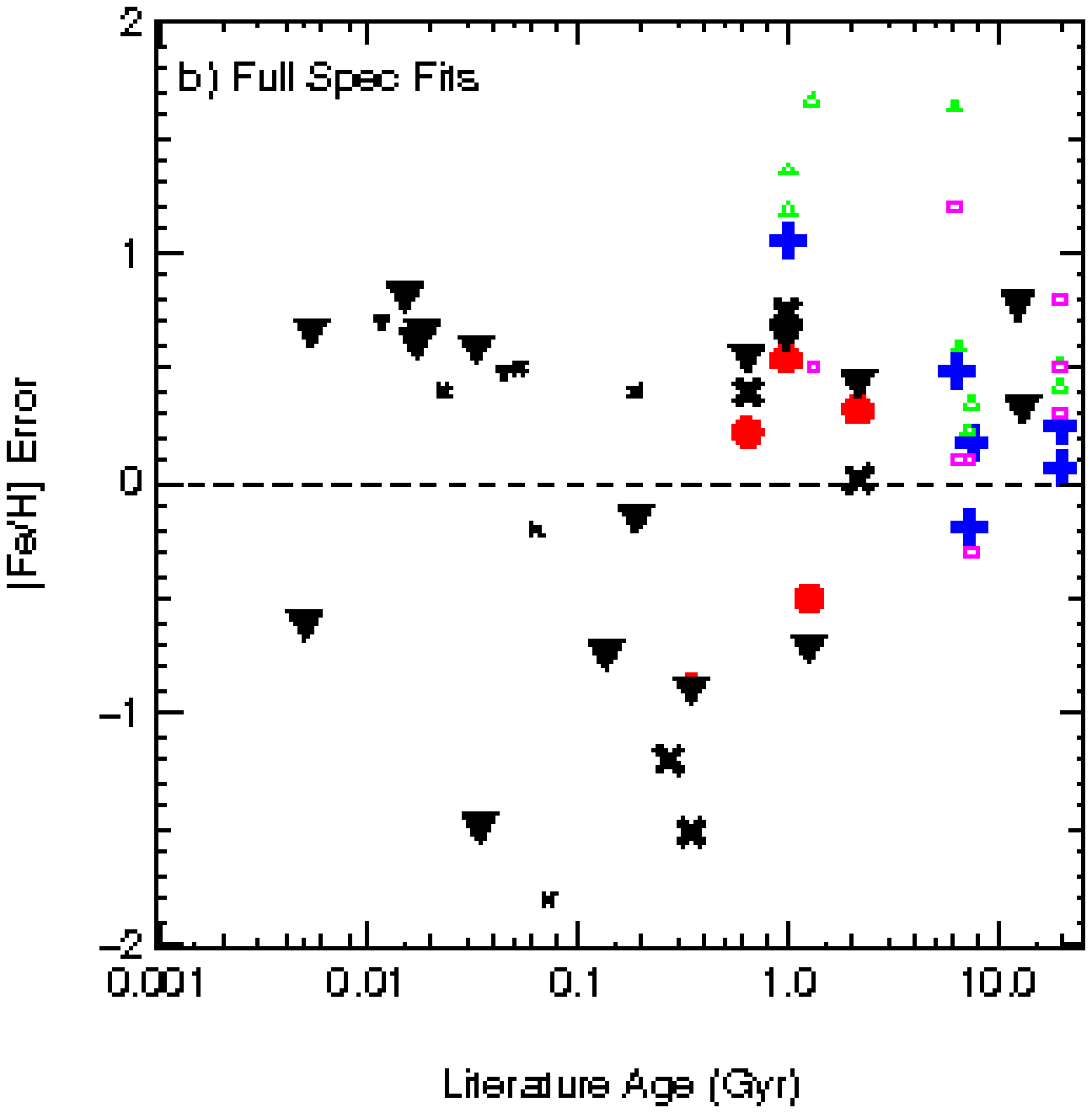}
    \plottwo{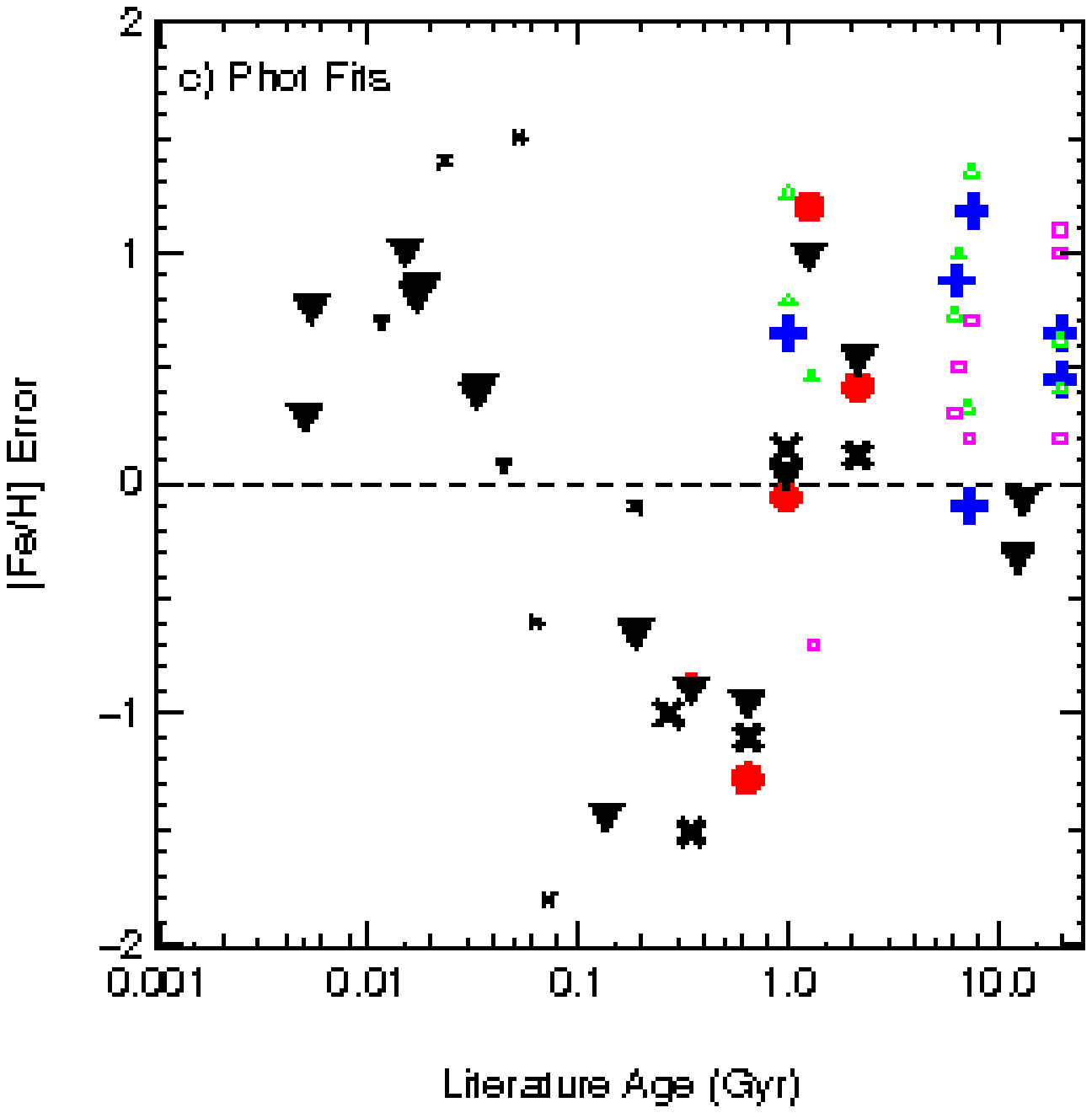}{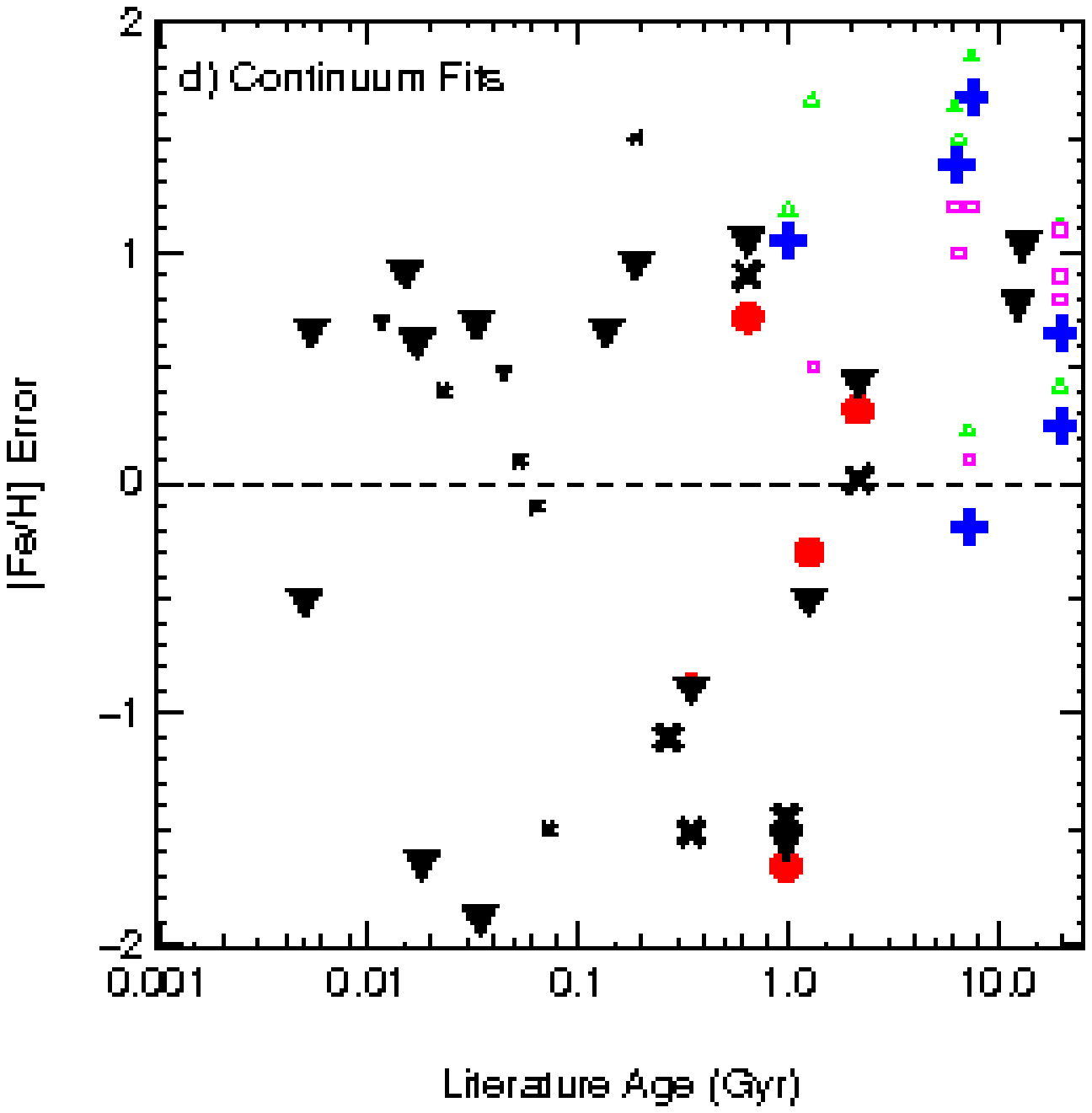}
  \caption{ [Fe/H] errors relative to literature values (fit -
  literature) for a) CN spectrum, b) full spectrum, c) photometry, and
  d) continuum fits, plotted against average literature ages. Symbols
  are the same as in Figure~\ref{fig8} with [Fe/H] from comparison
  values in \citet{leonardi03} in the MC (\textit{black crosses}),
  line indices by \citet{huchra91} in M31 (\textit{blue pluses}), line
  index ratios by \citet{leonardi03} in the MC (\textit{red circles}),
  VJK colors by \citet{cohen94} in M31 (\textit{green triangles}),
  photmetric fits to BC96 models by \citet{jiang03} in M31
  (\textit{magenta squares}), and literature [Fe/H] on a homogeneous
  scale by \citet{santos04} in the MC (\textit{black inverted
  triangles}). Large symbols denote literature [Fe/H] derived from
  spectra and small symbols are from colors. The dependence of [Fe/H]
  errors on cluster age can more easily be seen here for all types of
  fits. \label{fig9}}
\end{figure}

\begin{figure}
\figurenum{10}
    \centering \includegraphics[scale=.37]{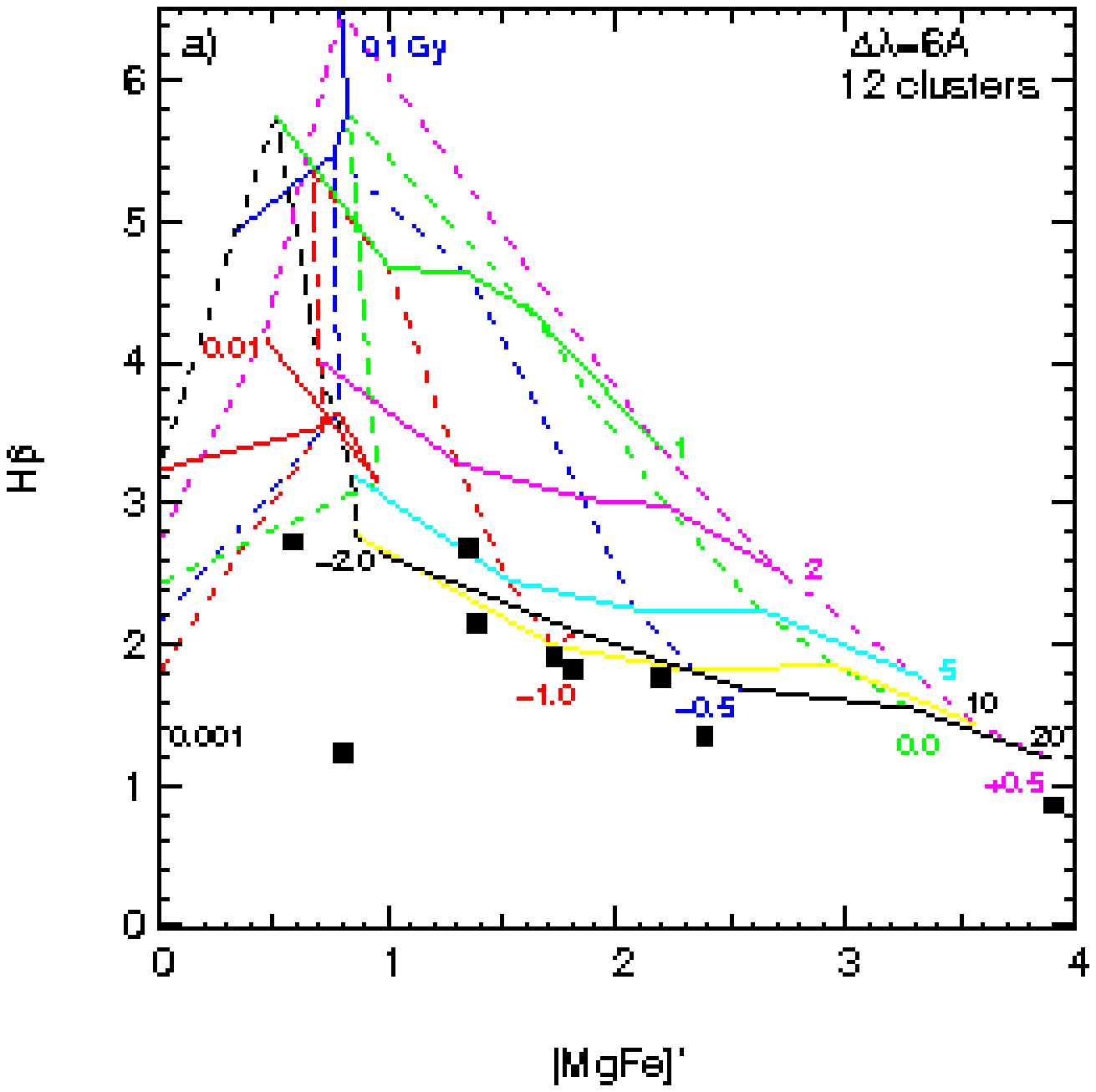}
    \centering \includegraphics[scale=.37]{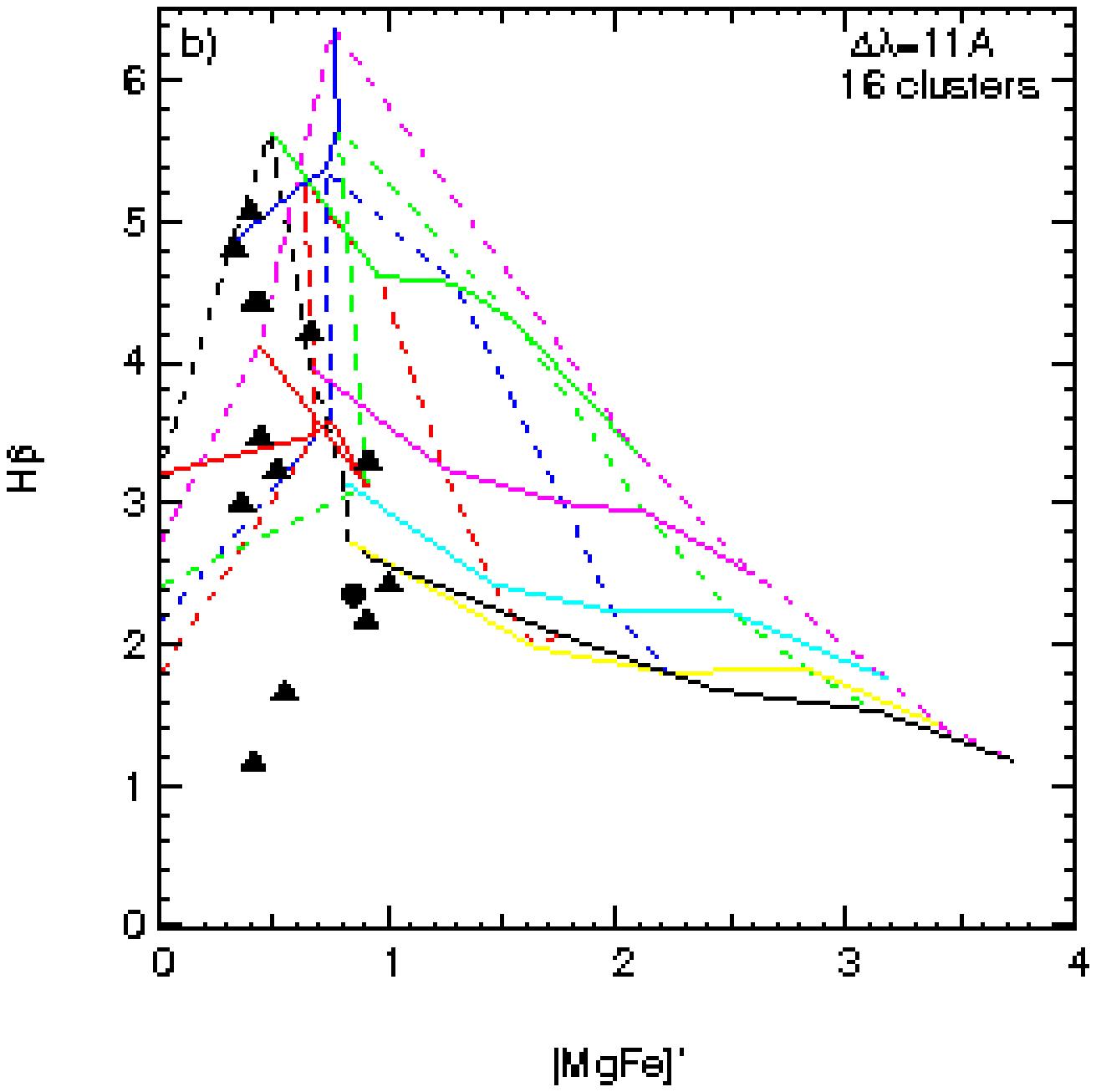}
    \centering \includegraphics[scale=.37]{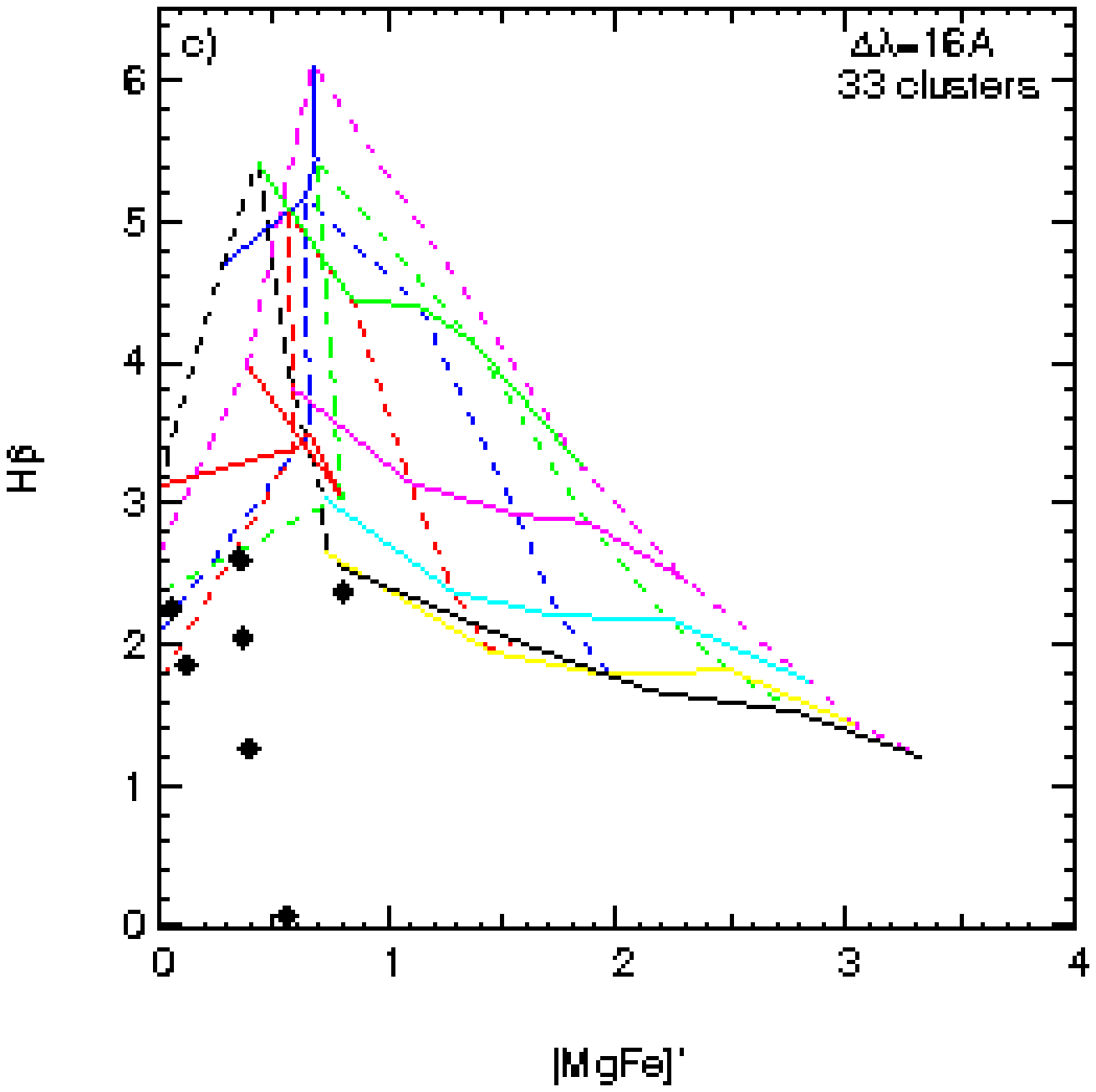}
    \centering \includegraphics[scale=.35]{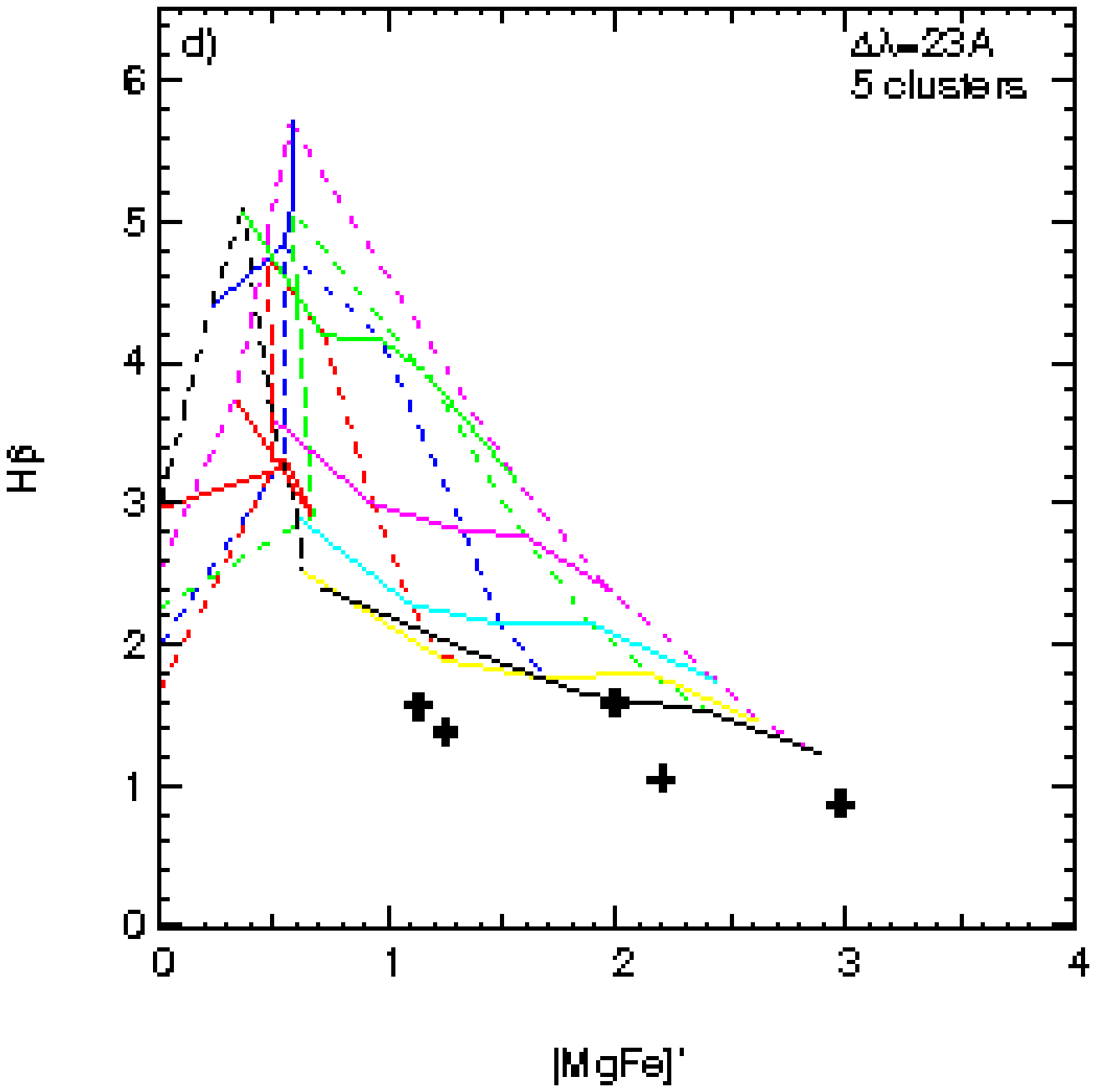}
    \centering \includegraphics[scale=.35]{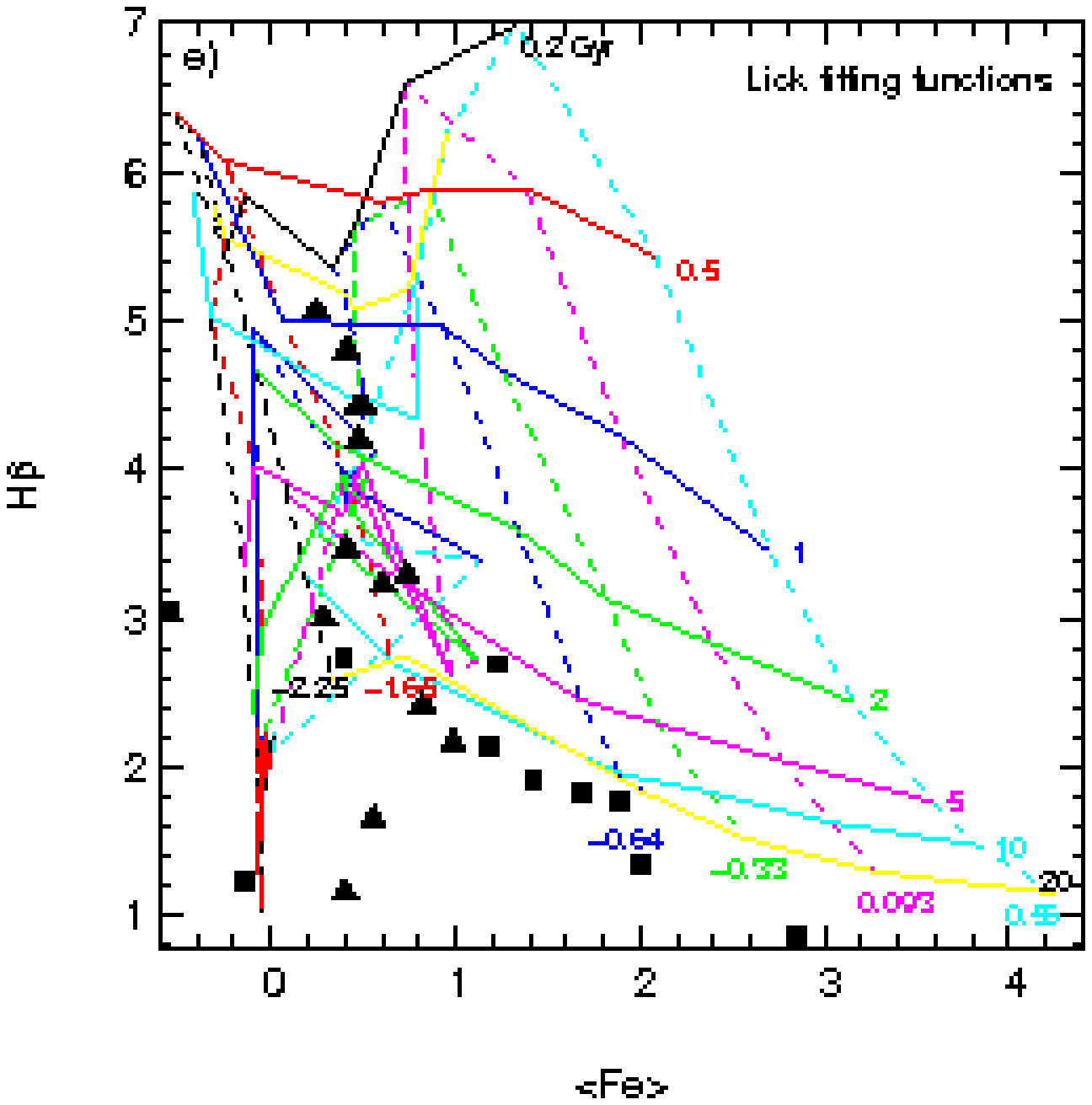}
    \centering \includegraphics[scale=.35]{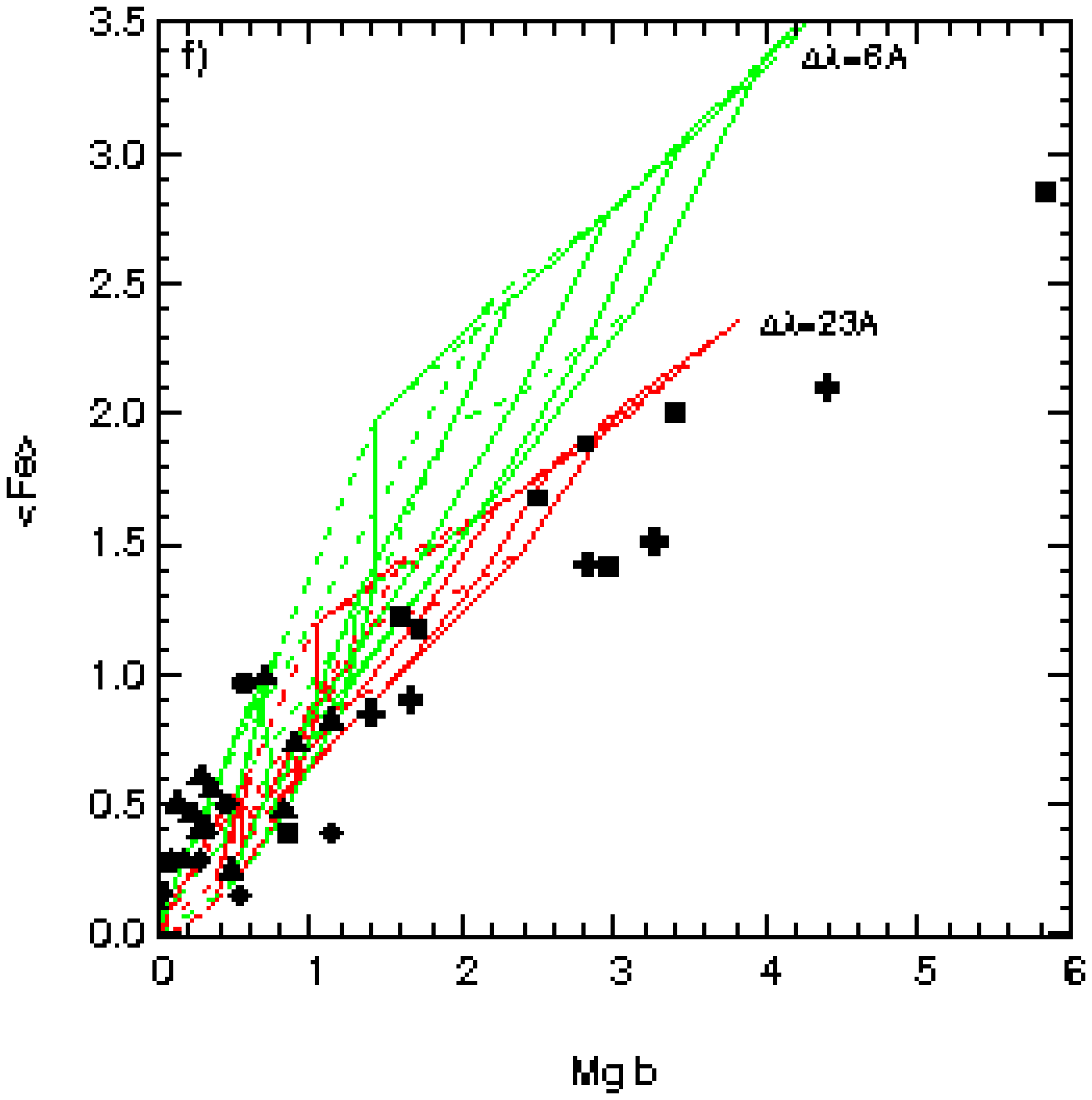}
  \caption{Line indices calculated from BC03 models. a-d)
  H$\beta$-[MgFe]$^\prime$ model indices calculated at different
  spectral resolutions. Grid age lines (\textit{solid}) from bottom to
  top are 20, 10, 5, 2, 1, 0.1, 0.01, and 0.001 Gyr, and [Fe/H] lines
  (\textit{dashed}) from right to left are +0.5, 0.0, -0.5, -1.0, and
  -2.0 (these are labeled in panel a). Note that the model indices
  become degenerate at $t_{age} \lesssim$ 100 Myr for -1.0 $<$ [Fe/H]
  $<$ +0.5, and at $t_{age} \lesssim$ 1 Gyr for [Fe/H] $<$
  -2.0. Cluster indices are plotted as symbols, differientiated by
  spectral resolution: 6~\AA~(\textit{squares}, M31),
  11~\AA~(\textit{triangles}, MC), 12~\AA~(\textit{circles}, MC),
  16~\AA~(\textit{diamonds}, MC), and 23~\AA~(\textit{crosses},
  M31). e) contains grids of the traditional Lick indices provided
  with the BC03 model package. They suffer from similar degeneracy at
  low ages and metallicities. f) shows an indication of [Mg/Fe]
  enhancement in some of the M31 clusters. Model grids for indices at
  the minimum and maximum spectral resolutions are
  included. \label{fig10}}
\end{figure}

\begin{figure}
\figurenum{11}
    \plottwo{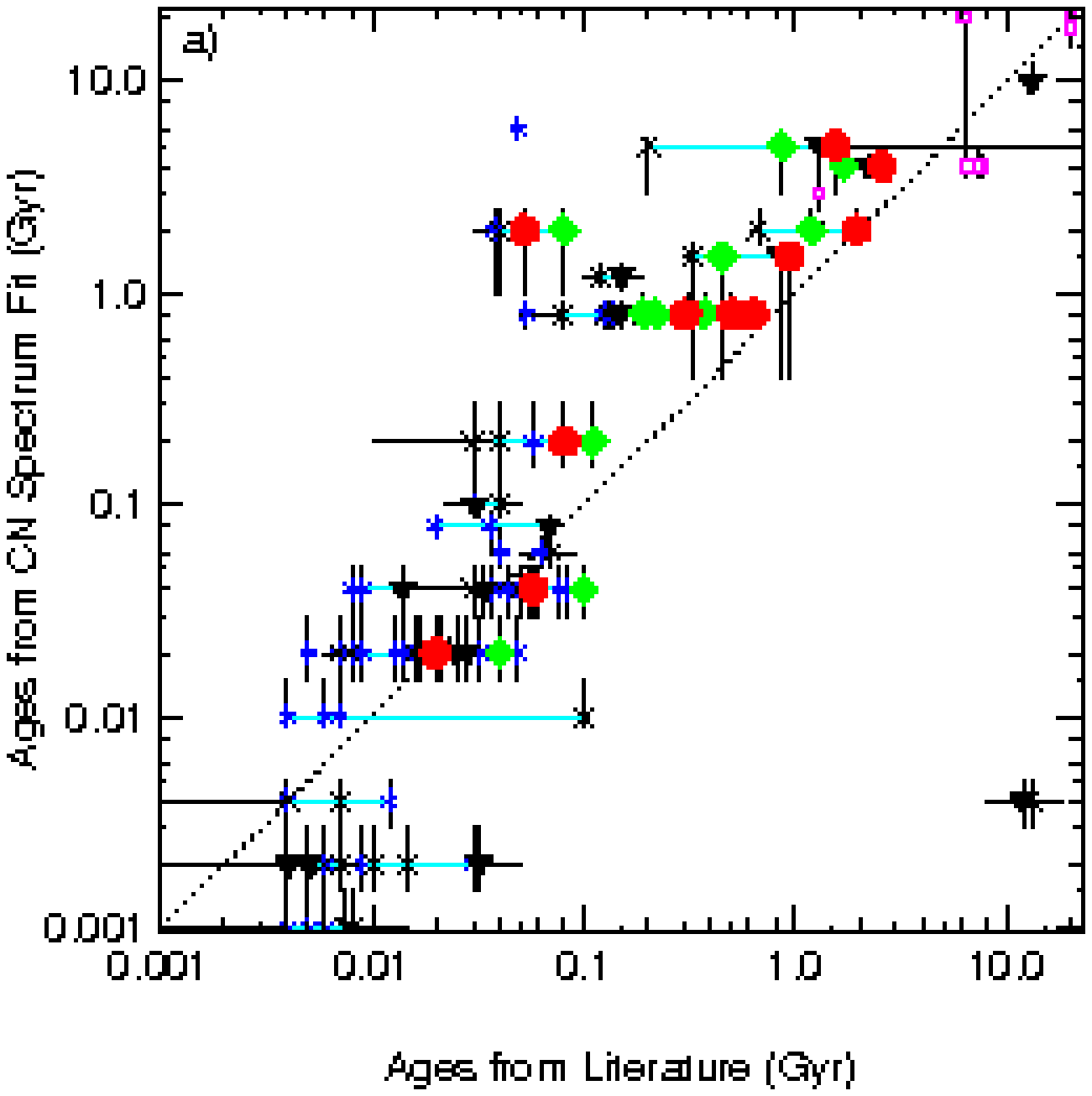}{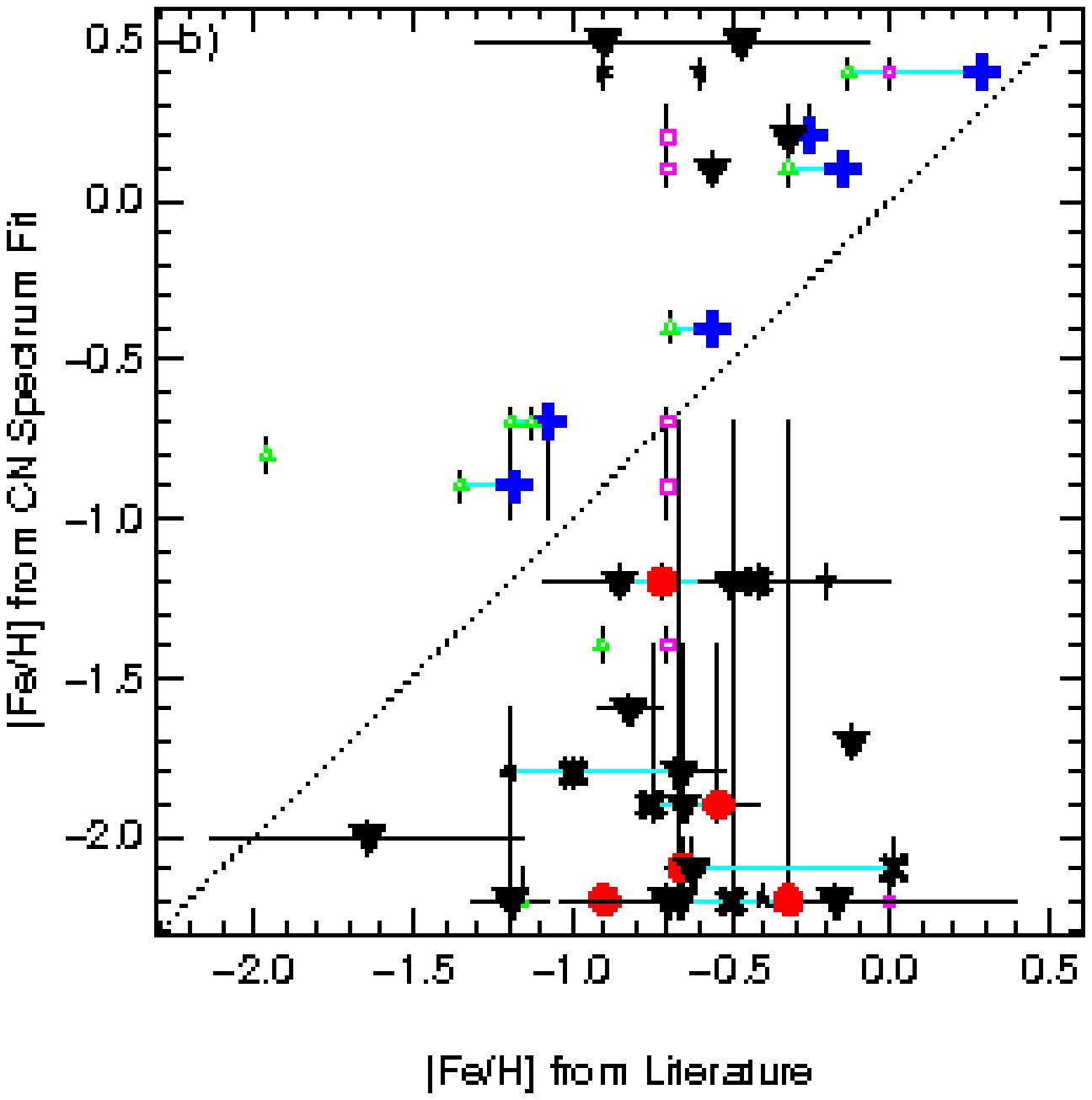}
    \plottwo{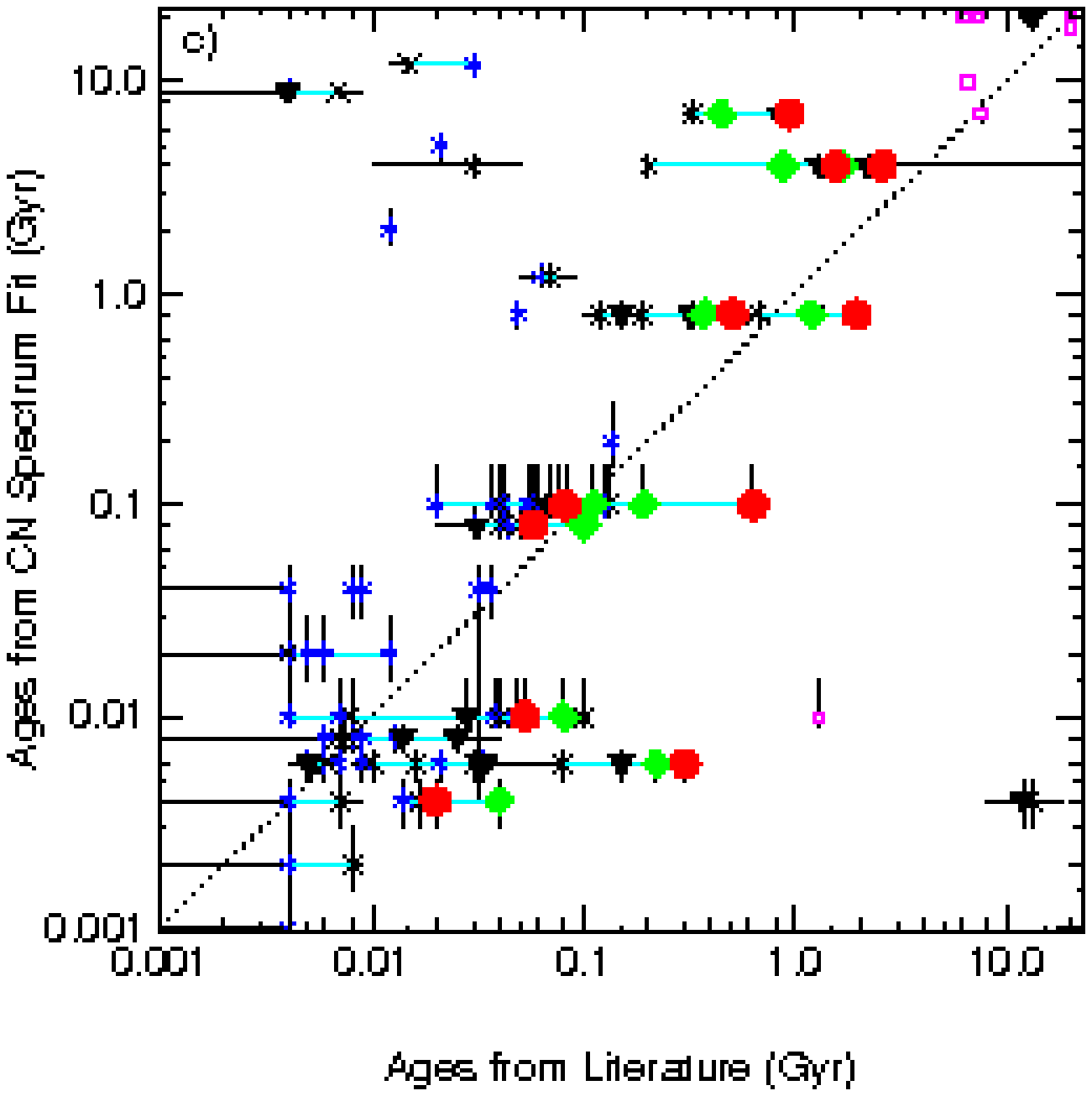}{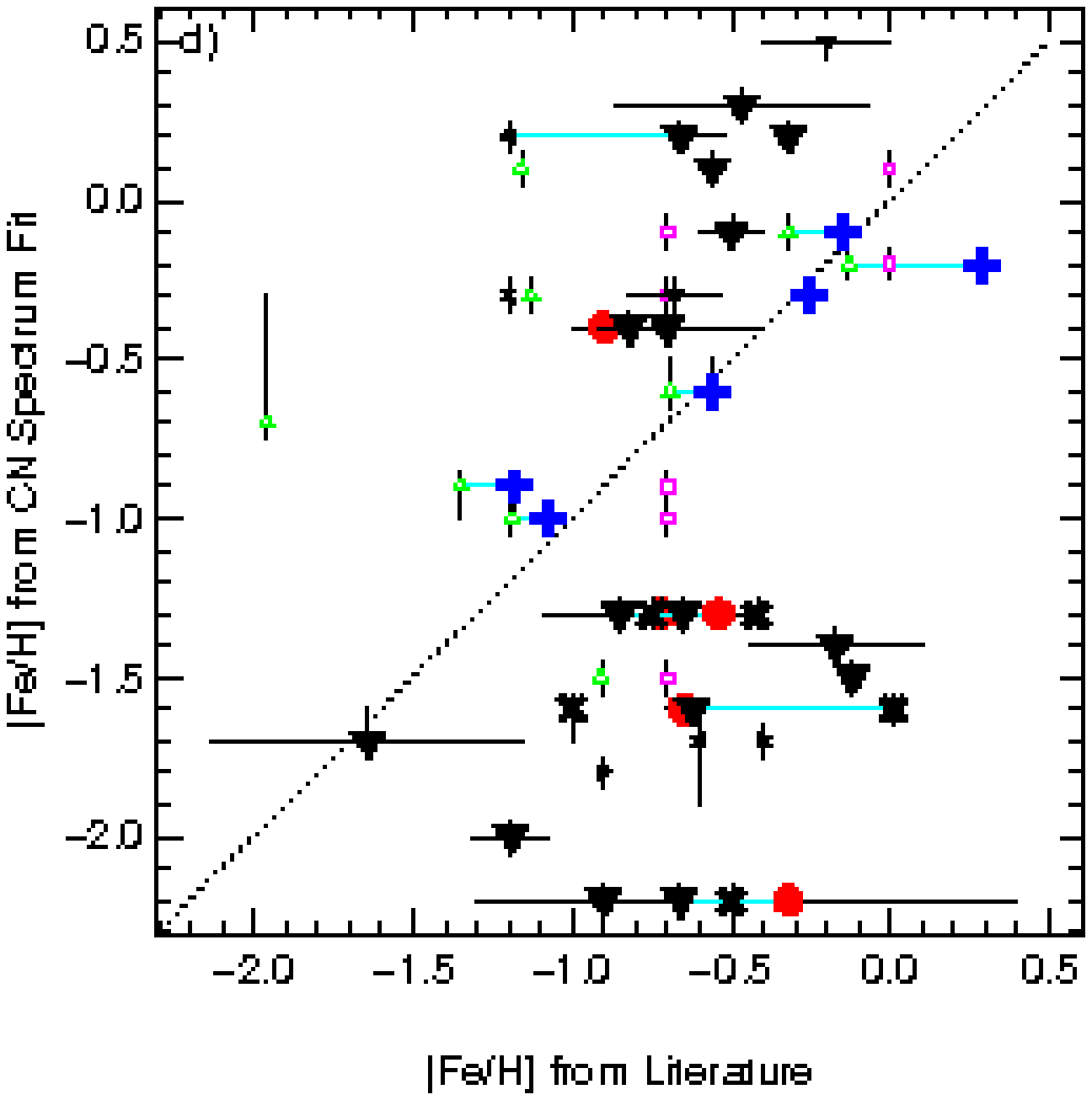}
  \caption{Ages and metallicities from CN spectrum fits to only the
  regions around typically used line indices and spectral features
  (a\&b) and to the regions excluding these features (c\&d). The
  features used were $D_n$(4000), H$\delta$, G4300, H$\gamma$,
  H$\beta$, Mg \textit{b}, Fe5270, and Fe5335. Refer to the captions
  on Figure~\ref{fig5} and \ref{fig8} for symbol
  description. \label{fig11}}
\end{figure}

\begin{figure}
\figurenum{12}
%    \plottwo{f12a.ps}{./fig_orig/f12b.ps}
    \includegraphics[scale=.55]{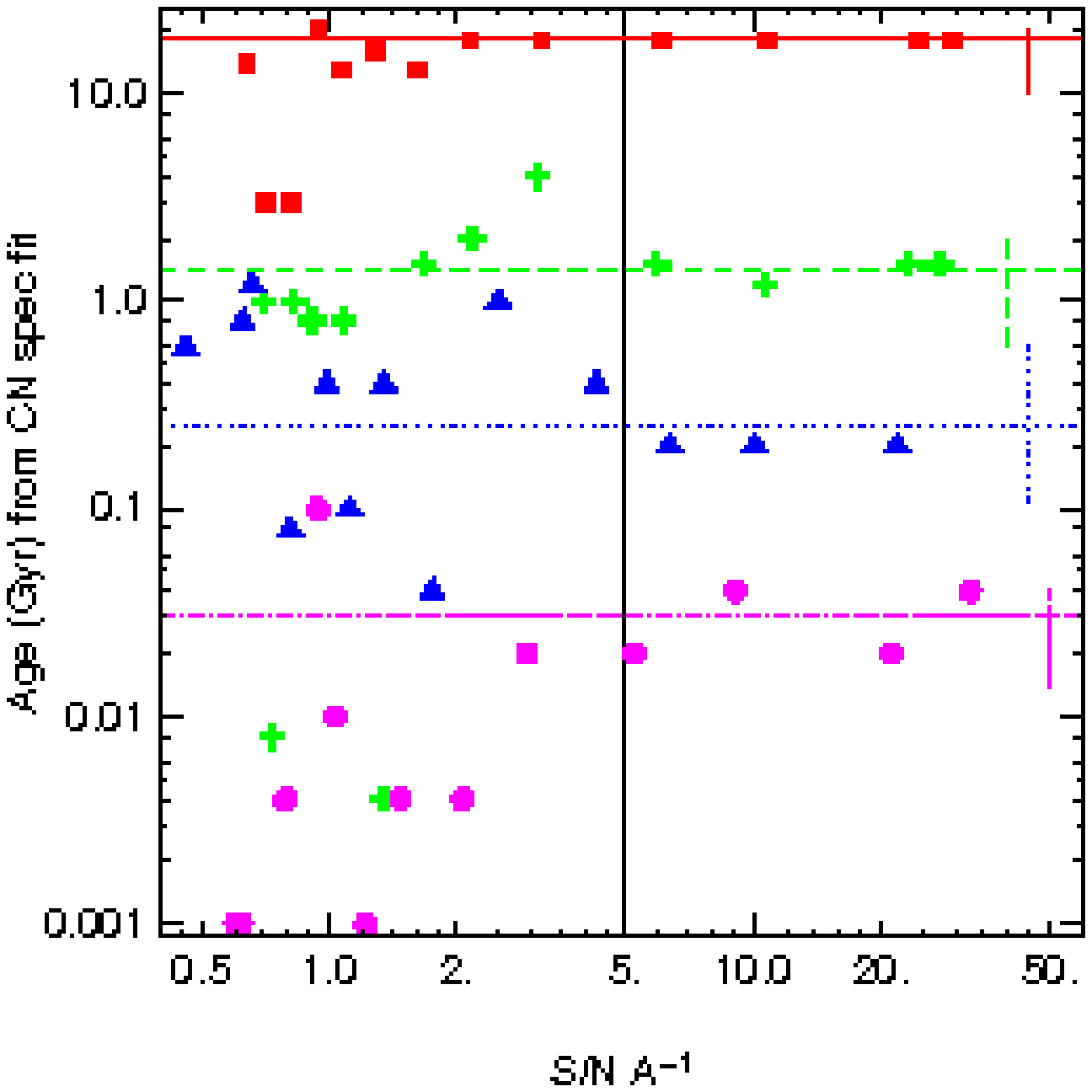}
    \includegraphics[scale=.55]{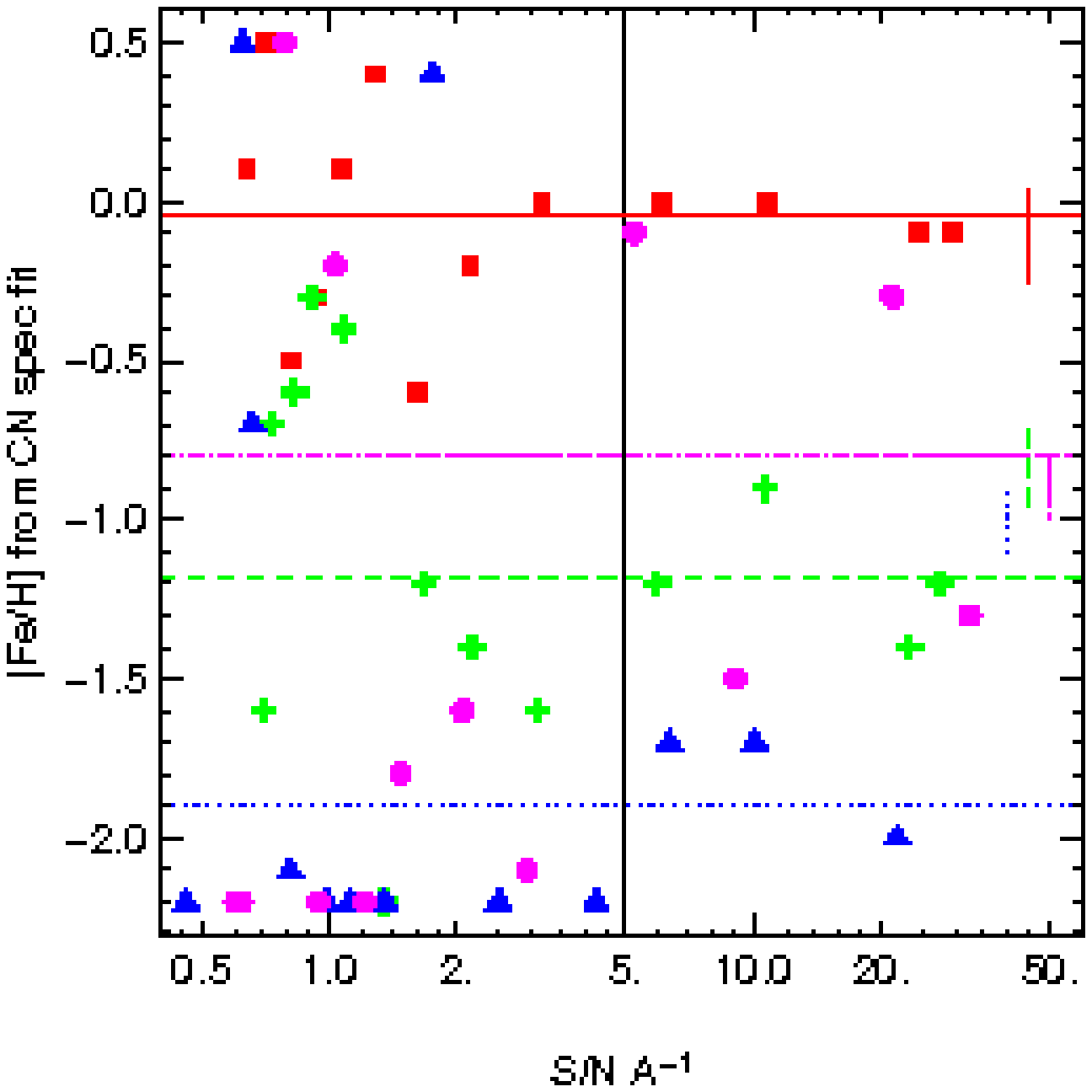}
  \caption{Dependence of ages and metallicities derived from CN
  spectrum fits on average S/N \AA$^{-1}$. S/N=5 is marked by vertical
  lines. The clusters shown are G158 (\textit{squares, solid red
  line}), NGC 419 (\textit{pluses, dashed green line}), NGC 2134
  (\textit{triangles, dotted blue line}), and NGC 1818
  (\textit{circles, dash-dot magenta line}). The horizontal lines mark
  the average age or metallicity of each cluster for the S/N$\geq$5
  points. The spectral resolution of these clusters is 23\AA~for G158
  and 13\AA~for the others. The vertical bars on the right show the
  spread of literature values for each cluster, in the same line types
  and colors. If the 0.45 dex [Fe/H] scale offset is subtracted from
  the MC cluster (NGC~419, 2134, 1818) literature values, then our
  derived [Fe/H] for NGC~419 falls within the literature range. Our
  [Fe/H] values for the two younger clusters do not.  \label{fig12}}
\end{figure}

\begin{figure}
\figurenum{13}
    \includegraphics[scale=.4,angle=-90]{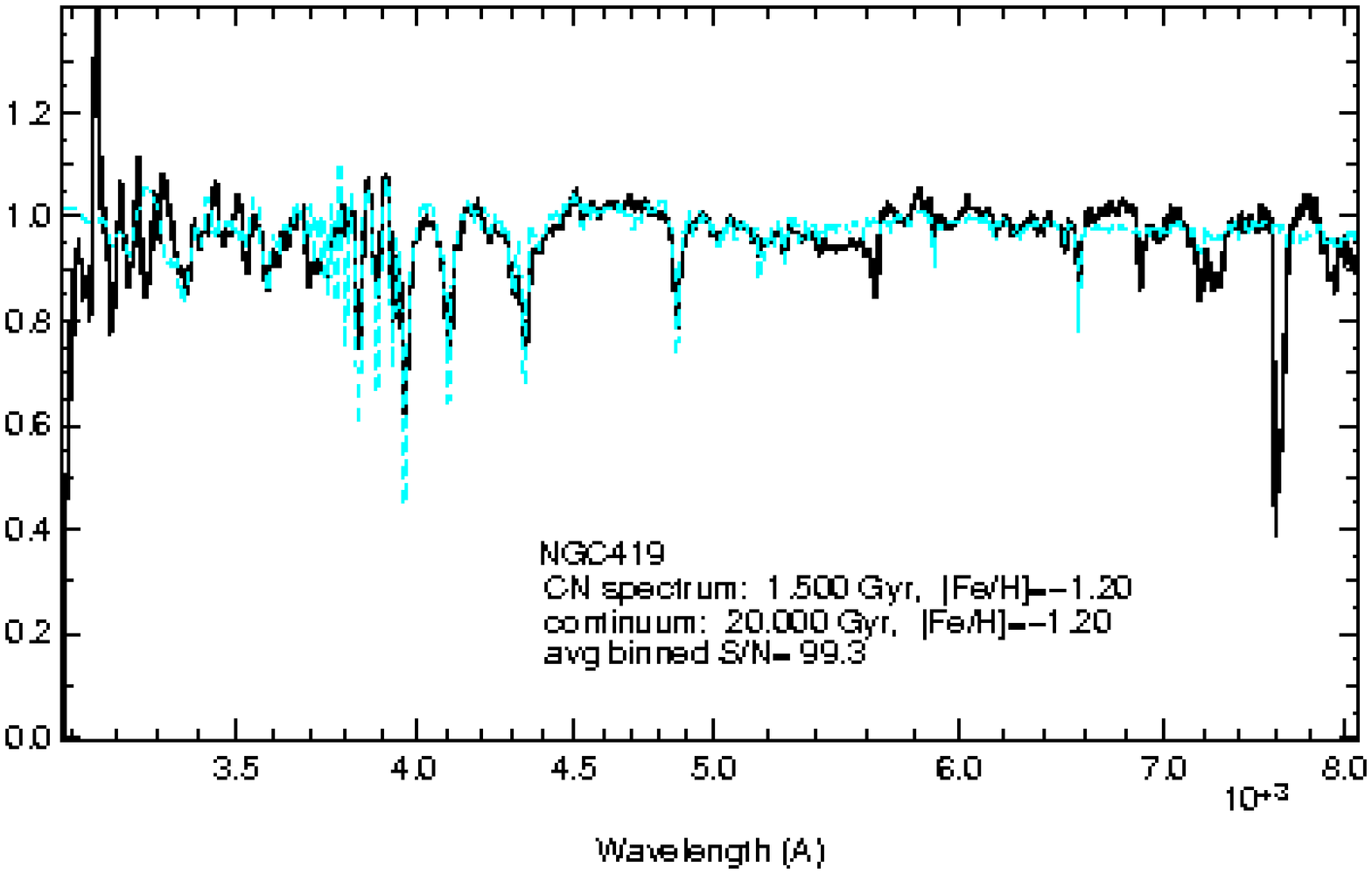}
    \includegraphics[scale=.4,angle=-90]{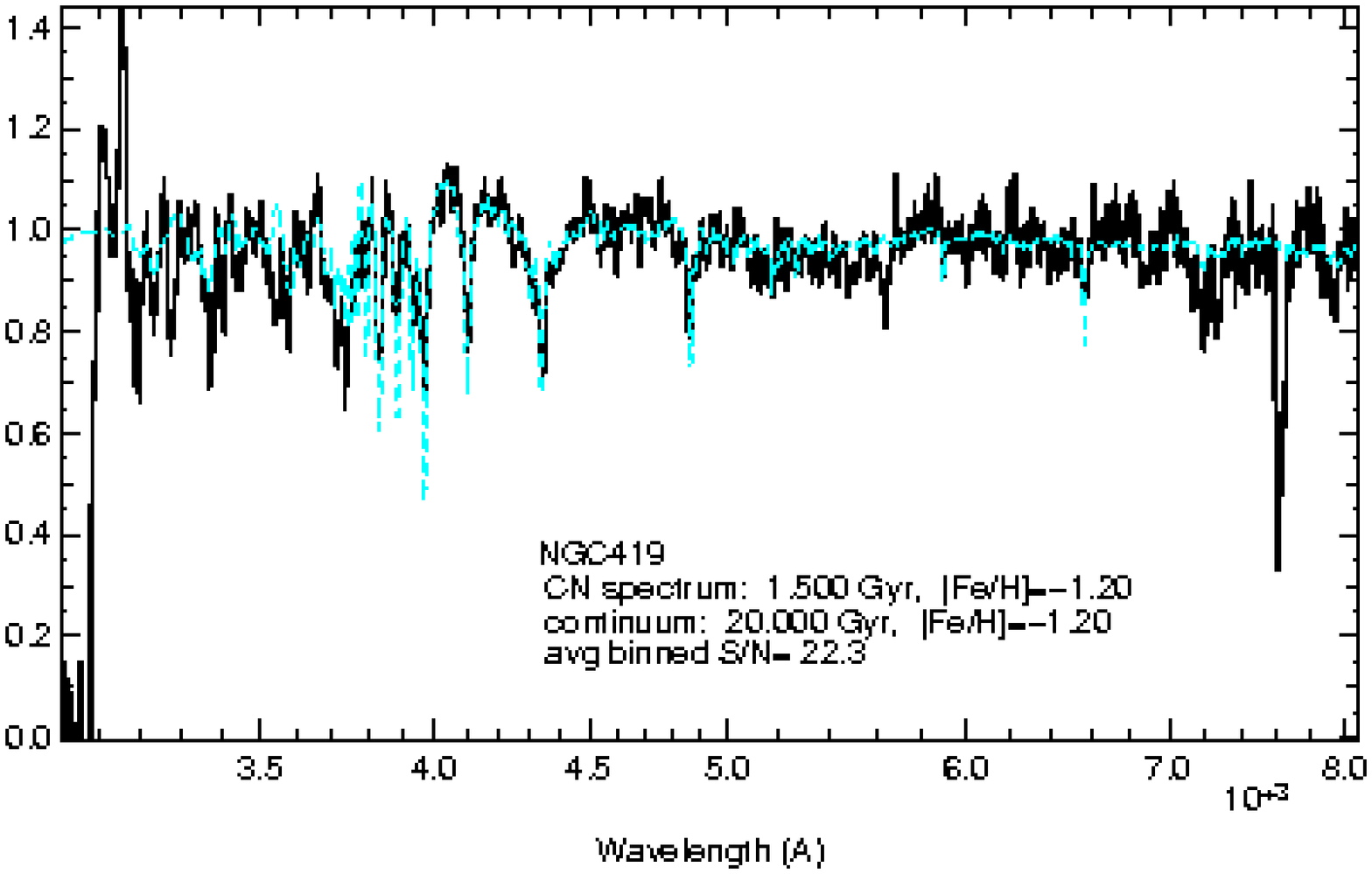}
    \includegraphics[scale=.4,angle=-90]{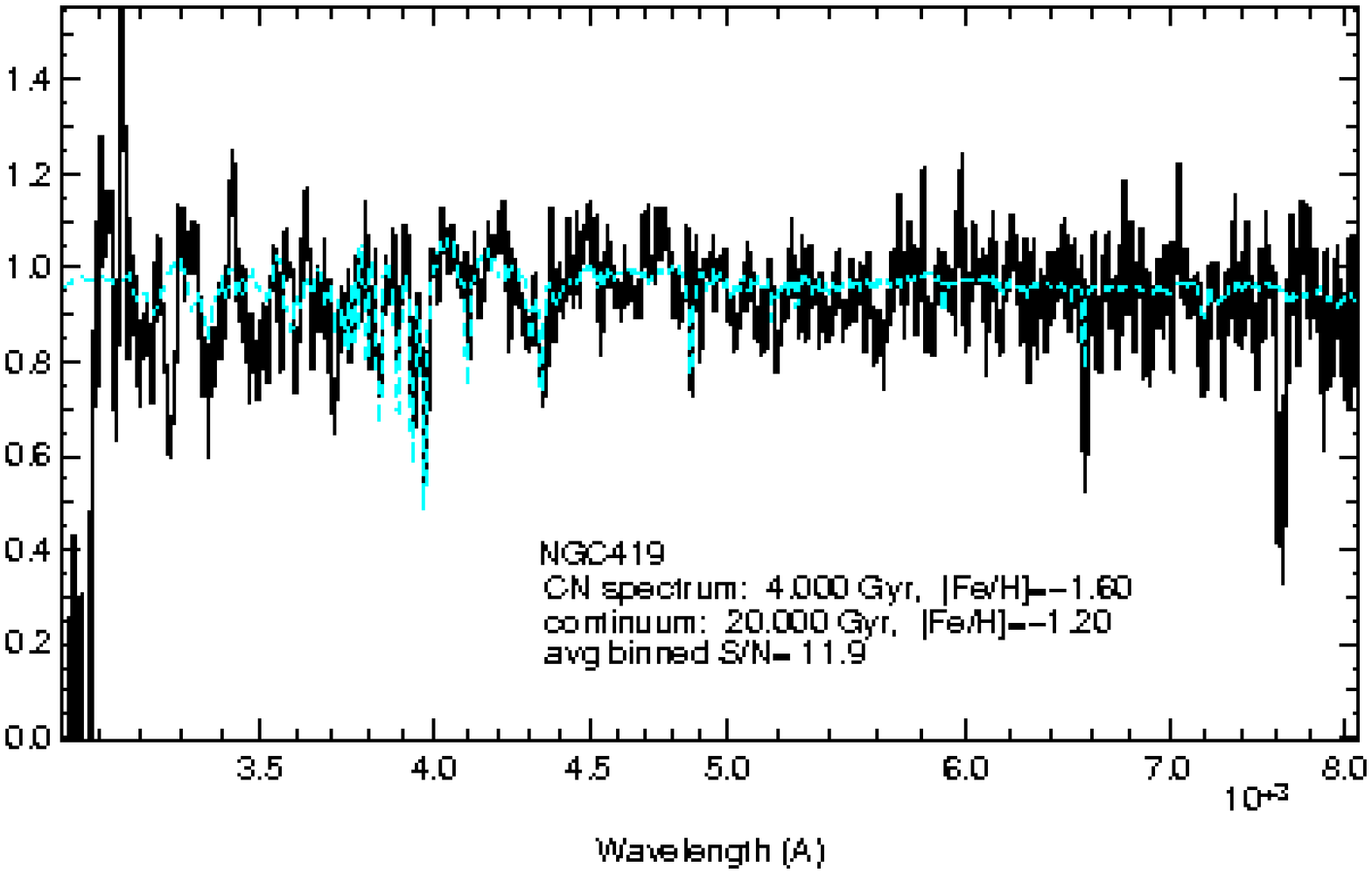}
  \caption{NGC 419 best-fitting spectra with artificially added noise
  resulting in S/N$\sim$99, S/N$\sim$22, and S/N$\sim$12. Between the
  last two S/N levels is where the model fits begin to give unstable
  ages for this cluster. \label{fig13}}
\end{figure}

Comparing the fitting methods (Figure~\ref{fig4}), we find a tighter
correlation between ages derived from the full spectrum and continuum
fits than from the full spectrum and CN spectrum fits for clusters
younger than 1~Gyr, suggesting that these derived ages are more
strongly driven by the continuum shape than by the spectral
lines. Clusters older than 1~Gyr show a tighter correlation between
ages derived from the CN spectrum and photometry fits. Because the
derived ages of young clusters are so strongly influenced by the
continua and we have seen signs of problems in the models matching the
cluster continua, we conclude that the best method for deriving
accurate ages for globular clusters, especially when simultaneously
determining metallicity, is fitting models to their
continuum-normalized spectra.

On metallicities of clusters older than 1 Gyr, we agree to 0.12~dex
with the literature values that are based on spectra. Those based on
colors have much more scatter, which suggests that CN spectrum
fits are more robust in estimating [Fe/H] than methods using
colors. We cannot accurately derive metallicity estimates for clusters
younger than $\sim$~1~Gyr.

Blueward of the 4000~\AA~break the continuum shape is more important
than the lines for deriving age. Redward of the break the lines alone
in a continuum-normalized spectrum are sufficient to derive the
age. Many of the globular clusters tested here have spectra out to
rest wavelengths of 1 $\mu$m. If fitting models to
continuum-normalized spectra of z$\sim$1 galaxies, this means that
near infrared spectra through K band will be important to provide
similar spectral coverage and constraints on age and metallicity. The
4000~\AA~break is a very important feature to be contained in the
spectrum. It will be contained in typical z$\sim$1 spectra, but higher
redshifts may require near infrared spectra to correctly constrain
ages and metallicites. We should note, however, that although a strong
age indicator for globular clusters and early type galaxies, the
4000~\AA~break can be significantly affected by dust in star forming
galaxies. For example, \citet{macarthur05} find dust imparted age
errors as large as 9.5~Gyr for a 0.5~Gyr BC03 model SSP when ages are
derived from the D$_n$(4000)-Fe4668 plane.

The spectral S/N required to derive ages by fitting the entire CN
spectrum is similar to the 30-40~\AA$^{-1}$ typically required when
using line indices on old objects. For a 10~Gyr old object at
$\Delta\lambda$~=~5-25~\AA~our required S/N is 15-30~\AA$^{-1}$ for a
20\% age uncertainty and 30-55~\AA$^{-1}$ for a 10\% uncertainty. A
1~Gyr year old object requires S/N~$\sim$~3-15~\AA$^{-1}$ for a 20\%
age uncertainty and 10-25~\AA$^{-1}$ for a 10\% uncertainty. At
$\Delta\lambda$~=~30~\AA~both ages require S/N~$\ga$~60~\AA. Fairly
low resolution, R$\sim$240-1200, spectra can be used, at the cost of
increasing required S/N with decreasing resolution.

We have successfully tested most of the age-metallicity space expected
to be occupied by galaxies at z$\sim$1, except for the highest
metallicities of [Fe/H]$>$+0.3. Fitting the continuum-normalized
spectra of these galaxies should work well for spectra without
accurate flux calibration, typical when faint galaxies are observed on
multiple masks over long periods of time during surveys. The required
S/N and spectral resolution for determining ages and metallicities
matches that typically achieved in recent galaxy surveys, as long as a
few like objects are coadded to boost the S/N. The techniques
presented here will be applied to galaxies in future work.

\acknowledgments

We wish to thank the anonymous referee for very useful comments and
suggestions. MJW was partially funded for this work by NASA GSRP grant
number NGT 5-50301.

%% Figures start here.

\clearpage

\clearpage

\clearpage

\clearpage

\clearpage

\clearpage

\clearpage

\clearpage

\clearpage

\clearpage

\clearpage

\clearpage

\clearpage

\clearpage

\begin{figure}
\figurenum{14}
    \includegraphics[scale=0.72,angle=-90]{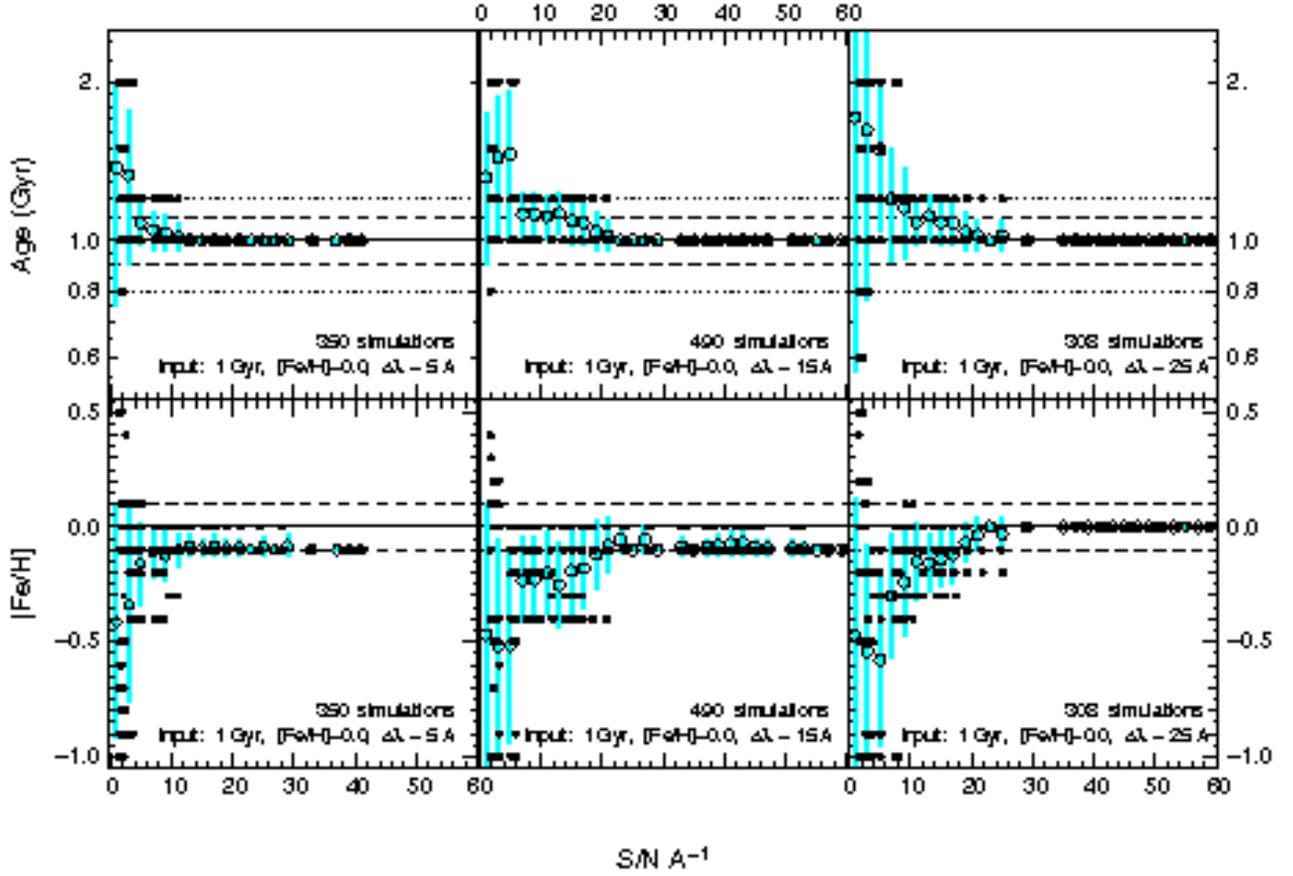}
 \caption{Noise simulations for 1 and 10~Gyr, [Fe/H] = 0.0 at
 $\Delta\lambda$~=~5, 15, 25~\AA. The top rows are derived age and
 bottom rows are derived [Fe/H]. Input age and metallicity are marked by
 solid lines, age uncertainties of 10\% by dashed lines, age
 uncertainties of 20\% by dotted lines, and [Fe/H] uncertainties of
 0.1 dex by dashed lines. Small filled circles represent individual
 simulations, open circles mark the average derived parameter in S/N
 bins of width 2, and vertical cyan bars mark the 1$\sigma$ variation
 in derived values within each bin. }
 \label{fig14}
\end{figure}

\begin{figure}
\figurenum{14}
    \includegraphics[scale=0.72,angle=-90]{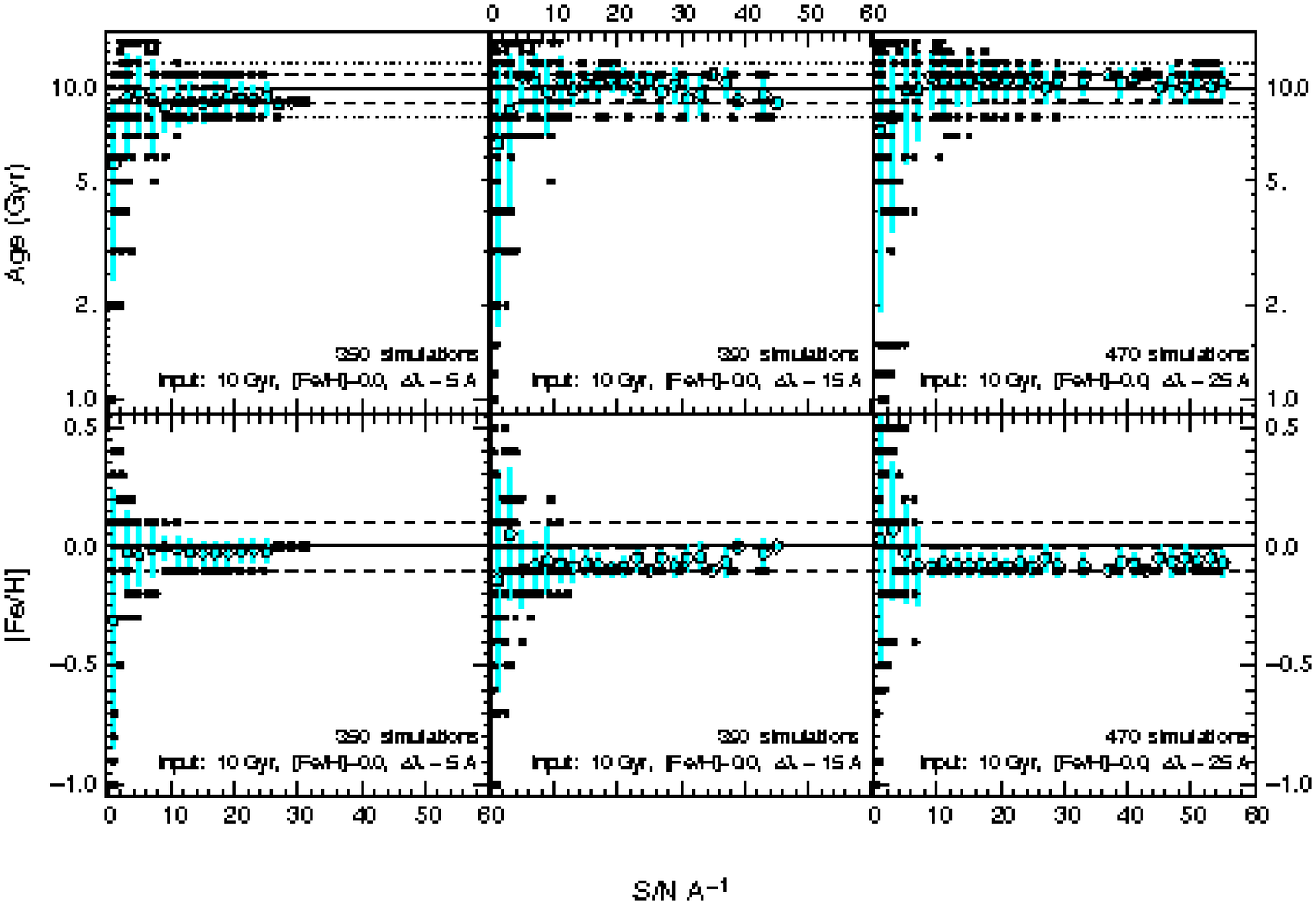}
 \caption{cont.}
\end{figure}

\begin{figure}
\figurenum{15}
    \includegraphics[scale=0.72,angle=-90]{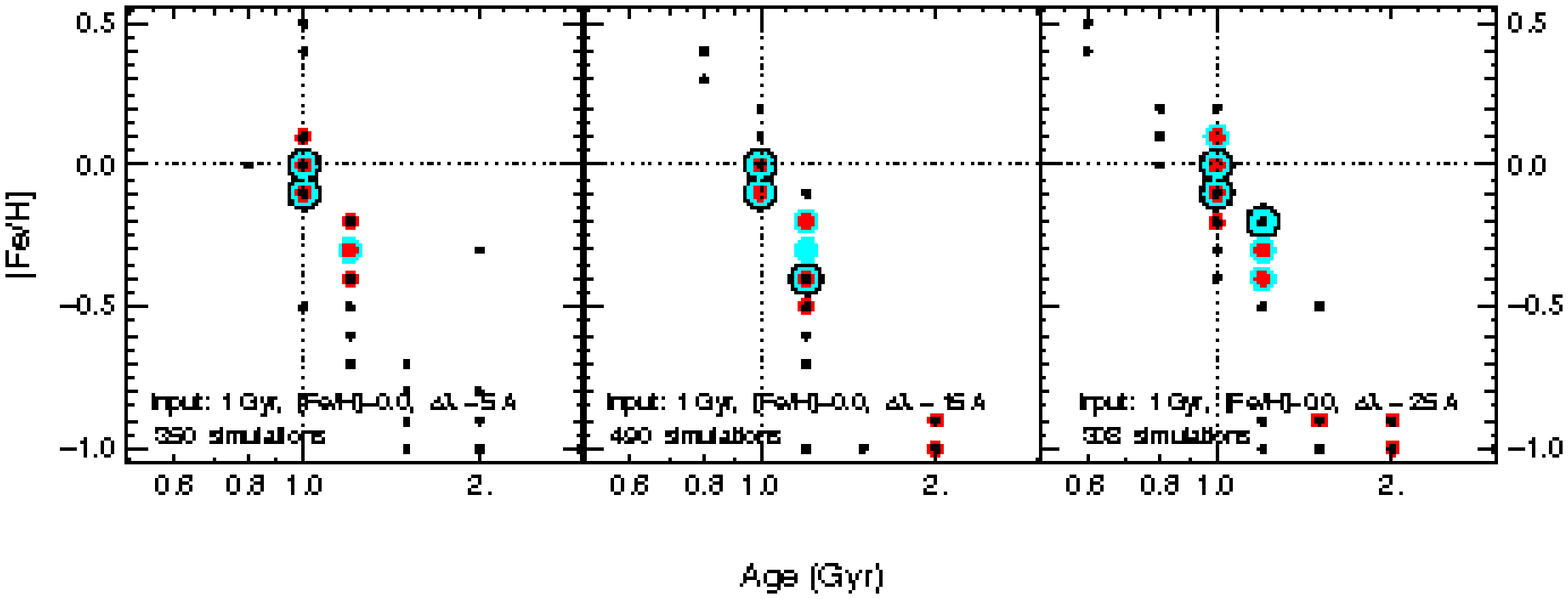}
    \includegraphics[scale=0.72,angle=-90]{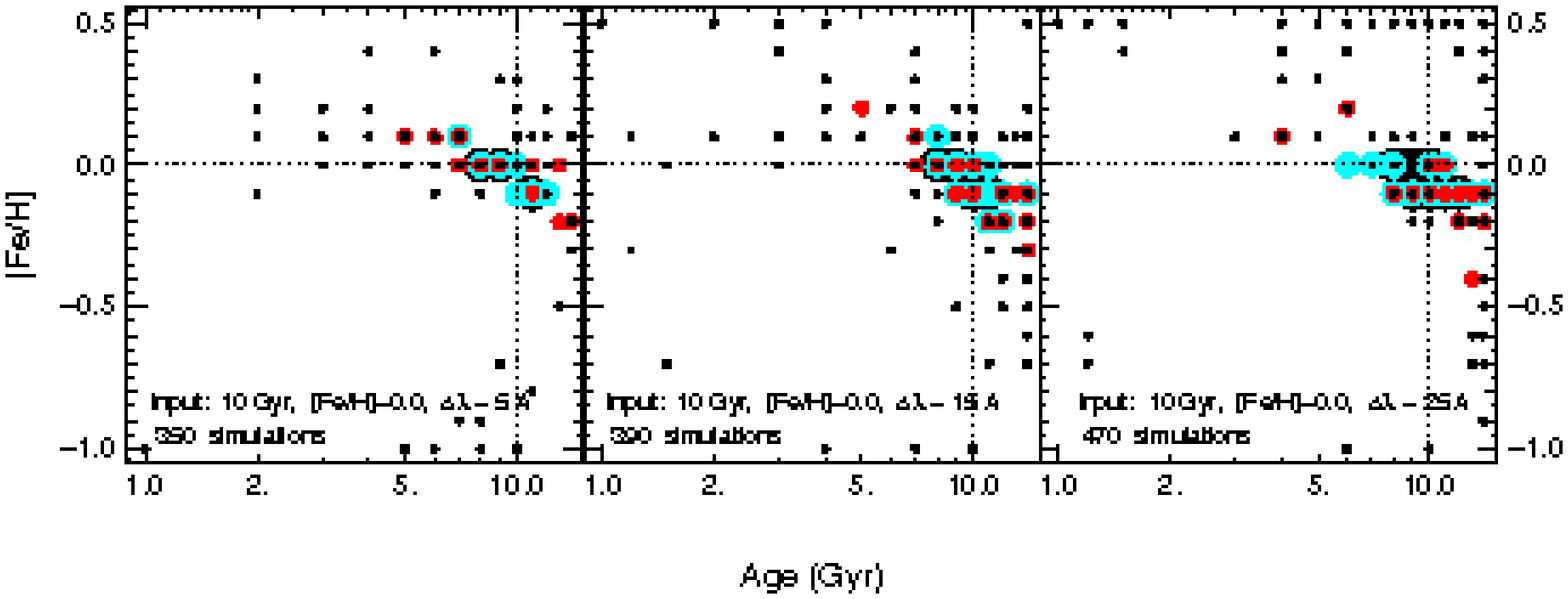}
 \caption{Age-metallicity plots of noise simulations with
 $\Delta\lambda$~=~5, 15, 25~\AA. The top row is 1~Gyr and bottom row
 is 10~Gyr. Symbol size denotes the S/N \AA$^{-1}$: largest black
 circles have S/N~$>$~20, next smaller cyan circles have
 10~$<$~S/N~$<$~20, next smaller red circles have 5~$<$~S/N~$<$~10,
 smallest black circles have S/N~$<$~5. Input age and metallicity are
 marked by dotted lines. }
 \label{fig15}
\end{figure}

\begin{figure}
\figurenum{16}
    \centering \includegraphics[scale=0.8,angle=-90]{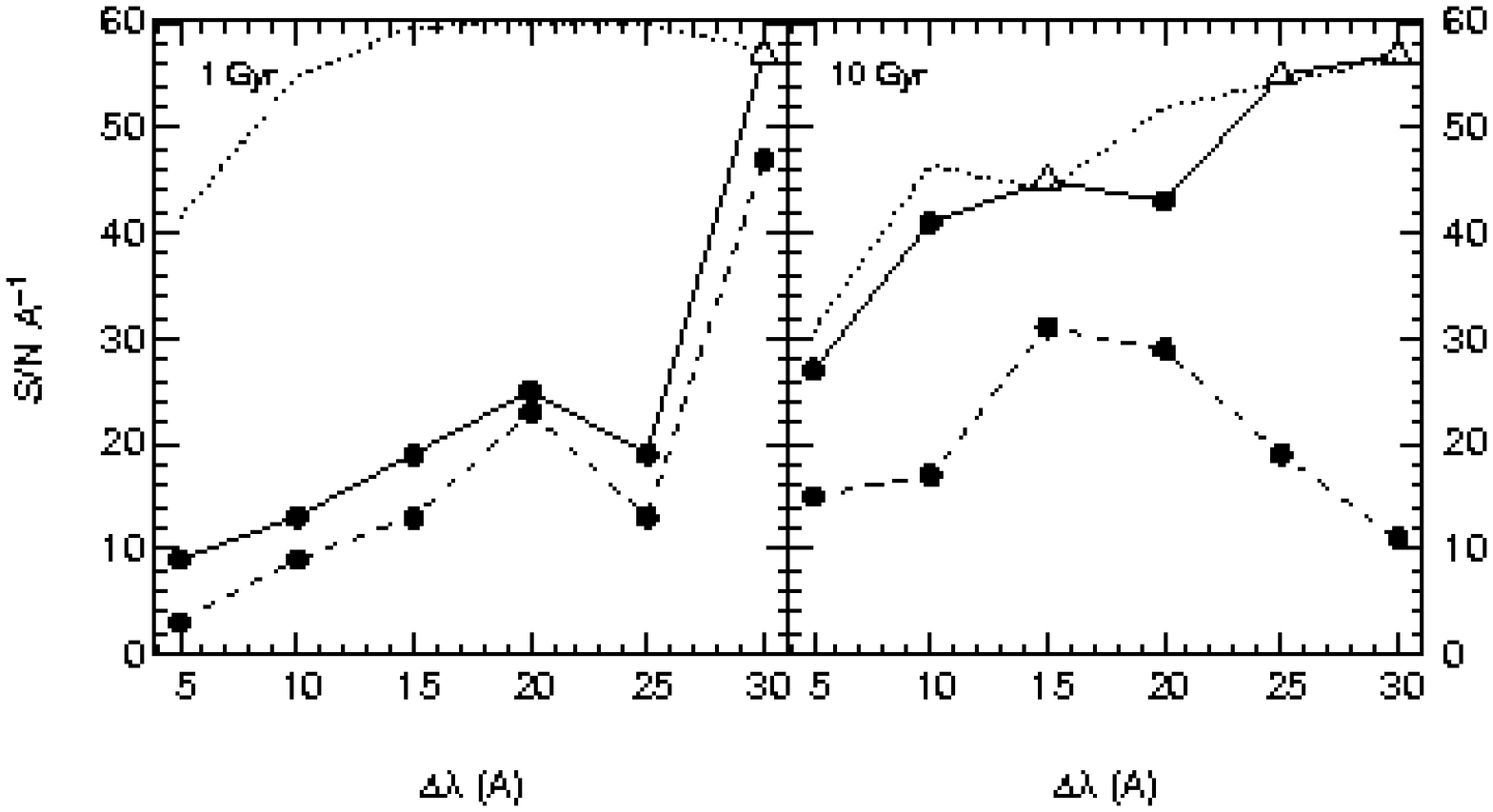}
 \caption{S/N~\AA$^{-1}$ required to derive ages within 10\% (solid
 lines) and 20\% (dashed lines) of a 1~Gyr input model on the left and
 a 10~Gyr model on the right (both with [Fe/H]~=~0) for a range of
 spectral resolutions. The dotted lines mark maximum S/N values at
 each resolution (see text). The upward facing open triangles are
 lower limits to the required S/N. }
 \label{fig16}
\end{figure}

\clearpage
\begin{figure}
\figurenum{17}
    \plottwo{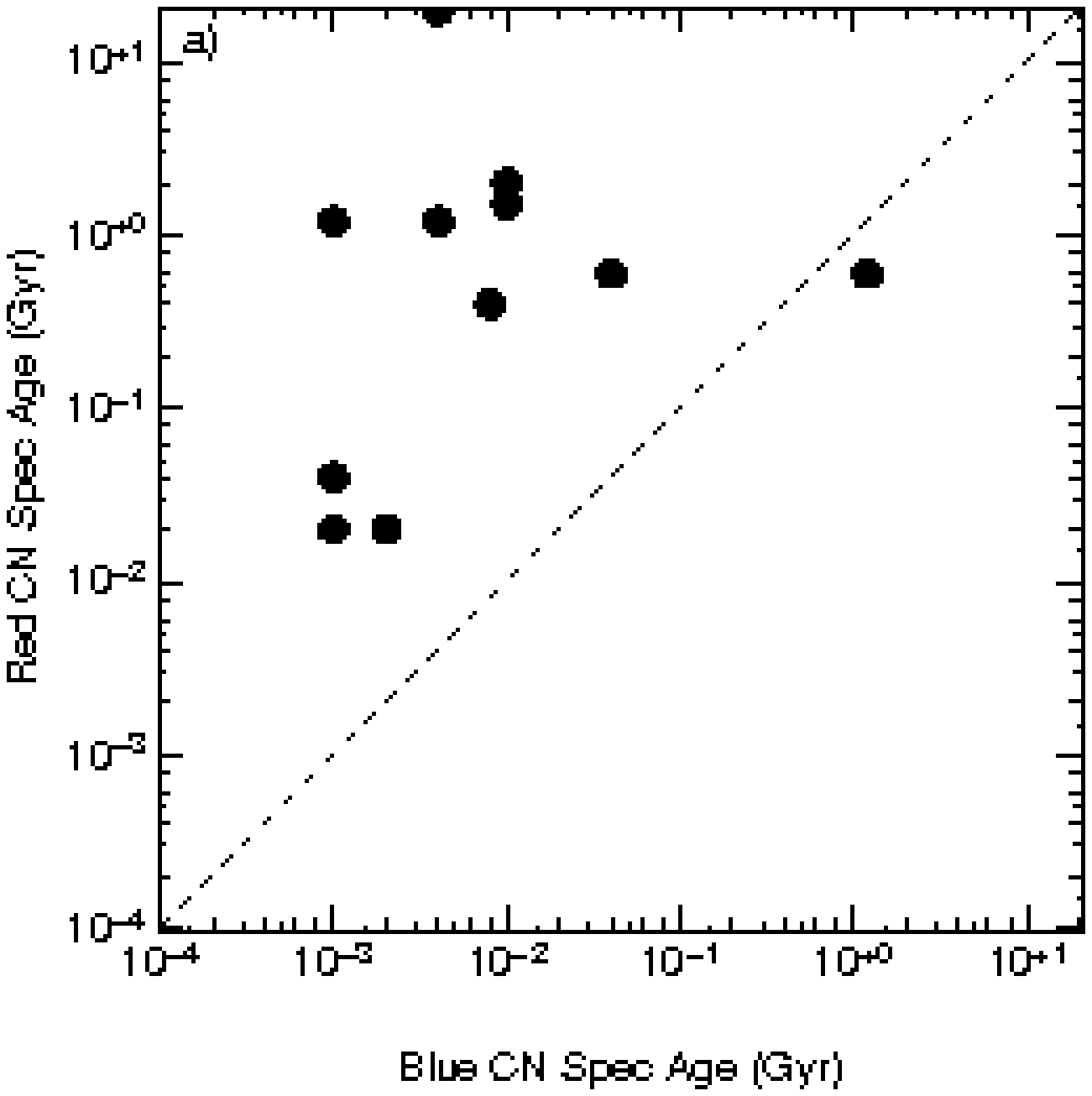}{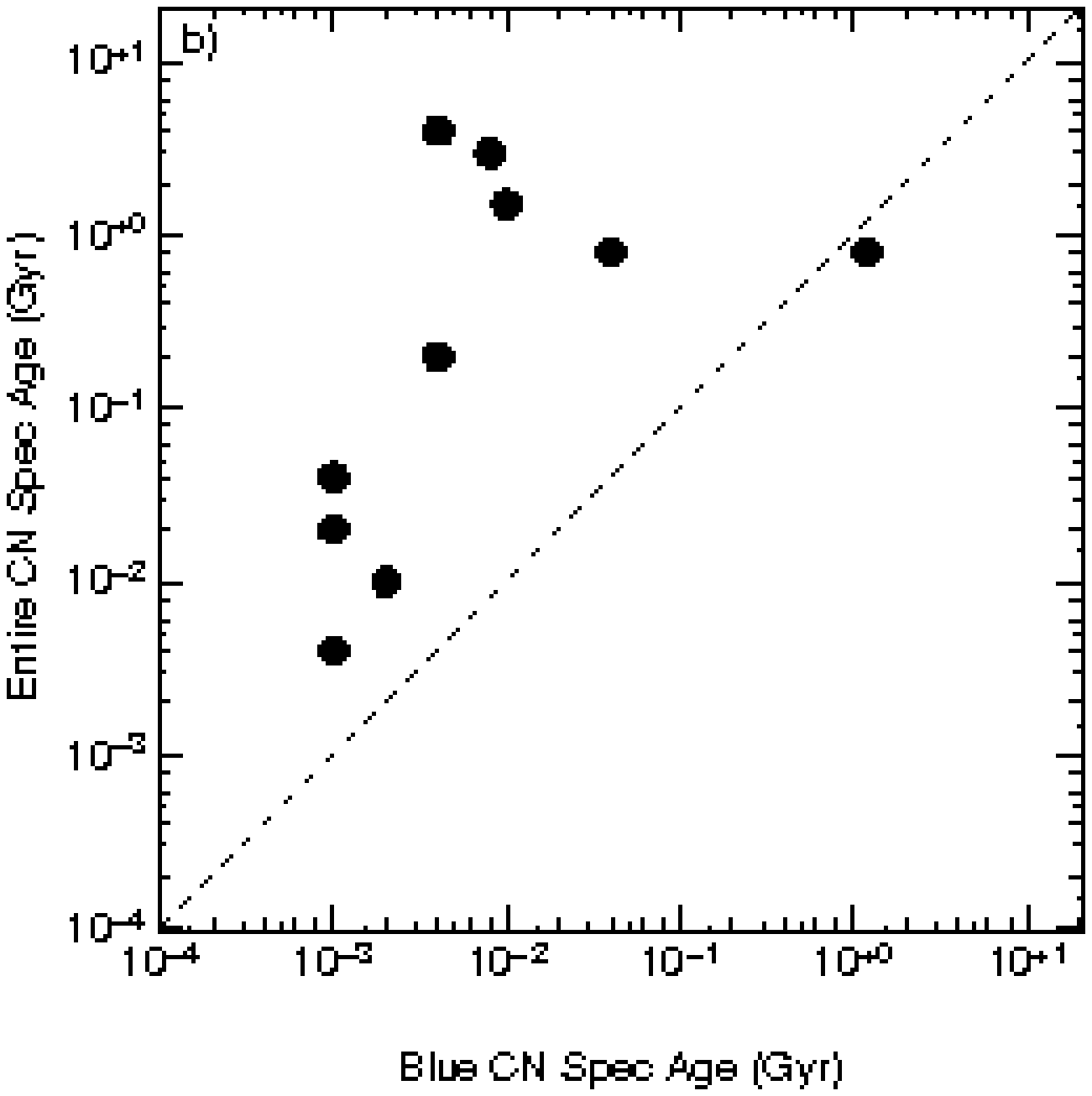}
    \plottwo{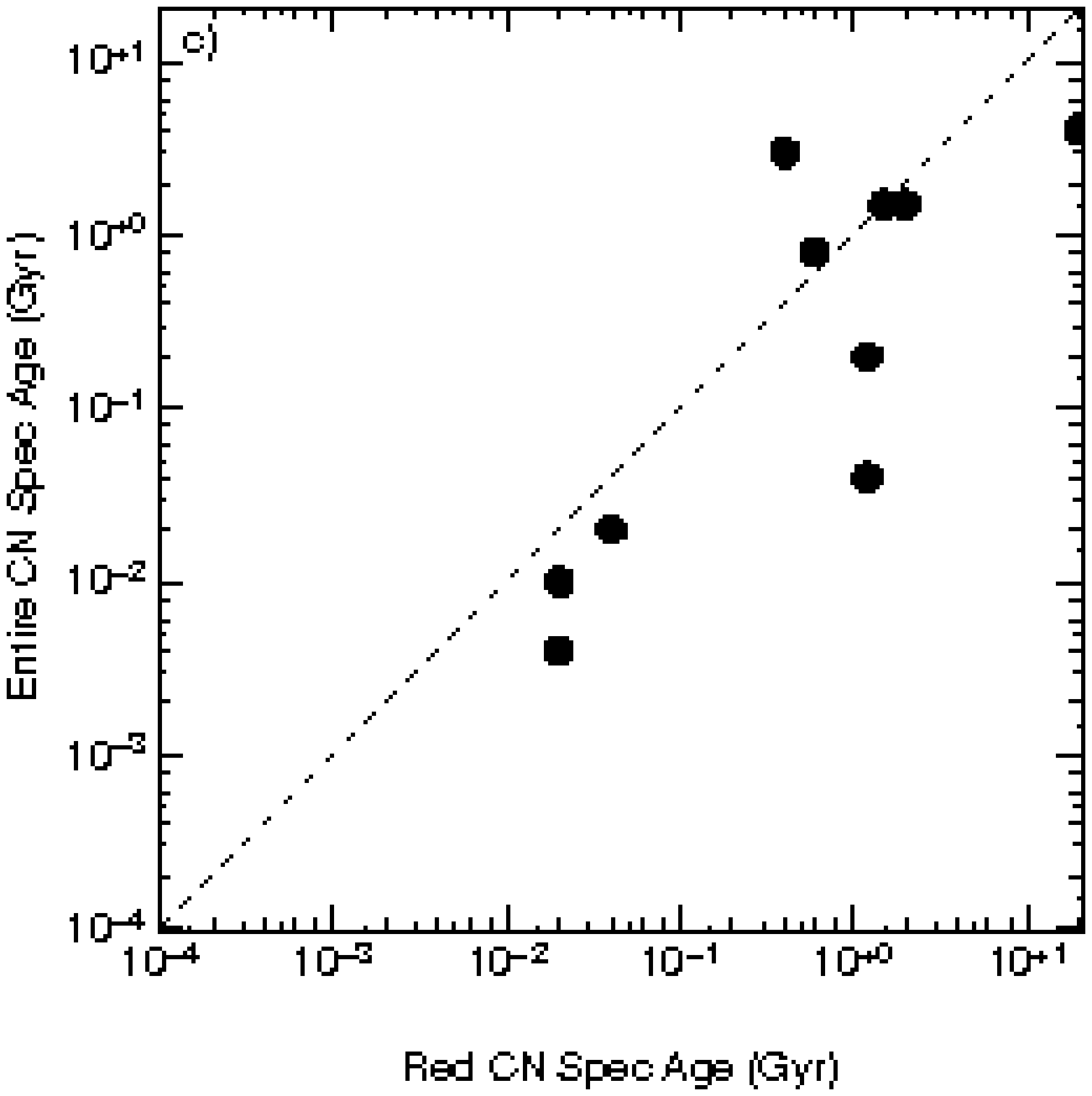}{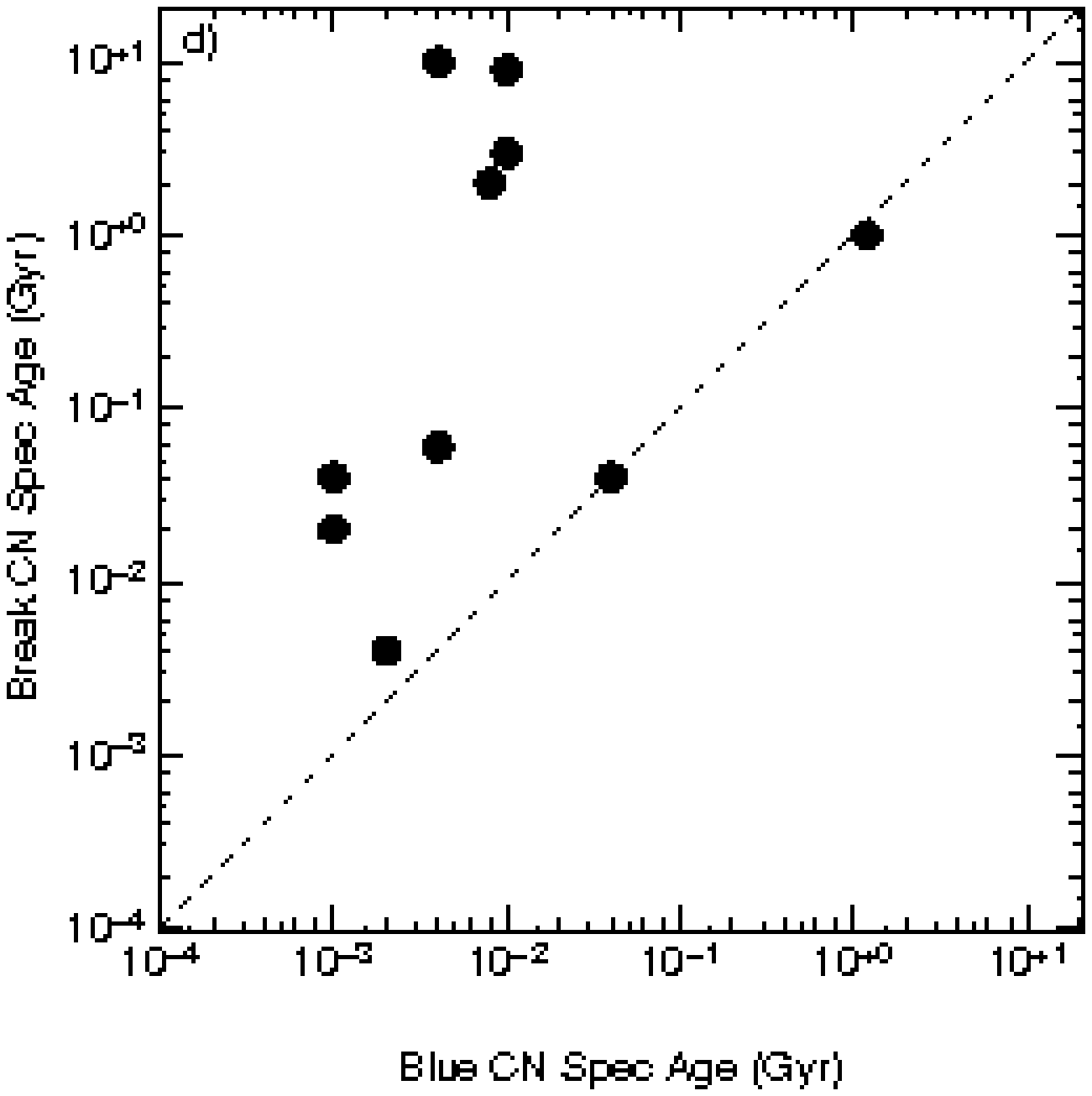}
  \caption{Age results from fitting the blue ($\lambda<$ 4000~\AA),
    red ($\lambda>$ 4000~\AA), and 4000~\AA~break (3750-4250~\AA)
    regions of the spectra. a-e) are CN spectrum fits, f-j) are full
    spectrum fits, k-n) are CN spectrum fit ages compared to the
    literature, and o-r) are full spectrum fit ages compared to the
    literature. h\&r) are from using the entire spectral
    range. \label{fig17}}
\end{figure}

\begin{figure}
\figurenum{17}
    \plottwo{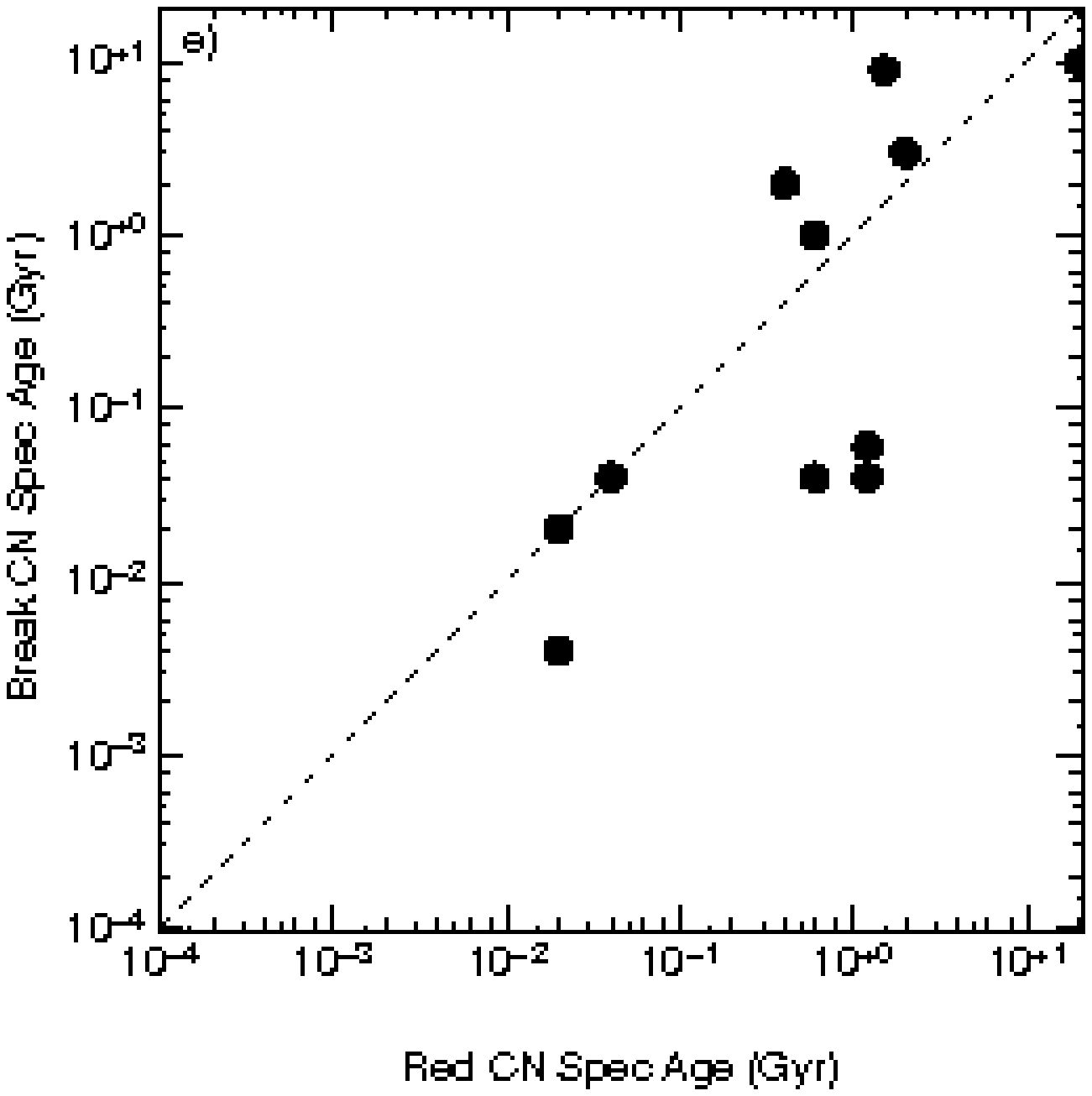}{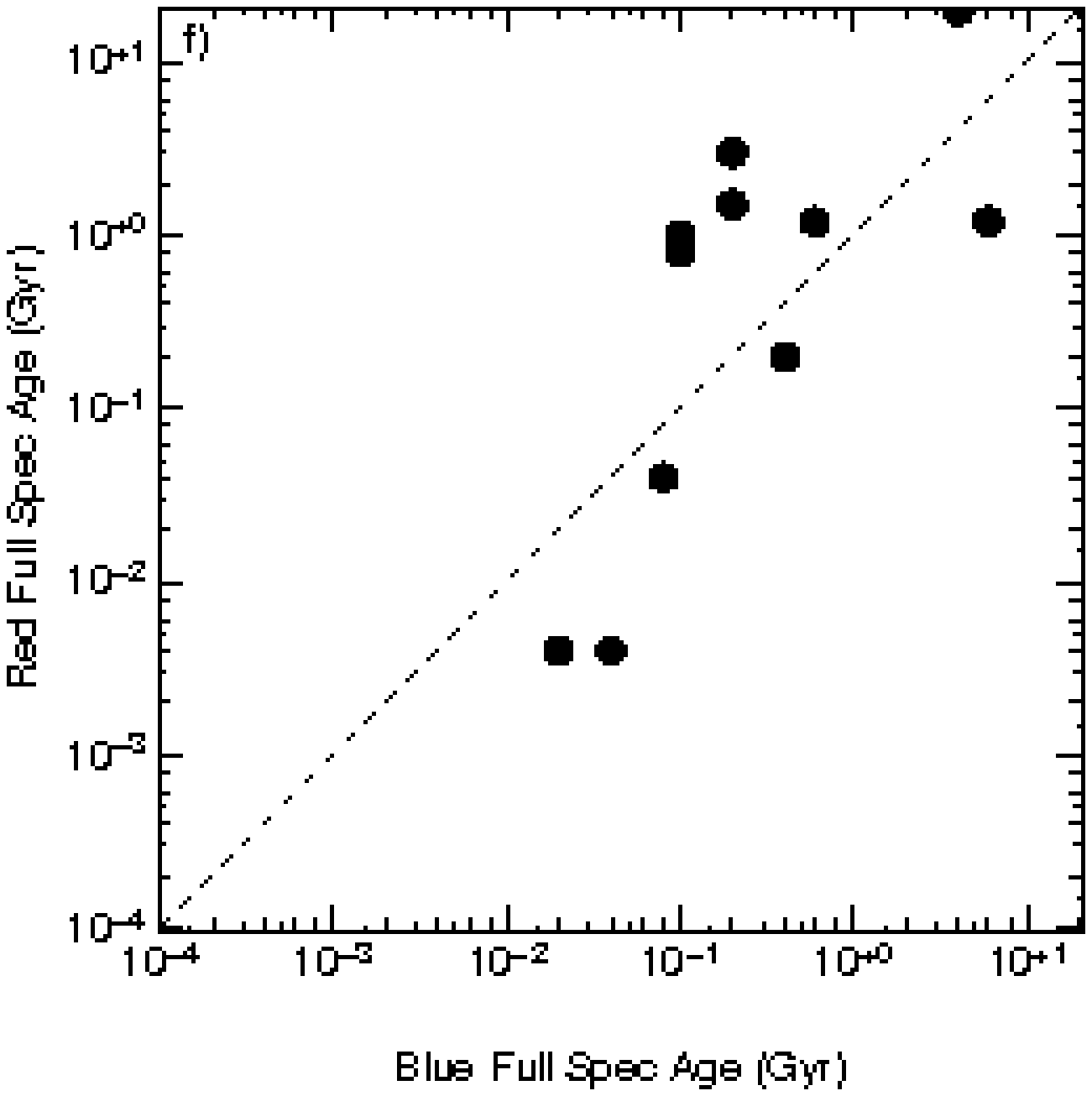}
    \plottwo{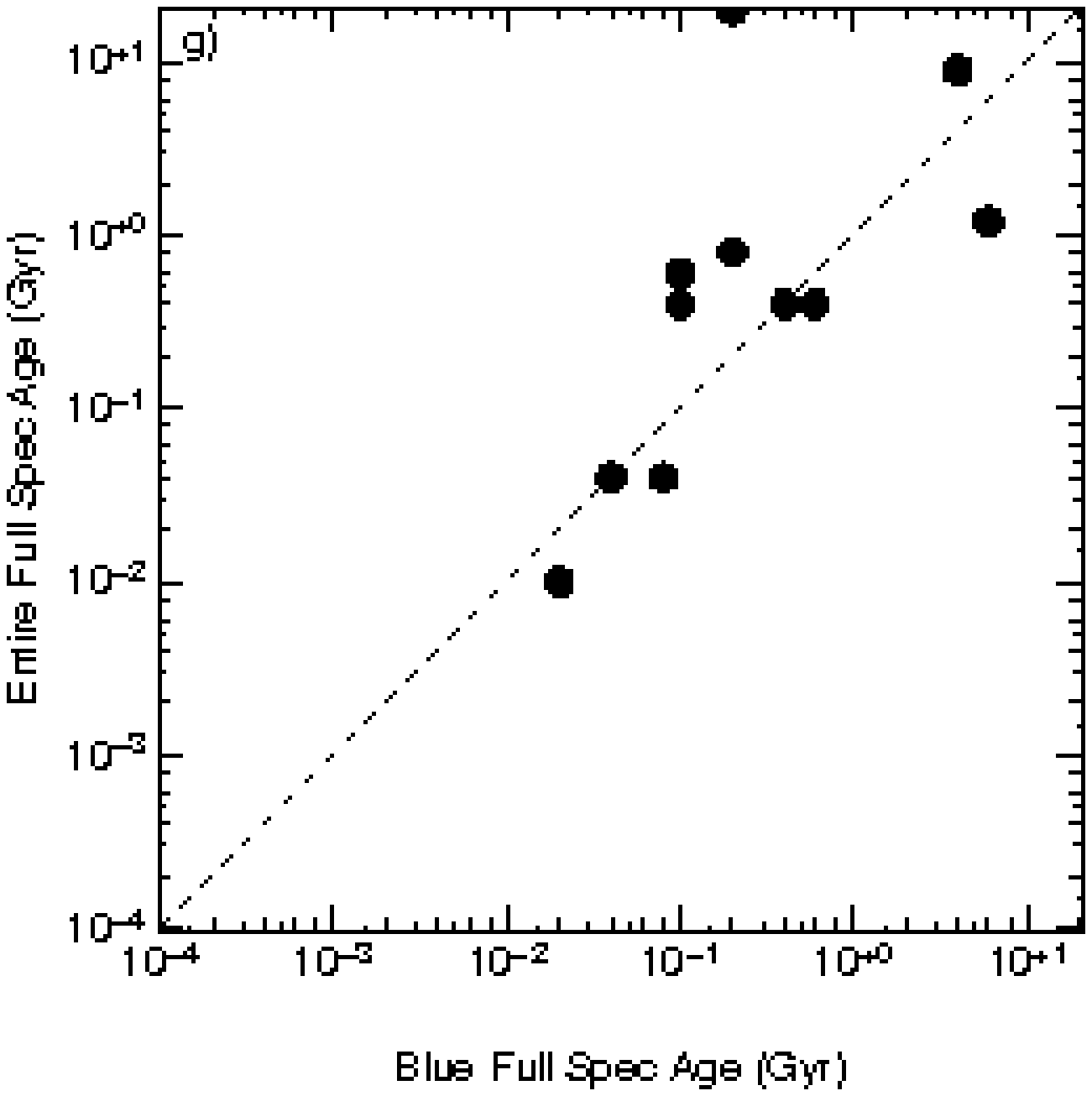}{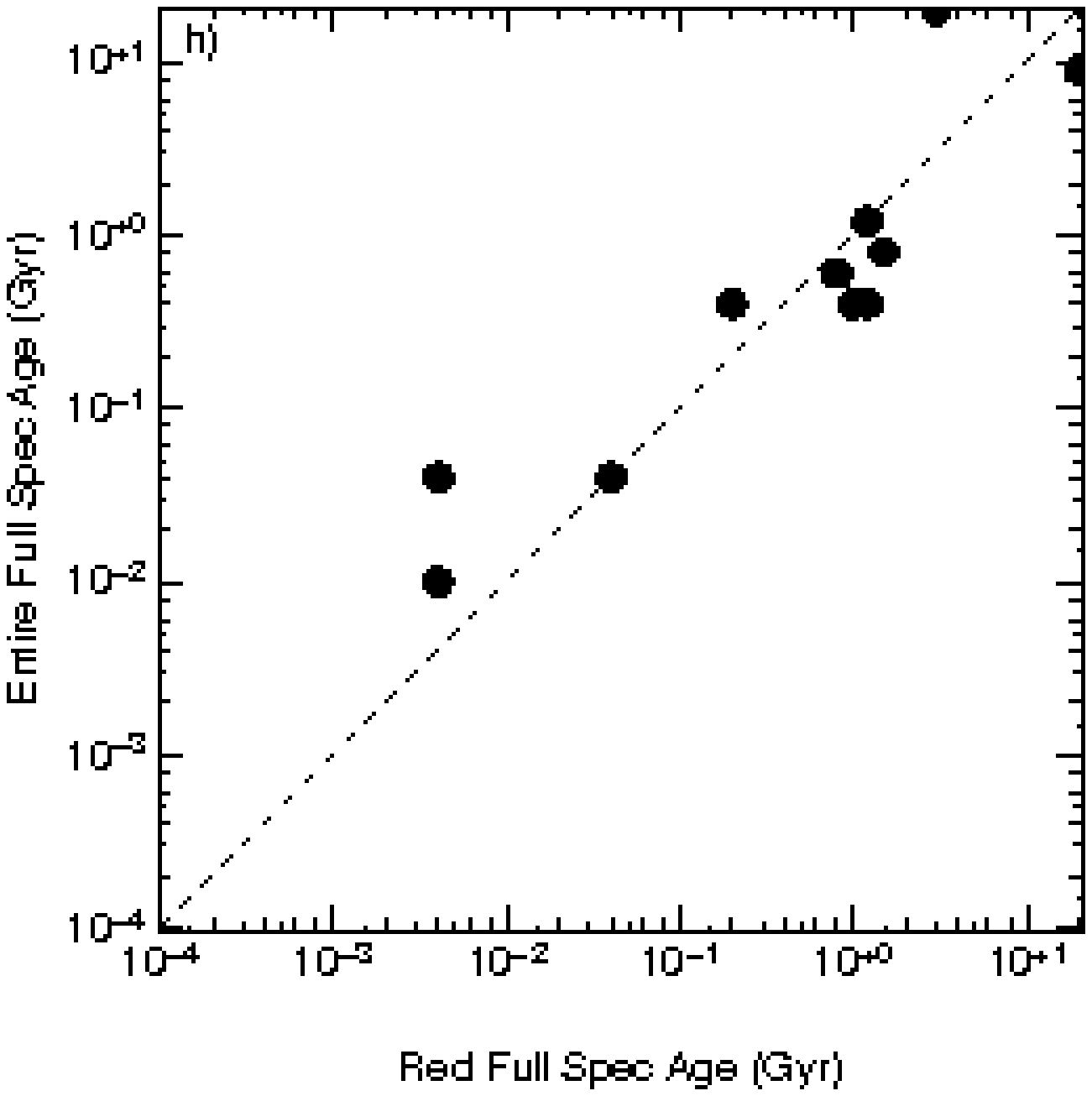}
\caption{cont.}
\end{figure}

\begin{figure}
\figurenum{17}
    \plottwo{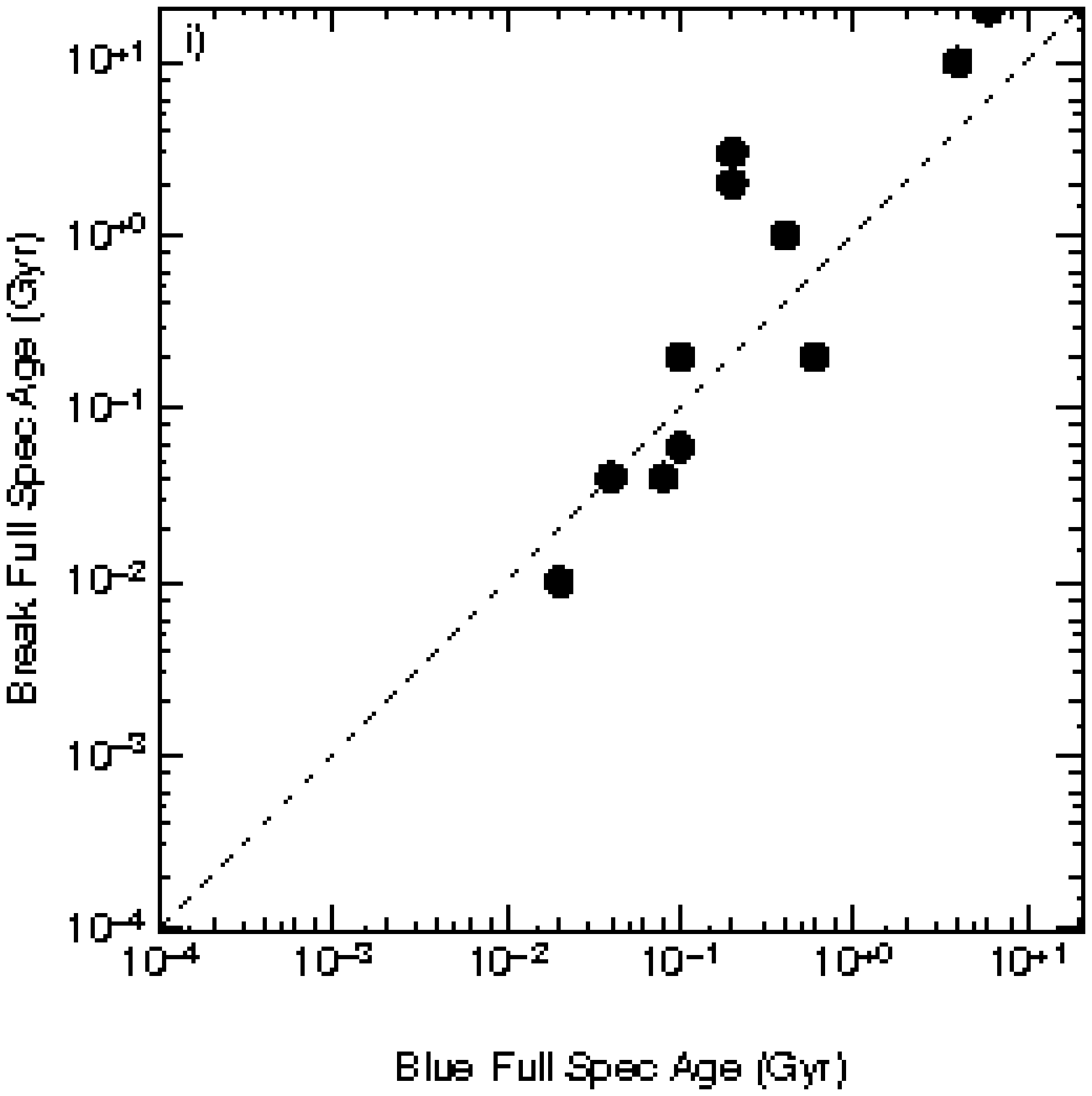}{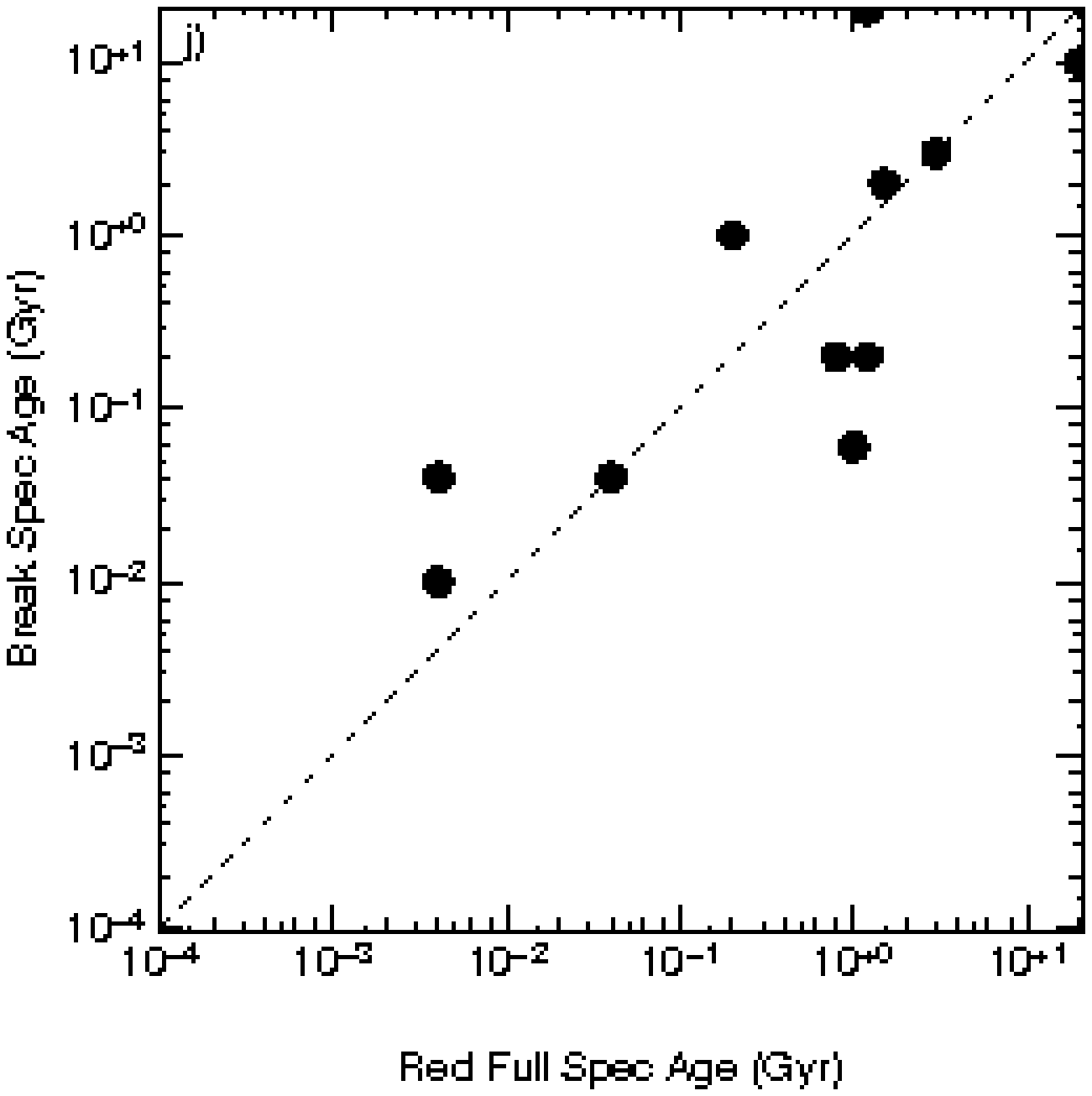}
    \plottwo{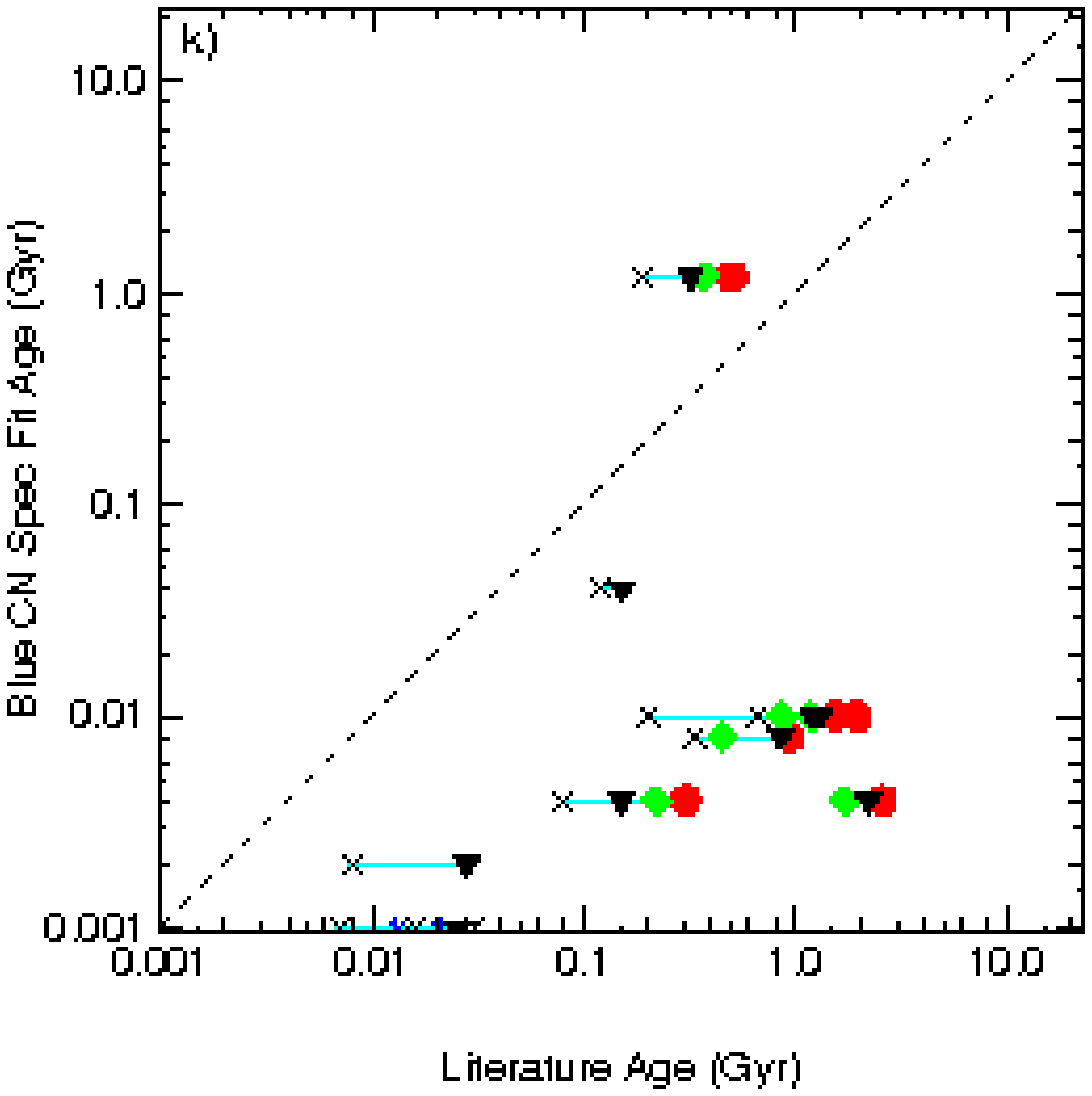}{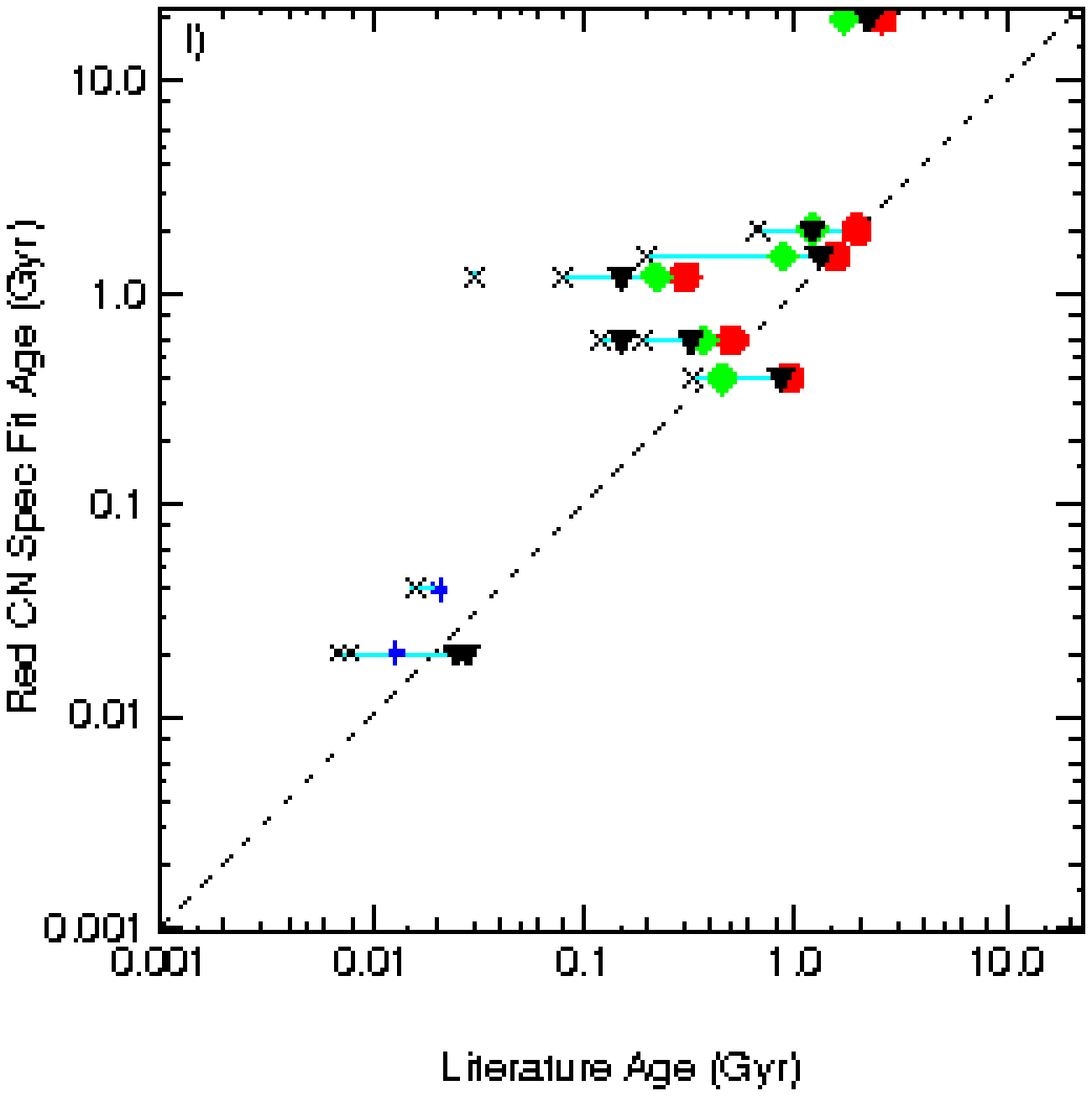}
\caption{cont.}
\end{figure}

\clearpage

\begin{figure}
\figurenum{17}
    \plottwo{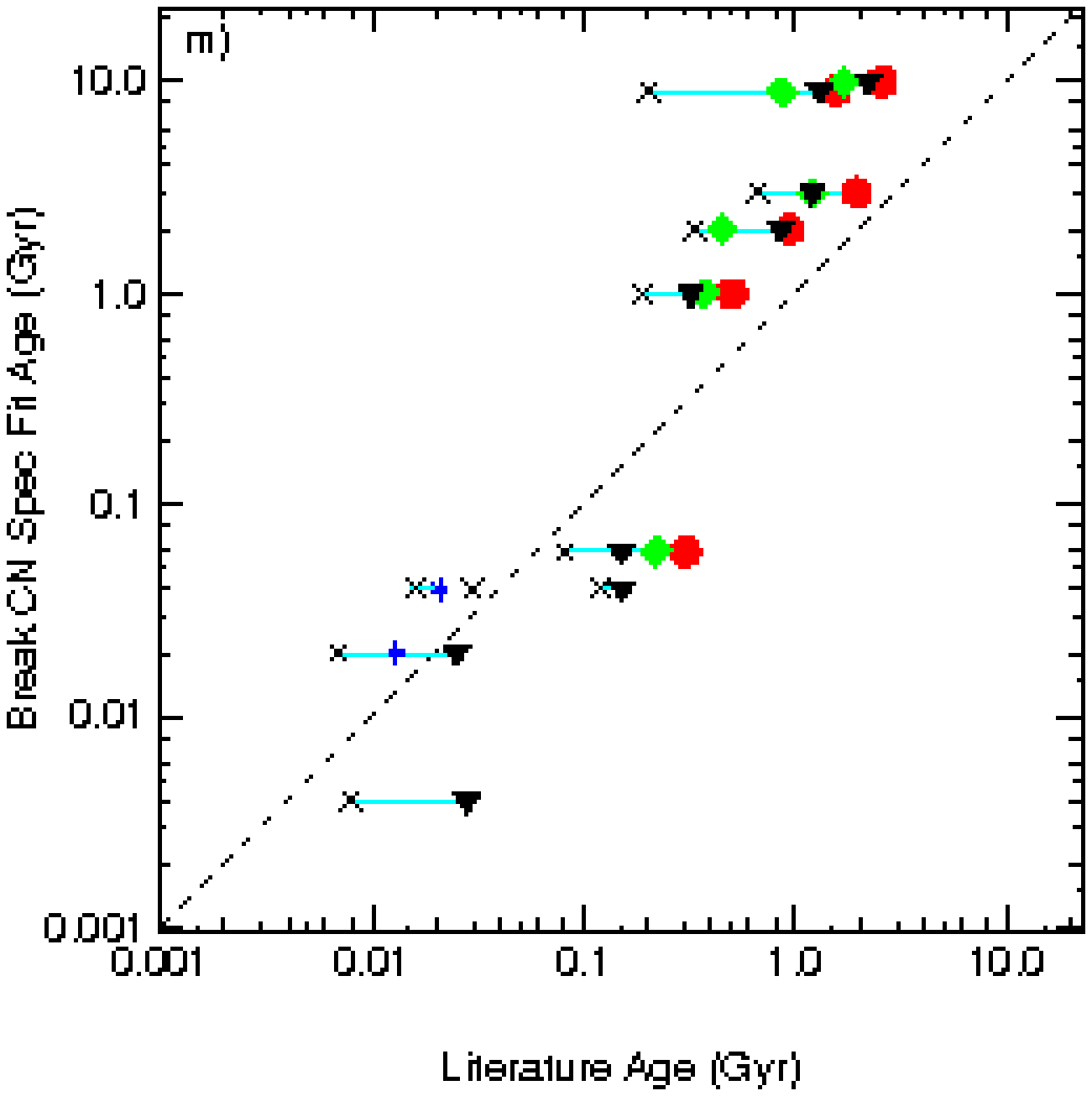}{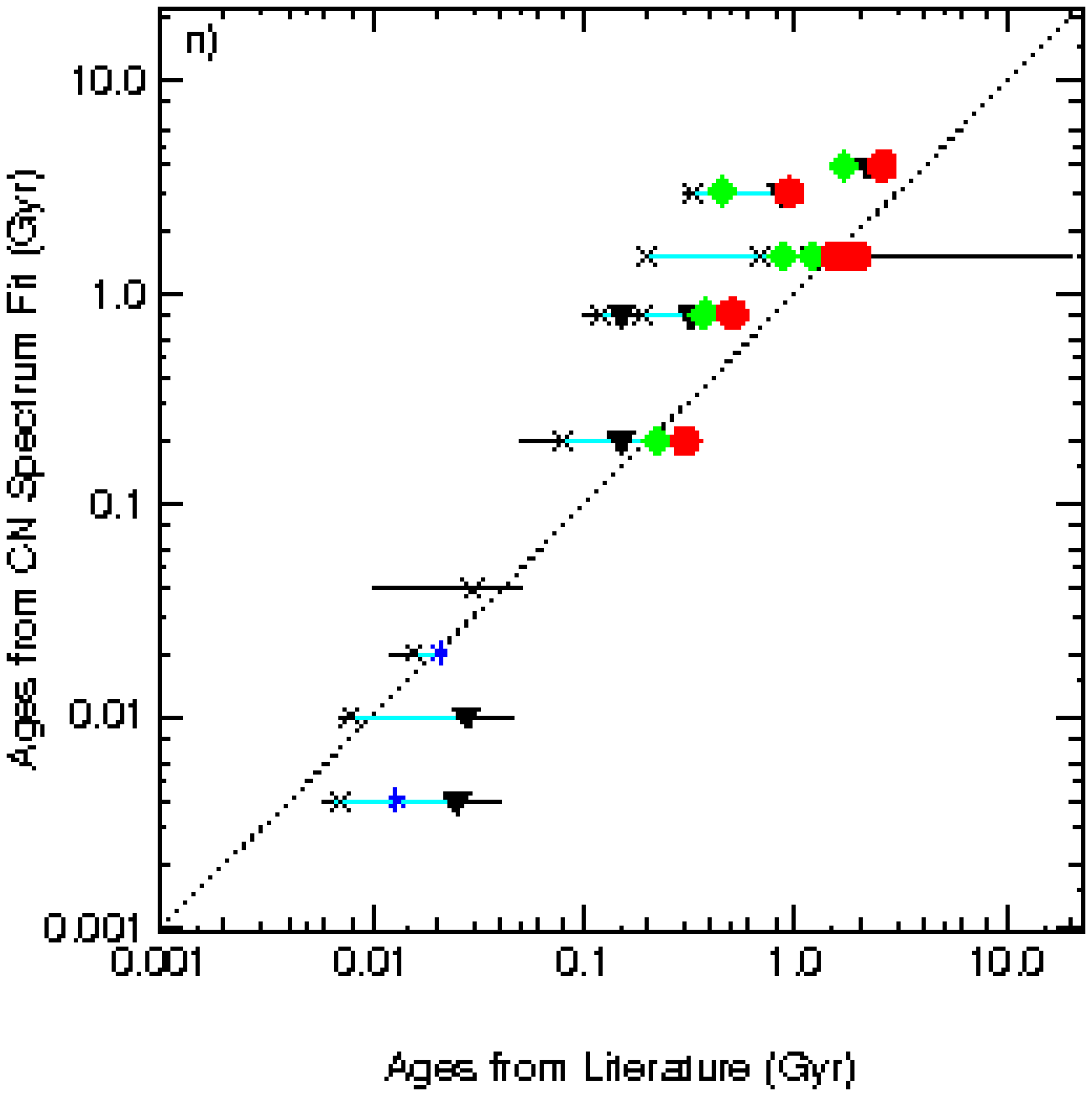}
    \plottwo{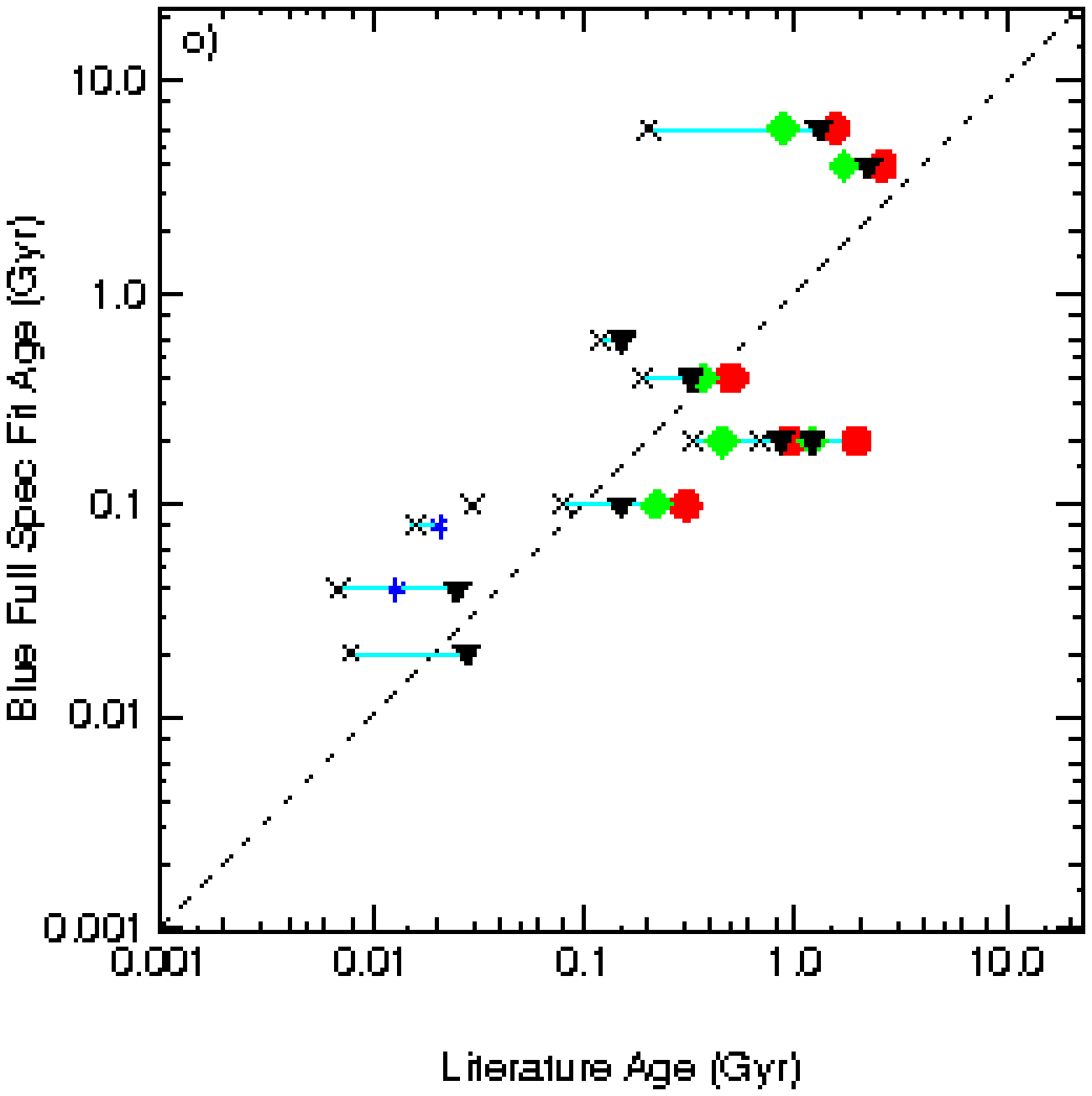}{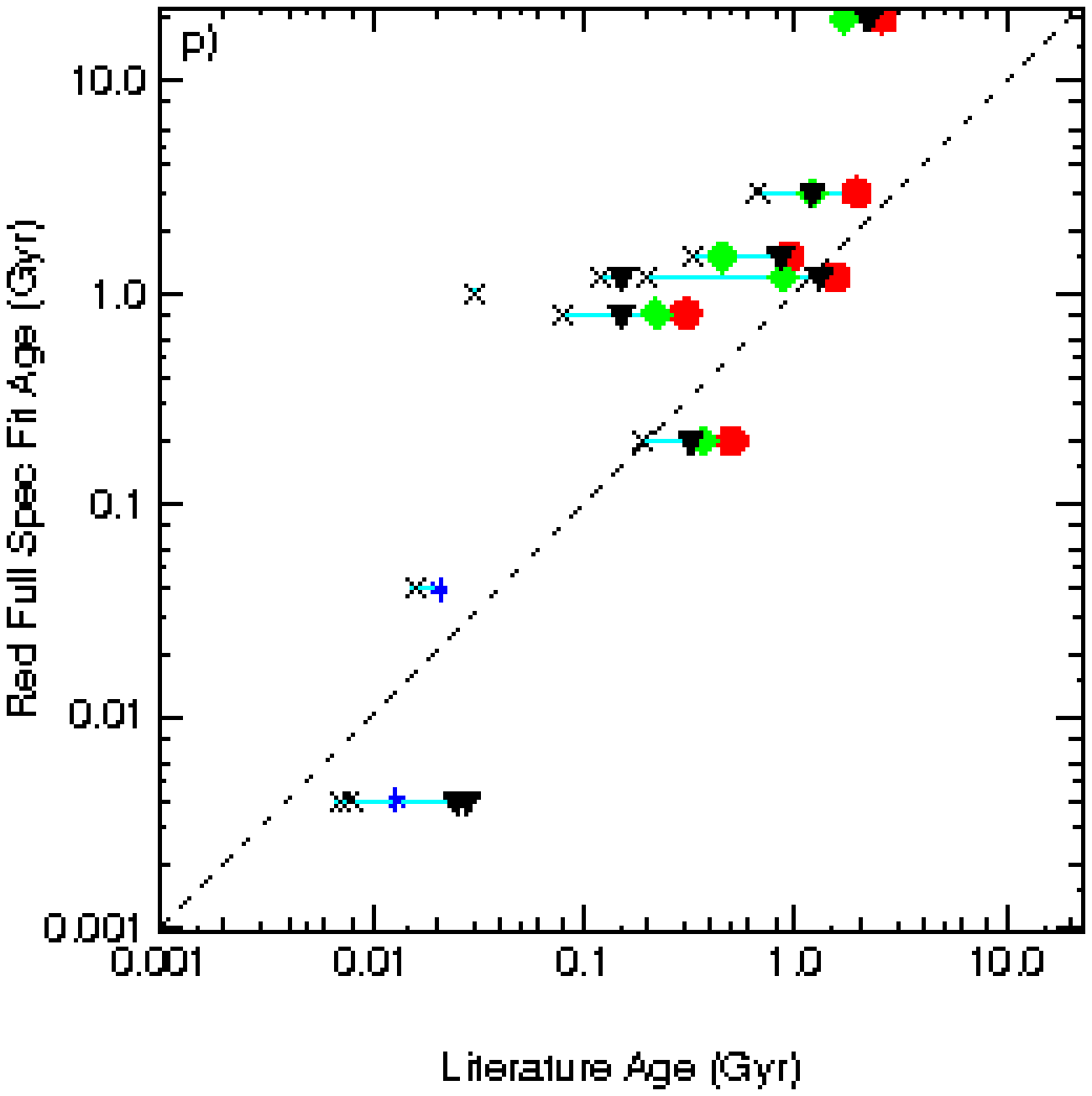}
\caption{cont.}
\end{figure}

\begin{figure}
\figurenum{17}
    \plottwo{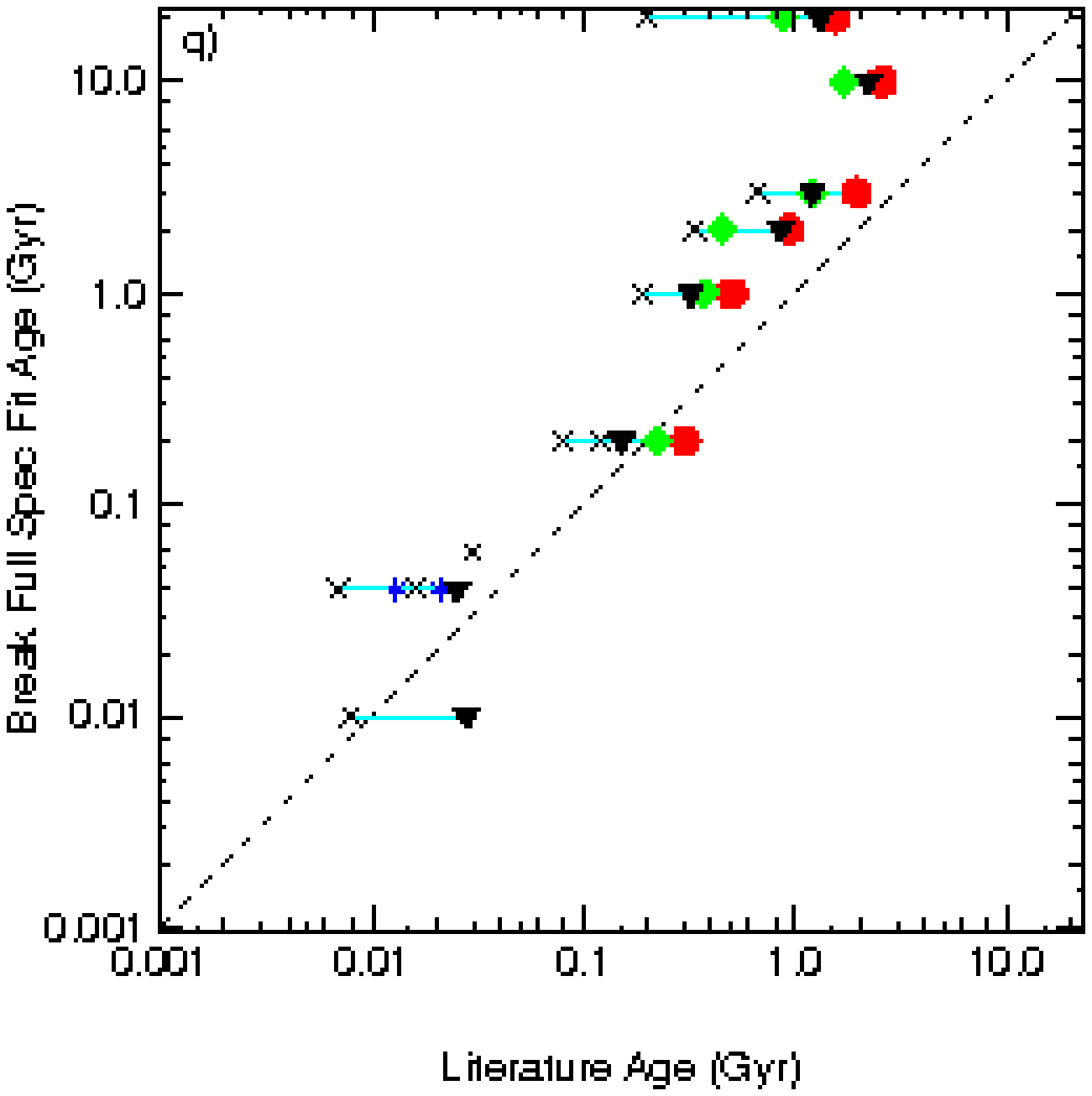}{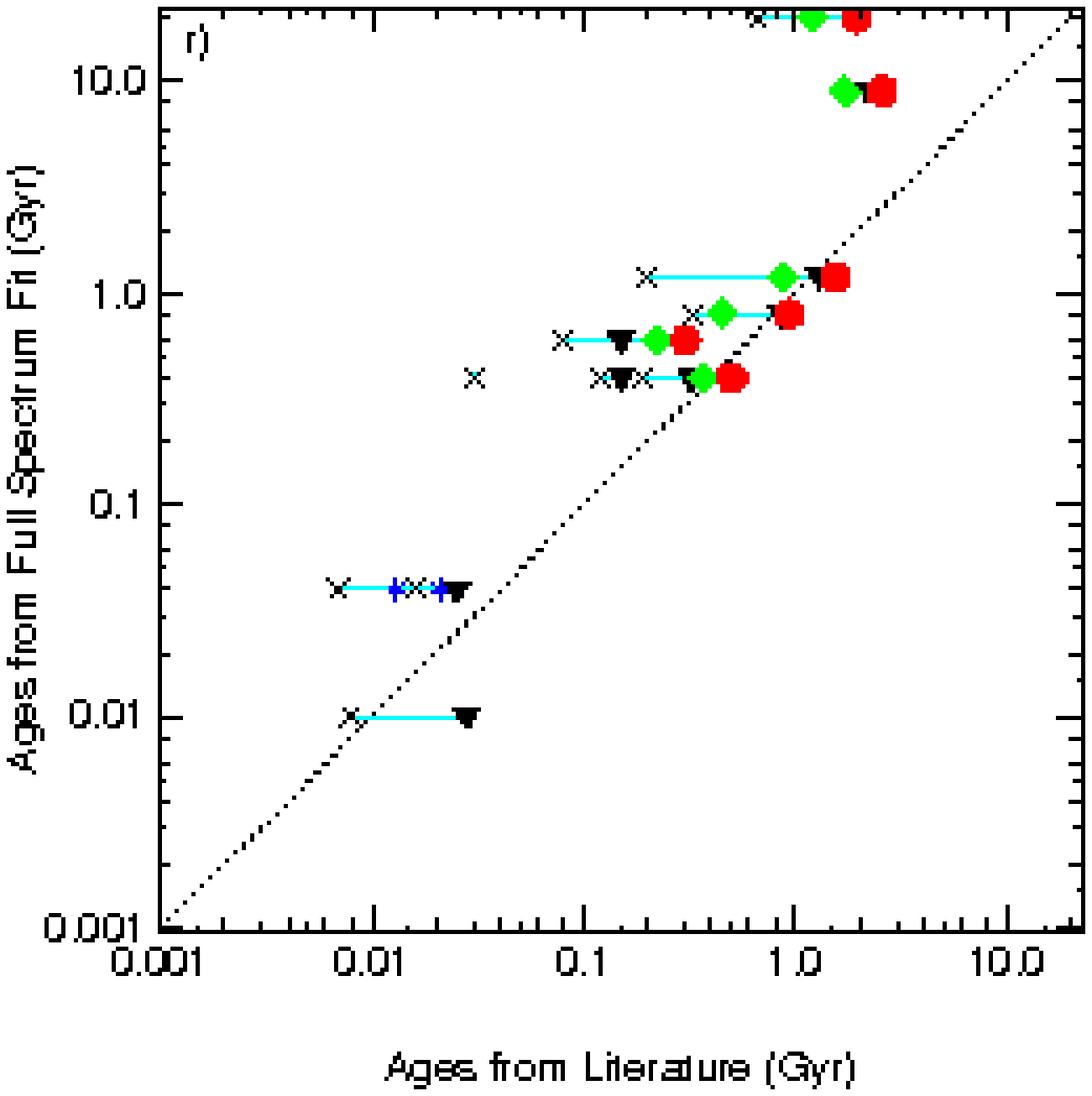}
\caption{cont.}
\end{figure}

\begin{figure}
\figurenum{18}
    \includegraphics[scale=.4,angle=-90]{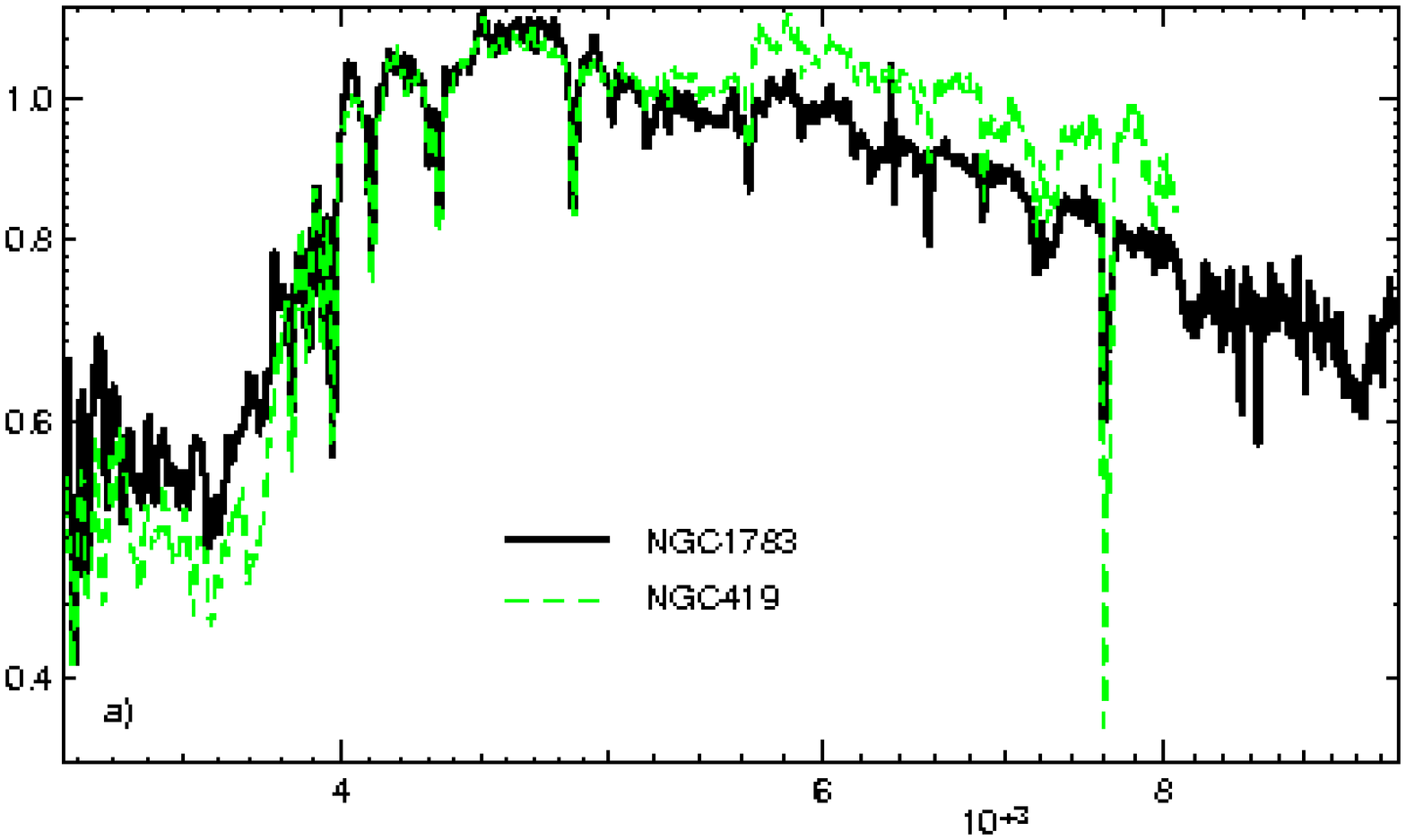}
    \includegraphics[scale=.4,angle=-90]{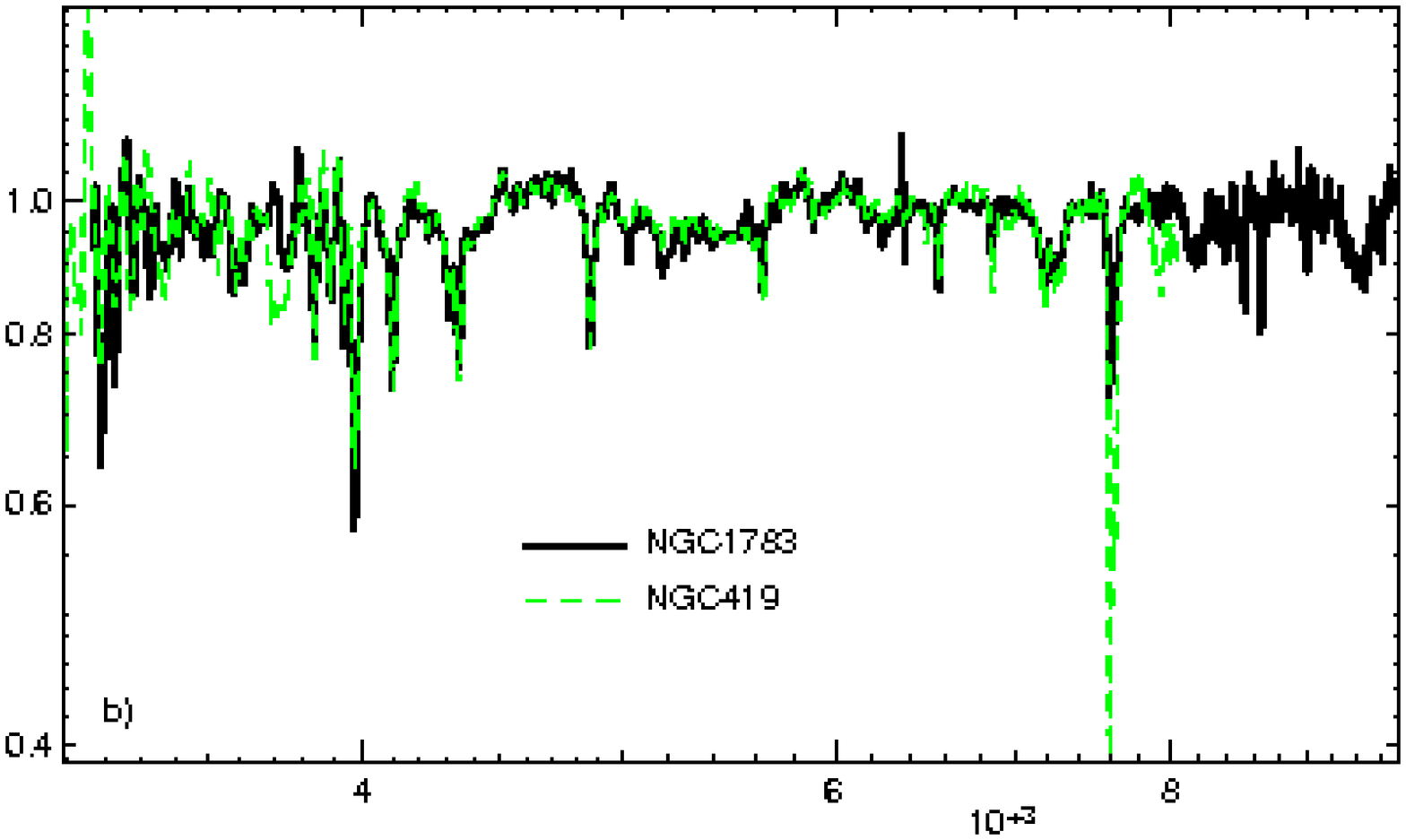}
    \includegraphics[scale=.4,angle=-90]{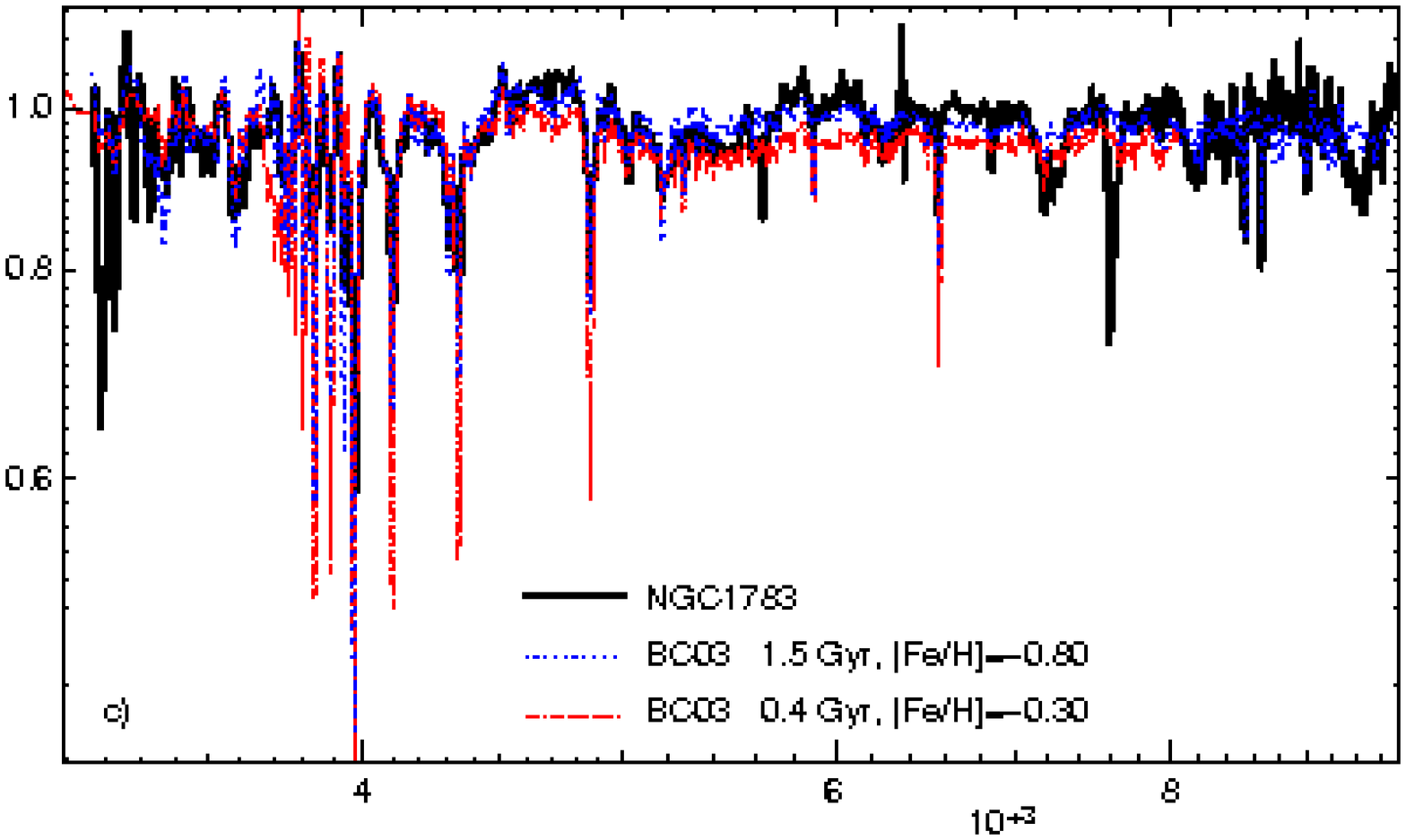}
    \includegraphics[scale=.4,angle=-90]{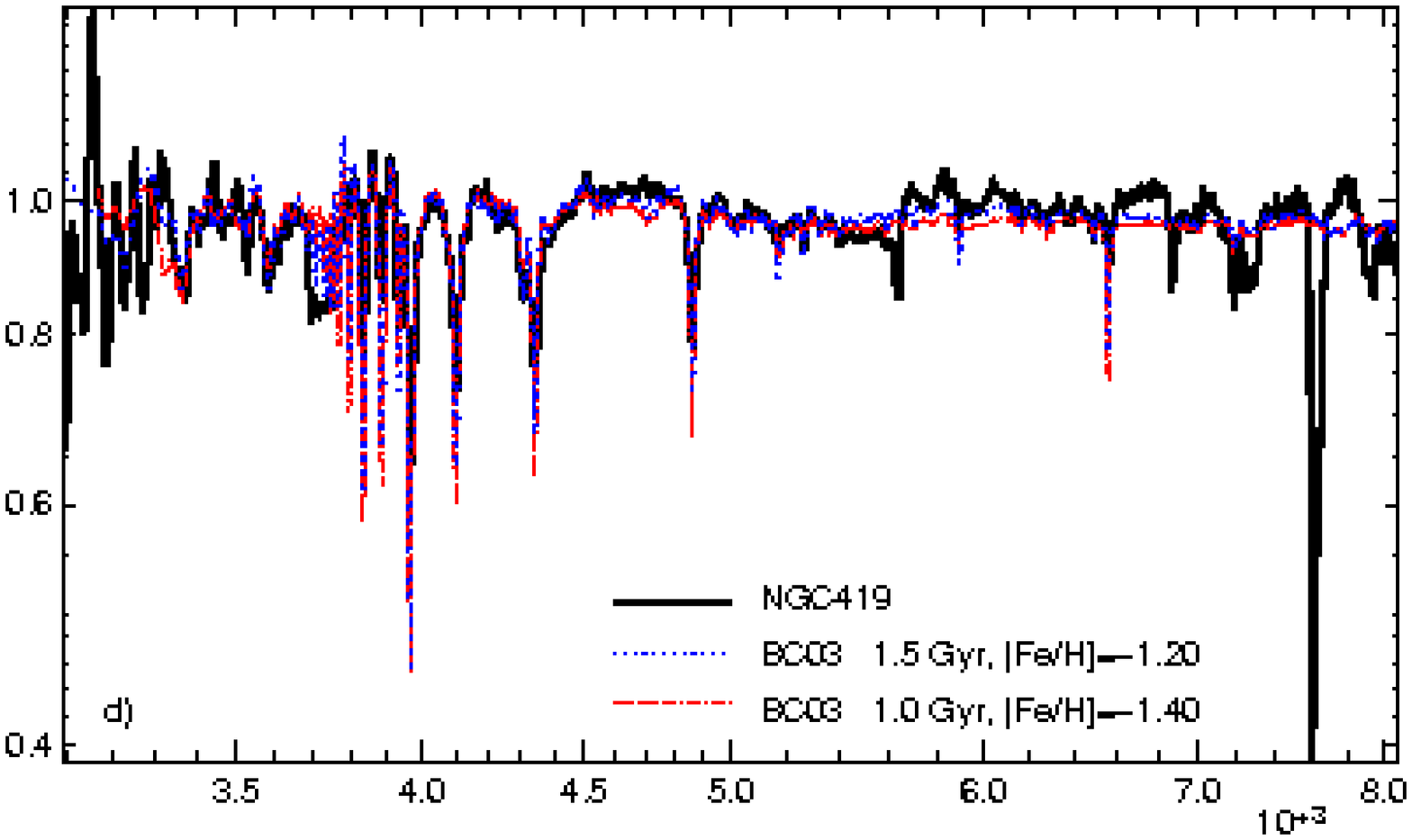}
  \caption{NGC 1783 and NGC 419 were two clusters in common with
  \citet{maraston05} who fit enhanced TP-AGB star models to the
  photometric SEDs. a) shows the two cluster spectra from
  \citet{santos02}, b) shows their CN spectra, and c\&d) include BC03
  model CN spectra. Our best fitting model to NGC 1783 was 1.5 Gyr and
  [Fe/H]=$-$0.8 and to NGC~419 was 1.5 Gyr and [Fe/H]=$-$1.2
  (\textit{blue dotted lines}). Maraston's best fit to the photometry
  of NGC 1783 was 0.3 Gyr and [Fe/H]=$-$0.33 (the nearest model in our
  grid was 0.4 Gyr and [Fe/H]=$-$0.3) and of NGC~419 was 1 Gyr and
  [Fe/H]=$-$1.35 (\textit{red dot-dashed lines}). \label{fig18}}
\end{figure}

%% Tables start here.

\clearpage
\begin{deluxetable}{lcccccccccc}
\tabletypesize{\scriptsize}
%% \rotate
\tablecaption{Cluster Wavelength Coverage and Resolution \label{res_tbl}}
\tablewidth{0pt}
\tablehead{
\colhead{} & \colhead{UV} & \colhead{UV} & \colhead{UV} & \colhead{Optical} & \colhead{Optical} & \colhead{Optical} & \colhead{IR} & \colhead{IR} & \colhead{IR} & \colhead{}  \\
\colhead{Object} & \colhead{$\lambda_{start}$} & \colhead{$\lambda_{end}$} & \colhead{Resolution} & \colhead{$\lambda_{start}$} & \colhead{$\lambda_{end}$} & \colhead{Resolution} & \colhead{$\lambda_{start}$} & \colhead{$\lambda_{end}$} & \colhead{Resolution} & \colhead{Refs}  \\

}
\startdata

\textbf{LMC} \\
\hline
H88-267 & \nodata & \nodata & \nodata & 3500 & 5870 & 16 & \nodata & \nodata & \nodata & 1 \\
HD269828 & \nodata & \nodata & \nodata & 3500 & 5870 & 16 & \nodata & \nodata & \nodata & 1 \\
HD32228 & \nodata & \nodata & \nodata & 3500 & 5812 & 16 & \nodata & \nodata & \nodata & 1 \\
IC2105 & \nodata & \nodata & \nodata & 3500 & 4704 & 12 & \nodata & \nodata & \nodata & 1 \\
IC2111 & \nodata & \nodata & \nodata & 3500 & 4704 & 12 & \nodata & \nodata & \nodata & 1 \\
IC2116 & \nodata & \nodata & \nodata & 3500 & 5878 & 16 & \nodata & \nodata & \nodata & 1 \\
KMHK1074 & \nodata & \nodata & \nodata & 3500 & 5878 & 16 & \nodata & \nodata & \nodata & 1 \\
NGC1466 & 3000 & 5652 & 15 & \nodata & \nodata & \nodata & \nodata & \nodata & \nodata & 2 \\
NGC1711 & \nodata & \nodata & \nodata & 3500 & 5878 & 16 & 5600 & 10000 & 14 & 1,3 \\
NGC1714 & \nodata & \nodata & \nodata & 3700 & 8200 & 11 & 6365 & 9809 & 12.5 & 1,4,5 \\
NGC1731 & \nodata & \nodata & \nodata & 3500 & 4704 & 12 & \nodata & \nodata & \nodata & 1 \\
NGC1735 & \nodata & \nodata & \nodata & 3500 & 4704 & 12 & 5600 & 10000 & 14 & 1,3 \\
NGC1736 & \nodata & \nodata & \nodata & 3500 & 4704 & 12 & \nodata & \nodata & \nodata & 1 \\
NGC1743 & \nodata & \nodata & \nodata & 3500 & 4740 & 12 & \nodata & \nodata & \nodata & 1 \\
NGC1755 & \nodata & \nodata & \nodata & 3500 & 4704 & 12 & 5600 & 10000 & 14 & 1,3 \\
NGC1761 & \nodata & \nodata & \nodata & 3500 & 5812 & 16 & \nodata & \nodata & \nodata & 1 \\
NGC1767 & \nodata & \nodata & \nodata & 3400 & 9700 & 12 & 5656 & 10112 & 14 & 1,3,6 \\
NGC1772 & \nodata & \nodata & \nodata & 3500 & 5870 & 16 & \nodata & \nodata & \nodata & 1 \\
NGC1774 & \nodata & \nodata & \nodata & 3500 & 4704 & 12 & 5600 & 10000 & 14 & 1,3 \\
NGC1782 & \nodata & \nodata & \nodata & 3500 & 4702 & 12 & 5600 & 10000 & 14 & 1,3 \\
NGC1783 & 3000 & 5654 & 15 & 3700 & 8200 & 11 & 6365 & 9809 & 12.5 & 2,4,5 \\
NGC1805 & \nodata & \nodata & \nodata & 3500 & 5878 & 16 & 5600 & 10000 & 14 & 1,3 \\
NGC1818 & \nodata & \nodata & \nodata & 3500 & 4704 & 12 & 5600 & 10000 & 14 & 1,3 \\
NGC1831 & 3000 & 5652 & 15 & 3700 & 8200 & 11 & 5656 & 9800 & 14 & 2,3,5 \\
NGC1847 & 3000 & 5652 & 15 & 3700 & 8200 & 11 & 5656 & 9796 & 14 & 2,1,3,5 \\
NGC1850 & \nodata & \nodata & \nodata & 3500 & 5878 & 16 & 5600 & 10000 & 14 & 1,3 \\
NGC1850A & \nodata & \nodata & \nodata & 3500 & 5878 & 16 & \nodata & \nodata & \nodata & 1 \\
NGC1854 & \nodata & \nodata & \nodata & 3500 & 5878 & 16 & 5600 & 10000 & 14 & 1,3 \\
NGC1856 & 3000 & 5654 & 15 & 3700 & 8200 & 11 & 5656 & 9800 & 14 & 2,3,5 \\
NGC1866 & 3000 & 5654 & 15 & 3700 & 8200 & 11 & 5656 & 9800 & 14 & 2,3,5 \\
NGC1868 & 3000 & 5654 & 15 & 3700 & 8200 & 11 & 5656 & 9800 & 14 & 2,3,5 \\
NGC1872 & \nodata & \nodata & \nodata & 3500 & 4704 & 12 & \nodata & \nodata & \nodata & 1 \\
NGC1895 & \nodata & \nodata & \nodata & 3700 & 8200 & 11 & 6365 & 9809 & 12.5 & 4,5 \\
NGC1903 & \nodata & \nodata & \nodata & 3500 & 4704 & 12 & 5600 & 10000 & 14 & 1,3 \\
NGC1921 & \nodata & \nodata & \nodata & 3500 & 4704 & 12 & \nodata & \nodata & \nodata & 1 \\
NGC1935 & \nodata & \nodata & \nodata & 3500 & 4704 & 12 & \nodata & \nodata & \nodata & 1 \\
NGC1936 & \nodata & \nodata & \nodata & 3500 & 4702 & 12 & \nodata & \nodata & \nodata & 1 \\
NGC1949 & \nodata & \nodata & \nodata & 3500 & 4704 & 12 & \nodata & \nodata & \nodata & 1 \\
NGC1951 & \nodata & \nodata & \nodata & 3500 & 4702 & 12 & 5600 & 10000 & 14 & 1,3 \\
NGC1967 & \nodata & \nodata & \nodata & 3500 & 5870 & 16 & \nodata & \nodata & \nodata & 1 \\
NGC1970 & \nodata & \nodata & \nodata & 3500 & 4704 & 12 & \nodata & \nodata & \nodata & 1 \\
NGC1978 & 3000 & 5654 & 15 & 3700 & 8200 & 11 & 6365 & 9811 & 12.5 & 2,4,5 \\
NGC1983 & \nodata & \nodata & \nodata & 3500 & 5874 & 16 & \nodata & \nodata & \nodata & 1 \\
NGC1984 & \nodata & \nodata & \nodata & 3500 & 5870 & 16 & \nodata & \nodata & \nodata & 1 \\
NGC1994 & \nodata & \nodata & \nodata & 3500 & 5870 & 16 & 5600 & 10000 & 14 & 1,3 \\
NGC2002 & \nodata & \nodata & \nodata & 3500 & 5870 & 16 & 5600 & 10000 & 14 & 1,3 \\
NGC2003 & \nodata & \nodata & \nodata & 3400 & 9700 & 12 & 5656 & 10112 &  & 3,6 \\
NGC2004 & 3000 & 5654 & 15 & 3700 & 8200 & 11 & 5656 & 9800 & 14 & 2,3,5 \\
NGC2006 & \nodata & \nodata & \nodata & 3500 & 5874 & 16 & 5600 & 10000 & 14 & 1,3 \\
NGC2009 & \nodata & \nodata & \nodata & 3500 & 5878 & 16 & \nodata & \nodata & \nodata & 1 \\
NGC2011 & \nodata & \nodata & \nodata & 3500 & 5870 & 16 & 5600 & 10000 & 14 & 1,3 \\
NGC2021 & \nodata & \nodata & \nodata & 3500 & 5878 & 16 & \nodata & \nodata & \nodata & 1 \\
NGC2025 & \nodata & \nodata & \nodata & 3500 & 4704 & 12 & 5600 & 10000 & 14 & 1,3 \\
NGC2032 & \nodata & \nodata & \nodata & 3500 & 4704 & 12 & \nodata & \nodata & \nodata & 1 \\
NGC2033 & \nodata & \nodata & \nodata & 3500 & 5878 & 16 & \nodata & \nodata & \nodata & 1 \\
NGC2035 & \nodata & \nodata & \nodata & 3500 & 4704 & 12 & \nodata & \nodata & \nodata & 1 \\
NGC2041 & \nodata & \nodata & \nodata & 3500 & 4704 & 12 & 5600 & 10000 & 14 & 1,3 \\
NGC2058 & \nodata & \nodata & \nodata & 3500 & 4704 & 12 & 5600 & 10000 & 14 & 1,3 \\
NGC2070 & 3000 & 5654 & 15 & 3700 & 8200 & 11 & \nodata & \nodata & \nodata & 2,5 \\
NGC2091 & \nodata & \nodata & \nodata & 3500 & 5878 & 16 & \nodata & \nodata & \nodata & 1 \\
NGC2092 & \nodata & \nodata & \nodata & 3500 & 5878 & 16 & \nodata & \nodata & \nodata & 1 \\
NGC2096 & \nodata & \nodata & \nodata & 3500 & 5874 & 16 & \nodata & \nodata & \nodata & 1 \\
NGC2098 & \nodata & \nodata & \nodata & 3500 & 5870 & 16 & 5600 & 10000 & 14 & 1,3 \\
NGC2100 & \nodata & \nodata & \nodata & 3700 & 8200 & 11 & 5656 & 10112 & 14 & 1,3,5 \\
NGC2102 & \nodata & \nodata & \nodata & 3500 & 5878 & 16 & \nodata & \nodata & \nodata & 1 \\
NGC2134 & \nodata & \nodata & \nodata & 3500 & 4704 & 12 & 5600 & 10000 & 14 & 1,3 \\
NGC2136 & \nodata & \nodata & \nodata & 3500 & 4704 & 12 & 5600 & 10000 & 14 & 1,3 \\
NGC2147 & \nodata & \nodata & \nodata & 3500 & 5878 & 16 & \nodata & \nodata & \nodata & 1 \\
NGC2156 & \nodata & \nodata & \nodata & 3500 & 4704 & 12 & \nodata & \nodata & \nodata & 1 \\
NGC2157 & 3000 & 5654 & 15 & 3700 & 8200 & 11 & 5656 & 9800 & 14 & 2,3,5 \\
NGC2159 & \nodata & \nodata & \nodata & 3500 & 4704 & 12 & \nodata & \nodata & \nodata & 1 \\
NGC2164 & \nodata & \nodata & \nodata & 3500 & 4704 & 12 & 5600 & 10000 & 14 & 1,3 \\
NGC2214 & \nodata & \nodata & \nodata & 3500 & 4740 & 12 & 5600 & 10000 & 14 & 1,3 \\
SL153 & \nodata & \nodata & \nodata & 3500 & 4704 & 12 & \nodata & \nodata & \nodata & 1 \\
SL492 & \nodata & \nodata & \nodata & 3500 & 5870 & 16 & \nodata & \nodata & \nodata & 1 \\
SL538 & \nodata & \nodata & \nodata & 3500 & 5870 & 16 & 5600 & 10000 & 14 & 1,3 \\
SL586 & \nodata & \nodata & \nodata & 3500 & 5878 & 16 & \nodata & \nodata & \nodata & 1 \\
SL601 & \nodata & \nodata & \nodata & 3500 & 4704 & 12 & \nodata & \nodata & \nodata & 1 \\
SL639 & \nodata & \nodata & \nodata & 3500 & 5882 & 16 & \nodata & \nodata & \nodata & 1 \\
\hline
\textbf{SMC} \\
\hline
NGC121 & \nodata & \nodata & \nodata & 3700 & 6800 & 11 & \nodata & \nodata & \nodata & 5 \\
NGC299 & \nodata & \nodata & \nodata & 3500 & 5877 & 16 & 5600 & 10000 & 14 & 1,3 \\
NGC330 & 3000 & 5656 & 15 & 3700 & 8200 & 11 & 5658 & 10116 & 14 & 2,1,3,5 \\
NGC419 & 3000 & 5654 & 15 & 3700 & 8200 & 11 & \nodata & \nodata & \nodata & 2,5 \\
\hline
\textbf{M31} \\
\hline
Bol104 & \nodata & \nodata & \nodata & 3200 & 6100 & 6 & \nodata & \nodata & \nodata & 7 \\
Bol118 & \nodata & \nodata & \nodata & 3200 & 6100 & 6 & \nodata & \nodata & \nodata & 7 \\
Bol119 & \nodata & \nodata & \nodata & 3200 & 6100 & 6 & \nodata & \nodata & \nodata & 7 \\
Bol124 & \nodata & \nodata & \nodata & 3200 & 6100 & 6 & \nodata & \nodata & \nodata & 7 \\
G1 & \nodata & \nodata & \nodata & 3600 & 9900 & 23 & \nodata & \nodata & \nodata & 8 \\
G78 & \nodata & \nodata & \nodata & 3600 & 9900 & 23 & \nodata & \nodata & \nodata & 8 \\
G158 & \nodata & \nodata & \nodata & 3600 & 9900 & 23 & \nodata & \nodata & \nodata & 7,8 \\
G165 & \nodata & \nodata & \nodata & 3200 & 6100 & 6 & \nodata & \nodata & \nodata & 7 \\
G168 & \nodata & \nodata & \nodata & 3200 & 6100 & 6 & \nodata & \nodata & \nodata & 7 \\
G169 & \nodata & \nodata & \nodata & 3200 & 6100 & 6 & \nodata & \nodata & \nodata & 7 \\
G170 & \nodata & \nodata & \nodata & 3600 & 9900 & 23 & \nodata & \nodata & \nodata & 8 \\
G174 & \nodata & \nodata & \nodata & 3200 & 6100 & 6 & \nodata & \nodata & \nodata & 7 \\
G175 & \nodata & \nodata & \nodata & 3200 & 6100 & 6 & \nodata & \nodata & \nodata & 7 \\
G177 & \nodata & \nodata & \nodata & 3600 & 9900 & 23 & \nodata & \nodata & \nodata & 7,8 \\
G184 & \nodata & \nodata & \nodata & 3200 & 6100 & 6 & \nodata & \nodata & \nodata & 7 \\
G185 & \nodata & \nodata & \nodata & 3200 & 6100 & 6 & \nodata & \nodata & \nodata & 7 \\
G190 & \nodata & \nodata & \nodata & 3200 & 6100 & 6 & \nodata & \nodata & \nodata & 7 \\
G222 & \nodata & \nodata & \nodata & 3600 & 7934 & 8 & \nodata & \nodata & \nodata & 8 \\

\enddata 

\tablerefs{(1) \citet{santos95}, (2) \citet{bica94}, (3)
\citet{bica90}, (4) \citet{bica87a}, (5) \citet{bica86a}, (6)
unpublished, \citet{santos02}, (7) \citet{jablonka98}, (8)
\citet{jablonka92}}

\end{deluxetable}

 \clearpage

 \begin{deluxetable}{lcccccccccc}
 \tabletypesize{\scriptsize} \rotate \tablecaption{Offsets and
 Dispersions of Derived Cluster Ages \label{rms_age_tbl}}
 \tablewidth{0pt} 

\tablehead{ 

  \colhead{} & \colhead{(1)} & \colhead{(2)} & \colhead{(3)} &
 \colhead{(4)} & \colhead{(5)} & \colhead{(6)} & \colhead{(7)} &
 \colhead{(8)} & \colhead{(9)} & \colhead{(10)} \\

 \colhead{} & \colhead{} & \colhead{All} & \colhead{} & \colhead{} &
  \colhead{UBV} & \colhead{Line Index} & \colhead{Homog.} &
 \colhead{BC96 Phot.} & \colhead{Old} & \colhead{Young} \\

 \colhead{} & \colhead{All} & \colhead{Good} & \colhead{MSTO} &
 \colhead{CMDs\tablenotemark{a}} & \colhead{Colors} & \colhead{Ratios}
 & \colhead{Scale} & \colhead{Fits} & \colhead{$\geq$ 1 Gyr} &
 \colhead{$<$ 1 Gyr}
 
}

 \startdata 

 Locations & MC, M31 & MC, M31  & MC & MC & MC & MC & MC  & M31 & MC,
 M31 & MC\\

 References & 1,2,3,4,5,6 & 1,2,3,4,5,6 & 1 & 3,5 & 2 & 5 & 6 & 4 &
 1,3,4,5,6 & 1,2,3,5,6 \\

 No. objects & 75 & 73 & 29 & 11 & 54 & 11 & 18 & 8 & 4 & 61 \\

 \hline
%% \sidehead{\textbf{CN Spectrum Fits}}
 \textbf{CN Spectrum Fits} \\
 Avg Offset (Gyr) & 0.84 & 0.19 & 0.12 & 0.56 & 0.03 & 0.26 & 0.59 &
 4.36 & 1.63 & 0.09 \\
 Avg Offset (dex) & 0.34 & 0.29 & 0.36 & 0.21 & 0.29 & 0.22 & 0.31 & 0.25 & 0.16 & 0.30 \\
 RMS (Gyr) & 3.18 & 1.00 & 0.82 & 1.06 & 0.12 & 0.79 & 1.82 & 8.34 &
 3.78 & 0.35 \\
 RMS (dex) & 0.57 & 0.36 & 0.45 & 0.31 & 0.37 & 0.27 & 0.40 & 0.46 & 0.17 & 0.37 \\

 \hline
%% \sidehead{\textbf{Full Spectrum Fits}} 
 \textbf{Full Spectrum Fits} \\
 Avg Offset (Gyr) & 0.93 & 0.54 & 0.71 & 2.54 & 0.42 & 2.24 & 0.70 &
 3.24 & 2.76 & 0.39  \\
 Avg Offset (dex) & 0.70 & 0.69 & 0.64 & 0.36 & 0.78 & 0.33 & 0.52 & 0.36 & 0.70 & 0.69 \\
 RMS (Gyr) & 4.42 & 3.20 & 3.68 & 6.08 & 1.56 & 5.78 & 5.43 & 9.61 &
 11.55 & 1.47 \\
 RMS (dex) & 0.93 & 0.90 & 0.71 & 0.51 & 0.98 & 0.44 & 0.64 & 0.42 & 0.79 & 0.91 \\

 \hline
%% \sidehead{\textbf{Photometry Fits}} 
 \textbf{Photometry Fits} \\
 Avg Offset (Gyr) & 0.99 & 0.48 & 0.73 & 0.67 & 0.34 & 0.37 & 0.74 &
 5.24 & 2.06 & 0.38 \\
 Avg Offset (dex) & 0.004 & 0.59 & 0.71 & 0.42 & 0.63 & 0.38 & 0.41 & 0.32 & 0.10 & 0.62 \\
 RMS (Gyr) & 3.29 & 1.19 & 1.55 & 0.85 & 1.04 & 0.61 & 1.42 & 9.47 &
 2.67 & 1.02 \\
 RMS (dex) & 0.52 & 0.76 & 0.90 & 0.52 & 0.80 & 0.48 & 0.49 & 0.49 & 0.11 & 0.78 \\

 \hline
%% \sidehead{\textbf{Continuum Fits}}
 \textbf{Continuum Fits} \\
 Avg Offset (Gyr) & 0.87 & 1.17 & 2.05 & 4.51 & 0.46 & 4.21 & 2.90 &
 -2.49 & 6.06 & 0.85 \\
 Avg Offset (dex) & 0.06 & 0.80 & 0.84 & 0.49 & 0.84 & 0.46 & 0.73 & 0.55 & 0.80 & 0.80 \\
 RMS (Gyr) & 5.44 & 4.30 & 5.51 & 8.66 & 1.57 & 8.29 & 7.71 & 10.88 &
 12.72 & 3.02 \\
 RMS (dex) & 0.74 & 1.03 & 1.05 & 0.68 & 1.04 & 0.58 & 0.93 & 0.65 & 0.73 & 1.04 \\

 \enddata 

 \tablecomments{Column 1 includes all clusters with literature ages
   and column 2 has excluded the two outliers in the upper left plot
   of Figure~\ref{fig5} and the objects that only have BC96
   photometric fits from the literature. All columns to the right,
   except for column 8, also exclude these points.}

\tablenotetext{a}{These age estimates were collected from
   \citet{girardi95,girardi98} and \citet{elson88} by
   \citet{leonardi03} for comparison to their estimates from line
   index ratios.}

\tablerefs{(1) \citet{hodge83},
   (2) \citet{santos95}, (3) \citet{girardi95,girardi98}, (4)
   \citet{jiang03}, (5) \citet{leonardi03}, (6) \citet{santos04}}

 \end{deluxetable}

 \clearpage

\begin{deluxetable}{lccccccccc}
 \tabletypesize{\scriptsize} \rotate \tablecaption{Offsets and
 Dispersions of Derived Cluster [Fe/H] \label{rms_met_tbl}}
 \tablewidth{0pt} 

\tablehead{ 

  \colhead{} & \colhead{(1)} & \colhead{(2)} & \colhead{(3)} &
 \colhead{(4)} & \colhead{(5)} & \colhead{(6)} & \colhead{(7)} &
 \colhead{(8)} & \colhead{(9)} \\

 \colhead{} & \colhead{} & \colhead{Line} & \colhead{VJK} & \colhead{BC96} &
  \colhead{LIR} & \colhead{Line Index} &
 \colhead{Homog.} & \colhead{Old} & \colhead{Young} \\

 \colhead{} & \colhead{All} & \colhead{Indices} & \colhead{Colors} &
 \colhead{Colors} & \colhead{Lit Compare\tablenotemark{a}} & \colhead{Ratios}
 & \colhead{Scale} & \colhead{$\geq$ 1 Gyr} &
 \colhead{$<$ 1 Gyr}
 
}

 \startdata 

 Locations & MC, M31 & M31 & M31 & M31 & MC & MC & MC  & MC, M31 & MC \\

 References & 1,2,3,4,5,6 & 1 & 2 & 3 & 4 & 5 & 6 & 1,2,3,4,5,6 &
 4,5,6 \\

 No. objects & 33 & 6 & 9 & 8 & 11 & 5 & 18 & 12 & 21  \\

 \hline
%% \sidehead{\textbf{CN Spectrum Fits}}
 \textbf{CN Spectrum Fits} \\
 Avg Offset (Gyr) & -0.23 & 0.11 & 0.25 & -0.14 & -0.68 & -0.69 & -0.37 &
 -0.10 & -0.29  \\
 RMS (Gyr) & 0.85 & 0.12 & 0.52 & 0.71 & 1.06 & 0.81 & 0.91 & 0.41 &
 1.02  \\

 \hline
%% \sidehead{\textbf{Full Spectrum Fits}} 
 \textbf{Full Spectrum Fits} \\
 Avg Offset (Gyr) & 0.23 & 0.31 & 0.88 & 0.40 & -0.27 & -0.05 & 0.16 &
 0.48 & 0.10  \\
 RMS (Gyr) & 0.84 & 0.50 & 1.03 & 0.59 & 0.89 & 0.53 & 0.71 & 0.78 &
 0.88  \\

 \hline
%% \sidehead{\textbf{Photometry Fits}} 
 \textbf{Photometry Fits} \\
 Avg Offset (Gyr) & 0.28 & 0.62 & 0.77 & 0.41 & -0.18 & -0.11 & 0.15 &
 0.51 & 0.14  \\
 RMS (Gyr) & 0.87 & 0.74 & 0.84 & 0.67 & 1.10 & 0.89 & 0.72 & 0.68 &
 0.96  \\

 \hline
%% \sidehead{\textbf{Continuum Fits}}
 \textbf{Continuum Fits} \\
 Avg Offset (Gyr) & 0.36 & 0.81 & 1.34 & 0.85 & -0.29 & -0.35 & 0.12 &
 0.85 & 0.08  \\
 RMS (Gyr) & 1.11 & 1.03 & 1.49 & 0.92 & 1.02 & 0.92 & 0.98 & 1.09 &
 1.13  \\

 \enddata 

\tablecomments{Column 1 includes all clusters with literature [Fe/H].}

\tablenotetext{a}{The \citet{leonardi03} comparison literature values
  come from \citet{olszewski91}, \citet{cohen82}, \citet{seggewiss89},
  \citet{sagar89}, and \citet{piatti02}.}

\tablerefs{(1) \citet{huchra91}, (2) \citet{cohen94}, (3)
  \citet{jiang03}, (4) \citet{leonardi03} literature comparison
  values, (5) \citet{leonardi03}, (6) \citet{santos04}}

\end{deluxetable}

\clearpage
\begin{deluxetable}{lccccccccccccc}
\tabletypesize{\scriptsize}
\rotate
\tablecaption{Cluster Ages and Metallicities \label{age_tbl}}
\tablewidth{0pt}
\tablehead{
\colhead{} & \colhead{CN Spec} & \colhead{CN Spec} & \colhead{Full Spec} & \colhead{UBV\tablenotemark{a}} & \colhead{MSTO\tablenotemark{b}} & \colhead{LIR\tablenotemark{c}} & \colhead{HLit\tablenotemark{d}} & \colhead{BC96\tablenotemark{e}} & \colhead{LI\tablenotemark{f}} & \colhead{VJK\tablenotemark{g}} & \colhead{LIR lit\tablenotemark{h}} & \colhead{LIR\tablenotemark{c}} & \colhead{HLit\tablenotemark{d}} \\
\colhead{Object} & \colhead{Age} & \colhead{[Fe/H]} & \colhead{[Fe/H]} & \colhead{Age} & \colhead{Age} & \colhead{Age} & \colhead{Age} & \colhead{Age} & \colhead{[Fe/H]} & \colhead{[Fe/H]} & \colhead{[Fe/H]} & \colhead{[Fe/H]} & \colhead{[Fe/H]} \\
}
\startdata

\textbf{LMC} \\
\hline
H88-267 & ${ 0.001}_{- 0.000}^{+ 0.001}$ & ${-2.20}_{-0.00}^{+0.05}$ & ${-1.80}_{-0.05}^{+0.05}$ &  0.005 &  \nodata &  \nodata &  \nodata &  \nodata & \nodata & \nodata & \nodata & \nodata & \nodata \\
HD32228 & ${ 0.002}_{- 0.001}^{+ 0.001}$ & ${0.50}_{-0.05}^{+0.00}$ & ${-2.20}_{-0.00}^{+0.05}$ &  0.004 &  \nodata &  \nodata &  \nodata &  \nodata & \nodata & \nodata & \nodata & \nodata & \nodata \\
HD269828 & ${ 0.002}_{- 0.001}^{+ 0.001}$ & ${-0.70}_{-0.10}^{+0.05}$ & ${-1.10}_{-0.05}^{+0.05}$ &  0.004 &  \nodata &  \nodata &  \nodata &  \nodata & \nodata & \nodata & \nodata & \nodata & \nodata \\
IC2105 & ${ 0.006}_{- 0.001}^{+ 0.034}$ & ${-2.20}_{-0.00}^{+0.10}$ & ${-2.00}_{-0.10}^{+0.05}$ &  \nodata &  \nodata &  \nodata &  \nodata &  \nodata & \nodata & \nodata & \nodata & \nodata & \nodata \\
IC2111 & ${ 0.006}_{- 0.001}^{+ 0.001}$ & ${-1.60}_{-0.05}^{+0.05}$ & ${-1.90}_{-0.05}^{+0.05}$ &  \nodata &  \nodata &  \nodata &  \nodata &  \nodata & \nodata & \nodata & \nodata & \nodata & \nodata \\
IC2116 & ${ 0.004}_{- 0.001}^{+ 0.001}$ & ${0.50}_{-0.05}^{+0.00}$ & ${-1.20}_{-0.05}^{+0.05}$ &  \nodata &  \nodata &  \nodata &  \nodata &  \nodata & \nodata & \nodata & \nodata & \nodata & \nodata \\
KMHK1074 & ${ 0.004}_{- 0.001}^{+ 0.001}$ & ${-0.10}_{-0.05}^{+0.05}$ & ${0.10}_{-0.05}^{+0.05}$ &  0.004 &  \nodata &  \nodata &  \nodata &  \nodata & \nodata & \nodata & \nodata & \nodata & \nodata \\
NGC1466 & ${20.000}_{- 0.500}^{+ 0.000}$ & ${-2.00}_{-0.05}^{+0.05}$ & ${-1.30}_{-0.05}^{+0.05}$ &  \nodata &  \nodata &  \nodata & 13.100 &  \nodata & \nodata & \nodata & \nodata & \nodata & -1.64 \\
NGC1711 & ${ 0.080}_{- 0.010}^{+ 0.010}$ & ${-1.40}_{-0.05}^{+0.20}$ & ${-0.20}_{-0.05}^{+0.05}$ &  0.020 &  \nodata &  \nodata &  0.068 &  \nodata & \nodata & \nodata & \nodata & \nodata & -0.68 \\
NGC1714 & ${20.000}_{- 1.000}^{+ 0.000}$ & ${0.40}_{-0.05}^{+0.05}$ & ${0.30}_{-0.05}^{+0.05}$ &  \nodata &  \nodata &  \nodata &  \nodata &  \nodata & \nodata & \nodata & \nodata & \nodata & \nodata \\
NGC1731 & ${ 0.001}_{- 0.000}^{+ 0.001}$ & ${-2.20}_{-0.00}^{+0.05}$ & ${-1.10}_{-0.05}^{+0.05}$ &  0.004 &  \nodata &  \nodata &  \nodata &  \nodata & \nodata & \nodata & \nodata & \nodata & \nodata \\
NGC1735 & ${ 0.040}_{- 0.010}^{+ 0.010}$ & ${-2.20}_{-0.00}^{+0.05}$ & ${-1.60}_{-0.05}^{+0.05}$ &  0.032 &  \nodata &  \nodata &  \nodata &  \nodata & \nodata & \nodata & \nodata & \nodata & \nodata \\
NGC1736 & ${ 0.002}_{- 0.001}^{+ 0.001}$ & ${-1.90}_{-0.10}^{+0.30}$ & ${-1.70}_{-0.10}^{+0.05}$ &  \nodata &  \nodata &  \nodata &  \nodata &  \nodata & \nodata & \nodata & \nodata & \nodata & \nodata \\
NGC1743 & ${ 0.010}_{- 0.008}^{+ 0.005}$ & ${-2.20}_{-0.00}^{+0.80}$ & ${-0.70}_{-0.10}^{+0.05}$ &  \nodata &  \nodata &  \nodata &  \nodata &  \nodata & \nodata & \nodata & \nodata & \nodata & \nodata \\
NGC1755 & ${ 0.100}_{- 0.010}^{+ 0.050}$ & ${-1.70}_{-0.05}^{+0.05}$ & ${-1.30}_{-0.10}^{+0.05}$ &  0.055 &  \nodata &  \nodata &  \nodata &  \nodata & \nodata & \nodata & \nodata & \nodata & \nodata \\
NGC1761 & ${ 0.010}_{- 0.001}^{+ 0.005}$ & ${-2.20}_{-0.00}^{+0.05}$ & ${0.50}_{-0.05}^{+0.00}$ &  \nodata &  \nodata &  \nodata &  \nodata &  \nodata & \nodata & \nodata & \nodata & \nodata & \nodata \\
NGC1767 & ${ 0.006}_{- 0.001}^{+ 0.001}$ & ${0.10}_{-0.05}^{+0.05}$ & ${0.20}_{-0.05}^{+0.05}$ &  0.007 &  \nodata &  \nodata &  \nodata &  \nodata & \nodata & \nodata & \nodata & \nodata & \nodata \\
NGC1772 & ${ 0.020}_{- 0.005}^{+ 0.010}$ & ${0.50}_{-0.05}^{+0.00}$ & ${-0.60}_{-0.05}^{+0.05}$ &  0.009 &  \nodata &  \nodata &  \nodata &  \nodata & \nodata & \nodata & \nodata & \nodata & \nodata \\
NGC1774 & ${ 0.100}_{- 0.010}^{+ 0.050}$ & ${-1.90}_{-0.10}^{+0.05}$ & ${-1.80}_{-0.05}^{+0.05}$ &  0.037 &  \nodata &  \nodata &  \nodata &  \nodata & \nodata & \nodata & \nodata & \nodata & \nodata \\
NGC1782 & ${ 0.040}_{- 0.010}^{+ 0.010}$ & ${-0.40}_{-0.05}^{+0.05}$ & ${-0.30}_{-0.05}^{+0.05}$ &  0.036 &  \nodata &  \nodata &  \nodata &  \nodata & \nodata & \nodata & \nodata & \nodata & \nodata \\
NGC1783 & ${ 1.500}_{- 0.150}^{+ 0.250}$ & ${-0.80}_{-0.10}^{+0.05}$ & ${0.00}_{-0.05}^{+0.05}$ &  \nodata &  0.200 &  1.548 &  1.300 &  \nodata & \nodata & \nodata & -0.75 & -0.54 & -0.65 \\
NGC1805 & ${ 0.008}_{- 0.001}^{+ 0.001}$ & ${0.50}_{-0.05}^{+0.00}$ & ${0.50}_{-0.05}^{+0.00}$ &  0.009 &  \nodata &  \nodata &  0.014 &  \nodata & \nodata & \nodata & \nodata & \nodata & -0.20 \\
NGC1818 & ${ 0.040}_{- 0.010}^{+ 0.010}$ & ${-1.30}_{-0.05}^{+0.05}$ & ${-0.50}_{-0.05}^{+0.05}$ &  0.014 &  0.017 &  0.020 &  \nodata &  \nodata & \nodata & \nodata & -0.90 & \nodata & \nodata \\
NGC1831 & ${ 0.800}_{- 0.100}^{+ 0.100}$ & ${-1.60}_{-0.10}^{+0.05}$ & ${-1.50}_{-0.05}^{+0.05}$ &  \nodata &  0.190 &  0.501 &  0.320 &  \nodata & \nodata & \nodata & 0.01 & -0.65 & -0.62 \\
NGC1847 & ${ 0.020}_{- 0.005}^{+ 0.010}$ & ${-1.70}_{-0.05}^{+0.05}$ & ${0.10}_{-0.05}^{+0.05}$ &  0.021 &  0.016 &  \nodata &  \nodata &  \nodata & \nodata & \nodata & \nodata & \nodata & \nodata \\
NGC1850 & ${ 0.100}_{- 0.010}^{+ 0.050}$ & ${-1.60}_{-0.05}^{+0.05}$ & ${-1.60}_{-0.05}^{+0.05}$ &  0.031 &  0.040 &  \nodata &  0.031 &  \nodata & \nodata & \nodata & \nodata & \nodata & -0.12 \\
NGC1850A & ${ 0.002}_{- 0.001}^{+ 0.001}$ & ${0.50}_{-0.10}^{+0.00}$ & ${-2.20}_{-0.00}^{+0.05}$ &  0.004 &  \nodata &  \nodata &  \nodata &  \nodata & \nodata & \nodata & \nodata & \nodata & \nodata \\
NGC1854 & ${ 0.060}_{- 0.010}^{+ 0.010}$ & ${-1.40}_{-0.05}^{+0.05}$ & ${0.10}_{-0.05}^{+0.05}$ &  0.034 &  0.030 &  \nodata &  0.034 &  \nodata & \nodata & \nodata & \nodata & \nodata & -0.50 \\
NGC1856 & ${ 0.800}_{- 0.100}^{+ 0.100}$ & ${-1.60}_{-0.05}^{+0.05}$ & ${-0.90}_{-0.05}^{+0.05}$ &  \nodata &  0.120 &  \nodata &  0.151 &  \nodata & \nodata & \nodata & \nodata & \nodata & -0.17 \\
NGC1866 & ${ 0.200}_{- 0.050}^{+ 0.100}$ & ${-2.20}_{-0.00}^{+0.05}$ & ${-0.80}_{-0.05}^{+0.05}$ &  \nodata &  0.080 &  0.302 &  0.150 &  \nodata & \nodata & \nodata & -1.20 & \nodata & -0.66 \\
NGC1868 & ${ 3.000}_{- 0.500}^{+ 0.500}$ & ${-1.70}_{-0.05}^{+0.05}$ & ${-0.10}_{-0.05}^{+0.05}$ &  \nodata &  0.330 &  0.933 &  0.850 &  \nodata & \nodata & \nodata & -0.50 & -0.32 & -0.66 \\
NGC1872 & ${ 1.000}_{- 0.100}^{+ 0.100}$ & ${-2.20}_{-0.00}^{+0.05}$ & ${0.00}_{-0.05}^{+0.05}$ &  0.136 &  \nodata &  \nodata &  \nodata &  \nodata & \nodata & \nodata & \nodata & \nodata & \nodata \\
NGC1895 & ${ 0.040}_{- 0.010}^{+ 0.010}$ & ${-0.70}_{-0.05}^{+0.05}$ & ${-0.90}_{-0.05}^{+0.05}$ &  \nodata &  \nodata &  \nodata &  \nodata &  \nodata & \nodata & \nodata & \nodata & \nodata & \nodata \\
NGC1903 & ${ 0.080}_{- 0.010}^{+ 0.010}$ & ${-2.00}_{-0.05}^{+0.05}$ & ${-1.70}_{-0.05}^{+0.05}$ &  0.044 &  \nodata &  \nodata &  \nodata &  \nodata & \nodata & \nodata & \nodata & \nodata & \nodata \\
NGC1921 & ${ 0.200}_{- 0.050}^{+ 0.100}$ & ${-2.20}_{-0.00}^{+0.05}$ & ${-2.10}_{-0.05}^{+0.10}$ &  0.041 &  \nodata &  \nodata &  \nodata &  \nodata & \nodata & \nodata & \nodata & \nodata & \nodata \\
NGC1935 & ${ 0.002}_{- 0.001}^{+ 0.001}$ & ${-2.10}_{-0.10}^{+0.40}$ & ${-1.50}_{-0.05}^{+0.05}$ &  \nodata &  \nodata &  \nodata &  \nodata &  \nodata & \nodata & \nodata & \nodata & \nodata & \nodata \\
NGC1936 & ${ 0.100}_{- 0.010}^{+ 0.050}$ & ${-2.00}_{-0.20}^{+0.10}$ & ${0.20}_{-0.10}^{+0.05}$ &  \nodata &  \nodata &  \nodata &  \nodata &  \nodata & \nodata & \nodata & \nodata & \nodata & \nodata \\
NGC1949 & ${ 0.006}_{- 0.001}^{+ 0.001}$ & ${-2.20}_{-0.00}^{+0.05}$ & ${-1.70}_{-0.30}^{+0.05}$ &  \nodata &  \nodata &  \nodata &  \nodata &  \nodata & \nodata & \nodata & \nodata & \nodata & \nodata \\
NGC1951 & ${ 0.020}_{- 0.005}^{+ 0.010}$ & ${-1.80}_{-0.10}^{+0.10}$ & ${-0.30}_{-0.05}^{+0.05}$ &  0.048 &  \nodata &  \nodata &  \nodata &  \nodata & \nodata & \nodata & \nodata & \nodata & \nodata \\
NGC1967 & ${ 0.010}_{- 0.001}^{+ 0.005}$ & ${-0.30}_{-0.05}^{+0.05}$ & ${-1.50}_{-0.05}^{+0.05}$ &  0.004 &  0.100 &  \nodata &  \nodata &  \nodata & \nodata & \nodata & \nodata & \nodata & \nodata \\
NGC1970 & ${ 0.002}_{- 0.001}^{+ 0.001}$ & ${-2.20}_{-0.00}^{+0.05}$ & ${-0.80}_{-0.05}^{+0.05}$ &  0.004 &  \nodata &  \nodata &  \nodata &  \nodata & \nodata & \nodata & \nodata & \nodata & \nodata \\
NGC1978 & ${ 4.000}_{- 0.500}^{+ 0.500}$ & ${-1.30}_{-0.05}^{+0.05}$ & ${-0.40}_{-0.05}^{+0.05}$ &  \nodata &  \nodata &  2.570 &  2.200 &  \nodata & \nodata & \nodata & -0.42 & -0.72 & -0.85 \\
NGC1983 & ${ 0.002}_{- 0.001}^{+ 0.001}$ & ${0.30}_{-0.05}^{+0.05}$ & ${-0.90}_{-0.05}^{+0.05}$ &  0.004 &  0.008 &  \nodata &  \nodata &  \nodata & \nodata & \nodata & \nodata & \nodata & \nodata \\
NGC1984 & ${ 0.004}_{- 0.001}^{+ 0.001}$ & ${0.50}_{-0.05}^{+0.00}$ & ${-1.50}_{-0.05}^{+0.05}$ &  0.004 &  0.007 &  \nodata &  0.004 &  \nodata & \nodata & \nodata & \nodata & \nodata & -0.90 \\
NGC1994 & ${ 0.008}_{- 0.001}^{+ 0.001}$ & ${0.30}_{-0.05}^{+0.05}$ & ${0.10}_{-0.05}^{+0.05}$ &  0.006 &  0.007 &  \nodata &  \nodata &  \nodata & \nodata & \nodata & \nodata & \nodata & \nodata \\
NGC2002 & ${ 0.008}_{- 0.001}^{+ 0.001}$ & ${0.40}_{-0.05}^{+0.05}$ & ${-2.20}_{-0.00}^{+0.05}$ &  0.008 &  \nodata &  \nodata &  \nodata &  \nodata & \nodata & \nodata & \nodata & \nodata & \nodata \\
NGC2003 & ${ 0.006}_{- 0.001}^{+ 0.001}$ & ${0.10}_{-0.05}^{+0.05}$ & ${0.20}_{-0.05}^{+0.05}$ &  \nodata &  \nodata &  \nodata &  \nodata &  \nodata & \nodata & \nodata & \nodata & \nodata & \nodata \\
NGC2004 & ${ 0.010}_{- 0.001}^{+ 0.005}$ & ${0.10}_{-0.05}^{+0.05}$ & ${0.10}_{-0.05}^{+0.05}$ &  \nodata &  0.008 &  \nodata &  0.028 &  \nodata & \nodata & \nodata & \nodata & \nodata & -0.56 \\
NGC2006 & ${ 0.006}_{- 0.001}^{+ 0.001}$ & ${0.00}_{-0.05}^{+0.05}$ & ${-0.60}_{-0.10}^{+0.05}$ &  0.005 &  \nodata &  \nodata &  \nodata &  \nodata & \nodata & \nodata & \nodata & \nodata & \nodata \\
NGC2009 & ${ 0.010}_{- 0.001}^{+ 0.005}$ & ${0.30}_{-0.05}^{+0.05}$ & ${-1.70}_{-0.05}^{+0.05}$ &  0.006 &  \nodata &  \nodata &  \nodata &  \nodata & \nodata & \nodata & \nodata & \nodata & \nodata \\
NGC2011 & ${ 0.006}_{- 0.001}^{+ 0.002}$ & ${0.30}_{-0.10}^{+0.05}$ & ${0.20}_{-0.05}^{+0.05}$ &  0.005 &  0.006 &  \nodata &  0.005 &  \nodata & \nodata & \nodata & \nodata & \nodata & -0.47 \\
NGC2021 & ${ 0.020}_{- 0.005}^{+ 0.010}$ & ${-0.80}_{-0.05}^{+0.10}$ & ${-2.20}_{-0.00}^{+0.05}$ &  0.004 &  \nodata &  \nodata &  \nodata &  \nodata & \nodata & \nodata & \nodata & \nodata & \nodata \\
NGC2025 & ${ 0.100}_{- 0.010}^{+ 0.050}$ & ${-1.90}_{-0.05}^{+0.10}$ & ${-1.50}_{-0.10}^{+0.05}$ &  0.083 &  \nodata &  \nodata &  \nodata &  \nodata & \nodata & \nodata & \nodata & \nodata & \nodata \\
NGC2032 & ${ 0.040}_{- 0.010}^{+ 0.010}$ & ${-2.20}_{-0.00}^{+0.05}$ & ${-1.70}_{-0.20}^{+0.05}$ &  \nodata &  \nodata &  \nodata &  \nodata &  \nodata & \nodata & \nodata & \nodata & \nodata & \nodata \\
NGC2033 & ${ 0.002}_{- 0.001}^{+ 0.001}$ & ${0.30}_{-0.05}^{+0.05}$ & ${-1.90}_{-0.05}^{+0.05}$ &  0.004 &  \nodata &  \nodata &  \nodata &  \nodata & \nodata & \nodata & \nodata & \nodata & \nodata \\
NGC2035 & ${ 0.002}_{- 0.001}^{+ 0.001}$ & ${-1.90}_{-0.30}^{+0.40}$ & ${-1.70}_{-0.50}^{+0.10}$ &  \nodata &  \nodata &  \nodata &  \nodata &  \nodata & \nodata & \nodata & \nodata & \nodata & \nodata \\
NGC2041 & ${ 0.100}_{- 0.010}^{+ 0.050}$ & ${-1.70}_{-0.05}^{+0.10}$ & ${-1.10}_{-0.05}^{+0.05}$ &  0.058 &  \nodata &  \nodata &  \nodata &  \nodata & \nodata & \nodata & \nodata & \nodata & \nodata \\
NGC2058 & ${ 0.200}_{- 0.050}^{+ 0.100}$ & ${-1.50}_{-0.30}^{+0.10}$ & ${-1.60}_{-0.05}^{+0.05}$ &  0.062 &  0.070 &  \nodata &  \nodata &  \nodata & \nodata & \nodata & \nodata & \nodata & \nodata \\
NGC2070 & ${ 0.060}_{- 0.010}^{+ 0.010}$ & ${0.50}_{-0.05}^{+0.00}$ & ${-2.20}_{-0.00}^{+0.05}$ &  \nodata &  \nodata &  \nodata &  \nodata &  \nodata & \nodata & \nodata & \nodata & \nodata & \nodata \\
NGC2091 & ${ 0.020}_{- 0.005}^{+ 0.010}$ & ${0.10}_{-0.05}^{+0.05}$ & ${-1.70}_{-0.05}^{+0.05}$ &  0.007 &  \nodata &  \nodata &  \nodata &  \nodata & \nodata & \nodata & \nodata & \nodata & \nodata \\
NGC2092 & ${10.000}_{- 1.000}^{+ 0.500}$ & ${-1.70}_{-0.10}^{+0.05}$ & ${-2.20}_{-0.00}^{+0.05}$ &  0.012 &  0.004 &  \nodata &  \nodata &  \nodata & \nodata & \nodata & \nodata & \nodata & \nodata \\
NGC2096 & ${ 6.000}_{- 0.500}^{+ 0.500}$ & ${-2.20}_{-0.00}^{+0.05}$ & ${-1.40}_{-0.05}^{+0.05}$ &  0.049 &  \nodata &  \nodata &  \nodata &  \nodata & \nodata & \nodata & \nodata & \nodata & \nodata \\
NGC2098 & ${ 0.020}_{- 0.005}^{+ 0.010}$ & ${0.00}_{-0.05}^{+0.05}$ & ${0.10}_{-0.05}^{+0.05}$ &  0.008 &  \nodata &  \nodata &  \nodata &  \nodata & \nodata & \nodata & \nodata & \nodata & \nodata \\
NGC2100 & ${ 0.006}_{- 0.001}^{+ 0.001}$ & ${0.20}_{-0.05}^{+0.05}$ & ${0.30}_{-0.05}^{+0.05}$ &  0.009 &  0.010 &  \nodata &  0.032 &  \nodata & \nodata & \nodata & \nodata & \nodata & -0.32 \\
NGC2102 & ${ 0.004}_{- 0.001}^{+ 0.001}$ & ${-0.10}_{-0.05}^{+0.05}$ & ${-2.20}_{-0.00}^{+0.05}$ &  0.004 &  0.007 &  \nodata &  \nodata &  \nodata & \nodata & \nodata & \nodata & \nodata & \nodata \\
NGC2134 & ${ 0.200}_{- 0.050}^{+ 0.100}$ & ${-2.00}_{-0.10}^{+0.20}$ & ${-2.20}_{-0.00}^{+0.05}$ &  0.124 &  0.130 &  0.631 &  \nodata &  \nodata & \nodata & \nodata & -1.00 & \nodata & \nodata \\
NGC2136 & ${ 0.100}_{- 0.010}^{+ 0.050}$ & ${-1.60}_{-0.10}^{+0.05}$ & ${-2.20}_{-0.00}^{+0.05}$ &  0.058 &  0.040 &  0.081 &  \nodata &  \nodata & \nodata & \nodata & -0.40 & \nodata & \nodata \\
NGC2147 & ${ 0.020}_{- 0.005}^{+ 0.010}$ & ${0.10}_{-0.05}^{+0.05}$ & ${-1.80}_{-0.05}^{+0.05}$ &  0.021 &  \nodata &  \nodata &  \nodata &  \nodata & \nodata & \nodata & \nodata & \nodata & \nodata \\
NGC2156 & ${ 0.100}_{- 0.010}^{+ 0.050}$ & ${-1.60}_{-0.05}^{+0.05}$ & ${-1.80}_{-0.05}^{+0.10}$ &  0.077 &  0.060 &  \nodata &  \nodata &  \nodata & \nodata & \nodata & \nodata & \nodata & \nodata \\
NGC2157 & ${ 0.040}_{- 0.010}^{+ 0.010}$ & ${-1.50}_{-0.05}^{+0.05}$ & ${-0.70}_{-0.05}^{+0.05}$ &  \nodata &  0.030 &  \nodata &  \nodata &  \nodata & \nodata & \nodata & \nodata & \nodata & \nodata \\
NGC2159 & ${ 0.100}_{- 0.010}^{+ 0.050}$ & ${-1.80}_{-0.05}^{+0.05}$ & ${-1.50}_{-0.10}^{+0.05}$ &  0.056 &  0.060 &  \nodata &  \nodata &  \nodata & \nodata & \nodata & \nodata & \nodata & \nodata \\
NGC2164 & ${ 0.100}_{- 0.010}^{+ 0.050}$ & ${-1.80}_{-0.05}^{+0.05}$ & ${-0.80}_{-0.05}^{+0.10}$ &  0.044 &  0.050 &  0.058 &  \nodata &  \nodata & \nodata & \nodata & -0.60 & \nodata & \nodata \\
NGC2214 & ${ 0.040}_{- 0.010}^{+ 0.010}$ & ${0.30}_{-0.05}^{+0.05}$ & ${-0.70}_{-0.05}^{+0.10}$ &  0.038 &  0.040 &  0.052 &  \nodata &  \nodata & \nodata & \nodata & -1.20 & \nodata & \nodata \\
SL153 & ${ 0.200}_{- 0.050}^{+ 0.100}$ & ${-2.20}_{-0.00}^{+0.10}$ & ${-1.00}_{-0.10}^{+0.05}$ &  0.054 &  \nodata &  \nodata &  \nodata &  \nodata & \nodata & \nodata & \nodata & \nodata & \nodata \\
SL492 & ${ 0.010}_{- 0.001}^{+ 0.005}$ & ${0.30}_{-0.05}^{+0.05}$ & ${-0.80}_{-0.05}^{+0.05}$ &  0.005 &  \nodata &  \nodata &  \nodata &  \nodata & \nodata & \nodata & \nodata & \nodata & \nodata \\
SL538 & ${ 0.006}_{- 0.001}^{+ 0.001}$ & ${0.10}_{-0.05}^{+0.05}$ & ${0.40}_{-0.05}^{+0.05}$ &  0.006 &  \nodata &  \nodata &  \nodata &  \nodata & \nodata & \nodata & \nodata & \nodata & \nodata \\
SL586 & ${ 0.002}_{- 0.001}^{+ 0.001}$ & ${0.50}_{-0.05}^{+0.00}$ & ${-2.20}_{-0.00}^{+0.05}$ &  \nodata &  \nodata &  \nodata &  \nodata &  \nodata & \nodata & \nodata & \nodata & \nodata & \nodata \\
SL601 & ${ 0.004}_{- 0.001}^{+ 0.001}$ & ${-0.30}_{-0.05}^{+0.10}$ & ${-1.70}_{-0.05}^{+0.05}$ &  0.004 &  \nodata &  \nodata &  \nodata &  \nodata & \nodata & \nodata & \nodata & \nodata & \nodata \\
SL639 & ${ 0.004}_{- 0.001}^{+ 0.001}$ & ${-1.90}_{-0.05}^{+0.05}$ & ${-2.20}_{-0.00}^{+0.05}$ &  0.030 &  0.015 &  \nodata &  \nodata &  \nodata & \nodata & \nodata & \nodata & \nodata & \nodata \\
\hline
\textbf{SMC} \\
\hline
NGC121 & ${10.000}_{- 0.500}^{+ 0.500}$ & ${-1.70}_{-0.05}^{+0.05}$ & ${-0.40}_{-0.05}^{+0.05}$ &  \nodata & 13.000 &  \nodata & 11.900 &  \nodata & \nodata & \nodata & \nodata & \nodata & -1.19 \\
NGC299 & ${ 0.008}_{- 0.001}^{+ 0.001}$ & ${-0.20}_{-0.05}^{+0.05}$ & ${-2.20}_{-0.00}^{+0.05}$ &  0.012 &  \nodata &  \nodata &  \nodata &  \nodata & \nodata & \nodata & \nodata & \nodata & \nodata \\
NGC330 & ${ 0.004}_{- 0.001}^{+ 0.001}$ & ${-1.50}_{-0.05}^{+0.05}$ & ${0.00}_{-0.05}^{+0.05}$ &  0.013 &  0.007 &  \nodata &  0.025 &  \nodata & \nodata & \nodata & \nodata & \nodata & -0.82 \\
NGC419 & ${ 1.500}_{- 0.150}^{+ 0.250}$ & ${-1.20}_{-0.05}^{+0.05}$ & ${-1.40}_{-0.05}^{+0.05}$ &  \nodata &  0.670 &  1.950 &  1.200 &  \nodata & \nodata & \nodata & -0.70 & -0.90 & -0.70 \\
\hline
\textbf{M31} \\
\hline
Bo104 & ${ 4.000}_{- 0.500}^{+ 0.500}$ & ${-1.20}_{-0.05}^{+0.05}$ & ${-0.90}_{-0.05}^{+0.05}$ & 10.000 &  \nodata &  \nodata &  \nodata &  \nodata & \nodata & \nodata & \nodata & \nodata & \nodata \\
Bo118 & ${13.000}_{- 0.500}^{+ 1.000}$ & ${-1.50}_{-0.05}^{+0.05}$ & ${0.40}_{-0.05}^{+0.05}$ & 10.000 &  \nodata &  \nodata &  \nodata &  \nodata & \nodata & \nodata & \nodata & \nodata & \nodata \\
Bo119 & ${10.000}_{- 1.000}^{+ 1.000}$ & ${-1.20}_{-0.05}^{+0.10}$ & ${0.50}_{-0.05}^{+0.00}$ & 10.000 &  \nodata &  \nodata &  \nodata &  \nodata & \nodata & \nodata & \nodata & \nodata & \nodata \\
Bo124 & ${10.000}_{- 0.500}^{+ 1.000}$ & ${-0.80}_{-0.05}^{+0.05}$ & ${0.40}_{-0.10}^{+0.05}$ & 10.000 &  \nodata &  \nodata &  \nodata &  \nodata & \nodata & \nodata & \nodata & \nodata & \nodata \\
G1 & ${ 3.000}_{- 0.500}^{+ 0.500}$ & ${-0.50}_{-0.05}^{+0.05}$ & ${-0.20}_{-0.05}^{+0.05}$ &  \nodata &  \nodata &  \nodata &  \nodata &  \nodata & \nodata & \nodata & \nodata & \nodata & \nodata \\
G78 & ${ 3.000}_{- 0.500}^{+ 0.500}$ & ${-0.50}_{-0.05}^{+0.05}$ & ${-0.10}_{-0.05}^{+0.05}$ &  \nodata &  \nodata &  \nodata &  \nodata &  \nodata & \nodata & \nodata & \nodata & \nodata & \nodata \\
G158 & ${18.000}_{- 0.500}^{+ 1.000}$ & ${-0.10}_{-0.05}^{+0.05}$ & ${-0.20}_{-0.05}^{+0.05}$ & 10.000 &  \nodata &  \nodata &  \nodata & 19.800 & -0.26 & \nodata & \nodata & \nodata & \nodata \\
G165 & ${15.000}_{- 1.000}^{+ 0.500}$ & ${-0.50}_{-0.05}^{+0.05}$ & ${0.50}_{-0.05}^{+0.00}$ & 10.000 &  \nodata &  \nodata &  \nodata &  \nodata & -0.56 & -0.69 & \nodata & \nodata & \nodata \\
G168 & ${20.000}_{- 1.000}^{+ 0.000}$ & ${-0.50}_{-0.05}^{+0.05}$ & ${0.50}_{-0.05}^{+0.00}$ & 10.000 &  \nodata &  \nodata &  \nodata &  6.200 & \nodata & -1.13 & \nodata & \nodata & \nodata \\
G169 & ${ 7.000}_{- 0.500}^{+ 0.500}$ & ${-1.00}_{-0.05}^{+0.05}$ & ${-1.00}_{-0.05}^{+0.05}$ & 10.000 &  \nodata &  \nodata &  \nodata &  7.500 & -1.18 & -1.35 & \nodata & \nodata & \nodata \\
G170 & ${20.000}_{- 0.500}^{+ 0.000}$ & ${0.00}_{-0.05}^{+0.05}$ & ${0.20}_{-0.05}^{+0.05}$ &  \nodata &  \nodata &  \nodata &  \nodata &  \nodata & \nodata & \nodata & \nodata & \nodata & \nodata \\
G174 & ${ 8.000}_{- 1.000}^{+ 0.500}$ & ${0.30}_{-0.05}^{+0.05}$ & ${0.10}_{-0.05}^{+0.05}$ & 10.000 &  \nodata &  \nodata &  \nodata &  7.200 & 0.29 & -0.13 & \nodata & \nodata & \nodata \\
G175 & ${20.000}_{- 0.500}^{+ 0.000}$ & ${-1.50}_{-0.10}^{+0.05}$ & ${0.50}_{-0.05}^{+0.00}$ & 10.000 &  \nodata &  \nodata &  \nodata &  1.300 & \nodata & -1.16 & \nodata & \nodata & \nodata \\
G177 & ${20.000}_{- 0.500}^{+ 0.000}$ & ${0.00}_{-0.05}^{+0.05}$ & ${0.10}_{-0.05}^{+0.05}$ & 10.000 &  \nodata &  \nodata &  \nodata & 19.800 & -0.15 & -0.32 & \nodata & \nodata & \nodata \\
G184 & ${20.000}_{- 0.500}^{+ 0.000}$ & ${-1.50}_{-0.05}^{+0.05}$ & ${-0.40}_{-0.05}^{+0.05}$ & 10.000 &  \nodata &  \nodata &  \nodata & 19.800 & \nodata & -0.91 & \nodata & \nodata & \nodata \\
G185 & ${10.000}_{- 0.500}^{+ 0.500}$ & ${-1.00}_{-0.05}^{+0.05}$ & ${-0.60}_{-0.05}^{+0.05}$ & 10.000 &  \nodata &  \nodata &  \nodata &  6.500 & -1.08 & -1.19 & \nodata & \nodata & \nodata \\
G190 & ${ 4.000}_{- 1.000}^{+ 1.000}$ & ${-0.90}_{-0.05}^{+0.10}$ & ${-0.60}_{-0.05}^{+0.05}$ & 10.000 &  \nodata &  \nodata &  \nodata &  \nodata & \nodata & -1.96 & \nodata & \nodata & \nodata \\
G222 & ${11.000}_{- 1.000}^{+ 0.500}$ & ${-0.50}_{-0.05}^{+0.05}$ & ${-0.40}_{-0.05}^{+0.05}$ &  \nodata &  \nodata &  \nodata &  \nodata &  \nodata & \nodata & \nodata & \nodata & \nodata & \nodata \\

\enddata

\tablecomments{All ages are in Gyr.}
\tablenotetext{a}{\citet{santos95}}
\tablenotetext{b}{\citet{hodge83}}
\tablenotetext{c}{The \citet{leonardi03} comparison literature values
  come from \citet{olszewski91}, \citet{cohen82}, \citet{seggewiss89},
  \citet{sagar89}, and \citet{piatti02}.}
\tablenotetext{d}{Homogeneous lit values \citep{santos04}}
\tablenotetext{e}{BC96 fits to colors in BATC filters \citep{jiang03}}
\tablenotetext{f}{$_{}$Line indices \citep{huchra91}}
\tablenotetext{g}{\citet{cohen94}}
\tablenotetext{h}{Equivalent width of Ca triplet \citep{olszewski91}}

\end{deluxetable}

\clearpage

\begin{deluxetable}{lccccccccccc}
\tabletypesize{\scriptsize} 
\rotate
\tablecaption{Clusters in Red vs. Blue Spectrum Effects Study
\label{red_blue_id} }
\tablewidth{0pt} 
\tablehead{ 
\colhead{} & \colhead{} & \colhead{Blue CN} &
\colhead{Red CN} & \colhead{Break CN} &
\colhead{Entire CN} & \colhead{Blue Full} &
\colhead{Red Full} & \colhead{Break Full} &
\colhead{Entire Full} & \colhead{Min Lit} & \colhead{Max
  Lit} \\
\colhead{ID} & \colhead{Loc} & \colhead{Spec Age} &
\colhead{Spec Age} & \colhead{Spec Age} &
\colhead{Spec Age} & \colhead{Spec Age} &
\colhead{Spec Age} & \colhead{Spec Age} &
\colhead{Spec Age} & \colhead{Age} & \colhead{Age}
 }

\startdata

NGC 1466 & LMC & 0.004 & 20.000 & 0.004 & 20.000 & 4.000 & 1.500 &
8.000 & 2.000 & 13.1 & 13.1 \\
NGC 1783 & LMC & 0.010 & 1.500 & 9.000 & 1.500 & 6.000 & 1.200 &
20.000 & 1.200 & 0.200 & 1.548 \\
NGC 1831 & LMC & 1.200 & 0.600 & 1.000 & 0.800 & 0.400 & 0.200 & 1.000
& 0.400 & 0.190 & 0.501 \\
NGC 1847 & LMC & 0.001 & 0.040 & 0.040 & 0.020 & 0.080 & 0.040 & 0.040
& 0.040 & 0.016 & 0.021 \\
NGC 1856 & LMC & 0.040 & 0.600 & 0.040 & 0.800 & 0.600 & 1.200 & 0.200
& 0.400 & 0.120 & 0.151\\
NGC 1866 & LMC & 0.004 & 1.200 & 0.060 & 0.200 & 0.100 & 0.800 & 0.200
& 0.600 & 0.080 &0.302 \\
NGC 1868 & LMC & 0.008 & 0.400 & 2.000 & 3.000 & 0.200 & 1.500 & 2.000
& 0.800 & 0.330 & 0.933 \\
NGC 1978 & LMC & 0.004 & 20.000 & 10.000 & 4.000 & 4.000 & 20.000 &
10.000 & 9.000 & 2.200 & 2.570 \\
NGC 2004 & LMC & 0.002 & 0.020 & 0.004 & 0.010 & 0.020 & 0.004 & 0.010
& 0.010 & 0.008 & 0.028 \\
NGC 2070 & LMC & 0.004 & 18.000 & 10.000 & 0.060 & 0.008 & 0.400 &
0.004 & 0.006 & \nodata & \nodata \\
NCG 2157 & LMC & 0.001 & 1.200 & 0.040 & 0.040 & 0.100 & 1.000 & 0.060
& 0.400 & 0.030 & 0.030 \\
NGC 330 & SMC & 0.001 & 0.020 & 0.020 & 0.004 & 0.040 & 0.004 & 0.040
& 0.040 & 0.007 & 0.025 \\
NGC 419 & SMC & 0.010 & 2.000 & 3.000 & 1.500 & 0.200 & 3.000 & 3.000
& 20.000 & 0.670 & 1.950 \\

\enddata

\tablecomments{All ages are in Gyr. Fitting ranges are: Blue =
  $\lambda<$4000~\AA, Red = $\lambda>$4000~\AA, Break = 3750-4250~\AA,
  Entire = full available range.}

\end{deluxetable}

\end{document}